\documentstyle[12pt,amsfonts]{article}
\newcommand{\svec}[1]{ \stackrel{\rightarrow}{#1} }
\newcommand{\dvec}[1]{ \stackrel{\leftrightarrow}{#1} }
\newcommand{\lvec}[1]{ \stackrel{\leftarrow}{#1} }
\newcommand{\swav}[1]{ \stackrel{\sim}{#1} }
\newcommand{\tdot}[1]{ \stackrel{\cdot}{#1} }
\newcommand{\uhat}{ \hat U }
\newcommand{\ehat}{ \hat U_{\epsilon} }
\newcommand{\mhat}[1]{ \hat U_{\epsilon_{#1}} }
\newcommand{\define}{ \stackrel{\triangle}{=} }

\def\be{\begin{equation}}
\def\ee{\end{equation}}
\def\ba{\begin{array}}
\def\ea{\end{array}}

\begin{document}
\title{\bf Renormalizable Quantum Gauge General Relativity }
\author{{Ning Wu}
\thanks{email address: wuning@mail.ihep.ac.cn}
\\
\\
{\small Institute of High Energy Physics, P.O.Box 918-1,
Beijing 100039, P.R.China}}
\maketitle
\vskip 0.8in

~~

\vskip 0.4in

\begin{abstract}

The quantum gauge general relativity is proposed in the framework of
quantum gauge theory of gravity. It is formulated based on
gauge principle which states that the correct symmetry for
gravitational interactions should be gravitational gauge
symmetry. The gravitational gauge group is studied
in the paper. Then gravitational gauge interactions of pure
gravitational gauge field is studied. It is found that
the field equation of gravitational gauge field is just the
Einstein's field equation. After that, the gravitational
interactions of scalar field, Dirac field and vector
fields are studied, and unifications of fundamental
interactions are discussed. Path integral quantization
of the theory is studied in the paper. The quantum
gauge general relativity discussed in this paper is a
perturbatively renormalizable quantum gravity,
which is one of the most important advantage of
the quantum gauge general relativity proposed in this paper.
A strict proof on the renormalizability of the
theory is also given in this paper. Another
important advantage of the quantum gauge general relativity
is that it can explain both classical tests of gravity
and quantum effects of gravitational
interactions, such as gravitational phase effects found
in COW experiments and gravitational shielding effects
found in Podkletnov experiments. For all classical effects
of gravitational interactions, such as classical tests of
gravity and cosmological model, quantum gauge general 
relativity gives out the same theoretical predictions
as that of the Einstein's general relaitvity.    
\\
\end{abstract}

\begin{center}
CONTENTS
\end{center}

\begin{enumerate}
\item Introduction
\item The Transcendental Foundations
\item Gravitational Gauge Group
\item Two Pictures of Gravity
\item Pure Gravitational Gauge fields
\item Gravitational Interactions of Scalar Fields
\item Gravitational Interactions of Dirac Field
\item Gravitational Interactions of Vector Field
\item GSU(N) Unification Model
\item Unification of Fundamental Interactions
\item Classical Limit of Quantum Gauge general Relativity
\item Path Integral Quantization of Gravitational Gauge Field
\item Renormalization
\item Classical Tests of Quantum Gauge General Relativity
\item Quantum Effects of Gravitational Interactions
\item Discussions

\end{enumerate}

\newpage

\Roman{section}

\section{Introduction}
\setcounter{equation}{0}

In 1686, Isaac Newton published his book {\it MATHEMATICAL
 PRINCIPLES OF NATURAL PHILOSOPHY}. In this
book, through studying the motion of planet in solar system, he
found that gravity obeys the inverse square law and the magnitude
of gravity is proportional to the mass of the object\cite{01}. The
Newton's classical theory of gravity is kept unchanged until 1916. At
that year, Einstein proposed General
Relativity\cite{02,03}. In this great work, he founded a
relativistic theory on gravity, which is based on principle of
general relativity and equivalence principle. Newton's classical
theory of gravity appears as
a classical limit of  general relativity.\\

One of the biggest revolution in human kind in the last century is
the foundation of quantum theory. The quantum hypothesis was first
introduced into physics by Max Plank in 1900.
In 1916, Albert Einstein points out that quantum effects must
lead to modifications in the theory of general relativity\cite{5}.
Soon after the foundation of  quantum mechanics,
physicists try to found a theory that could describe the
quantum behavior of the full gravitational field. In the 70
years attempts, physicists have found two theories based
on quantum mechanics that attempt to unify general
relativity and quantum mechanics, one is canonical quantum
gravity and another is superstring theory. But for quantum
field theory, there are different kinds of mathematical
infinities that naturally occur in quantum descriptions of
fields. These infinities should be removed by the technique
of perturbative renormalization. However, the perturbative
renormalization does not work for the quantization of Einstein's
theory of gravity, especially in canonical quantum gravity. In
superstring theory, in order to make perturbative renormalization
to work, physicists have to introduce six extra dimensions. But
up to now, none of the extra dimensions have been observed.
To found a consistent theory that can unify general relativity and
quantum mechanics is a long dream for physicists. \\

The "relativity revolution" and the "quantum revolution" are among
the greatest successes of twentieth century physics, yet two
theories appears to be fundamentally incompatible. General
relativity remains a purely classical theory which describes the
geometry of space and time as smooth and continuous, on the
contrary, quantum mechanics divides everything into discrete
quanta. The underlying theoretical incompatibility between two
theories arises from the way  that they treat the geometry of
space and time. This situation makes some physicists still wonder
whether quantum theory is a truly fundamental theory of Nature, or
just a convenient description of some aspects of the microscopic
world. Some physicists even consider the twentieth century as the
century of the incomplete revolution. To set up a consistent
quantum theory of gravity is considered to be the last challenge
of quantum theory\cite{50,51}. In other words, combining general
relativity with quantum mechanics is considered to be the last
hurdle to be overcome in the "quantum revolution".
\\

In 1921, H.Weyl introduced the concept of gauge transformation
into physics\cite{d01,d05}, which is one of the most important
concepts in modern physics, though his original theory is not
successful. Later, V.Fock, H.Weyl and W.Pauli found that quantum
electrodynamics is a gauge invariant theory\cite{d02,d03,d04}. In
1954, Yang and Mills proposed non-Abelian gauge field theory\cite{1}.
This theory was soon applied to elementary particle physics.
Unified electroweak theory\cite{2,3,4} and quantum chromodynamics
are all based on gauge field theory. The predictions of unified
electroweak theory have been confirmed in a large number of
experiments, and the intermediate gauge bosons $W^{\pm}$ and $Z^0$
which are predicted by unified electroweak model are also found in
experiments. The $U(1)$ part of the unified electroweak model,
quantum electrodynamics, now become one of the most accurate and
best-tested theories of modern physics. All these achievements of
gauge field theories suggest that gauge field theory is a
fundamental theory that describes fundamental interactions in Nature.
Now, it is generally  believed that four kinds of fundamental
interactions in Nature are all gauge interactions and they can be
described by gauge field theory. From theoretical point of view,
the principle of local gauge invariance plays a fundamental role
in particle's interaction theory.  \\

Gauge treatment of gravity was suggested immediately after the
gauge theory birth itself\cite{b1,b2,b3,b4,b5,b6}. In the
traditional gauge treatment of gravity, Lorentz group is localized,
and the gravitational field is not represented by gauge
potential. It is represented by metric field.
The theory has beautiful mathematical forms, but up to now, its
renormalizability is not proved. In other words,
it is conventionally considered to be non-renormalizable.
There are also some other attempts to use Yang-Mills theory
to reformulate quantum gravity\cite{c1,c2,c3,c4,cc5}. In these new
approaches, the importance of gauge fields is emphasized. Some
physicists also try to use gauge potential to represent
gravitational field, some suggest that we should pay more
attention on translation group.
\\

I will not talk too much on the history of quantum gravity and
the incompatibilities between quantum mechanics and general
relativity here. Materials on these subject can be widely found
in literatures. Now we want to ask that, except for traditional
canonical quantum gravity and superstring theory, whether exists
another approach to set up a fundamental theory, in which general
relativity and quantum mechanics are compatible? In this paper,
a new approach is proposed to consistently unify general relativity
and quantum mechanics.
\\

In 1991, a completely new formulation of quantum gauge
theory of gravity is proposed By N.Wu \cite{wu01,wu02,wu03,wu04}.
In this new attempt, the gravitational interactions are
treated as a kind of fundamental interactions in flat
Minkowski space-time, not space-time geometry. So,
in quantum gauge theory of gravity, the basic physical
space-time is always flat. In this new approach,
the unification of fundamental interactions can be
formulated in a simple and beautiful way\cite{12,13,14}.
If we use the mass generation mechanism which is proposed
in literature \cite{15}, we can propose a new
theory on gravity which contains massive graviton and
the introduction of massive graviton does not affect
the strict local gravitational gauge symmetry of the
Lagrangian and does not affect the traditional long-range
gravitational force\cite{16}. The existence of massive graviton
will help us to understand the possible origin of dark
energy and dark matter in the Universe.
Quantum gauge theory of gravity can be used to
explain quantum effects of gravitational interactions,
such as gravitational phase effects in COW
experiments\cite{q01,q02,q03}, and the gravitational
shielding effects in Podkletnov experiments\cite{q04,q05,q06,q07}.
  \\

In this paper, the quantum gauge general relativity is studied
in the  frame work of quantum gauge theory of gravity.
In all previous work of quantum gauge theory of gravity,
a very simple Lagrangian for pure gravitational gauge
field is selected, which is quite similar to that
of ordinary $SU(N)$ gauge field. As is discussed in
literature \cite{wu04}, the selection of the Lagrangian
for pure gravitational gauge field is not unique,
which is quite a special case in gauge field theory,
which originates from the fact that many gravitational
gauge invariant terms can be constructed from Minkowski
metric, pure gravitational gauge field and the field
strength of gravitational gauge field. Different selections
of the Lagrangian for pure gravitational gauge field
will result in different models of quantum gravity.
In this paper, another selection of the Lagrangian
for pure gravitational gauge field is adopted, that is,
the scalar "curvature" $R$ is selected to be the
Lagrangian for pure gravitational gauge field. In this
new quantum gauge model of gravity, one of the most
interesting thing is that the field equation of the
gravitational gauge field is just the Einstein's field
equation. So, this model is called quantum gauge general
relativity. In this paper, we try to give a systematic
formulation of the quantum gauge general relativity. So,
for the sake of integrity, we will repeat all important
discussions in quantum gauge theory of gravity, so
that those readers who are not familiar with quantum
gauge theory of gravity can understand the whole paper
without the help of other related literature. Some
discussions are just copied from literature \cite{wu04}.
The difference between the model proposed in this paper
and the model proposed in literature \cite{wu04}
is only that they have different Lagrangian for pure
gravitational gauge field, and therefore that the dynamics
for gravitational gauge field is different.
\\

\section{The Transcendental Foundations}
\setcounter{equation}{0}

It is known that action principle is one of the most important
fundamental principle in quantum field theory. Action principle
says that any quantum system is described by an action. The action
of the system contains all interaction information, contains all
information of the fundamental dynamics. The least of the action
gives out the classical equation of motion of a field. Action
principle is widely used in quantum field theory. We will accept
it as one of the most fundamental principles in this new quantum
gauge general relativity. The rationality of action principle will not be
discussed here, but it is well know that the rationality of the
action principle has already been tested by a huge amount of
experiments. However, this principle is not a special principle
for quantum gravity, it is a ubiquitous principle in quantum field
theory. Quantum gravity discussed in this paper is a kind of
quantum field theory, it's naturally to accept action principle as
one of its fundamental principles. \\

We need a special transcendental principle to introduce quantum
gravitational field, which should be the foundation of all kinds
of fundamental interactions in Nature. This special principle is
gauge principle. In order to introduce this important principle,
let's first study some fundamental laws in the fundamental
interactions other than gravitational interactions. We know that,
except for gravitational interactions, there are strong
interactions, electromagnetic interactions and weak interactions,
which are described by quantum chromodynamics,  quantum
electrodynamics and unified electroweak theory respectively. Let's
study these three fundamental interactions one by one. \\

Quantum electrodynamics (QED) is one of the most successful theory
in physics which has been tested by most accurate experiments. Let's
study some logic in QED. It is know that QED theory has $U(1)$
gauge symmetry. According to Noether's theorem, there is a
conserved charge corresponding to the global $U(1)$ gauge
transformations. This conserved charge is just the ordinary
electric charge. On the other hand, in order to keep local $U(1)$
gauge symmetry of the system, we had to introduce a $U(1)$ gauge
field, which transmits electromagnetic interactions. This $U(1)$
gauge field is just the well-know electromagnetic field. The
electromagnetic interactions between charged particles and the
dynamics of electromagnetic field are completely determined by the
requirement of local $U(1)$ gauge symmetry. The source of this
electromagnetic field is just the conserved charge which is given
by Noether's theorem. After quantization of the field, this
conserved charge becomes the generator of the quantum $U(1)$ gauge
transformations. The quantum $U(1)$ gauge transformation has only
one generator, it has no generator other than the quantum electric
charge.
\\

Quantum chromodynamics (QCD) is a prospective fundamental theory
for strong interactions. QCD theory has $SU(3)$ gauge symmetry.
The global $SU(3)$ gauge symmetry of the system gives out
conserved charges of the theory, which are usually called color
charges. The local $SU(3)$ gauge symmetry of the system requires
introduction of a set of $SU(3)$ non-Abelian gauge fields, and the
dynamics of non-Abelian gauge fields and the strong interactions
between color charged particles and gauge fields are completely
determined by the requirement of local $SU(3)$ gauge symmetry of
the system. These $SU(3)$ non-Abelian gauge fields are usually call
gluon fields. The sources of gluon fields are color charges. After
quantization, these color charges become generators of quantum
$SU(3)$ gauge transformation. Something which is different from
$U(1)$ Abelian gauge symmetry is that gauge fields themselves carry
color charges. \\

Unified electroweak model is the fundamental theory for
electroweak interactions. Unified electroweak model is usually
called the standard model. It has $SU(2)_L \times U(1)_Y$
symmetry. The global $SU(2)_L \times U(1)_Y$ gauge symmetry of the
system also gives out conserved charges of the system, The
requirement of local $SU(2)_L \times U(1)_Y$ gauge symmetry needs
introducing a set of $SU(2)$ non-Abelian gauge fields and one $U(1)$
Abelian gauge field. These gauge fields transmit weak interactions
and electromagnetic interactions, which correspond to intermediate
gauge bosons $W^{\pm}$, $Z^0$ and photon. The sources of these
gauge fields are just the conserved Noether charges. After
quantization, these conserved charges become generators of
quantum $SU(2)_L \times U(1)_Y$ gauge transformation. \\

QED, QCD and the standard model are three fundamental theories
for three kinds of fundamental interactions. Now we need to
summarize some fundamental laws of Nature on interactions.
Let's first ruminate over above discussions. Then we will find
that our formulations on three different fundamental interaction
theories are almost completely the same, that is the global gauge
symmetry of the system gives out conserved Noether charges,
in order to keep local gauge symmetry of the system, we have to
introduce gauge field or a set of gauge fields, these gauge
fields transmit interactions, the source of these gauge fields are
the conserved charges, and these conserved Noether charges
become generators of quantum gauge transformations after
quantization. These will be the main content of gauge principle.
\\

Before we formulate gauge principle formally, we need to study
something more on symmetry. It is know that not all symmetries can
be localized, and not all symmetries can be regarded as gauge
symmetries and have corresponding gauge fields. For example, time
reversal symmetry, space reflection symmetry, $\cdots$ are those
kinds of symmetries. We can not find any gauge fields or
interactions which correspond to these symmetries. It suggests
that symmetries can be divided into two different classes in
nature. Gauge symmetry is a special kind of symmetry which has the
following properties: 1) it can be localized; 2) it has some
conserved charges related to it; 3) it has a kind of interactions
related to it; 4) it is usually a continuous symmetry. This
symmetry can completely determine the dynamical behavior of a kind
of fundamental interactions. For the sake of simplicity, we call
this kind of symmetry dynamical symmetry or gauge symmetry. Any
kind of fundamental interactions has a gauge  symmetry
corresponding to it. In QED, the $U(1)$ symmetry is a gauge
symmetry, in QCD, the color $SU(3)$ symmetry is a gauge symmetry
and in the standard model, the $SU(2)_L \times U(1)_Y$ symmetry is
also a gauge symmetry. The gravitational gauge symmetry which we
will discuss below is also a kind of gauge symmetry. The time
reversal symmetry and space reflection symmetry are not gauge
symmetries. Those global symmetries which can not be localized are
not gauge symmetries either. Gauge symmetry is a fundamental
concept for gauge principle. \\

Gauge principle can be formulated as follows: Any kind of
fundamental interactions has a gauge symmetry corresponding to it;
the gauge symmetry completely determines the forms of
interactions. In principle, the gauge principle has the following
four different contents:
\begin{enumerate}

\item {\bf Conservation Law:} the global gauge symmetry  gives out
conserved current and conserved charge;

\item {\bf Interactions:} the requirement of the local gauge
symmetry requires introduction of gauge field or a set of gauge
fields; the interactions between gauge fields and matter fields
are completely determined by the requirement of local gauge
symmetry; these gauge fields transmit the corresponding
interactions;

\item {\bf Source:} qualitative speaking,
the conserved charge given by global gauge  symmetry is the source
of gauge field; for non-Abel gauge field, gauge field is also the
source of itself;

\item {\bf Quantum Transformation:} the conserved charges given
by global gauge symmetry become generators of quantum gauge
transformation after quantization, and for this kind  of
interactions, the quantum transformation can not have generators
other than quantum conserved charges given by global gauge
symmetry.

\end{enumerate}
It is known that conservation law is the objective origin of gauge
symmetry, so gauge symmetry is the exterior exhibition of the
interior conservation law. The conservation law is the law that
exists in fundamental interactions, so fundamental interactions
are the logic precondition and foundation of the conservation law.
Gauge principle tells us how to study conservation law and
fundamental interactions through symmetry. Gauge principle is one
of the most important transcendental fundamental principles for
all kinds of fundamental interactions in Nature; it reveals the
common nature of all kinds of fundamental interactions in Nature.
It is also the transcendental foundation of the quantum gravity
which is formulated in this paper. It will help us to select the
gauge symmetry for quantum gravitational theory and help us to
determine the Lagrangian of the system. In a meaning, we can say
that without gauge principle, we can not set up this new
renormalizable quantum gauge general relativity.
\\

Another transcendental principle that widely used in quantum field
theory is the microscopic causality principle. The central idea of the
causality principle is that any changes in the objective world have
their causation. Quantum field theory is a relativistic theory.
It is known that, in the special theory of relativity, the limit
spread speed is the speed
of light. It means that, in a definite reference system, the limit
spread speed
of the causation of some changes is the speed of light. Therefore,
the special theory of relativity exclude the possibility of the existence
of any kinds of non-local interactions in a fundamental theory. Quantum
field theory inherits this basic idea and calls it the microscopic causality
principle. There are several expressions of the microscopic causality
principle in quantum field theory. One expression says that two events
which happen at the same time but in different space position are two
independent events. The mathematical formulation for microscopic
causality principle is that
\be
[ O_1(\svec{x},t)~~,~~O_2(\svec{y},t) ] =0,
\label{2.1}
\ee
when $\svec{x} \not= \svec{y}$. In the above relation,
$ O_1(\svec{x},t)$ and $ O_2(\svec{y},t)$ are two different
arbitrary local bosonic operators. Another important expression
of the microscopic causality principle is that, in the Lagrangian
of a fundamental theory, all operators appear in the same point
of space-time. Gravitational interactions are a kind of physical
interactions, the fundamental theory of gravity should also obey
microscopic causality principle. This requirement is realized
in the construction of the Lagrangian for gravity. We will require
that all field operators in the Lagrangian should be at the same
point of space-time. \\

Because quantum field theory is a kind of relativistic theory, it
should obey some fundamental principles of the special theory of
relativity, such as principle of special relativity and principle
of invariance of light speed. These two principles conventionally
exhibit themselves through Lorentz invariance. So, in constructing
the Lagrangian of the quantum theory of gravity, we require that
it should have global Lorentz invariance. This is also a transcendental
requirement for the quantum theory of gravity. But what we treat
here that is different from that of the traditional quantum
gravity is that we
do not localize Lorentz transformation. Because gauge principle
forbids us to localize Lorentz transformation, asks us only to
localize gravitational gauge transformation. We will discuss this
topic in details later. However, it is important to remember that
global Lorentz invariance of the Lagrangian is a fundamental
requirement. The requirement of global Lorentz invariance can also
be treated as a transcendental principle of the quantum theory
of gravity.  \\

It is widely known that the {\it priori} foundations of
general relativity is the equivalence principle and the
principle of general covariance\cite{02,03,r01,r02}.
In the classical general  relativity,
it was generally believed that the fundamental
laws of nature should be invariant or covariant under a
general transformation of reference frame. In general
relativity, the equivalence principle tells us the existence
of a local inertial reference frame and the fundamental laws
in the local inertial reference frame which are the same as
those in inertial reference system. Then, after a general
transformation of reference frame, the fundamental laws of
nature in arbitrary reference frame or in arbitrary curved
space-time can be obtained. However, this point of view
is criticized in the literature \cite{r03}. In literature
\cite{r03}, it is argued that almost all basic physical
equations of nature are not invariant or covariant under
the most general transformations of reference frame. In
other words, almost all basic physical equations in
general relativity will change their forms under the
transformation from the local inertial reference frame
to a curved space-time. So, if we deduce basic physical
equations from a local inertial reference frame by a
transformation of reference frame as dictated by equivalence
principle and the principle of general covariance, what we
obtained are different from those in general relativity.
Therefore, the equivalence principle and the principle of
general covariance are not real {\it priori} foundations
of general relativity. So, the new renormalizable quantum
gauge general relativity will not be constructed based on
these two principles. We will find that gauge principle
is enough for us to set up a self-consistent quantum gauge
general relativity.
\\

Though we will not use the logic of classical general
relativity in the formulation of quantum gauge general relativity,
we should not forget the principle of general covariance.
It should be stated that, in the new quantum gauge theory of
gravity, the principle of general covariance appears in  another
way, that is, it realized itself through local gravitational gauge
symmetry. From mathematical point of view, the local gravitational
gauge invariance is just the general covariance in general
relativity. In this case, the general transformation in
the principle of general covariance can not be as big as
in classical general relativity, for it can not be the
most general transformation of reference frame, can only
be the first kind of general transformation of reference
frame\cite{r03}. In other words, the symmetry for
gravitational interactions is much smaller than that
in general relativity. In the new quantum gauge
general relativity, the equivalence principle
plays no role. In other words, we will not accept
the equivalence principle as a transcendental principle
of the new quantum gauge general relativity, for gauge
principle is enough for us to construct quantum
gauge general relativity. In fact, there are some
essential difficulties in using the equivalence principle
as a transcendental principle in quantum gauge general relativity.
It is known that, in quantum field theory, both inertial mass
and gravitational mass are not fundamental quantities. In quantum
field theory, mass is a paranotion from conservative
energy-momentum tensor. In other words, mass is not a
transcendental concepts in quantum field theory. So, it
can not be a constituent of a transcendental principle,
for all constituents of a transcendental principle should
be transcendental. On the other hand, inertial energy
momentum tensor and the gravitational energy-momentum tensor
have different origins, if gravity is considered, they
are essentially different. It is generally believed that
the transcendental principles should reveal the most deepest
essentially relationship of objective laws of nature,
not only a superficial quantitative relationship
of some quantities. And in a meaning, equivalence
principle has a meaning of quantitative relation
of two physical quantities, so it is not suitable
to be a transcendental principle of a fundamental
theory. In all, in quantum gauge general
relativity, the equivalence principle has no position.
\\

\section{Gravitational Gauge Group}
\setcounter{equation}{0}

Before we start our mathematical formulation of quantum gauge
general relativity, we have to determine  which group is the
symmetry of gravitational interactions, which is the starting point
of the whole theory. It is know that, in the traditional  gauge
theory of gravity, Lorentz group is localized. We will not follow
this way, for it contradicts with gauge principle. Now, we use
gauge principle to determine which group is the exact group for
quantum gauge general relativity. \\

Some of the most important properties of gravity can be obtained
by studying classical Newton's gravity. In the classical Newton's
theory of gravity, the gravitational force between two point
objects is given by:
\be
f = G \frac{m_1 m_2}{r^2}
\label{3.1}
\ee
with $m_1$ and $m_2$ masses of two objects, $r$ the distance
between two objects. So, gravity is proportional to the masses of
both objects, in other words, mass is the source of gravitational
field. In general relativity, the Einstein's field equation is
the equation which gives out the relation between energy-momentum
tensor and space-time curvature, which is essentially the relation
between energy-momentum tensor and gravitational field. In the
Einstein's field equation, energy-momentum is treated as the source
of gravity. This points of view is inherited in the  quantum gauge
general relativity. In other words, the starting point of
the new quantum gauge general relativity is that the energy-momentum is
the source of gravitational field. According to rule 3 and rule 1
of gauge principle, we know that, energy-momentum should be the conserved
charges of the corresponding global symmetry, which is just the
symmetry for gravity. According to quantum field theory,
energy-momentum is the conserved charge of global space-time
translation, the corresponding conserved current is
energy-momentum tensor. Therefore, the global space-time
translation is the global gravitational gauge transformation.
According to rule 4, we know that, after quantization, the
energy-momentum operator becomes the generator of gravitational
gauge transformation. It also states that, except for energy-momentum
operator, there is no other generator for gravitational
gauge transformation, such as, angular momentum operator $M_{\mu
\nu}$ can not be the generator of gravitational gauge
transformation. This is the reason why we do not localize Lorentz
transformation in the quantum gauge general relativity, for
the generator of Lorentz transformation is not energy-momentum
operator. According to rule 2 of gauge principle, the
gravitational interactions will be completely determined by the
requirement of the local gravitational gauge symmetry. These are
the basic ideas of the new quantum gauge general relativity, and
they are completely deductions of gauge principle. \\

We know that the generator of Lorentz group is angular momentum
operator $M_{\mu \nu}$. If we localize Lorentz group, according to
gauge principle, angular momentum will become source of a new
filed, which transmits direct spin interactions. This kind of
interactions does not belong to traditional Newton-Einstein
gravity. It is a new kind of interactions. Up to now, we do not
know  whether this kind of interactions exists in Nature or
not. Besides, spin-spin interaction is a kind of
non-renormalizable interaction. In other words, a quantum theory
which contains spin-spin interaction is a non-renormalizable
quantum theory. For these reasons, we will not localize Lorentz
group in this paper. We only localize translation group in this
paper. We will find that go along this way, we can set up a
consistent quantum gauge general relativity which is perturbatively
renormalizable. In other words, only localizing space-time
translation group is enough for us to set up a
consistent quantum gravity. \\

From above discussions, we know that, from mathematical point of
view,  gravitational gauge transformation is the inverse
transformation of  space-time translation, and gravitational gauge
group is space-time translation group. Suppose that there is an
arbitrary function $\phi(x)$ of space-time coordinates $x^{\mu}$.
The global space-time translation is:
\be
x^{\mu} \to x'^{\mu} = x^{\mu} + \epsilon^{\mu}.
\label{3.2}
\ee
The corresponding transformation for function $\phi(x)$ is
\be
\phi(x) \to \phi'(x')=\phi(x) = \phi(x' - \epsilon).
\label{3.3}
\ee
According to Taylor series expansion, we have:
\be
\phi(x - \epsilon) = \left(1 + \sum_{n=1}^{\infty} \frac{(-1)^n}{n!}
\epsilon ^{\mu_1} \cdots \epsilon^{\mu^n}
\partial_{\mu_1} \cdots \partial_{\mu_n}  \right) \phi(x),
\label{3.4}
\ee
where
\be
\partial_{\mu_i}=\frac{\partial}{\partial x^{\mu_i}}.
\label{3.5}
\ee
\\

Let's define a special exponential operation here. Define
\be
E^{a^{\mu} \cdot b_{\mu}} \define 1 +
\sum_{n=1}^{\infty} \frac{1}{n!}
a^{\mu_1} \cdots a^{\mu_n} \cdot
b_{\mu_1} \cdots b_{\mu_n}.
\label{3.6}
\ee
This definition is quite different from that of ordinary
exponential function. In general cases, operators $a^{\mu}$ and
$b_{\mu}$ do not commutate each other, so
\be
E^{a^{\mu} \cdot b_{\mu}} \not=
E^{b_{\mu} \cdot a^{\mu}},
\label{3.7}
\ee
\be
E^{a^{\mu} \cdot b_{\mu}} \not=
e^{a^{\mu} \cdot b_{\mu}},
\label{3.8}
\ee
where $ e^{a^{\mu} \cdot b_{\mu}}$ is the ordinary exponential
function whose definition is
\be
e^{a^{\mu} \cdot b_{\mu}} \equiv 1 +
\sum_{n=1}^{\infty} \frac{1}{n!}
(a^{\mu_1} \cdot b_{\mu_1}) \cdots
(a_{\mu_n} \cdot b_{\mu_n}).
\label{3.9}
\ee
If operators $a^{\mu}$ and $b_{\mu}$ commutate each other, we will
have
\be
E^{a^{\mu} \cdot b_{\mu}}  =
E^{b_{\mu} \cdot a^{\mu}},
\label{3.10}
\ee
\be
E^{a^{\mu} \cdot b_{\mu}} =
e^{a^{\mu} \cdot b_{\mu}}.
\label{3.11}
\ee
The translation operator $\ehat$ can be defined by
\be
\ehat \equiv 1 + \sum_{n=1}^{\infty} \frac{(-1)^n}{n!}
\epsilon^{\mu_1} \cdots \epsilon^{\mu_n}
\partial_{\mu_1} \cdots \partial_{\mu_n}.
\label{3.12}
\ee
Then we have
\be
\phi(x - \epsilon) = ( \uhat_{ \epsilon} \phi(x)).
\label{3.13}
\ee
In order to have a good form which is similar to ordinary gauge
transformation operators, the form of $\ehat$ can also be
written as
\be
\ehat =  E^{- i \epsilon^{\mu} \cdot \hat{P}_{\mu}},
\label{3.14}
\ee
where
\be
\hat{P}_{\mu} = -i \frac{\partial}{\partial x^{\mu}}.
\label{3.15}
\ee
$\hat{P}_{\mu}$ is just the energy-momentum operator in
space-time coordinate space. In the definition of $\ehat$ given by
eq.(\ref{3.14}), $\epsilon^{\mu}$ can be independent  of space-time
coordinate or a function of space-time coordinate, in a ward, it
can be an arbitrary function of space time coordinate $x^{\mu}$.  \\

Some operation properties of the translation operator $\ehat$ are
summarized below.
\begin{enumerate}

\item Operator $\ehat$ translate the space-time point of a field
from $x$ to $x - \epsilon$,
\be
\phi(x- \epsilon) = (\ehat \phi(x)),
\label{3.16}
\ee
where $\epsilon^{\mu}$ can be any function of space-time
coordinate. This relation can also be regarded as the definition
of the translation operator $\ehat$.

\item If $\epsilon$ is a function of space-time coordinate, that is
$\partial_{\mu} \epsilon^{\nu} \not= 0$, then
\be
\ehat =  E^{- i \epsilon^{\mu} \cdot \hat{P}_{\mu}}
\not= E^{- i \hat{P}_{\mu} \cdot \epsilon^{\mu}  },
\label{3.17}
\ee
and
\be
\ehat =  E^{- i \epsilon^{\mu} \cdot \hat{P}_{\mu}}
\not= e^{- i \epsilon^{\mu} \cdot \hat{P}_{\mu}}.
\label{3.18}
\ee
If $\epsilon$ is a constant, that is
$\partial_{\mu} \epsilon^{\nu} = 0$, then
\be
\ehat =  E^{- i \epsilon^{\mu} \cdot \hat{P}_{\mu}}
= E^{- i \hat{P}_{\mu} \cdot \epsilon^{\mu}  },
\label{3.19}
\ee
and
\be
\ehat =  E^{- i \epsilon^{\mu} \cdot \hat{P}_{\mu}}
= e^{- i \epsilon^{\mu} \cdot \hat{P}_{\mu}}.
\label{3.20}
\ee

\item Suppose that $\phi_1(x)$ and $\phi_2(x)$ are two arbitrary
functions of space-time coordinate, then we have
\be
\left( \ehat (\phi_1(x) \cdot \phi_2(x))\right)=
\left(\ehat \phi_1(x)  \right) \cdot
\left(\ehat \phi_2(x)  \right)
\label{3.21}
\ee

\item Suppose that $A^{\mu}$ and $B_{\mu}$ are two arbitrary
operators in Hilbert space, $\lambda$ is an arbitrary ordinary
c-number which is commutate with operators $A^{\mu}$ and
$B_{\mu}$, then we have
\be
\frac{\rm d}{{\rm d} \lambda}
E^{\lambda A^{\mu} \cdot B_{\mu}} =
A^{\mu} \cdot E^{\lambda A^{\mu} \cdot B_{\mu}} \cdot B_{\mu}.
\label{3.22}
\ee

\item Suppose that $\epsilon$ is an arbitrary function of space-time
coordinate, then
\be
(\partial_{\mu} \ehat) =
-i (\partial_{\mu} \epsilon^{\nu} ) \ehat \hat{P}_{\nu}.
\label{3.23}
\ee

\item Suppose that $A^{\mu}$ and $B_{\mu}$ are two arbitrary
operators in Hilbert space, then
\be
tr( E^{ A^{\mu} \cdot B_{\mu}} E^{ - B_{\mu} \cdot A^{\mu}} )=
tr {\bf I},
\label{3.24}
\ee
where $tr$ is the trace operation and ${\bf I}$ is the unit operator
in the Hilbert space.

\item Suppose that $A^{\mu}$, $B_{\mu}$ and $C^{\mu}$ are three
operators in Hilbert space, but operators $A^{\mu}$ and $C^{\nu}$
commutate each other:
\be
[ A^{\mu}~~,~~C^{\nu} ] = 0,
\label{3.25}
\ee
then
\be
tr( E^{ A^{\mu} \cdot B_{\mu}} E^{ B_{\nu} \cdot C^{\nu}} )=
tr( E^{ (A^{\mu}+ C^{\mu}) \cdot B_{\mu}}).
\label{3.26}
\ee

\item Suppose that $A^{\mu}$, $B_{\mu}$ and $C^{\mu}$ are three
operators in Hilbert space, they satisfy
\be
\begin{array}{rcl}
\lbrack A^{\mu}~~,~~C^{\nu} \rbrack & = &  0 , \\
\lbrack B_{\mu}~~,~~C^{\nu} \rbrack & = & 0,
\end{array}
\label{3.27}
\ee
then
\be
E^{ A^{\mu} \cdot B_{\mu}} E^{ C^{\nu} \cdot B_{\nu}}  =
E^{ (A^{\mu}+ C^{\mu}) \cdot B_{\mu}}.
\label{3.28}
\ee

\item Suppose that $A^{\mu}$, $B_{\mu}$ and $C^{\mu}$ are three
operators in Hilbert space, they satisfy
\be
\begin{array}{rcl}
\lbrack A^{\mu}~~,~~C^{\nu} \rbrack & = & 0, \\
\lbrack  \lbrack B_{\mu}~~,~~C^{\nu}
\rbrack ~~,~~ A^{\rho} \rbrack &=& 0,  \\
\lbrack  \lbrack B_{\mu}~~,~~C^{\nu}
\rbrack ~~,~~ C^{\rho} \rbrack  &=& 0,
\end{array}
\label{3.29}
\ee
then,
\be
E^{ A^{\mu} \cdot B_{\mu}} E^{ C^{\nu} \cdot B_{\nu}}  =
E^{ (A^{\mu}+ C^{\mu}) \cdot B_{\mu}}
+ [E^{ A^{\mu} \cdot B_{\mu}} ~~,~~C^{\sigma}]
E^{ C^{\nu} \cdot B_{\nu}} B_{\sigma}.
\label{3.30}
\ee

\item Suppose that $\mhat{1}$ and $\mhat{2}$ are two arbitrary
translation operators, define
\be
\mhat{3}=\mhat{2} \cdot \mhat{1},
\label{3.31}
\ee
then,
\be
\epsilon_3^{\mu}(x) = \epsilon_2^{\mu}(x) +
\epsilon_1^{\mu}(x- \epsilon_2(x) ).
\label{3.32}
\ee
This property means that the product of two translation operator
satisfy closure property, which is one of the conditions that any group
must satisfy.

\item Suppose that $\ehat$ is a non-singular translation operator,
then
\be
\ehat^{-1} = E^{ i \epsilon^{\mu}(f(x)) \cdot \hat{P}_{\mu}},
\label{3.33}
\ee
where $f(x)$ is defined by the following relations:
\be
f(x- \epsilon (x)) = x.
\label{3.34}
\ee
$\ehat^{-1}$ is the inverse operator of $\ehat$, so
\be
\ehat^{-1} \ehat = \ehat \ehat^{-1} = {\rm\bf 1},
\label{3.35}
\ee
where {\bf 1} is the unit element of the gravitational gauge group.

\item The product operation of translation also satisfies associative
law. Suppose that $\mhat{1}$ , $\mhat{2}$ and $\mhat{3}$ are three
arbitrary translation operators, then
\be
\mhat{3} \cdot ( \mhat{2} \cdot \mhat{1} )
 = ( \mhat{3} \cdot \mhat{2} ) \cdot \mhat{1} .
\label{3.36}
\ee

\item Suppose that $\ehat$ is an arbitrary translation operator and
$\phi(x)$ is an arbitrary function of space-time coordinate, then
\be
\ehat \phi(x) \ehat^{-1} = \phi (x- \epsilon(x)).
\label{3.37}
\ee
This relation is quite useful in the discussions of
gravitational gauge symmetry.

\item Suppose that $\ehat$ is an arbitrary translation operator.
Define
\be
\Lambda^{\alpha}_{~~\beta} =
\frac{\partial x^{\alpha}}{\partial ( x - \epsilon (x) )^{\beta}},
\label{3.38}
\ee
\be
\Lambda_{\alpha}^{~~\beta} =
\frac{\partial ( x - \epsilon (x))^{\beta}}{\partial x^{\alpha}}.
\label{3.39}
\ee
They satisfy
\be
\Lambda_{\alpha}^{~~ \mu} \Lambda^{\alpha}_{~~ \nu}
= \delta^{\mu}_{\nu},
\label{3.40}
\ee
\be
\Lambda_{\mu}^{ ~~\alpha} \Lambda^{\nu}_{~~ \alpha}
= \delta_{\mu}^{\nu}.
\label{3.41}
\ee
Then we have following relations:
\be
\ehat \hat{P}_{\alpha} \ehat ^{-1}
= \Lambda^{\beta}_{~~\alpha}  \hat{P}_{\beta},
\label{3.42}
\ee
\be
\ehat {\rm d}x^{\alpha} \ehat ^{-1}
= \Lambda_{\beta}^{~~\alpha}  {\rm d}x^{\beta},
\label{3.43}
\ee
which give out the the transformation laws of  $\hat{P}_{\alpha}$
and d$x^{\alpha}$ under local gravitational gauge transformations.

\end{enumerate}

Gravitational gauge group (GGG) is a transformation group which
consists of all non-singular translation operators $\ehat$. We can
easily see that gravitational gauge group is indeed a  group, for
\begin{enumerate}

\item the product of two arbitrary non-singular translation operators
is also a non-singular translation operator, which is also an element
of the gravitational gauge group. So, the product of the group satisfies
closure property which is expressed in eq(\ref{3.31});

\item the product of the gravitational gauge group also satisfies the
associative law which is expressed in eq(\ref{3.36});

\item the gravitational gauge group has
its unit element {\bf 1}, which satisfies
\be
{\rm\bf 1} \cdot \ehat = \ehat \cdot {\rm\bf 1} = \ehat;
\label{3.44}
\ee

\item every non-singular element $\ehat$ has its inverse element
which is given by eqs(\ref{3.33}) and (\ref{3.35}).

\end{enumerate}
According to gauge principle, the gravitational gauge group is the symmetry
of gravitational interactions. The global invariance of gravitational gauge
transformation will give out conserved charges which is just the ordinary
inertial energy-momentum; the requirement of local gravitational gauge
invariance needs introducing gravitational gauge field, and gravitational
interactions are completely determined by the local gravitational gauge
invariance. \\

The generators of gravitational gauge group is just the energy-momentum
operators $\hat{P}_{\alpha}$.  This is required by gauge principle. It can
also be seen from the form of infinitesimal transformations. Suppose that
$\epsilon$ is an infinitesimal quantity, then we have
\be
\ehat \simeq 1 - i \epsilon^{\alpha} \hat{P}_{\alpha}.
\label{3.45}
\ee
Therefore,
\be
i \frac{\partial \ehat}{\partial \epsilon^{\alpha}} \mid _{\epsilon = 0}
\label{3.46}
\ee
gives out  generators $\hat{P}_{\alpha}$ of gravitational gauge group.
It is known  that generators of gravitational gauge
group commute each other
\be
\lbrack \hat{P}_{\alpha} ~~,~~ \hat{P}_{\beta} \rbrack = 0.
\label{3.47}
\ee
However, the commutation property of generators does not mean that
gravitational gauge group is an Abelian group, because two general elements
of gravitational gauge group do not commute:
\be
\lbrack \mhat{1} ~~,~~ \mhat{2} \rbrack \not= 0.
\label{3.48}
\ee
Gravitational gauge group is a kind of non-Abelian gauge group. The
non-Able nature of  gravitational gauge group will cause
self-interactions of gravitational gauge field. \\

In order to avoid confusion, we need to pay some attention to some
differences between two concepts: space-time translation group and
gravitational gauge group. Generally speaking, space-time
translation is a kind of coordinates transformation, that is, the
objects or fields in space-time are fixed while the
space-time coordinates that describe the motion of objective
matter undergo transformation. But gravitational gauge
transformation is a kind of system transformation rather than a
kind of coordinates transformation. In system transformation, the
space-time coordinate system is fixed while objects or
fields undergo transformation. From mathematical point of view,
space-time translation and gravitational gauge transformation are
essentially the same, and the space-time translation is the
inverse transformation of the gravitational gauge transformation;
but from physical point of view, space-time translation and
gravitational gauge transformation are quite different, especially
when we discuss gravitational gauge transformation of
gravitational gauge field. For gravitational gauge field, its
gravitational gauge transformation is not the inverse
transformation of its space-time translation. In a meaning,
space-time translation is a kind of mathematical transformation,
which contains little dynamical information of interactions; while
gravitational gauge transformation is a kind of physical
transformation, which contains all dynamical information of
interactions and is convenient for us to study physical
interactions. Through gravitational gauge symmetry, we can
determine the whole gravitational interactions among various kinds
of fields. This is the reason why we do not call gravitational
gauge transformation space-time translation. This is important for
all of our discussions on gravitational
gauge transformations of various kinds of fields. \\

Suppose that $\phi(x)$ is an arbitrary scalar field. Its gravitational
gauge transformation is
\be
\phi(x) \to \phi'(x) = ( \ehat \phi(x)).
\label{3.49}
\ee
Similar to ordinary $SU(N)$ non-Abelian gauge field theory, there are
two kinds of scalars. For example, in chiral perturbative theory, the
ordinary $\pi$ mesons are scalar fields, but they are vector fields in
isospin space. Similar case exists in gravitational gauge field theory.
A Lorentz scalar can be a scalar or a vector or a tensor in the space
of gravitational gauge group. For the sake of simplicity, we call
this space gravitational gauge space or gravitational gauge Lie algebra.
If $\phi (x)$ is a scalar in the
gravitational gauge space, we just simply denote it as $\phi (x)$ in
gauge group space. If it is a vector in the  gravitational
gauge space, it can be expanded in the gravitational gauge
space in the following way:
\be
\phi(x)  =  \phi^{\alpha}(x) \cdot \hat{P}_{\alpha}.
\label{3.50}
\ee
The transformation of component field is
\be
\phi^{\alpha}(x)  \to \phi^{\prime  \alpha}(x) =
\Lambda^{\alpha}_{~~\beta} \ehat \phi^{\beta}(x) \ehat^{-1} .
\label{3.51}
\ee
The important thing that we must remember is that, the $\alpha$
index is not a Lorentz index, it is just a group index. For
gravitation gauge group, it is quite special that a group index looks
like a Lorentz index. We must be carefully on this important thing.
This will cause some fundamental changes on quantum gravity.
Lorentz scalar $\phi(x)$ can also be a tensor in gauge group
space. suppose that it is a $n$th order tensor in gauge group
space, then it can be expanded as
\be
\phi(x)  =  \phi^{\alpha_1 \cdots \alpha_n}(x)
\cdot \hat{P}_{\alpha_1}  \cdots \hat{P}_{\alpha_n}.
\label{3.52}
\ee
The transformation of component field is
\be
\phi^{\alpha_1 \cdots \alpha_n}(x)  \to
\phi ^{\prime \alpha_1 \cdots \alpha_n}(x) =
\Lambda^{\alpha_1}_{~~\beta_1} \cdots \Lambda^{\alpha_n}_{~~\beta_n}
\ehat \phi^{\beta_1 \cdots \beta_n}(x) \ehat^{-1} .
\label{3.53}
\ee
\\

If $\phi (x)$ is a spinor field, the above discussion is also valid.
That is, a spinor can also be a scalar or a vector or a tensor in
the space of gravitational gauge group. The gravitational
gauge transformations of the component fields are also given by
eqs.(\ref{3.49}-\ref{3.53}).
There is no transformations in spinor space, which
is different from  the Lorentz transformation of a spinor.  \\

Suppose that $A_{\mu} (x)$ is an arbitrary vector field. Here,
the index $\mu$ is a Lorentz index. Its gravitational gauge
transformation is:
\be
A_{\mu}(x) \to A'_{\mu}(x) = ( \ehat A_{\mu}(x)).
\label{3.54}
\ee
Please remember that there is no rotation in the space of Lorentz
index $\mu$, while in the general coordinates transformations in
general relativity, there is rotation in the space of Lorentz index
$\mu$. The reason is that gravitational gauge transformation is a
kind of system transformation, while in general relativity, the
general coordinates transformation is a kind of coordinates
transformation. If $A_{\mu}(x)$ is a scalar in the
gravitational gauge space, eq(\ref{3.54}) is all for its gauge
transformation. If $A_{\mu}(x)$ is a vector in the
gravitational gauge space, it can be expanded as:
\be
A_{\mu}(x)  =  A_{\mu}^{\alpha}(x) \cdot \hat{P}_{\alpha}.
\label{3.55}
\ee
The transformation of component field  is
\be
A_{\mu}^{\alpha}(x)  \to A_{ \mu}^{\prime\alpha}(x) =
\Lambda^{\alpha}_{~~\beta} \ehat A_{\mu}^{~~\beta}(x) \ehat^{-1} .
\label{3.56}
\ee
If $A_{\mu}(x)$ is a $n$th order tensor in the
gravitational gauge space, then
\be
A_{\mu}(x)  =  A_{\mu}^{\alpha_1 \cdots \alpha_n}(x)
\cdot \hat{P}_{\alpha_1}  \cdots \hat{P}_{\alpha_n}.
\label{3.57}
\ee
The transformation of component fields is
\be
A_{\mu}^{\alpha_1 \cdots \alpha_n}(x)  \to
A_{\mu}^{\prime\alpha_1 \cdots \alpha_n}(x) =
\Lambda^{\alpha_1}_{~~\beta_1} \cdots \Lambda^{\alpha_n}_{~~\beta_n}
\ehat A_{\mu}^{\beta_1 \cdots \beta_n}(x) \ehat^{-1} .
\label{3.58}
\ee
Therefore, under gravitational gauge transformations, the behavior
of a group index is quite different from that of a Lorentz index. However,
they have the same behavior in global Lorentz transformations. \\

Generally speaking, suppose that $T^{\mu_1 \cdots \mu_n}
_{\nu_1 \cdots \nu_m} (x)$ is an arbitrary tensor, its
gravitational gauge transformations are:
\be
T^{\mu_1 \cdots \mu_n}_{\nu_1 \cdots \nu_m} (x)
\to T'^{\mu_1 \cdots \mu_n}_{\nu_1 \cdots \nu_m} (x)
 = ( \ehat T^{\mu_1 \cdots \mu_n}_{\nu_1 \cdots \nu_m} (x)).
\label{3.59}
\ee
If it is a $p$th order tensor in group space, then
\be
T^{\mu_1 \cdots \mu_n}_{\nu_1 \cdots \nu_m} (x)
= T^{\mu_1 \cdots \mu_n ; \alpha_1 \cdots \alpha_p}
_{\nu_1 \cdots \nu_m} (x)
\cdot \hat{P}_{\alpha_1}  \cdots \hat{P}_{\alpha_p}.
\label{3.60}
\ee
The transformation of component fields is
\be
T^{\mu_1 \cdots \mu_n ; \alpha_1 \cdots \alpha_p}
_{\nu_1 \cdots \nu_m} (x)  \to
T'^{\mu_1 \cdots \mu_n ; \alpha_1 \cdots \alpha_p}
_{\nu_1 \cdots \nu_m} (x) =
\Lambda^{\alpha_1}_{~~\beta_1} \cdots \Lambda^{\alpha_p}_{~~\beta_p}
\ehat T^{\mu_1 \cdots \mu_n ; \beta_1 \cdots \beta_p}
_{\nu_1 \cdots \nu_m} (x)\ehat^{-1}.
\label{3.61}
\ee
\\

$\eta^{\mu \nu}$ is a second order Lorentz tensor, but it is
a scalar in gravitational gauge space. It is the metric
of the coordinate space. A Lorentz
index can be raised or descended by this metric tensor.
In a Minkowski space-time, it is selected to be:
\be
\begin{array}{rcl}
\eta^{0~0} &=& -1,  \\
\eta^{1~1} &=&  1,  \\
\eta^{2~2} &=&  1,  \\
\eta^{3~3} &=&  1,
\end{array}
\label{3.73}
\ee
and other components of $\eta^{\mu \nu}$ vanish. $\eta^{\mu \nu}$
is the traditional Minkowski metric.
\\

\section{Two Pictures of Gravity}
\setcounter{equation}{0}

As we have mentioned above, quantum gauge general relativity
is logically independent of traditional quantum gravity. It
is know that, there are at least two pictures of gravity.
In one picture, gravity is treated as space-time geometry.
In this picture, space-time is curved and there is no
physical gravitational interactions, for all effects of
gravity are represented by space-time metric. In another
picture, gravity is treated as a kind of fundamental
interactions. In this picture, space-time is always flat
and space-time metric is always selected to be the
Minkowski metric. For the sake of simplicity, we call
the first picture gemeotry picture of gravity and the
second picture physics picture of gravity. \\

The concepts of "physics picture of gravity" and "geometrical
picture of gravity" are key important to understand the
present theory. In order to understand these important
things, we use quantum mechanics as an example.
In quantum mechanics, there are many pictures, such Schrodinger
picture, Heisenberg picture, $\cdots$ etc. In Schrodinger
picture, operators of physical quantities are fixed and do not
change with time, but wave functions are evolve with time.
On the contrary, in Heisenberg picture, wave functions are
 fixed and do not change with time, but operators evolve
with time. If we want to know whether wave functions is  changed
with time or not, you must first determine in which picture
you study wave functions. If you do not know in which picture
you study wave functions, you will not know whether
wave functions should be changed with time or not.
Now, similar case happens in quantum gauge theory of
gravity. If you want to know whether space-time
is curved or not, you must first determine in which
picture gravity is studied. In physics picture of
gravity, space-time is flat, but in geometry picture
of gravity, space-time is curved.
They are two different space-times, i.e., the space-time
in physics picture of gravity is different from the
space-time in geometrical picture of gravity.
Quantum gauge theory
of gravity is formulated in the physics picture of gravity,
classical Newton's theory of gravity is also
formulated in the physics picture of gravity,
and the Einstein's general relativity is formulated in the
geometrical picture of gravity. Please do not discuss any
problem simultaneous in two pictures, which is dangerous.\\

Quantum gauge general relativity is foumulated in the
physics picture of gravity. So, in quantum gauge general
relativity, space-time is always flat and gravity is
treated as a kind of fundatmental interactions. In
order to avoid confusing, we do
not introduce any comcept of curved space-time
and we do not use any language of geometry
at present. It is suggest that anyone read this
paper do not try to find any geometrical meaning
of any physical quantities, do not use the language
of geometry to understand anything of this paper
and forget everything about the concept of fibre bundles,
connections, curved space-time metric, $\cdots$ etc,
for the present theory is not formulated in the
geometry picture of gravity. After we go into the
geometry picture of gravity and set up the geometrical
picture of quantum gauge theory of gravity, we
can use geometry language and study the geometry
meaning of the present theory. But at present, we
will not use the language of geometry.
\\

There are mainly the following four reasons to
introduce the physics picture of gravity and formulate
quantum gauge general relativity in the physics
picture of gravity:
\begin{enumerate}

\item   It has a clear interaction picture, so we
        can use perturbation theory to calculate
        the amplitudes of physical process.

\item   We can use traditional gauge field theory to
        study quantum behavior of gravitational interactions,
        and four different kinds of fundamental interactions
        in Nature can be formulated in the same manner, and
        four kinds of fundamental interactions can be unified
        in a simple and beautiful way.

\item   The perturbatively renormalizability of the theory can be
        easily proved in the physics picture of gravity.\\

\item   Quantum effects of gravitational interactions can be
        easily understood in the physics picture of gravity. \\

\end{enumerate}

Gravitational gauge transformation is different from space-time
translation. In gravitational gauge transformation, space-time
is fixed, space-time coordinates are not changed, only fields
and objects undergo some translation. In a meaning, gravitational
gauge transformation is a kind of physical transformation on
objects and fields. The traditional space-time translation
is a kind of transformation in which objects and fields are
kept unchanged while space-time coordinates undergo
translation. In a meaning, space-time translation is a kind
of geometrical transformation on space-time. Because quantum
gauge theory of gravity is set up in the physics picture
of gravity, we have to use gravitational gauge transformation
in our discussion, for physics picture needs physical
transformation. We do not discuss translation transformation
of space-time and gauge translations in physics picture of
gravity. In a meaning, space-time translation
is a kind of geometrical transformation,
which contains  geometrical information of
space-time structure and is convenient for us to study
space-time geometry; while
gravitational gauge transformation is a kind of physical
transformation, which contains all dynamical information of
gravitational interactions and is convenient
for us to study physical interactions.
Though from mathematical point of view, for global transformations,
space-time translation is the inverse transformation of
the gravitational gauge transformation. But from physical
point of view and for local transformation, they are not
the same. In the quantum gauge general relativity,
we do not gauge translation group,
but gauge gravitational gauge group, for translation
group is different from gravitational gauge group.
Translation group is the symmetry of space-time, but
 gravitational gauge group is the symmetry group
of physical fields and objects. They have essential difference
from physical point of view.
\\

\section{Pure Gravitational Gauge Fields}
\setcounter{equation}{0}

Before we study gravitational field, we must determine which field
represents gravitational field. In the traditional gravitational
gauge theory, gravitational field is represented by space-time
metric tensor. If there is gravitational field in space-time, the
space-time metric will not be equivalent to Minkowski metric, and
space-time will become curved. In other words, in the traditional
gravitational gauge theory, quantum gravity is formulated in
curved space-time. In this paper, we will not follow this way. The
underlying point of view of this new quantum gauge general relativity
is that it is formulated in the framework of traditional
quantum field theory, gravity is treated as a kind of physical
interactions in flat space-time and the
gravitational field is represented by gauge
potential.  In other words, if we put gravity into the structure
of space-time, the space-time will become curved and there will be
no physical gravity in space-time, because all gravitational
effects are put into space-time metric and gravity is geometrized.
But if we study physical gravitational interactions, it is better
to rescue gravity from space-time metric and treat gravity as a
kind of physical interactions. In this case, space-time is
flat and there is physical gravity in Minkowski space-time.
For this reason, we will not introduce the concept of curved
space-time to study quantum gravity in the most part of
this paper. So, in most chapters of this paper,
the space-time is always flat, the gravitational field is
represented by gauge potential, and gravitational interactions
are always treated as physical interactions.
In fact, what gravitational field is represented by gauge
potential is required by gauge principle.
\\

Now, let's begin to construct the Lagrangian of quantum gauge
general relativity. For the sake of simplicity,
let's suppose that  $\phi (x)$ is a Lorentz scalar and
gauge group scalar. According to above discussions,
its gravitational gauge transformation is
\be
\phi (x) \to \phi '(x) = (\ehat \phi (x)).
\label{4.1}
\ee
Because
\be
(\partial_{\mu} \ehat ) \not= 0,
\label{4.2}
\ee
partial derivative of $\phi (x)$ does not transform covariantly
under gravitational gauge transformation
\be
\partial_{\mu} \phi (x) \to \partial_{\mu} \phi '(x)
\not= ( \ehat \partial_{\mu} \phi (x) ).
\label{4.3}
\ee
In order to construct an action which is invariant under local
gravitational gauge transformation, gravitational gauge
covariant derivative is highly necessary. The gravitational gauge
covariant derivative is defined by
\be
D_{\mu} = \partial_{\mu} - i g C_{\mu} (x),
\label{4.4}
\ee
where $C_{\mu} (x)$ is the gravitational gauge field and $g$ is the
coupling constant of gravitational gauge interactions. It is a Lorentz
vector. Under gravitational gauge transformations, it transforms as
\be
C_{\mu}(x) \to  C'_{\mu}(x) =
\ehat (x) C_{\mu} (x) \ehat^{-1} (x)
+ \frac{i}{g} \ehat (x) (\partial_{\mu} \ehat^{-1} (x)).
\label{4.5}
\ee
Using the original definition of $\ehat$, we can strictly proved
that
\be
\lbrack \partial_{\mu} ~~,~~ \ehat \rbrack
= (\partial_{\mu} \ehat).
\label{4.6}
\ee
Therefor, we have
\be
\ehat \partial_{\mu} \ehat^{-1} =
\partial_{\mu}  + \ehat (\partial_{\mu} \ehat^{-1}),
\label{4.7}
\ee
\be
\ehat D_{\mu} \ehat^{-1} =
\partial_{\mu} - i g  C'_{\mu} (x).
\label{4.8}
\ee
So, under local gravitational gauge transformations,
\be
D_{\mu} \phi (x) \to
D'_{\mu} \phi '(x) =
(\ehat D_{\mu} \phi (x)) ,
\label{4.9}
\ee
\be
D_{\mu} (x) \to D'_{\mu} (x)
= \ehat D_{\mu} (x) \ehat^{-1}.
\label{4.10}
\ee
\\

Gravitational gauge field $C_{\mu} (x)$ is vector field, it is a
Lorentz vector. It is also a vector in gravitational gauge space,
so it can be expanded as linear combinations of generators of
gravitational gauge group
\be
C_{\mu} (x) = C_{\mu}^{\alpha}(x) \cdot \hat{P}_{\alpha}.
\label{4.11}
\ee
$C_{\mu}^{\alpha}$ are component fields of gravitational gauge
field. It looks like a second rank tensor. But according to our
previous discussion, it is not a tensor field, it is a vector
field. The index $\alpha$ is not a Lorentz index, it is just a gauge
group index. Gravitational gauge field $C_{\mu}^{\alpha}$ has only
one Lorentz index, so it is a kind of vector field. This is a
result of gauge principle. The gravitational gauge transformation  of
component field is
\be
C_{\mu}^{\alpha}(x) \to C_{\mu}^{\prime \alpha}(x)=
\Lambda ^{\alpha}_{~~\beta} (\ehat C_{\mu}^{\beta}(x))
- \frac{1}{g} (\ehat  \partial_{\mu} \epsilon^{\alpha}(y)),
\label{4.12}
\ee
where $y$ is a function of space-time coordinates which satisfy
\be
( \ehat y(x) )=x.
\label{4.13}
\ee
\\

Define matrix $G$ as
\be  \label{4.701}
G = (G_{\mu}^{\alpha}) = ( \delta_{\mu}^{\alpha} - g C_{\mu}^{\alpha}).
\ee
 A simple form for matrix $G$ is
\be  \label{4.702}
G = I - gC,
\ee
where $I$ is a unit matrix and $C= (C_{\mu}^{\alpha})$. Therefore,
\be  \label{4.703}
G^{-1} = \frac{1}{I-gC}.
\ee
$G^{-1}$ is the inverse matrix of $G$, it satisfies
\be  \label{4.704}
(G^{-1})^{\mu}_{\beta} G^{\alpha}_{\mu}
= \delta_{\beta}^{\alpha},
\ee
\be  \label{4.705}
G^{\alpha}_{\mu} (G^{-1})^{\nu}_{\alpha}
= \delta^{\nu}_{\mu}.
\ee
Define
\be  \label{4.706}
g^{\alpha \beta} \define \eta^{\mu \nu}
G_{\mu}^{\alpha} G_{\nu}^{\beta}.
\ee
\be  \label{4.707}
g_{\alpha \beta} \define \eta_{\mu \nu}
(G^{-1})_{\alpha}^{\mu} (G^{-1})_{\beta}^{\nu}.
\ee
It can be easily proved that
\be  \label{4.708}
g_{\alpha \beta} g^{\beta \gamma} = \delta_{\alpha}^{\gamma},
\ee
\be  \label{4.7081}
g^{\alpha \beta} g_{\beta \gamma} = \delta^{\alpha}_{\gamma}.
\ee
Under gravitational gauge transformations, they transform
as
\be  \label{4.7082}
G^{\alpha}_{ \mu}(x)  \to G^{\prime \alpha}_{ \mu}(x)
= \Lambda^{\alpha}_{~\alpha_1}
(\ehat G^{\alpha_1}_{ \mu}(x)),
\ee
\be  \label{4.7083}
G_{\alpha}^{-1 \mu }(x)  \to G_{\alpha}^{\prime  -1 \mu }(x)
= \Lambda_{\alpha}^{~\alpha_1}
(\ehat G_{\alpha_1}^{-1 \mu}(x)),
\ee
\be  \label{4.709}
g_{\alpha \beta}(x)  \to g'_{\alpha \beta}(x)
= \Lambda_{\alpha}^{~\alpha_1} \Lambda_{\beta}^{~\beta_1}
(\ehat g_{\alpha_1 \beta_1}(x)),
\ee
\be  \label{4.710}
g^{\alpha \beta}(x)  \to g^{\prime \alpha \beta}(x)
= \Lambda^{\alpha}_{~\alpha_1} \Lambda^{\beta}_{~\beta_1}
(\ehat g^{\alpha_1 \beta_1}(x)),
\ee
\\

The strength of gravitational gauge field is defined by
\be
F_{\mu \nu} = \frac{1}{-i g}
\lbrack D_{\mu} ~~,~~ D_{\nu} \rbrack,
\label{4.14}
\ee
or
\be
F_{\mu \nu} = \partial_{\mu} C_{\nu}(x)
- \partial_{\nu} C_{\mu}(x)
- i g C_{\mu}(x) C_{\nu}(x)
+ i g  C_{\nu}(x) C_{\mu}(x).
\label{4.15}
\ee
$F_{\mu \nu}$ is a second order Lorentz tensor. It is a vector is
group space, so it can be expanded in group space,
\be
F_{\mu \nu} (x) = F_{\mu \nu}^{\alpha} (x) \cdot \hat{P}_{\alpha}.
\label{4.16}
\ee
The explicit form of component field strengths is
\be
F_{\mu \nu}^{\alpha} = \partial_{\mu} C_{\nu}^{\alpha}
- \partial_{\nu} C_{\mu}^{\alpha}
- g C_{\mu}^{\beta} \partial_{\beta} C_{\nu}^{\alpha}
+ g C_{\nu}^{\beta} \partial_{\beta} C_{\mu}^{\alpha}
\label{4.17}
\ee
The strength of gravitational gauge field transforms covariantly
under gravitational gauge transformation
\be
F_{\mu \nu} \to F'_{\mu \nu} =
\ehat F_{\mu \nu} \ehat^{-1}.
\label{4.18}
\ee
The gravitational gauge transformation of the
component field strength  is
\be
F_{\mu \nu}^{\alpha} \to F_{\mu \nu}^{\prime \alpha} =
\Lambda^{\alpha}_{~~\beta} (\ehat F_{\mu \nu}^{\beta}).
\label{4.19}
\ee
\\

In physics picture of gravity,
$g_{\alpha\beta}$ defined by (\ref{4.707}) and
$g^{\alpha\beta}$ defined by (\ref{4.706}) are not space-time metric,
for the space-time metric are always Minkowski metric.
They are only two composite field operators which are composed of
gravitational gauge field. Using these two operators,
we can calculate another important operator
\be
\Gamma^{\gamma}_{\alpha \beta}
= \frac{1}{2} g^{\gamma \delta}
\left( \frac{\partial g_{\alpha \delta}}{\partial x^{\beta}}
+ \frac{\partial g_{\beta \delta}}{\partial x^{\alpha}}
-\frac{\partial g_{\alpha \beta}}{\partial x^{\delta}} \right).
\label{4.1901}
\ee
From above definition, we can see that
$\Gamma^{\gamma}_{\alpha \beta}$ looks like
the affine connection in general relativity. But now, it
has no geometric meaning, it is not the affine connection in
the curved space-time in geometrical picture of gravity,
for now we are in the physics picture of gravity.
It is only a composite field operator.
Using the following relation,
\be
- g F^{\gamma}_{\mu \nu}
= G^{\alpha}_{\mu} G^{\beta}_{\nu}
\lbrack ( G^{-1} \partial_{\alpha} G)^{\gamma}_{\beta}
- ( G^{-1} \partial_{\beta} G)^{\gamma}_{\alpha} \rbrack,
\label{4.1902}
\ee
where $ F^{\gamma}_{\mu \nu}$ is the component field strength of
gravitational gauge field, we get
\be
\begin{array}{rcl}
\Gamma^{\gamma}_{\alpha \beta}
&=& - \frac{1}{2}
\lbrack  G^{-1 \mu}_{\beta} \partial_{\alpha} G^{\gamma}_{\mu}
+  G^{-1 \mu}_{\alpha} \partial_{\beta} G^{\gamma}_{\mu} \rbrack \\
&&\\
&& + \frac{g}{2}  g^{\gamma \delta}
(g_{\beta \alpha_1} G^{-1 \mu}_{\alpha}
+ g_{\alpha \alpha_1} G^{-1 \mu}_{\beta} )
G^{-1 \nu}_{\delta}
F^{\alpha_1}_{\mu \nu} .
\end{array}
\label{4.1903}
\ee
\\

Operator $R^{\delta}_{\alpha \beta \gamma}$ is defined by
\be
R^{\delta}_{\alpha \beta \gamma}
\define \partial_{\gamma} \Gamma^{\delta}_{\alpha \beta}
-\partial_{\beta} \Gamma^{\delta}_{\alpha \gamma}
+\Gamma^{\eta}_{\alpha \beta} \Gamma^{\delta}_{\gamma \eta}
- \Gamma^{\eta}_{\alpha \gamma} \Gamma^{\delta}_{\beta \eta},
\label{4.1904}
\ee
the operator $R_{\alpha \gamma}$ is defined by
\be
R_{\alpha \gamma} \define
R^{\beta}_{\alpha \beta \gamma},
\label{4.1905}
\ee
and the scalar operator $R$ is defined by
\be
R \define g^{\alpha \gamma} R_{\alpha \gamma}.
\label{4.1906}
\ee
Operator $R_{\alpha \gamma}$ can also be calculated from the
following relation
\be
\ba{rcl}
R_{\alpha \gamma}
& = & \frac{1}{2} g^{\beta \delta}
(\partial_{\beta} \partial_{\delta} g_{\alpha \gamma}
- \partial_{\alpha} \partial_{\delta} g_{\beta \gamma}
-\partial_{\beta} \partial_{\gamma} g_{\alpha \delta}
+\partial_{\alpha} \partial_{\gamma} g_{\beta \delta} ) \\
&&\\
&& + g^{\beta \delta} g_{\alpha_1 \beta_1}
(\Gamma^{\alpha_1}_{\alpha \gamma} \Gamma^{\beta_1}_{\beta \delta}
- \Gamma^{\alpha_1}_{\alpha \delta} \Gamma^{\beta_1}_{\beta \gamma}).
\ea
\label{4.19061}
\ee
The explicit expression for the operator $R_{\alpha \gamma}$ is
\be
\begin{array}{rcl}
R_{\alpha \gamma} &=&
-(\partial_{\gamma} \partial_{\alpha} G\cdot G^{-1}) ^{\mu}_{\mu}
+ 2 ( \partial_{\gamma} G \cdot G^{-1}
\cdot \partial_{\alpha} G \cdot G^{-1} )^{\mu}_{\mu}  \\
&& + \eta^{\rho \sigma} \eta_{\mu \nu}
( \partial_{\alpha} G \cdot G^{-1} )^{\mu}_{\rho}
( \partial_{\gamma} G \cdot G^{-1} )^{\nu}_{\sigma} \\
&&+ \frac{1}{2} g^{\beta \delta} \eta_{\mu \nu}
( G^{-1} \cdot \partial_{\beta} G \cdot G^{-1} \cdot \partial_{\delta} G
\cdot G^{-1} )^{\mu}_{\alpha} G^{-1 \nu}_{\gamma}  \\
&&  - \frac{1}{2} g^{\beta \delta} \eta_{\mu \nu}
( G^{-1} \cdot \partial_{\beta} \partial_{\delta} G
\cdot G^{-1} )^{\mu}_{\alpha} G^{-1 \nu}_{\gamma} \\
&& + \frac{1}{2} g^{\beta \delta} \eta_{\mu \nu}
( G^{-1} \cdot \partial_{\delta}G \cdot G^{-1} \cdot
\partial_{\beta} G \cdot G^{-1} ) ^{\mu}_{\alpha} G^{-1 \nu}_{\gamma} \\
&&+ \frac{1}{2} g^{\beta \delta} \eta_{\mu \nu}
G^{-1 \mu }_{\alpha}
( G^{-1} \cdot \partial_{\beta} G \cdot G^{-1} \cdot \partial_{\delta} G
\cdot G^{-1} )^{\nu}_{\gamma} \\
&& -\frac{1}{2} g^{\beta \delta} \eta_{\mu \nu}
G^{-1 \mu }_{\alpha}
( G^{-1} \cdot \partial_{\beta} \partial_{\delta}G
\cdot G^{-1} )^{\nu}_{\gamma} \\
&& + \frac{1}{2} g^{\beta \delta} \eta_{\mu \nu}
G^{-1 \mu }_{\alpha}
( G^{-1} \cdot \partial_{\delta} G \cdot G^{-1} \cdot \partial_{\beta} G
\cdot G^{-1} )^{\nu}_{\gamma}  \\
&&+  g^{\beta \delta} \eta_{\mu \nu}
( G^{-1} \cdot \partial_{\beta} G \cdot G^{-1} ) ^{\mu}_{\alpha}
( G^{-1} \cdot \partial_{\delta} G \cdot G^{-1} )^{\nu}_{\gamma} \\
&& -\frac{1}{2} ( G^{-1} \cdot \partial_{\gamma} G \cdot G^{-1}
\cdot \partial_{\beta} G ) ^{\beta}_{\alpha}
-\frac{1}{2} ( G^{-1} \cdot \partial_{\beta} G \cdot G^{-1}
\cdot \partial_{\gamma} G ) ^{\beta}_{\alpha}
+ \frac{1}{2} ( G^{-1} \cdot \partial_{\gamma}
\partial_{\beta} G ) ^{\beta}_{\alpha} \\
&& - \frac{1}{2} \eta^{\rho \sigma} \eta_{\mu \nu} G^{\delta}_{\rho}
( G^{-1} \cdot \partial_{\delta} G \cdot G^{-1} )^{\mu}_{\alpha}
(\partial_{\gamma} G \cdot G^{-1})^{ \nu}_{\sigma} \\
&& - \frac{1}{2} \eta^{\rho \sigma} \eta_{\mu \nu} G^{\delta}_{\rho}
( G^{-1} \cdot \partial_{\gamma} G \cdot G^{-1} )^{\mu}_{\alpha}
(\partial_{\delta} G \cdot G^{-1})^{ \nu}_{\sigma} \\
&& - \frac{1}{2} \eta^{\rho \sigma} \eta_{\mu \nu}
G^{\delta}_{\rho} G^{-1 \mu }_{\alpha}
 ( \partial_{\gamma} G \cdot G^{-1} \cdot \partial_{\delta} G
\cdot G^{-1} ) ^{\nu}_{\sigma}  \\
&& + \frac{1}{2}\eta^{\rho \sigma} \eta_{\mu \nu} G^{\delta}_{\rho}
G^{-1 \mu }_{\alpha}
( \partial_{\gamma} \partial_{\delta} G \cdot G^{-1} )^{\nu}_{\sigma}  \\
&& - \frac{1}{2} \eta^{\rho \sigma} \eta_{\mu \nu}
G^{\delta}_{\rho} G^{-1 \mu }_{\alpha}
 ( \partial_{\delta} G \cdot G^{-1} \cdot \partial_{\gamma} G
\cdot G^{-1} ) ^{\nu}_{\sigma}  \\
&&-\frac{1}{2} ( G^{-1} \cdot \partial_{\beta} G \cdot G^{-1}
\cdot \partial_{\alpha} G ) ^{\beta}_{\gamma}
-\frac{1}{2} ( G^{-1} \cdot \partial_{\alpha} G \cdot G^{-1}
\cdot \partial_{\beta} G ) ^{\beta}_{\gamma}
+ \frac{1}{2} ( G^{-1} \cdot \partial_{\beta}
\partial_{\alpha} G ) ^{\beta}_{\gamma} \\
&&- \eta^{\rho \sigma} \eta_{\mu \nu} G^{\beta}_{\sigma}
( G^{-1} \cdot \partial_{\alpha} G \cdot G^{-1} )^{\mu}_{\gamma}
(\partial_{\beta} G \cdot G^{-1})^{ \nu}_{\rho} \\
&&- \frac{1}{2} \eta^{\rho \sigma} \eta_{\mu \nu}
G^{\beta}_{\sigma} G^{-1 \mu }_{\gamma}
 ( \partial_{\beta} G \cdot G^{-1} \cdot \partial_{\alpha} G
\cdot G^{-1} ) ^{\nu}_{\rho}  \\
&& + \frac{1}{2}\eta^{\rho \sigma} \eta_{\mu \nu} G^{\beta}_{\sigma}
G^{-1 \mu }_{\gamma}
( \partial_{\beta} \partial_{\alpha} G \cdot G^{-1} )^{\nu}_{\rho}
- \frac{1}{2} \eta^{\rho \sigma} \eta_{\mu \nu}
G^{\beta}_{\sigma} G^{-1 \mu }_{\gamma}
 ( \partial_{\alpha} G \cdot G^{-1} \cdot \partial_{\beta} G
\cdot G^{-1} ) ^{\nu}_{\rho}  \\
&& + \frac{1}{2} \eta_{\mu \nu} \eta^{\mu_1 \nu_1}
G^{\beta}_{\nu_1} ( \partial_{\beta} G \cdot G^{-1})^{\mu}_{\mu_1}
( G^{-1} \cdot \partial_{\alpha} G \cdot G^{-1})^{\nu}_{\gamma} \\
&&+ \frac{1}{2} \eta_{\mu \nu} \eta^{\mu_1 \nu_1}
G^{\beta}_{\nu_1} ( \partial_{\beta} G \cdot G^{-1})^{\mu}_{\mu_1}
( G^{-1} \cdot \partial_{\gamma} G \cdot G^{-1})^{\nu}_{\alpha}  \\
&& - \frac{1}{4} \eta_{\mu \nu} \eta^{\mu_1 \nu_1}
( \partial_{\gamma} G \cdot G^{-1})^{\mu}_{\mu_1}
( \partial_{\alpha} G \cdot G^{-1})^{\nu}_{\nu_1} \\
&&- \frac{1}{4} \eta_{\mu \nu} \eta^{\mu_1 \nu_1} G^{\beta}_{\nu_1}
( \partial_{\gamma} G \cdot G^{-1})^{\mu}_{\mu_1}
( G^{-1} \cdot \partial_{\beta} G \cdot G^{-1})^{\nu}_{\alpha}\\
&& - \frac{1}{4} \eta_{\mu \nu} \eta^{\mu_1 \nu_1} G^{\delta}_{\mu_1}
(G^{-1} \cdot \partial_{\delta} G \cdot G^{-1})^{\mu}_{\gamma}
( \partial_{\alpha} G \cdot G^{-1})^{\nu}_{\nu_1} \\
&& - \frac{1}{4} \eta_{\mu \nu} \eta^{\mu_1 \nu_1}
G^{\delta}_{\mu_1}G^{\beta}_{\nu_1}
( G^{-1} \cdot \partial_{\delta} G \cdot G^{-1})^{\mu}_{\gamma}
( G^{-1} \cdot \partial_{\beta} G \cdot G^{-1})^{\nu}_{\alpha}\\
&& - \frac{g}{2} \eta_{\mu \nu} \eta^{\mu_3 \nu_3}
F^{\delta}_{\mu_1 \nu_1} G^{\beta}_{\nu_3} G^{-1 \mu}_{\delta}
G^{-1 \nu}_{\gamma} G^{-1 \mu_1}_{\alpha}
( \partial_{\beta} G \cdot G^{-1} )^{\nu_1}_{\mu_3}  \\
&& - \frac{g}{2} \eta_{\mu \nu} \eta^{\mu_3 \nu_3}
F^{\delta}_{\mu_1 \nu_1} G^{\beta}_{\nu_3} G^{-1 \mu}_{\delta}
G^{-1 \nu}_{\alpha} G^{-1 \mu_1}_{\gamma}
( \partial_{\beta} G \cdot G^{-1} )^{\nu_1}_{\mu_3} \\
&& - \frac{g}{2} F^{\beta}_{\mu \nu} G^{-1 \mu}_{\beta}
( G^{-1} \cdot \partial_{\alpha} G \cdot G^{-1})^{\nu}_{\gamma}
- \frac{g}{2} F^{\beta}_{\mu \nu} G^{-1 \mu}_{\beta}
( G^{-1} \cdot \partial_{\gamma} G \cdot G^{-1})^{\nu}_{\alpha}\\
&& + \frac{g}{4} \eta^{\mu_3 \mu_1} \eta_{\mu \nu}
F^{\beta}_{\mu_1 \nu_1} G^{-1 \mu}_{\beta} G^{-1 \nu}_{\alpha}
(\partial_{\gamma}G \cdot G^{-1} )^{\nu_1}_{\mu_3}
+ \frac{g}{4} F^{\beta}_{\mu \nu}
(G^{-1} \cdot \partial_{\gamma} G \cdot G^{-1} )^{\nu}_{\beta}
G^{-1 \mu}_{\alpha}  \\
&&+ \frac{g}{4} \eta^{\mu_3 \mu_1} \eta_{\mu \nu}
F^{\beta}_{\mu_1 \nu_1} G^{-1 \mu}_{\beta} G^{-1 \nu}_{\alpha}
G^{\delta}_{\mu_3}
(G^{-1} \cdot \partial_{\delta}G \cdot G^{-1} )^{\nu_1}_{\gamma} \\
&&+ \frac{g}{4} F^{\beta}_{\mu \nu}
(G^{-1} \cdot \partial_{\beta} G \cdot G^{-1} )^{\nu}_{\gamma}
G^{-1 \mu}_{\alpha}
+ \frac{g}{4} F^{\beta}_{\mu \nu}
(G^{-1} \cdot \partial_{\beta} G \cdot G^{-1} )^{\nu}_{\alpha}
G^{-1 \mu}_{\gamma} \\
&&+ \frac{g}{4} \eta^{\mu_3 \mu_1} \eta_{\mu \nu}
F^{\beta}_{\mu_1 \nu_1} G^{-1 \mu}_{\beta} G^{-1 \nu}_{\gamma}
 (\partial_{\alpha}G \cdot G^{-1} )^{\nu_1}_{\mu_3}
+ \frac{g}{4} F^{\beta}_{\mu \nu}
(G^{-1} \cdot \partial_{\alpha} G \cdot G^{-1} )^{\nu}_{\beta}
G^{-1 \mu}_{\gamma}  \\
&&+ \frac{g}{4} \eta^{\mu_3 \mu_1} \eta_{\mu \nu}
F^{\beta}_{\mu_1 \nu_1} G^{-1 \mu}_{\beta} G^{-1 \nu}_{\gamma}
G^{\delta}_{\mu_3}
(G^{-1} \cdot \partial_{\delta}G \cdot G^{-1} )^{\nu_1}_{\alpha} \\
&& + \frac{g^2}{2} \eta^{\nu \nu_1} \eta_{\mu_2 \nu_2}
F^{\beta}_{\mu \nu} F^{\beta_2}_{\mu_1 \nu_1}
G^{-1 \mu}_{\beta} G^{-1 \mu_2}_{\beta_2}
( G^{-1 \nu_2}_{\alpha} G^{-1 \mu_1}_{\gamma}
+ G^{-1 \nu_2}_{\gamma} G^{-1 \mu_1}_{\alpha}  )  \\
&& - \frac{g^2}{4} \eta_{\mu \nu} \eta_{\mu_2 \nu_2}
\eta^{\mu_1 \nu_1} \eta^{\rho_1 \sigma_1}
F^{\beta}_{\rho_1 \nu_1} F^{\beta_1}_{\sigma_1 \mu_1}
G^{-1 \mu }_{\beta}  G^{-1 \mu_2}_{\beta_1}
G^{-1 \nu}_{\gamma}  G^{-1 \nu_2}_{\alpha} \\
&&- \frac{g^2}{4} \eta_{\mu \nu} \eta^{\mu_1 \nu_1}
F^{\beta}_{\mu_2 \nu_1} F^{\beta_1}_{\mu_3 \mu_1}
G^{-1 \mu }_{\beta}  G^{-1 \mu_2}_{\beta_1}
G^{-1 \nu}_{\gamma}  G^{-1 \mu_3}_{\alpha}\\
&& - \frac{g^2}{4}  \eta_{\mu_2 \nu_2} \eta^{\mu_1 \nu_1}
F^{\beta}_{\nu \nu_1} F^{\beta_1}_{\mu \mu_1}
G^{-1 \mu }_{\beta}  G^{-1 \mu_2}_{\beta_1}
G^{-1 \nu}_{\gamma}  G^{-1 \nu_2}_{\alpha} \\
&&- \frac{g^2}{4} \eta_{\mu \nu} \eta^{\mu_2 \nu_2}
F^{\beta}_{\mu_1 \nu_2} F^{\beta_1}_{\mu_3 \mu_2}
G^{-1 \mu }_{\beta}  G^{-1 \nu}_{\beta_1}
G^{-1 \mu_1}_{\gamma}  G^{-1 \mu_3}_{\alpha}
\end{array}
\label{4.1907}
\ee
Operator $R$ can be calculated from the following relation
\be
R = g^{\alpha \gamma}  g^{\beta \delta}
(\partial_{\beta} \partial_{\delta} g_{\alpha \gamma}
-\partial_{\beta} \partial_{\gamma} g_{\alpha \delta} )
 + g^{\alpha \gamma} g^{\beta \delta} g_{\alpha_1 \beta_1}
(\Gamma^{\alpha_1}_{\alpha \gamma} \Gamma^{\beta_1}_{\beta \delta}
- \Gamma^{\alpha_1}_{\alpha \delta} \Gamma^{\beta_1}_{\beta \gamma}).
\label{4.19071}
\ee
The explicit expression for the operator $R$ is
\be
\begin{array}{rcl}
R & = &
4 g^{\alpha \gamma} ( \partial_{\alpha} G \cdot G^{-1}
\cdot \partial_{\gamma} G \cdot G^{-1} ) ^{\mu}_{\mu}
- 2 g^{\alpha \gamma} ( \partial_{\alpha} \partial_{\gamma}G
 \cdot G^{-1} ) ^{\mu}_{\mu} \\
 &&\\
&& + \frac{3}{2} \eta^{\rho \sigma } \eta_{\mu \nu} g^{\alpha \gamma}
( \partial_{\alpha} G  \cdot G^{-1} ) ^{\mu}_{\rho}
( \partial_{\gamma} G  \cdot G^{-1} ) ^{\nu}_{\sigma}
- 2 \eta^{\mu \nu} G^{\gamma}_{\nu}
( \partial_{\gamma} G \cdot G^{-1} \cdot
\partial_{\alpha} G )^{\alpha}_{\mu}\\
&&\\
&&- 2 \eta^{\mu \nu} G^{\gamma}_{\nu}
( \partial_{\alpha} G \cdot G^{-1} \cdot
\partial_{\gamma} G )^{\alpha}_{\mu}
 + 2 \eta^{\mu \nu} G^{\gamma}_{\nu}
( \partial_{\gamma} \partial_{\alpha} G )^{\alpha}_{\mu}\\
&&\\
&& - \frac{3}{2} \eta_{\mu \nu} \eta^{\rho \sigma} \eta^{\mu_1 \nu_1}
G^{\gamma}_{\nu_1} G^{\alpha}_{\rho}
( \partial_{\alpha} G  \cdot G^{-1} ) ^{\mu}_{\mu_1}
( \partial_{\gamma} G  \cdot G^{-1} ) ^{\nu}_{\sigma} \\
&&\\
&& -  \eta_{\mu \nu} \eta^{\rho \sigma} \eta^{\mu_1 \nu_1}
G^{\gamma}_{\nu_1} G^{\alpha}_{\rho}
( \partial_{\gamma} G  \cdot G^{-1} ) ^{\mu}_{\mu_1}
( \partial_{\alpha} G  \cdot G^{-1} ) ^{\nu}_{\sigma} \\
&&\\
&& + \eta_{\mu \nu} \eta^{\mu_2 \nu_2} \eta^{\mu_1 \nu_1}
G^{\beta}_{\nu_2} G^{\alpha}_{\mu_1}
( \partial_{\beta} G  \cdot G^{-1} ) ^{\mu}_{\mu_2}
( \partial_{\alpha} G  \cdot G^{-1} ) ^{\nu}_{\nu_1 } \\
&&\\
&&- 2 g \eta^{\mu \nu} F^{\alpha}_{\mu_1 \nu_1}
G^{\beta}_{\nu} G^{-1 \mu_1}_{\alpha}
( \partial_{\beta} G  \cdot G^{-1} ) ^{\nu_1}_{\mu} \\
&&\\
&& + g \eta^{\lambda \mu} F^{\alpha}_{\mu \nu}
 ( \partial_{\alpha} G  \cdot G^{-1} ) ^{\nu}_{\lambda} \\
 &&\\
&& + g \eta^{\rho \mu} F^{\alpha}_{\rho \sigma}
G^{\gamma}_{\mu}  (G^{-1} \cdot
\partial_{\gamma} G  \cdot G^{-1} ) ^{\sigma}_{\alpha} \\
&&\\
&& + g^2 \eta^{\nu \nu_1} F^{\alpha}_{\mu \nu}
F^{\alpha_1}_{\mu_1 \nu_1} G^{-1 \mu}_{\alpha}
G ^{-1 \mu_1}_{\alpha_1} \\
&&\\
&& - \frac{g^2}{2} \eta^{\mu \rho} \eta^{\nu \sigma}
g_{\alpha \beta}  F^{\alpha}_{\mu \nu}
F^{\beta}_{\rho \sigma }  \\
&&\\
&&- \frac{g^2}{2} \eta^{\nu \nu_1 } F^{\alpha}_{\mu \nu}
F^{\alpha_1}_{\mu_1 \nu_1} G^{-1 \mu_1}_{\alpha}
G ^{-1 \mu}_{\alpha_1}
\end{array}
\label{4.1909}
\ee
Please note that the second last line in the above expression
is just the Lagrangian that we used in literature \cite{wu04}.
In order to set up the quantum gauge general relativity, we will not
select it to be the Lagrangian of the model.
\\

It can be strictly proved that, under gravitational gauge
transformations, these operators transforms as
\be
\Gamma^{\gamma}_{\alpha \beta} (x)  \to
\Gamma^{\prime \gamma}_{\alpha \beta} (x)
= \Lambda_{\alpha}^{~\alpha_1} \Lambda_{\beta}^{~\beta_1}
\Lambda^{\gamma}_{~\gamma_1}
( \ehat \Gamma^{\gamma_1}_{\alpha_1 \beta_1} (x))
- \Lambda_{\alpha}^{~\alpha_1} \Lambda_{\beta}^{~\beta_1}
(\partial_{\beta_1} \Lambda^{\gamma}_{~\alpha_1})
\label{4.1910}
\ee
\be
R_{\alpha \beta \gamma \delta}(x)  \to
R^{\prime}_{ \alpha \beta \gamma \delta}(x)
= \Lambda_{\alpha}^{~\alpha_1} \Lambda_{\beta}^{~\beta_1}
\Lambda_{\gamma}^{~\gamma_1} \Lambda_{\delta}^{~\delta_1}
 (\ehat R_{\alpha_1 \beta_1\gamma_1 \delta_1}(x)) ,
\label{4.1911}
\ee
\be
R_{\alpha \beta }(x)  \to
R'_{ \alpha \beta }(x)
= \Lambda_{\alpha}^{~\alpha_1} \Lambda_{\beta}^{~\beta_1}
( \ehat R_{\alpha_1 \beta_1}(x)),
\label{4.1912}
\ee
\be
R(x) \to R'(x) = ( \ehat R(x)).
\label{4.1913}
\ee
\\

Now, we need to determine the Lagrangian of the model. In order
to set up the model for quantum gauge general relativity, we need to
select $R$ in eq.(\ref{4.1909}) to be the Lagrangian. However,
its expression is too complicated to be used in further calculation.
So, we need to changed it into a simpler form. After rather lengthy
and complicated calculations, it is found that it can be written into
another form
\be
R = R_0 + \frac{1}{J(C)} \partial_{\beta}
\left \lbrack
2 J(C) g^{\alpha \beta} G^{-1 \mu}_{\alpha}
(\partial_{\gamma} G_{\mu}^{\gamma})
- 2 J(C) g^{\beta \delta} G^{-1 \mu}_{\alpha}
(\partial_{\delta} G_{\mu}^{\alpha})
\right \rbrack,
\label{4.1914}
\ee
where
\be
\ba{rcl}
R_0 & = & \frac{g^2}{4} \eta^{\mu \rho}
\eta^{\nu \sigma} g_{\alpha \beta}
F^{\alpha}_{\mu \nu} F^{\beta}_{\rho \sigma} \\
&&\\
&& + \frac{g^2}{2} \eta^{\mu \rho}
G^{-1 \nu}_{\beta} G^{-1 \sigma}_{\alpha}
F^{\alpha}_{\mu \nu} F^{\beta}_{\rho \sigma} \\
&&\\
&& - g^2 \eta^{\mu \rho}
G^{-1 \nu}_{\alpha} G^{-1 \sigma}_{\beta}
F^{\alpha}_{\mu \nu} F^{\beta}_{\rho \sigma},
\ea
\label{4.1915}
\ee
and
\be
J(C) = \sqrt{- {\rm det} g_{\alpha \beta} }.
\label{4.1916}
\ee
It is quite surprising that operator $R_0$ has very simple
and beautiful form. It has an obvious symmetric form
under gravitational gauge transformation. The second term
in eq.(\ref{4.1914}) is only a surface term in action, so
it has no contribution to the action, and if we select
$R$ or $R_0$ to be the Lagrangian of the system, they will
give out the same action and the same field equation for
gravitational gauge field. Because $R_0$ is much simpler
and more symmetric, we select it to be the Lagrangian
of the  quantum gauge general relativity
\be
\ba{rcl}
{\cal L}_0 & = &- \frac{1}{16 \pi G} R_0 = - \frac{1}{4 g^2} R_0\\
&&\\
& = & - \frac{1}{16} \eta^{\mu \rho}
\eta^{\nu \sigma} g_{\alpha \beta}
F^{\alpha}_{\mu \nu} F^{\beta}_{\rho \sigma} \\
&&\\
&& - \frac{1}{8} \eta^{\mu \rho}
G^{-1 \nu}_{\beta} G^{-1 \sigma}_{\alpha}
F^{\alpha}_{\mu \nu} F^{\beta}_{\rho \sigma} \\
&&\\
&& + \frac{1}{4} \eta^{\mu \rho}
G^{-1 \nu}_{\alpha} G^{-1 \sigma}_{\beta}
F^{\alpha}_{\mu \nu} F^{\beta}_{\rho \sigma}.
\ea
\label{4.20}
\ee
If we define
\be
C^{\mu\nu\rho\sigma}_{\alpha\beta}
= \frac{1}{4} \eta^{\mu \rho}
\eta^{\nu \sigma} g_{\alpha \beta}
+ \frac{1}{2} \eta^{\mu \rho}
G^{-1 \nu}_{\beta} G^{-1 \sigma}_{\alpha}
- \eta^{\mu \rho}
G^{-1 \nu}_{\alpha} G^{-1 \sigma}_{\beta},
\label{4.20a}
\ee
then the above lagrangian ${\cal L}_0$ can be written into
a simpler form:
\be
{\cal L}_0 =  - \frac{1}{4} C^{\mu\nu\rho\sigma}_{\alpha\beta}
F^{\alpha}_{\mu \nu} F^{\beta}_{\rho \sigma} .
\label{4.20b}
\ee
The first term in the Lagrangian ${\cal L}_0$ of (\ref{4.20}) is just
the Lagrangian in the previous work\cite{wu04}. In fact, every
term in the above Lagrangian has gravitational gauge symmetry.
So, the most general Lagrangian for quantum gauge theory of
gravity is
\be
\ba{rcl}
{\cal L}_0 & = & c_1 \eta^{\mu \rho}
\eta^{\nu \sigma} g_{\alpha \beta}
F^{\alpha}_{\mu \nu} F^{\beta}_{\rho \sigma} \\
&&\\
&& + c_2 \eta^{\mu \rho}
G^{-1 \nu}_{\beta} G^{-1 \sigma}_{\alpha}
F^{\alpha}_{\mu \nu} F^{\beta}_{\rho \sigma} \\
&&\\
&& + c_3 \eta^{\mu \rho}
G^{-1 \nu}_{\alpha} G^{-1 \sigma}_{\beta}
F^{\alpha}_{\mu \nu} F^{\beta}_{\rho \sigma}.
\ea
\label{4.2001}
\ee
Because the action given by this Lagrangian has strict
local gravitational gauge symmetry, the model based on
this Lagrangian is perturbatively renormalizable. The strict
formal proof on the renormalizability of the quantum gauge general
relativity is suitable to the model based on this most
general Lagrangian. For quantum gauge general relativity, the
parameters $c_1$, $c_2$ and $c_3$ are
\be
c_1 = - \frac{1}{16}, ~~~
c_2 = - \frac{1}{8}, ~~~
c_3 = \frac{1}{4}.
\label{4.2002}
\ee
For the model discussed in the literature \cite{wu04},
the parameters $c_1$, $c_2$ and $c_3$ are
\be
c_1 = - \frac{1}{4}, ~~~
c_2 = 0, ~~~
c_3 = 0.
\label{4.2003}
\ee
Different selection of the  parameters  $c_1$, $c_2$ and $c_3$
gives out different model for quantum gravity, and therefore
different dynamics of gravitational field. This situation
is quite special for gravitational interactions. We know that,
for ordinary $SU(N)$ gauge field theory, using gauge field
strength, we can only construct one gauge invariant Lagrangian
which is a quadratic form of field strength. In other words,
the Lagrangian for pure gauge field in ordinary $SU(N)$ gauge
field theory is unique. However, in quantum gauge theory of
gravity, we can construct three different gauge invariant
terms which are quadratic forms of field strength
of gravitational gauge field. Indeed, in many aspects,
quantum gauge theory of gravity is different from the
traditional gauge field theory. \\

Using relations (\ref{4.7083}), (\ref{4.709}) and (\ref{4.19}),
we can easily prove that the Lagrangian defined by eq.(\ref{4.20})
transforms covariantly  under gravitational gauge transformation
\be
{\cal L}_0 \to {\cal L}'_0 = ( \ehat {\cal L}_0).
\label{4.21}
\ee
In order to resume the gravitational gauge symmetry of the action,
we introduce an extremely important factor $J(C)$ which is
defined by eq.(\ref{4.1916}).
The gravitational gauge transformations of $g_{\alpha\beta}$
is given by eq.(\ref{4.709}). Then $J(C)$ transforms as
\be
J(C) \to J'(C') = J \cdot (\ehat J(C)),
\label{4.22}
\ee
where $J$ is the Jacobian of the transformation,
\be
J = det \left(\frac{\partial (x - \epsilon)^{\mu}}
{\partial x^{\nu}} \right).
\label{4.2301}
\ee
The Lagrangian for gravitational gauge field is selected as
\be
{\cal L} = J(C) {\cal L}_0 =  \sqrt{- {\rm det} g_{\alpha \beta} }
\cdot {\cal L}_0,
\label{4.24}
\ee
and the action for gravitational gauge field is
\be
S = \int {\rm d}^4 x {\cal L}.
\label{4.25}
\ee
It can be proved that this action has gravitational gauge symmetry.
In other words, it is invariant under gravitational gauge
transformation,
\be
S \to S' = S.
\label{4.26}
\ee
In order to prove the gravitational gauge symmetry of the action,
the following relation is important,
\be
\int {\rm d}^4 x J \left(\ehat f(x) \right)
= \int {\rm d}^4 x f(x),
\label{4.27}
\ee
where $f(x)$ is an arbitrary function of space-time coordinate.
\\

According to gauge principle, the global gauge symmetry will give out
conserved charges. Now, let's discuss the conserved charges of
global gravitational gauge transformation. Suppose that
$\epsilon^{\alpha}$ is an infinitesimal constant 4-vector. Then,
in the first order approximation, we have
\be
\ehat = 1 - \epsilon^{\alpha} \partial_{\alpha} + o(\epsilon^2).
\label{4.29}
\ee
The first order variation of the gravitational gauge field is
\be
\delta C_{\mu}^{\alpha} (x) =
- \epsilon^{\nu} \partial_{\nu} C_{\mu}^{\alpha},
\label{4.30}
\ee
and the first order variation of action is:
\be
\delta S = \int {\rm d}^4 x ~ \epsilon^{\alpha} \partial_{\mu}
T_{i \alpha}^{\mu},
\label{4.31}
\ee
where $T_{i \alpha}^{\mu}$ is the inertial energy-momentum
tensor, whose definition is
\be
T_{i \alpha}^{\mu} \equiv J(C)
\left( - \frac{\partial {\cal L}_0}
{\partial \partial_{\mu} C_{\nu}^{\beta}}
\partial_{\alpha} C_{\nu}^{\beta}
+ \delta^{\mu}_{\alpha} {\cal L}_0 \right).
\label{4.32}
\ee
Its explicit expression is
\be
\ba{rcl}
T_{i \alpha}^{\mu} &= J(C) \lbrack &
\frac{1}{4} \eta^{\lambda \rho}
\eta^{\nu \sigma} g_{\beta \gamma}
G^{\mu}_{\lambda} F_{\rho \sigma}^{\gamma}
(\partial_{\alpha} C_{\nu}^{\beta}) \\
&&\\
&&- \frac{1}{4} \eta^{\nu\rho} G^{-1 \sigma}_{\beta}
F_{\rho\sigma}^{\mu}
(\partial_{\alpha} C_{\nu}^{\beta}) \\
&&\\
&& + \frac{1}{4} \eta^{\lambda \rho} G^{-1 \nu}_{\gamma}
G^{-1 \sigma}_{\beta} G^{\mu}_{\lambda}
F_{\rho \sigma}^{\gamma}
(\partial_{\alpha} C_{\nu}^{\beta}) \\
&&\\
&& - \frac{1}{2} \eta^{\lambda \rho} G^{-1 \nu}_{\beta}
G^{-1 \sigma}_{\gamma} G^{\mu}_{\lambda}
F_{\rho \sigma}^{\gamma}
(\partial_{\alpha} C_{\nu}^{\beta}) \\
&&\\
&& + \frac{1}{2} \eta^{\nu\rho} \delta^{\mu}_{\beta}
G^{-1 \sigma}_{\gamma} F_{\rho \sigma}^{\gamma}
(\partial_{\alpha} C_{\nu}^{\beta}) \\
&&\\
&& - \frac{1}{16} \delta^{\mu}_{\alpha}
\eta^{\lambda \rho} \eta^{\nu\sigma}
g_{\beta\gamma}
F^{\gamma}_{\lambda\nu} F^{\beta}_{\rho\sigma} \\
&&\\
&& - \frac{1}{8} \delta^{\mu}_{\alpha}
\eta^{\lambda \rho} G^{-1 \nu}_{\beta}
G^{-1 \sigma}_{\gamma}
F^{\gamma}_{\lambda\nu} F^{\beta}_{\rho\sigma} \\
&&\\
&& + \frac{1}{4} \delta^{\mu}_{\alpha}
\eta^{\lambda \rho} G^{-1 \nu}_{\gamma}
G^{-1 \sigma}_{\beta} F^{\gamma}_{\lambda\nu} F^{\beta}_{\rho\sigma}
\rbrack .
\ea
\label{4.3201}
\ee
The global gravitational gauge symmetry of the system gives
out the conservation law of inertial energy-momentum tensor
\be
\partial_{\mu} T_{i \alpha}^{\mu} = 0.
\label{4.33}
\ee
Except for the factor $J(C)$, the form of the inertial
energy-momentum tensor is almost completely the same as that
in the traditional quantum field theory. It means that gravitational
interactions will change energy-momentum of matter fields, which
is what we expected in Einstein's general relativity.
\\

The Euler-Lagrange equation for gravitational gauge field is
\be
\partial_{\mu} \frac{\partial \cal L}
{\partial \partial_{\mu} C_{\nu}^{\alpha}}
= \frac{\partial \cal L}{\partial C_{\nu}^{\alpha}}.
\label{4.34}
\ee
This form is completely the same as what we have ever seen in
quantum field theory. But if we insert eq.(\ref{4.24}) into it,
we will get
\be
\partial_{\mu} \frac{\partial {\cal L }_0}
{\partial \partial_{\mu} C_{\nu}^{\alpha}}
= \frac{\partial {\cal L}_0}{\partial C_{\nu}^{\alpha}}
+g G_{\alpha}^{-1 \nu} {\cal L}_0
- g G_{\alpha}^{-1 \nu}
(  \partial_{\mu} C_{\nu}^{\alpha})
\frac{\partial {\cal L}_0}{\partial \partial_{\mu} C_{\nu}^{\alpha}}.
\label{4.35}
\ee
Eq.(\ref{4.17}) can be changed into
\be
F_{\mu \nu}^{\alpha} =
(D_{\mu} C_{\nu}^{\alpha})
- (D_{\nu} C_{\mu}^{\alpha}) ,
\label{4.36}
\ee
so the Lagrangian ${\cal L}_0$ depends on gravitational gauge fields
through its covariant derivative, $g_{\alpha \beta}$ and
$G^{-1 \mu}_{\alpha}$. Therefore, we have
\be
\ba{rcl}
\frac{\partial {\cal L}_0}{\partial C_{\nu}^{\alpha}}
&=& \frac{g}{4} \eta^{\nu\rho} \eta^{\lambda\sigma}
g_{\beta\gamma} F_{\rho \sigma}^{\beta}
(\partial_{\alpha} C_{\lambda}^{\gamma}) \\
&&\\
&& + \frac{g}{4} \eta^{\nu\rho}
G^{-1 \lambda}_{\beta} G^{-1 \sigma}_{\gamma}
F_{\rho \sigma}^{\beta}
(\partial_{\alpha} C_{\lambda}^{\gamma}) \\
&&\\
&& - \frac{g}{2} \eta^{\nu\rho}
G^{-1 \lambda}_{\gamma} G^{-1 \sigma}_{\beta}
F_{\rho \sigma}^{\beta}
(\partial_{\alpha} C_{\lambda}^{\gamma}) \\
&&\\
&& - \frac{g}{4} \eta^{\lambda\rho}
G^{-1 \nu}_{\beta} G^{-1 \sigma}_{\gamma}
F_{\rho \sigma}^{\beta}
(\partial_{\alpha} C_{\lambda}^{\gamma}) \\
&&\\
&& + \frac{g}{2} \eta^{\lambda\rho}
G^{-1 \nu}_{\gamma} G^{-1 \sigma}_{\beta}
F_{\rho \sigma}^{\beta}
(\partial_{\alpha} C_{\lambda}^{\gamma}) \\
&&\\
&& + \frac{g}{8} \eta^{\lambda\rho}
\eta^{\kappa\sigma} g_{\alpha\gamma}
G^{-1 \nu}_{\beta} F_{\rho \sigma}^{\beta}
F^{\gamma}_{\kappa\lambda} \\
&&\\
&&
- \frac{g}{4} \eta^{\kappa\rho}
G^{-1 \nu}_{\beta} G^{-1 \lambda}_{\alpha}
G^{-1 \sigma}_{\gamma}
F_{\rho \sigma}^{\beta}
F^{\gamma}_{\kappa\lambda} \\
&&\\
&&
+ \frac{g}{2} \eta^{\kappa\rho}
G^{-1 \nu}_{\gamma} G^{-1 \lambda}_{\alpha}
G^{-1 \sigma}_{\beta}
F_{\rho \sigma}^{\beta}
F^{\gamma}_{\kappa\lambda},
\ea
\label{4.37}
\ee
 and
\be
\ba{rcl}
\frac{\partial {\cal L}_0}{\partial \partial_{\mu} C_{\nu}^{\alpha}}
&=&- \frac{1}{4} \eta^{\lambda \rho}
\eta^{\nu \sigma} g_{\alpha \beta }
G^{\mu}_{\lambda} F_{\rho \sigma}^{\beta} \\
&&\\
&& + \frac{1}{4} \eta^{\nu\rho} G^{-1 \sigma}_{\alpha}
F_{\rho\sigma}^{\mu} \\
&&\\
&& - \frac{1}{4} \eta^{\lambda \rho} G^{-1 \nu}_{\beta}
G^{-1 \sigma}_{\alpha} G^{\mu}_{\lambda}
F_{\rho \sigma}^{\beta} \\
&&\\
&& + \frac{1}{2} \eta^{\lambda \rho} G^{-1 \nu}_{\alpha}
G^{-1 \sigma}_{\beta} G^{\mu}_{\lambda}
F_{\rho \sigma}^{\beta} \\
&&\\
&& - \frac{1}{2} \eta^{\nu\rho} \delta^{\mu}_{\alpha}
G^{-1 \sigma}_{\beta} F_{\rho \sigma}^{\beta}.
\ea
\label{4.38}
\ee
The above  field equation of gravitational gauge fields
are changed into
\be
\ba{rl}
\partial_{\mu} ( & \frac{1}{4} \eta^{\mu \rho}
\eta^{\nu \sigma} g_{\alpha \beta}
F_{\rho \sigma}^{\beta}
- \frac{1}{4} \eta^{\nu \rho} F^{\mu}_{\rho\alpha}
+ \frac{1}{4} \eta^{\mu \rho} F^{\nu}_{\rho\alpha} \\
&\\
& - \frac{1}{2} \eta^{\mu\rho} \delta^{\nu}_{\alpha}
F^{\beta}_{\rho\beta}
+ \frac{1}{2} \eta^{\nu\rho} \delta^{\mu}_{\alpha}
F^{\beta}_{\rho\beta})  = - g T_{g \alpha}^{\nu},
\label{4.41}
\ea
\ee
where
\be  \label{3.25}
\ba{rcl}
T_{g \alpha}^{\nu} & = &
 \frac{1}{4} \eta^{\nu\rho} \eta^{\lambda\sigma}
g_{\beta\gamma} F^{\beta}_{\rho\sigma}
(\partial_{\alpha} C_{\lambda}^{\gamma}) \\
&& + \frac{1}{4} \eta^{\nu\rho} G^{-1 \lambda}_{\beta}
G^{-1 \sigma}_{\gamma} F^{\beta}_{\rho\sigma}
(\partial_{\alpha} C_{\lambda}^{\gamma}) \\
&& - \frac{1}{2} \eta^{\nu\rho} G^{-1 \lambda}_{\gamma}
G^{-1 \sigma}_{\beta} F^{\beta}_{\rho\sigma}
(\partial_{\alpha} C_{\lambda}^{\gamma}) \\
&& - \frac{1}{4} \eta^{\lambda\rho} G^{-1 \nu}_{\beta}
G^{-1 \sigma}_{\gamma} F^{\beta}_{\rho\sigma}
(\partial_{\alpha} C_{\lambda}^{\gamma}) \\
&& + \frac{1}{2} \eta^{\lambda\rho} G^{-1 \nu}_{\gamma}
G^{-1 \sigma}_{\beta} F^{\beta}_{\rho\sigma}
(\partial_{\alpha} C_{\lambda}^{\gamma}) \\
&& + \frac{1}{4} \eta^{\lambda\rho} \eta^{\nu\sigma}
g_{\alpha\beta}
G^{-1 \kappa}_{\gamma} F^{\beta}_{\rho\sigma}
(D_{\lambda} C_{\kappa}^{\gamma}) \\
&& - \frac{1}{4} \eta^{\nu\rho} G^{-1 \sigma}_{\alpha}
G^{-1 \kappa}_{\gamma} F^{\mu}_{\rho\sigma}
(\partial_{\mu} C_{\kappa}^{\gamma}) \\
&& + \frac{1}{2} \eta^{\nu\rho} G^{-1 \sigma}_{\beta}
G^{-1 \kappa}_{\gamma} F^{\beta}_{\rho\sigma}
(\partial_{\alpha} C_{\kappa}^{\gamma}) \\
&& + \frac{1}{4} \eta^{\lambda\rho} G^{-1 \nu}_{\beta}
G^{-1 \sigma}_{\alpha}
G^{-1 \kappa}_{\gamma} F^{\beta}_{\rho\sigma}
(D_{\lambda} C_{\kappa}^{\gamma}) \\
&& - \frac{1}{2} \eta^{\lambda\rho} G^{-1 \nu}_{\alpha}
G^{-1 \sigma}_{\beta}
G^{-1 \kappa}_{\gamma} F^{\beta}_{\rho\sigma}
(D_{\lambda} C_{\kappa}^{\gamma}) \\
&& - \frac{1}{4} \eta^{\lambda\rho} \eta^{\nu\sigma}
\partial_{\mu} (g_{\alpha \beta} C^{\mu}_{\lambda}
 F^{\beta}_{\rho\sigma}) \\
&& - \frac{1}{4} \eta^{\nu\rho}
\partial_{\beta} (C^{\sigma}_{\lambda} G^{-1 \lambda}_{\alpha}
 F^{\beta}_{\rho\sigma}) \\
&&  + \frac{1}{2} \eta^{\nu\rho}
\partial_{\alpha} (C^{\sigma}_{\lambda} G^{-1 \lambda}_{\beta}
 F^{\beta}_{\rho\sigma}) \\
&& + \frac{1}{4g} \eta^{\lambda\rho}
\partial_{\mu} \lbrack ( G^{-1 \nu}_{\beta} G^{-1 \sigma}_{\alpha}
G^{\mu}_{\lambda} - \delta^{\nu}_{\beta} \delta^{\sigma}_{\alpha}
\delta^{\mu}_{\lambda}) F^{\beta}_{\rho\sigma}
\rbrack  \\
&& - \frac{1}{2 g} \eta^{\lambda\rho}
\partial_{\mu} \lbrack ( G^{-1 \nu}_{\alpha} G^{-1 \sigma}_{\beta}
G^{\mu}_{\lambda} - \delta^{\nu}_{\alpha} \delta^{\sigma}_{\beta}
\delta^{\mu}_{\lambda}) F^{\beta}_{\rho\sigma}
\rbrack  \\
&& - \frac{1}{4} \eta^{\kappa \rho} G^{-1 \nu}_{\beta}
G^{-1 \lambda}_{\alpha} G^{-1 \sigma}_{\gamma}
F^{\beta}_{\rho\sigma} F^{\gamma}_{\kappa\lambda}  \\
&& + \frac{1}{2} \eta^{\kappa \rho} G^{-1 \nu}_{\gamma}
G^{-1 \lambda}_{\alpha} G^{-1 \sigma}_{\beta}
F^{\beta}_{\rho\sigma} F^{\gamma}_{\kappa\lambda}  \\
&& - \frac{1}{8} \eta^{\mu \rho} \eta^{\lambda \sigma}
g_{\alpha \gamma} G^{-1 \nu}_{\beta}
F^{\beta}_{\rho\sigma} F^{\gamma}_{\mu\lambda}  \\
&& - \frac{1}{16} \eta^{\mu \rho} \eta^{\lambda \sigma}
g_{\beta \gamma} G^{-1 \nu}_{\alpha}
F^{\beta}_{\rho\sigma} F^{\gamma}_{ \mu\lambda}  \\
&& - \frac{1}{8} \eta^{\mu \rho} G^{-1 \nu}_{\alpha}
G^{-1 \lambda}_{\beta} G^{-1 \sigma}_{\gamma}
F^{\beta}_{\rho\sigma} F^{\gamma}_{\mu\lambda}  \\
&& + \frac{1}{4} \eta^{\mu \rho} G^{-1 \nu}_{\alpha}
G^{-1 \lambda}_{\gamma} G^{-1 \sigma}_{\beta}
F^{\beta}_{\rho\sigma} F^{\gamma}_{\mu\lambda}.
\ea
\ee
$ T_{g \alpha}^{\nu}$ is also a conserved current, that is
\be
\partial_{\nu} T_{g \alpha}^{\nu} = 0,
\label{4.42}
\ee
because of the following identity
\be
\ba{rl}
\partial_{\nu} \partial_{\mu} ( & \frac{1}{4} \eta^{\mu \rho}
\eta^{\nu \sigma} g_{\alpha \beta}
F_{\rho \sigma}^{\beta}
- \frac{1}{4} \eta^{\nu \rho} F^{\mu}_{\rho\alpha}
+ \frac{1}{4} \eta^{\mu \rho} F^{\nu}_{\rho\alpha} \\
&\\
& - \frac{1}{2} \eta^{\mu\rho} \delta^{\nu}_{\alpha}
F^{\beta}_{\rho\beta}
+ \frac{1}{2} \eta^{\nu\rho} \delta^{\mu}_{\alpha}
F^{\beta}_{\rho\beta})  = 0.
\label{4.43}
\ea
\ee
$ T_{g \alpha}^{\nu}$ is called gravitational energy-momentum
tensor, which is the source of gravitational gauge field. Now we get
two different energy-momentum tensors, one is the inertial energy-momentum
tensor $ T_{i \alpha}^{\nu}$ and another is the gravitational
energy-momentum tensor $ T_{g \alpha}^{\nu}$. They are similar,
but they are different. The inertial energy-momentum tensor
$ T_{i \alpha}^{\nu}$ is given by conservation law which is associate
with global gravitational gauge symmetry, it gives out an
energy-momentum 4-vector:
\be
P_{i \alpha} = \int {\rm d}^3 \svec{x} T_{i  \alpha}^{0}.
\label{4.44}
\ee
It is a conserved charges,
\be
\frac{\rm d}{{\rm d} t}P_{i \alpha} =  0.
\label{4.45}
\ee
The time component of $P_{i \alpha}$, that is
$P_{i 0}$,  gives out the Hamiltonian $H$ of the system,
\be
H = - P_{i~0} =  \int {\rm d}^3 \svec{x} J(C)
\left (\pi_{\alpha}^{\mu} \tdot{C} _{\mu}^{\alpha} -
{\cal L}_0 \right ).
\label{4.46}
\ee
According to our conventional belief, $H$ should be the inertial
energy of the system, therefore $P_{i \alpha}$ is the inertial
energy-momentum of the system. The gravitational energy-momentum
is given by the field equation of gravitational gauge field, it is
also a conserved current. The space integration of the time component
of it gives out a conserved energy-momentum 4-vector,
\be
P_{g \alpha} = \int {\rm d}^3 \svec{x} T_{g  \alpha}^{0}.
\label{4.47}
\ee
It is also a conserved charge,
\be
\frac{\rm d}{{\rm d} t}P_{g \alpha} =  0.
\label{4.48}
\ee
Its time component  just gives out the gravitational energy
of the system, which is the source of  gravitational gauge field.
This can be easily seen. For a static system, if
we Set $\nu$  and $\alpha$
in eq.(\ref{4.41}) to $0$, select harmonic gauge and make
the leading term approximation, we can get(details on this
deducing can be found in the chapter on classical limit
of quantum gauge general relativity)
\be
\nabla^2 C_0^0   = - g T_{g 0}^{0}.
\label{4.49}
\ee
Define
\be
E^i = - \partial_i C_0^0.
\label{4.50}
\ee
$E^i$ is just the field strength of gravitational gauge
field for a static system.
The space integration of eq.(\ref{4.49}) gives out
\be
\oint {\rm d}\svec{\sigma} \cdot \svec{E}
=   g \int {\rm d}^3 \svec{x} T_{g  0}^0.
\label{4.51}
\ee
According to Newton's classical theory of gravity,
$\int {\rm d}^3 \svec{x} T_{g  0}^0$ in the right hand term is
just the gravitational mass of the system. Denote the gravitational
mass of the system as $M_g$, that is
\be
M_g
= - \int {\rm d}^3 \svec{x} T_{g  0}^0.
\label{4.52}
\ee
Then eq(\ref{4.51}) is changed into
\be
\oint {\rm d}\svec{\sigma} \cdot \svec{E}
= - g M_g.
\label{4.53}
\ee
This is just the classical Newton's law of universal gravitation.
It can be strictly proved that gravitational mass is different from
inertial mass. They are not equivalent. But their difference is at least
first order infinitesimal quantity if the gravitational field
$ g C_{\mu}^{\alpha}$, for this difference is
proportional to $g C_{\mu}^{\alpha}$. So, this difference is
too small to be detected in experiments. But in the environment
of strong gravitational field, the difference will become relatively
larger and will be easier to be detected. Much more highly precise
measurement of this difference is strongly needed to test this
prediction and to test the validity of the equivalence principle.
In the chapter of classical limit of quantum gauge theory of
gravity, we will return to discuss this problem again. \\

As we have stated before, the model given by the lagrangian
(\ref{4.20}) is quantum gauge general relativity, so the field
equation of gravitational gauge field given by this Lagrangian
should be the Einstein's field equation. Indeed, this is true.
That is, the field equation (\ref{4.41}) is just the Einstein's
field equation. If we use periodic boundary conditions,
all surface terms in the action must vanish, that is
\be
\int {\rm d}^4 x ~J(C) \cdot \frac{1}{J(C)} \partial_{\beta}
\left \lbrack
2 J(C) g^{\alpha \beta} G^{-1 \mu}_{\alpha}
(\partial_{\gamma} G_{\mu}^{\gamma})
- 2 J(C) g^{\beta \delta} G^{-1 \mu}_{\alpha}
(\partial_{\delta} G_{\mu}^{\alpha})
\right \rbrack=0.
\label{4.5301}
\ee
Combine the above equation with the equation (\ref{4.1914}),
we have
\be
\int {\rm d}^4 x ~J(C) \cdot R_0 =\int {\rm d}^4 x ~J(C) \cdot R.
\label{4.5302}
\ee
So, the action of the model can be written into
\be
S = - \frac{1}{16 \pi G} \int {\rm d}^4 x ~
\sqrt{- {\rm det} g_{\alpha \beta} } \cdot R.
\label{4.5303}
\ee
Denote the action of matter field as $S_M$. If we consider
the gravitational interactions of matter fields, the total
action of the system should be
\be
S = - \frac{1}{16 \pi G} \int {\rm d}^4 x ~
\sqrt{- {\rm det} g_{\alpha \beta} } \cdot R + S_M.
\label{4.5304}
\ee
Make a variation of gravitational gauge field $C_{\mu}^{\alpha}$,
the variation of the operator $g_{\alpha \beta}$ is
\be
\delta g_{\alpha \beta} =
g (g_{\gamma \beta} G^{-1 \mu}_{\alpha}
+ g_{\gamma \alpha} G^{-1 \mu}_{\beta} )
\delta C_{\mu}^{\gamma},
\label{4.5305}
\ee
and the change in the action is
\be
\delta S = \frac{g}{16 \pi G} \int {\rm d}^4 x ~
\sqrt{- {\rm det} g_{\alpha_1 \beta_1} }
\left \lbrack
R^{\alpha \beta} - \frac{1}{2} g^{\alpha \beta} R
+ 8 \pi G T^{\alpha \beta}
\right \rbrack
(g_{\gamma \beta} G^{-1 \mu}_{\alpha}
+ g_{\gamma \alpha} G^{-1 \mu}_{\beta} )
\delta C_{\mu}^{\gamma},
\label{4.5306}
\ee
where $T^{\alpha \beta}$ is the symmetric energy-momentum
tensor which is defined by
\be
T^{\alpha \beta} = \frac{2}{\sqrt{- {\rm det} g_{\alpha_1 \beta_1} }}
\frac{\delta S_M}{\delta g_{\alpha \beta} }.
\label{4.5307}
\ee
The total action $S$ is stationary with respect to arbitrary
variations in $C_{\mu}^{\gamma}$ if and only if
\be
\left \lbrack
R^{\alpha_1 \beta_1} - \frac{1}{2} g^{\alpha_1 \beta_1} R
+ 8 \pi G T^{\alpha_1 \beta_1}
\right \rbrack
(g_{\gamma \beta_1} G^{-1 \mu}_{\alpha_1}
+ g_{\gamma \alpha_1} G^{-1 \mu}_{\beta_1} )
=0.
\label{4.5308}
\ee
Multiply both side of the above equation with $G_{\mu}^{\alpha}$,
we get
\be
\left \lbrack
R^{\alpha \beta_1} - \frac{1}{2} g^{\alpha \beta_1} R
+ 8 \pi G T^{\alpha \beta_1}
\right \rbrack
g_{\gamma \beta_1}
=0.
\label{4.5309}
\ee
Then multiply both side of the above equation with
$g^{\gamma \beta}$, we get
\be
R^{\alpha \beta} - \frac{1}{2} g^{\alpha \beta} R
+ 8 \pi G T^{\alpha \beta}
=0,
\label{4.5310}
\ee
which is just the Einstein's field equation. Now, from the
same action, the least action principle gives out two equations
(\ref{4.41}) and (\ref{4.5306}). They are different in forms,
but they are essentially the same, for one action can only give
out one field equation. So, in quantum gauge general relativity,
the field equation of gravitational gauge field is the Einstein's
field equation.
\\

Now, let's discuss self-coupling of gravitational
field. The Lagrangian of gravitational gauge field is given by
eq(\ref{4.24}). Because
\be
J(C) = 1 + \sum_{m=1}^{\infty} \frac{1}{m!}
\left( \sum_{n=1}^{\infty}
\frac{g^n}{n}{\rm tr} (C^n)
\right)^m
\label{4.54}
\ee
there are vertexes of  $n$ gravitational gauge fields in tree diagram
where $n$ can be arbitrary integer number that is greater than 3. This
property is important for renormalization of the theory. Because
the coupling constant of the gravitational gauge interactions has
negative mass dimension, any kind of regular vertex exists divergence.
In order to cancel these divergences, we need to introduce the
corresponding counterterms. Because of the existence of
the vertex of $n$ gravitational gauge fields in tree diagram in the
non-renormalized Lagrangian, we need not introduce any new
counterterm which does not exist in the non-renormalized
Lagrangian, what we need to do is to redefine gravitational coupling
constant $g$ and gravitational gauge field $C_{\mu}^{\alpha}$
in renormalization.
If there is no  $J(C)$ term in the original Lagrangian, then
we will have to introduce infinite counterterms in
renormalization, and therefore the theory is non-renormalizable.
Because of the existence of the factor $J(C)$, though
quantum gauge theory of gravity looks like a non-renormalizable
theory according to the power counting law, it is indeed
renormalizable. In a word, the factor $J(C)$ is highly important
for the quantum gauge general relativity. \\

\section{Gravitational Interactions of Scalar Fields}
\setcounter{equation}{0}

Now, let's start to discuss gravitational interactions of matter
fields. First, we discuss gravitational interactions of scalar fields.
For the sake of simplicity, we first discuss real scalar field. Suppose
that $\phi(x)$ is a real scalar field. The traditional Lagrangian for the
real scalar field is
\be
- \frac{1}{2} \eta^{\mu \nu} \partial_{\mu} \phi(x)
\partial_{\nu} \phi(x) - \frac{m^2}{2} \phi^2 (x),
\label{5.1}
\ee
where $m$ is the mass of scalar field.
This is the Lagrangian for a free real scalar field. Its Euler-Lagrangian
equation of motion  is
\be
( \eta^{\mu \nu} \partial_{\mu} \partial_{\nu} - m^2 ) \phi(x) =0,
\label{5.2}
\ee
which is the famous Klein-Gordan equation. \\

Now, replace the ordinary partial derivative $\partial_{\mu}$
with gauge covariant derivative $D_{\mu}$, and add into the
Lagrangian of pure gravitational gauge field, we get
\be
\ba{rcl}
{\cal L}_0 & = & -\frac{1}{2} \eta^{\mu \nu} (D_{\mu} \phi)( D_{\nu} \phi)
-\frac{m^2}{2} \phi^2 \\
&&\\
&&- \frac{1}{16} \eta^{\mu \rho}
\eta^{\nu \sigma} g_{\alpha \beta}
F^{\alpha}_{\mu \nu} F^{\beta}_{\rho \sigma} \\
&&\\
&& - \frac{1}{8} \eta^{\mu \rho}
G^{-1 \nu}_{\beta} G^{-1 \sigma}_{\alpha}
F^{\alpha}_{\mu \nu} F^{\beta}_{\rho \sigma} \\
&&\\
&& + \frac{1}{4} \eta^{\mu \rho}
G^{-1 \nu}_{\alpha} G^{-1 \sigma}_{\beta}
F^{\alpha}_{\mu \nu} F^{\beta}_{\rho \sigma}.
\ea
\label{5.3}
\ee
The full Lagrangian is selected to be
\be
{\cal L} = J(C) {\cal L}_0,
\label{5.4}
\ee
and the action $S$ is defined by
\be
S = \int {\rm d}^4 x ~ {\cal L}.
\label{5.5}
\ee
\\

Using our previous definitions of gauge covariant derivative $D_{\mu}$
and strength of gravitational gauge field $F_{\mu \nu}^{\alpha}$, we can
obtain an explicit form of Lagrangian ${\cal L}$,
\be
{\cal L} = {\cal L}_F  +  {\cal  L}_I,
\label{5.6}
\ee
with ${\cal L}_F$ the free Lagrangian and ${\cal L}_I$  the
interaction Lagrangian. Their explicit expressions are
\be
\ba{rcl}
{\cal L}_F &=&
- \frac{1}{2} \eta^{\mu \nu} \partial_{\mu} \phi(x)
\partial_{\nu} \phi(x) - \frac{m^2}{2} \phi^2 (x)\\
&&\\\
&& - \frac{1}{16} \eta^{\mu \rho} \eta^{\nu \sigma} \eta_{\alpha \beta }
F_{0 \mu \nu}^{\alpha} F_{0 \rho \sigma}^{\beta}
- \frac{1}{8} \eta^{\mu \rho}
F^{\alpha}_{0 \mu \beta} F^{\beta}_{0 \rho \alpha}
+ \frac{1}{4} \eta^{\mu \rho}
F^{\alpha}_{0 \mu \alpha} F^{\beta}_{0 \rho \beta}
\ea
\label{5.7}
\ee
\be
\begin{array}{rcl}
{\cal L}_I &=&
 ( J(C) - 1 ) \cdot
 ( - \frac{1}{2} \eta^{\mu \nu} \partial_{\mu} \phi(x)
\partial_{\nu} \phi(x) - \frac{m^2}{2} \phi^2 (x) ) \\
&&\\
&& + g J(C) \eta^{\mu \nu} C_{\mu}^{\alpha}
(\partial_{\alpha} \phi)(\partial_{\nu} \phi)
- \frac{g^2}{2} J(C) \eta^{\mu \nu} C_{\mu}^{\alpha}
C_{\nu}^{\beta} (\partial_{\alpha} \phi)(\partial_{\beta} \phi)  \\
&&\\
&&  + { self ~interaction~ terms ~of ~Gravitational~gauge~field},
\end{array}
\label{5.8}
\ee
where,
\be
F_{0 \mu \nu}^{\alpha} = \partial_{\mu} C_{\nu}^{\alpha}
- \partial_{\nu}  C_{\mu}^{\alpha}.
\label{5.9}
\ee
From eq.(\ref{5.8}), we can see that scalar field can directly couples
to any number of gravitational gauge fields. This is one of the most
important interaction properties of gravity. Other kinds of interactions,
such as strong interactions, weak interactions and electromagnetic
interactions do not have this kind of interaction properties. Because
the gravitational coupling constant has negative mass dimension,
renormalization of theory needs this kind of interaction properties.
In other words, if matter field can not directly couple to any
number of gravitational gauge fields, the theory will be
non-renormalizable. \\

The symmetries of the theory can be easily seen from eq.(\ref{5.3}).
First, let's discuss Lorentz symmetry. In eq.(\ref{5.3}), some indexes
are Lorentz indexes and some are group indexes. Lorentz indexes
and group indexes have different transformation law under
gravitational gauge transformation, but they have the same
transformation law under Lorentz transformation. Therefor,
it can be easily seen that both ${\cal L}_0$ and $J(C)$ are
Lorentz scalars, the Lagrangian ${\cal  L}$ and action $S$
are invariant under global Lorentz transformation. \\

Under gravitational gauge transformations, real scalar field
$\phi (x)$ transforms as
\be
\phi (x) \to \phi '(x) = ( \ehat \phi (x) ),
\label{5.10}
\ee
therefore,
\be
D_{\mu} \phi (x) \to D'_{\mu} \phi '(x) = ( \ehat D_{\mu} \phi (x) ).
\label{5.11}
\ee
Using above relations and relations (\ref{4.7083}),
(\ref{4.709}) and (\ref{4.19})
we can easily prove that ${\cal L}_0$ transforms covariantly
\be
{\cal L}_0 \to {\cal L}'_0  = (\ehat {\cal L}_0),
\label{5.12}
\ee
and the action eq.(\ref{5.5}) of the system is invariant,
\be
S \to S' =  S.
\label{5.13}
\ee
Please remember that eq.(\ref{4.27}) and eq.(\ref{4.22})
are an important relations to be used
in the proof of the gravitational gauge symmetry of the action. \\

Global gravitational gauge symmetry gives out conserved charges.
Suppose that $\ehat$ is an infinitesimal gravitational gauge
transformation, it will have the form of eq.(\ref{4.29}). The first order
variations of fields are
\be
\delta C_{\mu}^{\alpha} (x) =
- \epsilon^{\nu} (\partial_{\nu} C_{\mu}^{\alpha} (x)),
\label{5.14}
\ee
\be
\delta \phi (x) =
- \epsilon^{\nu} (\partial_{\nu} \phi (x)),
\label{5.15}
\ee
Using Euler-Lagrange equation of motions for scalar fields and
field equation for gravitational gauge fields, we can obtain that
\be
\delta S = \int {\rm d}^4 x
\epsilon^{\alpha} \partial_{\mu} T_{i \alpha}^{\mu},
\label{5.16}
\ee
where
\be
T_{i \alpha}^{\mu} \equiv J(C)
\left( - \frac{\partial {\cal L}_0}
{\partial \partial_{\mu} \phi} \partial_{\alpha} \phi
- \frac{\partial {\cal L}_0}{\partial \partial_{\mu} C_{\nu}^{\beta}}
\partial_{\alpha} C_{\nu}^{\beta}
+ \delta^{\mu}_{\alpha} {\cal L}_0 \right).
\label{5.17}
\ee
Because action is invariant under global gravitational gauge
transformation,
\be
\delta S = 0,
\label{5.18}
\ee
and $\epsilon^{\alpha}$ is an  arbitrary  infinitesimal
constant 4-vector,   we obtain,
\be
\partial_{\mu} T_{i \alpha}^{\mu} = 0.
\label{5.19}
\ee
This is the conservation equation for inertial energy-momentum
tensor. $T_{i \alpha}^{\mu}$ is the conserved current which
corresponds to the global gravitational gauge symmetry.
The space integration of the time component of inertial
energy-momentum tensor gives out the conserved charge,
which is just the inertial energy-momentum of the system.
The time component of the conserved charge is the Hamilton of
the system, which is
\be
H = - P_{i~0} =  \int {\rm d}^3 \svec{x} J(C)
( \pi_{\phi} \tdot{\phi} +
\pi_{\alpha}^{\mu} \tdot{C} _{\mu}^{\alpha} - {\cal L}_0),
\label{5.20}
\ee
where $\pi_{\phi}$ and $\pi_{\alpha}^{\mu}$ are  canonical conjugate
momenta of the real scalar field and gravitational field
\be
\pi_{\phi} =  \frac{\partial {\cal L}_0 }{\partial \tdot{\phi} } ,
\label{5.2001}
\ee
\be
\pi_{\alpha}^{\mu} =
\frac{\partial {\cal L}_0 }{\partial \tdot{C_{\mu}^{\alpha}} } .
\label{5.2002}
\ee
The inertial space momentum of the system is given by
\be
P^i  = P_{i ~i} =  \int {\rm d}^3 \svec{x} J(C)
( - \pi_{\phi} \partial_i {\phi} -
\pi_{\alpha}^{\mu} \partial_i {C} _{\mu}^{\alpha} ).
\label{5.21}
\ee
According to gauge principle, after quantization, they will become
generators of quantum gravitational gauge transformation. \\

Using the difinion (\ref{4.706}), we can change the
Lagrangian given by eq.(\ref{5.3}) into
\be
{\cal L}_0 = -\frac{1}{2} g^{\alpha \beta}
( \partial_{\alpha} \phi)
(\partial_{\beta} \phi ) -  \frac{m^2}{2} \phi^2
- \frac{1}{16\pi G} R_0.
\label{5.23}
\ee
$g^{\alpha \beta}$ is the metric tensor of curved group space-time.
We can easily see that, when there is no gravitational field
in space-time, that is,
\be
C_{\mu}^{\alpha} = 0,
\label{5.25}
\ee
the group space-time will be flat
\be
g^{ \alpha \beta} = \eta^{\alpha \beta}.
\label{5.26}
\ee
This is what we expected in Einstein's general relativity.   \\

Euler-Lagrange equations of motion can be easily deduced from
action principle. Keep gravitational gauge field $C_{\mu}^{\alpha}$
fixed and let real scalar field vary infinitesimally, then the first order
infinitesimal variation of action is
\be
\delta S =
\int {\rm d}^4 x  J(C)
\left( \frac{\partial {\cal L}_0}{\partial \phi}
- \partial_{\mu} \frac{\partial {\cal L}_0}{\partial \partial_{\mu} \phi}
- g G^{-1 \nu}_{\alpha} (\partial_{\mu} C_{\nu}^{\alpha} )
\frac{\partial {\cal L}_0}
{\partial \partial_{\mu} \phi} \right) \delta \phi.
\label{5.27}
\ee
Because $\delta \phi$ is an arbitrary variation of scalar field, according
to action principle, we get
\be
\frac{\partial {\cal L}_0}{\partial \phi}
- \partial_{\mu} \frac{\partial {\cal L}_0}{\partial \partial_{\mu} \phi}
- g G^{-1 \nu}_{\alpha} (\partial_{\mu} C_{\nu}^{\alpha} )
\frac{\partial {\cal L}_0}{\partial \partial_{\mu} \phi} = 0.
\label{5.28}
\ee
 Because of the existence of the factor $J(C)$,
the equation of motion for scalar field is quite different from the
traditional form in quantum field theory. But the difference is
a second order infinitesimal quantity if we suppose that both gravitational
coupling constant and gravitational gauge field are first order
infinitesimal quantities. Because
\be
\frac{\partial {\cal L}_0}{\partial \partial_{\alpha} \phi}
= - g^{\alpha \beta} \partial_{\beta} \phi,
\label{5.29}
\ee
\be
\frac{\partial {\cal L}_0}{\partial \phi}
= - m^2 \phi,
\label{5.30}
\ee
the explicit form of the equation of motion of scalar field is
\be
g^{\alpha \beta} \partial_{\alpha} \partial_{\beta} \phi
- m^2 \phi + (\partial_{\alpha} g^{\alpha \beta})
\partial_{\beta} \phi
+ g g^{\alpha \beta} (\partial_{\beta} \phi )
G^{-1 \nu}_{\gamma} (\partial_{\alpha} C_{\nu}^{\gamma} )
= 0.
\label{5.31}
\ee
The field equation for gravitational gauge field is:
\be
\ba{rl}
\partial_{\mu} ( & \frac{1}{4} \eta^{\mu \rho}
\eta^{\nu \sigma} g_{\alpha \beta}
F_{\rho \sigma}^{\beta}
- \frac{1}{4} \eta^{\nu \rho} F^{\mu}_{\rho\alpha}
+ \frac{1}{4} \eta^{\mu \rho} F^{\nu}_{\rho\alpha} \\
&\\
& - \frac{1}{2} \eta^{\mu\rho} \delta^{\nu}_{\alpha}
F^{\beta}_{\rho\beta}
+ \frac{1}{2} \eta^{\nu\rho} \delta^{\mu}_{\alpha}
F^{\beta}_{\rho\beta})  = - g T_{g \alpha}^{\nu},
\label{5.32}
\ea
\ee
where $ T_{g \alpha}^{\nu}$ is the gravitational energy-momentum
tensor, whose definition is:
\be
\begin{array}{rcl}
T_{g \alpha}^{\nu}&=&
 - \frac{\partial {\cal L}_0}{\partial D_{\nu} C_{\mu}^{\beta}}
\partial_{\alpha} C_{\mu}^{\beta}
- \frac{\partial {\cal L}_0}{\partial D_{\nu} \phi}
\partial_{\alpha} \phi
+ G_{\alpha}^{-1 \nu} {\cal L}_0
- G^{-1 \lambda}_{\beta} (\partial_{\mu} C_{\lambda}^{\beta} )
\frac{\partial {\cal L}_0}
{\partial \partial_{\mu} C_{\nu}^{\alpha}}\\
&&\\
&&- \frac{1}{4} \eta^{\lambda\rho} \eta^{\nu\sigma}
\partial_{\mu} (g_{\alpha \beta} C^{\mu}_{\lambda}
 F^{\beta}_{\rho\sigma}) \\
 &&\\
&& - \frac{1}{4} \eta^{\nu\rho}
\partial_{\beta} (C^{\sigma}_{\lambda} G^{-1 \lambda}_{\alpha}
 F^{\beta}_{\rho\sigma}) \\
 &&\\
&&  + \frac{1}{2} \eta^{\nu\rho}
\partial_{\alpha} (C^{\sigma}_{\lambda} G^{-1 \lambda}_{\beta}
 F^{\beta}_{\rho\sigma}) \\
 &&\\
&& + \frac{1}{4g} \eta^{\lambda\rho}
\partial_{\mu} \lbrack ( G^{-1 \nu}_{\beta} G^{-1 \sigma}_{\alpha}
G^{\mu}_{\lambda} - \delta^{\nu}_{\beta} \delta^{\sigma}_{\alpha}
\delta^{\mu}_{\lambda}) F^{\beta}_{\rho\sigma}
\rbrack  \\
&&\\
&& - \frac{1}{2 g} \eta^{\lambda\rho}
\partial_{\mu} \lbrack ( G^{-1 \nu}_{\alpha} G^{-1 \sigma}_{\beta}
G^{\mu}_{\lambda} - \delta^{\nu}_{\alpha} \delta^{\sigma}_{\beta}
\delta^{\mu}_{\lambda}) F^{\beta}_{\rho\sigma}
\rbrack  \\
&&\\
&& - \frac{1}{4} \eta^{\kappa \rho} G^{-1 \nu}_{\beta}
G^{-1 \lambda}_{\alpha} G^{-1 \sigma}_{\gamma}
F^{\beta}_{\rho\sigma} F^{\gamma}_{\kappa\lambda}  \\
&&\\
&& + \frac{1}{2} \eta^{\kappa \rho} G^{-1 \nu}_{\gamma}
G^{-1 \lambda}_{\alpha} G^{-1 \sigma}_{\beta}
F^{\beta}_{\rho\sigma} F^{\gamma}_{\kappa\lambda}  \\
&&\\
&& - \frac{1}{8} \eta^{\mu \rho} \eta^{\lambda \sigma}
g_{\alpha \gamma} G^{-1 \nu}_{\beta}
F^{\beta}_{\rho\sigma} F^{\gamma}_{\mu\lambda}.
\end{array}
\label{5.33}
\ee
We can see again that, for matter field, its inertial energy-momentum
tensor is also different from the gravitational energy-momentum
tensor, this difference completely originate from the influences
of gravitational gauge field. Compare eq.(\ref{5.33}) with eq.(\ref{5.17}),
and set gravitational gauge field to zero, that is
\be
D_{\mu} \phi = \partial_{\mu} \phi ,
\label{5.34}
\ee
\be
J(C) = 1,
\label{5.35}
\ee
then we find that two energy-momentum tensors are completely
the same:
\be
T_{i \alpha}^{\mu } = T_{g \alpha}^{\mu }.
\label{5.36}
\ee
It means that the equivalence principle only strictly hold in a space-time
where there is no gravitational field. In the environment of strong
gravitational field, such as in black hole, the equivalence principle
will be strongly violated. \\

Field equation of gravitational gauge field can be written into
another form. Starting from eq. (\ref{5.23}) and making a
variation of gravitational gauge field, we get
\be
\delta S = \frac{g}{16 \pi G} \int {\rm d}^4 x ~
J(C)
\left \lbrack
R^{\alpha \beta} - \frac{1}{2} g^{\alpha \beta} R
+ 8 \pi G T^{\alpha \beta}
\right \rbrack
(g_{\gamma \beta} G^{-1 \mu}_{\alpha}
+ g_{\gamma \alpha} G^{-1 \mu}_{\beta} )
\delta C_{\mu}^{\gamma},
\label{5.3601}
\ee
where $T^{\alpha\beta}$ is the energy-momentum tensor
of the scalar field
\be
T^{\alpha\beta} = g^{\alpha \alpha_1} g^{\beta \beta_1}
( \partial_{\alpha_1} \phi  )
(\partial_{\beta_1} \phi) + g^{\alpha \beta}
\left\lbrack  -\frac{1}{2} g^{\alpha_1 \beta_1}
( \partial_{\alpha_1} \phi)
(\partial_{\beta_1} \phi ) -  \frac{m^2}{2} \phi^2
\right\rbrack.
\label{5.3602}
\ee
Then action principle gives out the Einstein's field equation
\be
R^{\alpha \beta} - \frac{1}{2} g^{\alpha \beta} R
+ 8 \pi G T^{\alpha \beta} =0.
\label{5.3603}
\ee
The Einstein's field equation is equivalent to the field
equation (\ref{5.32}). \\

Define
\be
{\bf L} = \int {\rm d}^3 \svec{x} {\cal L}
= \int {\rm d}^3 \svec{x} J(C) {\cal L}_0.
\label{5.37}
\ee
Then, we can easily prove that
\be
\frac{\delta {\bf L} }{\delta \phi} =
J(C) \left( \frac{\partial {\cal L}_0 }{\partial \phi}
- \partial_i \frac{\partial {\cal L}_0}{\partial \partial_i \phi}
-g G^{-1 \mu}_{\alpha} (\partial_i C_{\mu}^{\alpha} )
\frac{\partial {\cal L}_0}{\partial \partial_i \phi} \right),
\label{5.38}
\ee
\be
\frac{\delta {\bf L} }{\delta \tdot{\phi} } =
J(C) \frac{\partial {\cal L}_0 }{\partial \tdot{\phi} } ,
\label{5.39}
\ee
\be
\frac{\delta {\bf L} }{\delta C_{\nu}^{\alpha}} =
J(C) \left( \frac{\partial {\cal L}_0 }{\partial C_{\nu}^{\alpha}}
- \partial_i \frac{\partial
{\cal L}_0}{\partial \partial_i C_{\nu}^{\alpha}}
+ g G_{\alpha}^{-1 \nu} {\cal L}_0
-g G^{-1 \mu}_{\beta} (\partial_i C_{\mu}^{\beta} )
\frac{\partial {\cal L}_0}{\partial \partial_i C_{\nu}^{\alpha} } \right),
\label{5.40}
\ee
\be
\frac{\delta {\bf L} }{\delta \tdot{ C_{\nu}^{\alpha}} } =
J(C) \frac{\partial {\cal L}_0 }{\partial \tdot{ C_{\nu}^{\alpha}} } .
\label{5.41}
\ee
Then, Hamilton's action principle gives out the following equations
of motion:
 \be
\frac{\delta {\bf L} }{\delta \phi}
- \frac{\rm d}{{\rm d}t}
\frac{\delta {\bf L} }{\delta \tdot{\phi}} = 0 ,
\label{5.42}
\ee
\be
\frac{\delta {\bf L} }{\delta C_{\nu}^{\alpha} }
- \frac{\rm d}{{\rm d}t}
\frac{\delta {\bf L} }{\delta \tdot{ C_{\nu}^{\alpha} }} = 0 .
\label{5.43}
\ee
These two equations of motion are essentially the same as the
Euler-Lagrange equations of motion which we have obtained before.
But these two equations have more beautiful forms. \\

The Hamiltonian of the system is given by a Legendre transformation,
\be
\begin{array}{rcl}
H  &= & \int {\rm d}^3 \svec{x}
\left (\frac{\delta {\bf L} }{\delta \tdot{ \phi} }  \tdot{\phi}
+ \frac{\delta {\bf L} }{\delta \tdot{ C_{\mu}^{\alpha}} }
\tdot{ C_{\mu}^{\alpha}} \right ) - {\bf L}  \\
&=&  \int {\rm d}^3 \svec{x} J(C)
\left ( \pi_{\phi} \tdot{\phi} +
\pi_{\alpha}^{\mu} \tdot{C} _{\mu}^{\alpha} - {\cal L}_0 \right ),
\end{array}
\label{5.44}
\ee
where $\pi_{\phi}$ and $\pi_{\alpha}^{\mu}$ are
canonical conjugate momenta whose definitions are
given by (\ref{5.2001}) and (\ref{5.2002}).
It can be easily seen that the Hamiltonian given by Legendre
transformation is completely the same as that given by
inertial energy-momentum tensor. After Legendre transformation,
$\phi$, $C_{\mu}^{\alpha}$, $J(C) \pi_{\phi}$ and
$J(C) \pi_{\alpha}^{\mu}$ are canonical independent
variables. Let these variables vary infinitesimally, we can get
\be
\frac{\delta H }{\delta \phi}
= - \frac{\delta {\bf L} }{\delta \phi} ,
\label{5.47}
\ee
\be
\frac{\delta H }{\delta ( J(C) \pi_{\phi})}
= \tdot{\phi} ,
\label{5.48}
\ee
\be
\frac{\delta H }{\delta C_{\nu}^{\alpha}}
= - \frac{\delta {\bf L} }{\delta C_{\nu}^{\alpha}} ,
\label{5.49}
\ee
\be
\frac{\delta H }{\delta ( J(C) \pi_{\alpha}^{\nu})}
= \tdot{ C_{\nu}^{\alpha}}.
\label{5.50}
\ee
Then, Hamilton's equations of motion read:
\be
\frac{\rm d}{{\rm d}t} \phi =
\frac{\delta H }{\delta ( J(C) \pi_{\phi})} ,
\label{5.51}
\ee
\be
\frac{\rm d}{{\rm d}t} (J(C) \pi_{\phi} ) =
- \frac{\delta H }{\delta \phi} ,
\label{5.52}
\ee
\be
\frac{\rm d}{{\rm d}t} C_{\nu}^{\alpha} =
\frac{\delta H }{\delta ( J(C) \pi_{\alpha}^{\nu})} ,
\label{5.53}
\ee
\be
\frac{\rm d}{{\rm d}t} (J(C) \pi_{\alpha}^{\nu} ) =
- \frac{\delta H }{\delta C_{\nu}^{\alpha}} .
\label{5.54}
\ee
The  forms of the Hamilton's equations of motion are completely
the same as those appears in usual quantum field theory and usual
classical analytical mechanics. Therefor, the introduction of the
factor $J(C)$ does not affect the forms of Lagrange equations
of motion and Hamilton's equations of motion. \\

The Poisson brackets of two general functional of canonical arguments
can be defined by
\be
\begin{array}{rcl}
\lbrace  A ~~,~~ B \rbrace  &=& \int {\rm d}^3 \svec{x}
\left
( \frac{\delta A}{\delta \phi} \frac{\delta B}{\delta (J(C) \pi_{\phi})}
- \frac{\delta A}{\delta (J(C) \pi_{\phi})}
\frac{\delta B}{\delta \phi} \right . \\
&&\\
&& \left . +\frac{\delta A}{\delta C_{\nu}^{\alpha}}
\frac{\delta B}{\delta (J(C) \pi_{\alpha}^{\nu})}
- \frac{\delta A}{\delta (J(C) \pi_{\alpha}^{\nu})}
\frac{\delta B}{\delta C_{\nu}^{\alpha}} \right).
\end{array}
\label{5.55}
\ee
According to this definition, we have
\be
\lbrace  \phi(\svec{x},t) ~~,
~~ (J(C) \pi_{\phi})(\svec{y},t) \rbrace
= \delta^3(\svec{x} - \svec{y}),
\label{5.56}
\ee
\be
\lbrace  C_{\nu}^{\alpha}(\svec{x},t) ~~,
~~ (J(C) \pi_{\beta}^{\mu})(\svec{y},t) \rbrace
= \delta_{\nu}^{\mu} \delta^{\alpha}_{\beta}
\delta^3(\svec{x} - \svec{y}).
\label{5.57}
\ee
These two relations can be used as the starting point of
canonical quantization of quantum gravity.\\

Using Poisson brackets, the Hamilton's equations of motion can
be expressed in other forms,
\be
\frac{\rm d}{{\rm d}t} \phi (\svec{x},t) =
\lbrace  \phi(\svec{x},t) ~~,~~ H \rbrace  ,
\label{5.58}
\ee
\be
\frac{\rm d}{{\rm d}t} (J(C) \pi_{\phi}) (\svec{x},t) =
\lbrace  (J(C) \pi_{\phi}) (\svec{x},t) ~~,~~ H \rbrace  ,
\label{5.59}
\ee
\be
\frac{\rm d}{{\rm d}t} C_{\nu}^{\alpha} (\svec{x},t) =
\lbrace C_{\nu}^{\alpha}(\svec{x},t) ~~,~~ H \rbrace  ,
\label{5.60}
\ee
\be
\frac{\rm d}{{\rm d}t} (J(C) \pi_{\alpha}^{\nu}) (\svec{x},t) =
\lbrace  (J(C) \pi_{\alpha}^{\nu} ) (\svec{x},t) ~~,~~ H \rbrace  .
\label{5.61}
\ee
Therefore, if $A$ is an arbitrary functional of the canonical arguments
$\phi$, $C_{\mu}^{\alpha}$, $J(C) \pi_{\phi}$ and
$J(C) \pi_{\alpha}^{\mu}$, then we have
\be
\tdot{A} =
\lbrace  A ~~,~~ H \rbrace  .
\label{5.62}
\ee
After quantization, this equation will become the Heisenberg equation.
\\

If $\phi(x)$ is a complex scalar field, its traditional Lagrangian is
\be
-  \eta^{\mu \nu} \partial_{\mu} \phi(x)
\partial_{\nu} \phi^{*}(x) - m^2 \phi(x) \phi^{*}(x).
\label{5.63}
\ee
Replace ordinary partial derivative with gauge covariant derivative,
and add into the Lagrangian for pure gravitational gauge field,
we get,
\be
\ba{rcl}
{\cal L}_0 &=& - \eta^{\mu \nu}( D_{\mu} \phi)  (D_{\nu} \phi)^*
- m^2\phi \phi^{*}  \\
&&\\
&& - \frac{1}{16} \eta^{\mu \rho}
\eta^{\nu \sigma} g_{\alpha \beta}
F^{\alpha}_{\mu \nu} F^{\beta}_{\rho \sigma} \\
&&\\
&& - \frac{1}{8} \eta^{\mu \rho}
G^{-1 \nu}_{\beta} G^{-1 \sigma}_{\alpha}
F^{\alpha}_{\mu \nu} F^{\beta}_{\rho \sigma} \\
&&\\
&& + \frac{1}{4} \eta^{\mu \rho}
G^{-1 \nu}_{\alpha} G^{-1 \sigma}_{\beta}
F^{\alpha}_{\mu \nu} F^{\beta}_{\rho \sigma}.
\ea
\label{5.64}
\ee
Repeating all above discussions, we can get the whole theory
for gravitational interactions of complex scalar fields. We will not
repeat this discussion here.\\

\section{Gravitational Interactions of Dirac Field}
\setcounter{equation}{0}

In the usual quantum field theory, the Lagrangian for Dirac field is
\be
- \bar{\psi} (\gamma^{\mu} \partial_{\mu} + m) \psi.
\label{6.1}
\ee
Replace ordinary partial derivative with gauge covariant derivative,
and add into the Lagrangian of pure gravitational gauge field,
we get,
\be
\ba{rcl}
{\cal L}_0 &=&
- \bar{\psi} (\gamma^{\mu} D_{\mu} + m) \psi\\
&&\\
&&  - \frac{1}{16} \eta^{\mu \rho}
\eta^{\nu \sigma} g_{\alpha \beta}
F^{\alpha}_{\mu \nu} F^{\beta}_{\rho \sigma} \\
&&\\
&& - \frac{1}{8} \eta^{\mu \rho}
G^{-1 \nu}_{\beta} G^{-1 \sigma}_{\alpha}
F^{\alpha}_{\mu \nu} F^{\beta}_{\rho \sigma} \\
&&\\
&& + \frac{1}{4} \eta^{\mu \rho}
G^{-1 \nu}_{\alpha} G^{-1 \sigma}_{\beta}
F^{\alpha}_{\mu \nu} F^{\beta}_{\rho \sigma}.
\ea
\label{6.2}
\ee
The full Lagrangian of the system is
\be
{\cal L} = J(C) {\cal L}_0 ,
\label{6.3}
\ee
and the corresponding action is
\be
S =  \int {\rm d}^4 x {\cal L}
=\int {\rm d}^4 x  ~~ J(C) {\cal L}_0 .
\label{6.4}
\ee
This Lagrangian can be separated into two parts,
\be
{\cal L} = {\cal L}_F + {\cal L}_I ,
\label{6.5}
\ee
with ${\cal L}_F$ the free Lagrangian and ${\cal L}_I$ the interaction
Lagrangian. Their explicit forms are
\be
\ba{rcl}
{\cal L}_F &=&
- \bar{\psi} (\gamma^{\mu} \partial_{\mu} + m) \psi \\
&&\\
&& - \frac{1}{16} \eta^{\mu \rho} \eta^{\nu \sigma} \eta_{\alpha \beta }
F_{0 \mu \nu}^{\alpha} F_{0 \rho \sigma}^{\beta}
- \frac{1}{8} \eta^{\mu \rho}
F^{\alpha}_{0 \mu \beta} F^{\beta}_{0 \rho \alpha}
+ \frac{1}{4} \eta^{\mu \rho}
F^{\alpha}_{0 \mu \alpha} F^{\beta}_{0 \rho \beta},
\ea
\label{6.6}
\ee
\be
\begin{array}{rcl}
{\cal L}_I &=&
 (J(C) - 1 ) \cdot \left\lbrack
- \bar{\psi} (\gamma^{\mu} \partial_{\mu} + m) \psi
\right\rbrack
+  g J(C) \bar{\psi} \gamma^{\mu} (\partial_{\alpha}
\psi) C_{\mu}^{\alpha}  \\
&&  \\
&&  + { self ~interaction~ terms ~of ~Gravitational~gauge~field}.
\end{array}
\label{6.7}
\ee
From ${\cal L}_I$, we can see that Dirac field can directly
couple to any number of gravitational gauge fields, the mass
term of Dirac field also take part in gravitational interactions. All these
interactions are completely determined by the requirement of
gravitational gauge symmetry. The Lagrangian function before
renormalization almost contains all kind of divergent vertex,
which is important in the renormalization of the theory.
Besides, from eq.(\ref{6.7}), we can directly write out Feynman
rules of the corresponding interaction vertexes.
\\

Because the traditional Lagrangian function eq.(\ref{6.1}) is invariant
under global Lorentz transformation, which is already proved in the
traditional quantum field theory, and the covariant derivative
has the same behavior as partial derivative under global Lorentz
transformation, the first two terms of Lagrangian ${\cal L}$ are
global Lorentz invariant. We have already prove that the Lagrangian
function for pure gravitational gauge field is invariant under
global Lorentz transformation. Therefor, ${\cal L}$ has global
Lorentz symmetry.  \\

The gravitational gauge transformation of Dirac field is
\be
\psi(x) \to \psi ' (x) = (\ehat \psi (x) ).
\label{6.8}
\ee
$\bar{\psi}$ transforms similarly,
\be
\bar{\psi}(x) \to \bar{\psi} ' (x) = (\ehat \bar{\psi} (x) ).
\label{6.9}
\ee
Dirac $\gamma$-matrices is not a physical field, so it keeps unchanged
under gravitational gauge transformation,
\be
\gamma^{\mu} \to \gamma^{\mu}.
\label{6.10}
\ee
Using above relations and relations  (\ref{4.7083}),
(\ref{4.709}) and (\ref{4.19})
we can  prove that, under gravitational gauge transformation,
${\cal L}_0$ transforms as
\be
{\cal L}_0 \to {\cal L}'_0 = (\ehat {\cal L}_0 ).
\label{6.11}
\ee
So,
\be
{\cal L} \to {\cal L}'  =  J (\ehat {\cal L}_0 ),
\label{6.12}
\ee
where $J$ is the Jacobi of the corresponding space-time
translation. Then using eq.(\ref{4.27}), we can prove that the action $S$
has gravitational gauge symmetry. \\

Suppose that $\ehat$ is an infinitesimal global transformation, then
the first order infinitesimal variations of Dirac field are
\be
\delta \psi = - \epsilon^{\nu} \partial_{\nu} \psi,
\label{6.13}
\ee
\be
\delta \bar{\psi}  = - \epsilon^{\nu} \partial_{\nu} \bar{\psi}.
\label{6.14}
\ee
The first order variation of action is
\be
\delta S = \int {\rm d}^4 x
\epsilon^{\alpha} \partial_{\mu} T_{i \alpha}^{\mu},
\label{6.15}
\ee
where $ T_{i \alpha}^{\mu}$ is the inertial energy-momentum
tensor whose definition is,
\be
T_{i \alpha}^{\mu} \equiv J(C)
\left( - \frac{\partial {\cal L}_0}{\partial \partial_{\mu} \psi}
\partial_{\alpha} \psi
- \frac{\partial {\cal L}_0}{\partial \partial_{\mu} C_{\nu}^{\beta}}
\partial_{\alpha} C_{\nu}^{\beta}
+ \delta^{\mu}_{\alpha} {\cal L}_0 \right).
\label{6.16}
\ee
The global gravitational gauge symmetry of action gives out conservation
equation of the inertial energy-momentum tensor,
\be
\partial_{\mu} T_{i \alpha}^{\mu} = 0.
\label{6.17}
\ee
The inertial energy-momentum tensor is the conserved current
which expected by gauge principle. The space integration of its
time component gives out the conserved energy-momentum of the
system,
\be
H = - P_{i~0} =  \int {\rm d}^3 \svec{x} J(C)
\left( \pi_{\psi} \tdot{\psi} +
\pi_{\alpha}^{\mu} \tdot{C} _{\mu}^{\alpha}
 - {\cal L}_0 \right),
\label{6.18}
\ee
\be
P^i  = P_{i~i} =  \int {\rm d}^3 \svec{x} J(C)
\left( - \pi_{\psi} \partial_i {\psi} -
\pi_{\alpha}^{\mu} \partial_i {C} _{\mu}^{\alpha} \right),
\label{6.19}
\ee
where
\be
\pi_{\psi} = \frac{\partial {\cal L}_0}{\partial \tdot{\psi}}.
\label{6.20}
\ee

The equation of motion of Dirac field is
\be
(\gamma^{\mu} D_{\mu} + m) \psi = 0.
\label{6.21}
\ee
From this expression, we can see that the factor $J(C)$ does not
affect the equation of motion of Dirac field. This is caused by the
asymmetric form of the Lagrangian. If we use a symmetric form of
Lagrangian, the factor $J(C)$ will also affect the equation
of motion of Dirac field, which will be discussed later. \\

The field equation  of gravitational gauge field can be
easily deduced,
\be
\ba{rl}
\partial_{\mu} ( & \frac{1}{4} \eta^{\mu \rho}
\eta^{\nu \sigma} g_{\alpha \beta}
F_{\rho \sigma}^{\beta}
- \frac{1}{4} \eta^{\nu \rho} F^{\mu}_{\rho\alpha}
+ \frac{1}{4} \eta^{\mu \rho} F^{\nu}_{\rho\alpha} \\
&\\
& - \frac{1}{2} \eta^{\mu\rho} \delta^{\nu}_{\alpha}
F^{\beta}_{\rho\beta}
+ \frac{1}{2} \eta^{\nu\rho} \delta^{\mu}_{\alpha}
F^{\beta}_{\rho\beta})  = - g T_{g \alpha}^{\nu},
\label{6.22}
\ea
\ee
where $ T_{g \alpha}^{\nu}$ is the gravitational energy-momentum
tensor, whose definition is:
\be
\begin{array}{rcl}
T_{g \alpha}^{\nu}&=&
 - \frac{\partial {\cal L}_0}{\partial D_{\nu} C_{\mu}^{\beta}}
\partial_{\alpha} C_{\mu}^{\beta}
- \frac{\partial {\cal L}_0}{\partial D_{\nu} \psi}
\partial_{\alpha} \psi
+ G_{\alpha}^{-1 \nu} {\cal L}_0
- G^{-1 \lambda}_{\beta}
(\partial_{\mu} C_{\lambda}^{\beta} )
\frac{\partial {\cal L}_0}
{\partial \partial_{\mu} C_{\nu}^{\alpha}}  \\
&&\\
&& - \frac{1}{4} \eta^{\lambda\rho} \eta^{\nu\sigma}
\partial_{\mu} (g_{\alpha \beta} C^{\mu}_{\lambda}
 F^{\beta}_{\rho\sigma}) \\
 &&\\
&& - \frac{1}{4} \eta^{\nu\rho}
\partial_{\beta} (C^{\sigma}_{\lambda} G^{-1 \lambda}_{\alpha}
 F^{\beta}_{\rho\sigma}) \\
 &&\\
&&  + \frac{1}{2} \eta^{\nu\rho}
\partial_{\alpha} (C^{\sigma}_{\lambda} G^{-1 \lambda}_{\beta}
 F^{\beta}_{\rho\sigma}) \\
 &&\\
&& + \frac{1}{4g} \eta^{\lambda\rho}
\partial_{\mu} \lbrack ( G^{-1 \nu}_{\beta} G^{-1 \sigma}_{\alpha}
G^{\mu}_{\lambda} - \delta^{\nu}_{\beta} \delta^{\sigma}_{\alpha}
\delta^{\mu}_{\lambda}) F^{\beta}_{\rho\sigma}
\rbrack  \\
&&\\
&& - \frac{1}{2 g} \eta^{\lambda\rho}
\partial_{\mu} \lbrack ( G^{-1 \nu}_{\alpha} G^{-1 \sigma}_{\beta}
G^{\mu}_{\lambda} - \delta^{\nu}_{\alpha} \delta^{\sigma}_{\beta}
\delta^{\mu}_{\lambda}) F^{\beta}_{\rho\sigma}
\rbrack  \\
&&\\
&& - \frac{1}{4} \eta^{\kappa \rho} G^{-1 \nu}_{\beta}
G^{-1 \lambda}_{\alpha} G^{-1 \sigma}_{\gamma}
F^{\beta}_{\rho\sigma} F^{\gamma}_{\kappa\lambda}  \\
&&\\
&& + \frac{1}{2} \eta^{\kappa \rho} G^{-1 \nu}_{\gamma}
G^{-1 \lambda}_{\alpha} G^{-1 \sigma}_{\beta}
F^{\beta}_{\rho\sigma} F^{\gamma}_{\kappa\lambda}  \\
&&\\
&& - \frac{1}{8} \eta^{\mu \rho} \eta^{\lambda \sigma}
g_{\alpha \gamma} G^{-1 \nu}_{\beta}
F^{\beta}_{\rho\sigma} F^{\gamma}_{\mu\lambda}.
\end{array}
\label{6.23}
\ee
We see again that the gravitational energy-momentum tensor is
different from the  inertial energy-momentum tensor. This field
equation is also equivalent to the Einstein's field equation
which has the following form
\be
R^{\alpha \beta} - \frac{1}{2} g^{\alpha \beta} R
+ 8 \pi G T^{\alpha \beta} =0,
\label{6.2301}
\ee
where $T^{\alpha\beta}$ is the symmetric energy-momentum
tensor of Dirac field
\be
T^{\alpha\beta} = \frac{1}{2} (\bar{\psi} \gamma^{\mu}
\partial_{\gamma} \psi)(G_{\mu}^{\alpha} g^{\beta\gamma}
+ G_{\mu}^{\beta} g^{\alpha\gamma} )
- g^{\alpha\beta} \bar{\psi} (\gamma^{\mu} D_{\mu}
+ m) \psi
\label{6.2302}
\ee
\\

In usual quantum field theory, the Lagrangian for Dirac field
has a more symmetric form, which is
\be
- \bar{\psi} (\gamma^{\mu} \dvec{\partial}_{\mu} + m) \psi,
\label{6.24}
\ee
where
\be
 \dvec{\partial}_{\mu} =
\frac{\partial_{\mu} - \lvec{\partial_{\mu}}}{2}.
\label{6.25}
\ee
The Euler-Lagrange equation of motion of eq.(\ref{6.24}) also
gives out the conventional Dirac equation. \\

Now replace ordinary space-time partial derivative with
covariant derivative, and add into the Lagrangian of pure
gravitational gauge field, we get,
\be
\ba{rcl}
{\cal L}_0 &=&
- \bar{\psi} (\gamma^{\mu} \dvec{D}_{\mu} + m) \psi\\
&&\\
&&- \frac{1}{16} \eta^{\mu \rho}
\eta^{\nu \sigma} g_{\alpha \beta}
F^{\alpha}_{\mu \nu} F^{\beta}_{\rho \sigma} \\
&&\\
&& - \frac{1}{8} \eta^{\mu \rho}
G^{-1 \nu}_{\beta} G^{-1 \sigma}_{\alpha}
F^{\alpha}_{\mu \nu} F^{\beta}_{\rho \sigma} \\
&&\\
&& + \frac{1}{4} \eta^{\mu \rho}
G^{-1 \nu}_{\alpha} G^{-1 \sigma}_{\beta}
F^{\alpha}_{\mu \nu} F^{\beta}_{\rho \sigma},
\ea
\label{6.26}
\ee
where $\dvec{D}_{\mu}$ is defined by
\be
 \dvec{D}_{\mu} =
\frac{D_{\mu} - \lvec{D}_{\mu}}{2}.
\label{6.27}
\ee
Operator $\lvec{D}_{\mu}$ is understood in the following way
\be
f(x) \lvec{D}_{\mu} g(x) =
(D_{\mu} f(x)) g(x),
\label{6.28}
\ee
with $f(x)$ and $g(x)$ two arbitrary functions. The Lagrangian density
${\cal L}$ and action $S$ are also defined by
eqs.(\ref{6.3}-\ref{6.4}). In this case,
the free Lagrangian ${\cal L}_F$ and interaction Lagrangian ${\cal L}_I$
are given by
\be
\ba{rcl}
{\cal L}_F &=&
- \bar{\psi} (\gamma^{\mu} \dvec{\partial}_{\mu} + m) \psi \\
&&\\
&& - \frac{1}{16} \eta^{\mu \rho} \eta^{\nu \sigma} \eta_{\alpha \beta }
F_{0 \mu \nu}^{\alpha} F_{0 \rho \sigma}^{\beta}
- \frac{1}{8} \eta^{\mu \rho}
F^{\alpha}_{0 \mu \beta} F^{\beta}_{0 \rho \alpha}
+ \frac{1}{4} \eta^{\mu \rho}
F^{\alpha}_{0 \mu \alpha} F^{\beta}_{0 \rho \beta},
\ea
\label{6.29}
\ee
\be
\begin{array}{rcl}
{\cal L}_I &=&
(J(C) - 1 ) \cdot \left\lbrack
- \bar{\psi} (\gamma^{\mu} \dvec{\partial}_{\mu} + m) \psi
\right\rbrack
 + g J(C)( \bar{\psi} \gamma^{\mu} \dvec{\partial}_{\alpha}
\psi) C_{\mu}^{\alpha}\\
&&\\
&& + { self ~interaction~ terms ~of ~Gravitational~gauge~field}.
\end{array}
\label{6.30}
\ee
\\

The Euler-Lagrange equation of motion for Dirac field is
\be
\frac{\partial {\cal L}_0}{\partial \bar{\psi}}
- \partial_{\mu} \frac{\partial {\cal L}_0}{\partial
\partial_{\mu} \bar{\psi}}
- g G^{-1 \nu}_{\alpha} ( \partial_{\mu} C_{\nu}^{\alpha})
\frac{\partial {\cal L}_0}{\partial
\partial_{\mu} \bar{\psi}} = 0.
\label{6.31}
\ee
Because
\be
\frac{\partial {\cal L}_0}{\partial \bar{\psi}}
= -\frac{1}{2} \gamma^{\mu} D_{\mu} \psi - m \psi ,
\label{6.32}
\ee
\be
\frac{\partial {\cal L}_0}{\partial \partial_{\mu} \bar{\psi}}
=   \frac{1}{2} \gamma^{\alpha} G_{\alpha}^{\mu} \psi ,
\label{6.33}
\ee
eq.(\ref{6.31}) will be changed into
\be
(\gamma^{\mu} D_{\mu} + m) \psi =
- \frac{1}{2}\gamma^{\mu}
(\partial_{\alpha} G_{\mu}^{\alpha}) \psi
-  \frac{1}{2} g \gamma^{\mu} \psi
G_{\beta}^{-1 \nu} (D_{\mu}  C_{\nu}^{\beta} ).
\label{6.34}
\ee
If gravitational gauge field vanishes, this equation of motion
will return to the traditional Dirac equation. \\

The inertial energy-momentum tensor now becomes
\be
T_{i \alpha}^{\mu} = J(C)
\left( - \frac{\partial {\cal L}_0}{\partial \partial_{\mu} \psi}
\partial_{\alpha} \psi
- (\partial_{\alpha} \bar{\psi})
\frac{\partial {\cal L}_0}{\partial \partial_{\mu} \bar{\psi}}
- \frac{\partial {\cal L}_0}{\partial \partial_{\mu} C_{\nu}^{\beta}}
\partial_{\alpha} C_{\nu}^{\beta}
+ \delta^{\mu}_{\alpha} {\cal L}_0 \right),
\label{6.35}
\ee
and the gravitational energy-momentum tensor becomes
\be
\begin{array}{rcl}
T_{g \alpha}^{\nu}&=&
 - \frac{\partial {\cal L}_0}{\partial D_{\nu} C_{\mu}^{\beta}}
\partial_{\alpha} C_{\mu}^{\beta}
- \frac{\partial {\cal L}_0}{\partial D_{\nu} \psi}
\partial_{\alpha} \psi
- (\partial_{\alpha} \bar{\psi})
\frac{\partial {\cal L}_0}{\partial D_{\nu} \bar{\psi}}
+ G_{\alpha}^{-1 \nu} {\cal L}_0  \\
&&\\
&& -  G^{-1 \lambda}_{\beta} (\partial_{\mu} C_{\lambda}^{\beta} )
\frac{\partial {\cal L}_0}{\partial \partial_{\mu} C_{\nu}^{\alpha}}  \\
&&\\
&& - \frac{1}{4} \eta^{\lambda\rho} \eta^{\nu\sigma}
\partial_{\mu} (g_{\alpha \beta} C^{\mu}_{\lambda}
 F^{\beta}_{\rho\sigma}) \\
 &&\\
&& - \frac{1}{4} \eta^{\nu\rho}
\partial_{\beta} (C^{\sigma}_{\lambda} G^{-1 \lambda}_{\alpha}
 F^{\beta}_{\rho\sigma}) \\
 &&\\
&&  + \frac{1}{2} \eta^{\nu\rho}
\partial_{\alpha} (C^{\sigma}_{\lambda} G^{-1 \lambda}_{\beta}
 F^{\beta}_{\rho\sigma}) \\
 &&\\
&& + \frac{1}{4g} \eta^{\lambda\rho}
\partial_{\mu} \lbrack ( G^{-1 \nu}_{\beta} G^{-1 \sigma}_{\alpha}
G^{\mu}_{\lambda} - \delta^{\nu}_{\beta} \delta^{\sigma}_{\alpha}
\delta^{\mu}_{\lambda}) F^{\beta}_{\rho\sigma}
\rbrack  \\
&&\\
&& - \frac{1}{2 g} \eta^{\lambda\rho}
\partial_{\mu} \lbrack ( G^{-1 \nu}_{\alpha} G^{-1 \sigma}_{\beta}
G^{\mu}_{\lambda} - \delta^{\nu}_{\alpha} \delta^{\sigma}_{\beta}
\delta^{\mu}_{\lambda}) F^{\beta}_{\rho\sigma}
\rbrack  \\
&&\\
&& - \frac{1}{4} \eta^{\kappa \rho} G^{-1 \nu}_{\beta}
G^{-1 \lambda}_{\alpha} G^{-1 \sigma}_{\gamma}
F^{\beta}_{\rho\sigma} F^{\gamma}_{\kappa\lambda}  \\
&&\\
&& + \frac{1}{2} \eta^{\kappa \rho} G^{-1 \nu}_{\gamma}
G^{-1 \lambda}_{\alpha} G^{-1 \sigma}_{\beta}
F^{\beta}_{\rho\sigma} F^{\gamma}_{\kappa\lambda}  \\
&&\\
&& - \frac{1}{8} \eta^{\mu \rho} \eta^{\lambda \sigma}
g_{\alpha \gamma} G^{-1 \nu}_{\beta}
F^{\beta}_{\rho\sigma} F^{\gamma}_{\mu\lambda}.
\end{array}
\label{6.36}
\ee
Both of them are conserved energy-momentum tensor.  But they
are not equivalent.
The field equation of gravitational gauge field still are
(\ref{6.22}), but should replace the gravitational energy-momentum
tensor $T^{\nu}_{g \alpha}$ with (\ref{6.36}). The Einstein's
field equation is still (\ref{6.2301}), but the energy-momentum
tensor $T^{\alpha \beta}$ should be
\be
T^{\alpha\beta} = \frac{1}{2} (\bar{\psi} \gamma^{\mu}
\dvec{\partial}_{\gamma} \psi)(G_{\mu}^{\alpha} g^{\beta\gamma}
+ G_{\mu}^{\beta} g^{\alpha\gamma} )
- g^{\alpha\beta} \left\lbrack
\bar{\psi} (\gamma^{\mu} \dvec{D}_{\mu}
+ m) \psi \right \rbrack
\label{6.37}
\ee
\\

\section{Gravitational Interactions of Vector Field}
\setcounter{equation}{0}

The traditional Lagrangian for vector field is
\be
- \frac{1}{4} \eta^{\mu \rho} \eta^{\nu \sigma}
A_{\mu \nu} A_{\rho \sigma}
- \frac{m^2}{2}  \eta^{\mu \nu} A_{\mu} A_{\nu},
\label{7.1}
\ee
where $A_{\mu \nu}$ is the strength of vector field which is
given by
\be
 \partial_{\mu} A_{\nu}
- \partial_{\nu} A_{\mu}.
\label{7.2}
\ee
The Lagrangian ${\cal L}_0$ that describes gravitational interactions
between vector field and gravitational fields is
\be
\ba{rcl}
{\cal L}_0 &=&
- \frac{1}{4} \eta^{\mu \rho} \eta^{\nu \sigma}
A_{\mu \nu} A_{\rho \sigma}
- \frac{m^2}{2}  \eta^{\mu \nu} A_{\mu} A_{\nu}  \\
&&\\
&&  - \frac{1}{16} \eta^{\mu \rho}
\eta^{\nu \sigma} g_{\alpha \beta}
F^{\alpha}_{\mu \nu} F^{\beta}_{\rho \sigma} \\
&&\\
&& - \frac{1}{8} \eta^{\mu \rho}
G^{-1 \nu}_{\beta} G^{-1 \sigma}_{\alpha}
F^{\alpha}_{\mu \nu} F^{\beta}_{\rho \sigma} \\
&&\\
&& + \frac{1}{4} \eta^{\mu \rho}
G^{-1 \nu}_{\alpha} G^{-1 \sigma}_{\beta}
F^{\alpha}_{\mu \nu} F^{\beta}_{\rho \sigma}.
\ea
\label{7.3}
\ee
In eq.(\ref{7.3}), the definition of strength $A_{\mu \nu}$ is not
given by eq.(\ref{7.2}), it is given by
\be
\begin{array}{rcl}
A_{\mu \nu} &=& D_{\mu} A_{\nu}
- D_{\nu} A_{\mu}  \\
&=&  \partial_{\mu} A_{\nu} - \partial_{\nu} A_{\mu}
- g C_{\mu}^{\alpha} \partial_{\alpha} A_{\nu}
+ g C_{\nu}^{\alpha} \partial_{\alpha} A_{\mu},
\end{array}
\label{7.4}
\ee
where $D_{\mu}$ is the gravitational gauge
 covariant derivative, whose definition
is given by eq.(\ref{4.4}). The full Lagrangian ${\cal L}$ is given by,
\be
{\cal L} = J(C) {\cal L}_0.
\label{7.5}
\ee
The action $S$ is defined by
\be
S = \int {\rm d}^4 x ~ {\cal L}.
\label{7.6}
\ee
\\

The Lagrangian ${\cal L}$ can be separated into two parts:
the free Lagrangian ${\cal L}_F$ and interaction Lagrangian
${\cal L}_I$. The explicit forms of them are
\be
\ba{rcl}
{\cal L}_F &=&
- \frac{1}{4} \eta^{\mu \rho} \eta^{\nu \sigma}
A_{0 \mu \nu} A_{0 \rho \sigma}
- \frac{m^2}{2}  \eta^{\mu \nu} A_{\mu} A_{\nu} \\
&&\\
&&  - \frac{1}{16} \eta^{\mu \rho} \eta^{\nu \sigma} \eta_{\alpha \beta }
F_{0 \mu \nu}^{\alpha} F_{0 \rho \sigma}^{\beta}
- \frac{1}{8} \eta^{\mu \rho}
F^{\alpha}_{0 \mu \beta} F^{\beta}_{0 \rho \alpha}
+ \frac{1}{4} \eta^{\mu \rho}
F^{\alpha}_{0 \mu \alpha} F^{\beta}_{0 \rho \beta},
\ea
\label{7.7}
\ee
\be
\begin{array}{rcl}
{\cal L}_I &=&
( J(C) - 1 ) \cdot ( - \frac{1}{4} \eta^{\mu \rho} \eta^{\nu \sigma}
A_{0 \mu \nu} A_{0 \rho \sigma}
- \frac{m^2}{2}  \eta^{\mu \nu} A_{\mu} A_{\nu}  ) \\
&&\\
&& + g J(C) \eta^{\mu \rho} \eta^{\nu \sigma}
A_{0 \mu \nu} C_{\rho}^{\alpha} \partial_{\alpha} A_{\sigma} \\
&&\\
&& - \frac{g^2}{2} J(C) \eta^{\mu \rho} \eta^{\nu \sigma}
( C_{\mu}^{\alpha} C_{\rho}^{\beta} (\partial_{\alpha} A_{\nu} )
(\partial_{\beta} A_{\sigma} )
- C_{\nu}^{\alpha} C_{\rho}^{\beta} (\partial_{\alpha} A_{\mu} )
(\partial_{\beta} A_{\sigma} )  )  \\
&&\\
&& + {~ self ~interaction~ terms ~of ~Gravitational~gauge~field},
\end{array}
\label{7.8}
\ee
where
$A_{0 \mu \nu} = \partial_{\mu} A_{\nu} - \partial_{\nu} A_{\mu}$.
The first three lines of ${\cal L}_I$ contain interactions between
vector field and gravitational gauge fields. It can be seen that
the vector field can also directly couple to arbitrary number of
gravitational gauge fields, which is one of the most important
properties of gravitational gauge interactions. This interaction
property is required and determined by local gravitational
gauge symmetry.  \\

Under Lorentz transformations, group index and Lorentz index have
the same behavior. Therefor every term in the Lagrangian ${\cal L}$
are Lorentz scalar, and the whole Lagrangian ${\cal L}$ and
action $S$ have Lorentz symmetry. \\

Under gravitational gauge transformations, vector field $A_{\mu}$
transforms as
\be
A_{\mu} (x) \to A'_{\mu}(x) = (\ehat A_{\mu} (x)).
\label{7.9}
\ee
$D_{\mu} A_{\nu}$ and $A_{\mu \nu}$ transform covariantly,
\be
D_{\mu} A_{\nu} \to D'_{\mu} A'_{\nu} =
(\ehat D_{\mu} A_{\nu}) ,
\label{7.10}
\ee
\be
A_{\mu \nu} \to A'_{\mu \nu} =
(\ehat A_{\mu \nu}).
\label{7.11}
\ee
So, the gravitational gauge transformations of ${\cal L}_0$
and ${\cal L}$ respectively are
\be
{\cal L}_0 \to {\cal L}'_0 = (\ehat {\cal L}_0 ),
\label{7.12}
\ee
\be
{\cal L} \to {\cal L}'  =  J (\ehat {\cal L}_0 ).
\label{7.13}
\ee
The action of the system is gravitational gauge invariant.
\\

The global gravitational gauge transformation gives out conserved
current of gravitational gauge  symmetry.
Under infinitesimal global gravitational gauge
transformation, the vector field $A_{\mu}$ transforms as
\be
\delta A_{\mu} = - \epsilon^{\alpha} \partial_{\alpha} A_{\mu}.
\label{7.14}
\ee
The first order variation of action is
\be
\delta S = \int {\rm d}^4 x
\epsilon^{\alpha} \partial_{\mu} T_{i \alpha}^{\mu},
\label{7.15}
\ee
where $ T_{i \alpha}^{\mu}$ is the inertial energy-momentum
tensor whose definition is,
\be
T_{i \alpha}^{\mu} = J(C)
\left( - \frac{\partial {\cal L}_0}{\partial \partial_{\mu} A_{\nu}}
\partial_{\alpha} A_{\nu}
- \frac{\partial {\cal L}_0}{\partial \partial_{\mu} C_{\nu}^{\beta}}
\partial_{\alpha} C_{\nu}^{\beta} +
\delta^{\mu}_{\alpha} {\cal L}_0 \right).
\label{7.16}
\ee
$T_{i \alpha}^{\mu}$ is a conserved current. The space integration
of its time component gives out inertial energy-momentum of the
system,
\be
H = - P_{i~0} =  \int {\rm d}^3 \svec{x} J(C)
\left( \pi^{\mu} \tdot{A_{\mu}} +
\pi_{\alpha}^{\mu} \tdot{C _{\mu}^{\alpha}}
 - {\cal L}_0 \right),
\label{7.17}
\ee
\be
P^i  = P_{i~i} =  \int {\rm d}^3 \svec{x} J(C)
\left ( - \pi^{\mu} \partial_i A_{\mu} -
\pi_{\alpha}^{\mu} \partial_i {C} _{\mu}^{\alpha} \right ),
\label{7.18}
\ee
where
\be
\pi^{\mu} = \frac{\partial {\cal L}_0}{\partial \tdot{A_{\mu}}}.
\label{7.19}
\ee
\\

The equation of motion for vector field is
\be
\frac{\partial {\cal L}_0}{\partial A_{\nu}}
- \partial_{\mu}
\frac{\partial {\cal L}_0}{\partial \partial_{\mu} A_{\nu}}
- g G_{\alpha}^{-1 \lambda}
( \partial_{\mu} C_{\lambda}^{\alpha})
\frac{\partial {\cal L}_0}{\partial \partial_{\mu} A_{\nu}} = 0.
\label{7.20}
\ee
From eq.(\ref{7.3}), we can obtain
\be
\frac{\partial {\cal L}_0}{\partial \partial_{\mu} A_{\nu}}
= - \eta^{\lambda \rho} \eta^{\nu \sigma}
G_{\lambda}^{\mu} A_{\rho \sigma},
\label{7.21}
\ee
\be
\frac{\partial {\cal L}_0}{\partial A_{\nu}}
= - m^2 \eta^{\lambda \nu} A_{\lambda}.
\label{7.22}
\ee
Then, eq.(\ref{7.20}) is changed into
\be
\eta^{\mu \rho} \eta^{\nu \sigma} D_{\mu} A_{\rho \sigma}
- m^2 \eta^{\mu \nu} A_{\mu}
= - \eta^{\lambda \rho} \eta^{\nu \sigma}
( \partial_{\mu} G_{\lambda}^{\mu} ) A_{\rho \sigma}
- g \eta^{\mu \rho} \eta^{\nu \sigma}
 A_{\rho \sigma} G_{\alpha}^{-1 \mu}
( D_{\mu} C_{\mu}^{\alpha}).
\label{7.23}
\ee
The equation of motion of gravitational gauge field is
\be
\ba{rl}
\partial_{\mu} ( & \frac{1}{4} \eta^{\mu \rho}
\eta^{\nu \sigma} g_{\alpha \beta}
F_{\rho \sigma}^{\beta}
- \frac{1}{4} \eta^{\nu \rho} F^{\mu}_{\rho\alpha}
+ \frac{1}{4} \eta^{\mu \rho} F^{\nu}_{\rho\alpha} \\
&\\
& - \frac{1}{2} \eta^{\mu\rho} \delta^{\nu}_{\alpha}
F^{\beta}_{\rho\beta}
+ \frac{1}{2} \eta^{\nu\rho} \delta^{\mu}_{\alpha}
F^{\beta}_{\rho\beta})  = - g T_{g \alpha}^{\nu},
\label{7.24}
\ea
\ee
where $ T_{g \alpha}^{\nu}$ is the gravitational energy-momentum
tensor,
\be
\begin{array}{rcl}
T_{g \alpha}^{\nu}&=&
 - \frac{\partial {\cal L}_0}{\partial D_{\nu} C_{\mu}^{\beta}}
\partial_{\alpha} C_{\mu}^{\beta}
- \frac{\partial {\cal L}_0}{\partial D_{\nu} A_{\mu}}
\partial_{\alpha} A_{\mu}
+ G_{\alpha}^{-1 \nu} {\cal L}_0
 -  G_{\gamma}^{-1 \rho} (\partial_{\mu} C_{\rho}^{\gamma})
\frac{\partial {\cal L}_0}{\partial \partial_{\mu} C_{\nu}^{\alpha}} \\
&&\\
&&- \frac{1}{4} \eta^{\lambda\rho} \eta^{\nu\sigma}
\partial_{\mu} (g_{\alpha \beta} C^{\mu}_{\lambda}
 F^{\beta}_{\rho\sigma}) \\
 &&\\
&& - \frac{1}{4} \eta^{\nu\rho}
\partial_{\beta} (C^{\sigma}_{\lambda} G^{-1 \lambda}_{\alpha}
 F^{\beta}_{\rho\sigma}) \\
 &&\\
&&  + \frac{1}{2} \eta^{\nu\rho}
\partial_{\alpha} (C^{\sigma}_{\lambda} G^{-1 \lambda}_{\beta}
 F^{\beta}_{\rho\sigma}) \\
 &&\\
&& + \frac{1}{4g} \eta^{\lambda\rho}
\partial_{\mu} \lbrack ( G^{-1 \nu}_{\beta} G^{-1 \sigma}_{\alpha}
G^{\mu}_{\lambda} - \delta^{\nu}_{\beta} \delta^{\sigma}_{\alpha}
\delta^{\mu}_{\lambda}) F^{\beta}_{\rho\sigma}
\rbrack  \\
&&\\
&& - \frac{1}{2 g} \eta^{\lambda\rho}
\partial_{\mu} \lbrack ( G^{-1 \nu}_{\alpha} G^{-1 \sigma}_{\beta}
G^{\mu}_{\lambda} - \delta^{\nu}_{\alpha} \delta^{\sigma}_{\beta}
\delta^{\mu}_{\lambda}) F^{\beta}_{\rho\sigma}
\rbrack  \\
&&\\
&& - \frac{1}{4} \eta^{\kappa \rho} G^{-1 \nu}_{\beta}
G^{-1 \lambda}_{\alpha} G^{-1 \sigma}_{\gamma}
F^{\beta}_{\rho\sigma} F^{\gamma}_{\kappa\lambda}  \\
&&\\
&& + \frac{1}{2} \eta^{\kappa \rho} G^{-1 \nu}_{\gamma}
G^{-1 \lambda}_{\alpha} G^{-1 \sigma}_{\beta}
F^{\beta}_{\rho\sigma} F^{\gamma}_{\kappa\lambda}  \\
&&\\
&& - \frac{1}{8} \eta^{\mu \rho} \eta^{\lambda \sigma}
g_{\alpha \gamma} G^{-1 \nu}_{\beta}
F^{\beta}_{\rho\sigma} F^{\gamma}_{\mu\lambda} .
\end{array}
\label{7.25}
\ee
$T_{g \alpha}^{\nu}$ is also a conserved current. The space integration
of its time component gives out the gravitational energy-momentum
which is the source of gravitational interactions.
It can be also seen that inertial energy-momentum tensor and
gravitational energy-momentum tensor are not equivalent.   \\

\section{GSU(N) Unification Model}
\setcounter{equation}{0}

Now, let's discuss how to unify traditional $SU(N)$ gauge
field  theory with gravitational gauge field theory\cite{13},
which is the foundation of the unification of fundamental
interactions. As an example, we discuss gravitational gauge
interactions and $SU(N)$ non-Abelian gauge interactions
of Dirac field.
The generators of $SU(N)$ group is denoted as $T_a$, they
satisfies
\be \label{19.1}
\lbrack T_a~~,~~ T_b \rbrack = i f_{abc} T_c,
\ee
\be \label{19.2}
Tr( T_a  T_b )  = K \delta_{ab}.
\ee
The $SU(N)$ non-Abelian gauge field is denoted as $A_{\mu}$, which is
an element of $SU(N)$ Lie algebra,
\be \label{19.3}
A_{\mu}(x) = A_{\mu}^{a}(x) T_a,
\ee
where $A_{\mu}^{a}(x)$ are component fields. \\

Because an arbitrary element $U(x)$ of $SU(N)$ group does not commute
with an arbitrary element $\ehat$ of gravitational gauge group,
\be \label{19.4}
\lbrack U(x)~~,~~ \ehat \rbrack \not= 0.
\ee
The product group of $SU(N)$ group and gravitational group is not
direct product group, but semi-direct product group, which we will
denoted as $GSU(N)$ group
\be \label{19.5}
GSU(N) \define
SU(N) \otimes_s Gravitational~Gauge~Group.
\ee
An arbitrary element of $GSU(N)$ is denoted as $g(x)$, which is
defined by
\be \label{19.6}
g(x) \define \ehat \cdot U(x).
\ee
The gauge covariant derivative of $GSU(N)$ group is
\be \label{19.7}
{\mathbb D}_{\mu} \define \partial_{\mu}
- i g C_{\mu} - i g_s A_{\mu} = D_{\mu} -ig_s A_{\mu},
\ee
where $g_s$ is the coupling constant of non-Abelian $SU(N)$ gauge
interactions, $C_{\mu}$ is the gravitational gauge field
and $D_{\mu}$ is the gravitational gauge covariant derivative.
\\

The field strength of non-Abelian gauge field $A_{\mu}$ is
\be \label{19.8}
A_{\mu\nu} = (D_{\mu} A_{\nu}) - (D_{\nu} A_{\mu})
-ig_s \lbrack A_{\mu} ~~,~~ A_{\nu} \rbrack.
\ee
$A_{\mu\nu}$ is also an element of $SU(N)$ Lie algebra,
\be \label{19.9}
A_{\mu\nu}(x) = A_{\mu\nu}^{a}(x) T_a,
\ee
where
\be \label{19.10}
A^a_{\mu\nu} = (D_{\mu} A^a_{\nu}) - (D_{\nu} A^a_{\mu})
+ g_s f_{abc} A^b_{\mu}  A^c_{\nu}.
\ee
$A_{\mu\nu}$ is not a $SU(N)$ gauge covariant field strength.
\\

$SU(N)$ Gauge covariant field strength is defined by
\be \label{19.11}
{\mathbb A}_{\mu\nu} = A_{\mu\nu} + g G^{-1 \lambda}_{\sigma}
A_{\lambda} F_{\mu\nu}^{\sigma} =
{\mathbb A}^a_{\mu\nu} T_a,
\ee
where
\be \label{19.12}
{\mathbb A}^a_{\mu\nu} = A^a_{\mu\nu} + g G^{-1 \lambda}_{\sigma}
A^a_{\lambda} F_{\mu\nu}^{\sigma} .
\ee
The lagrangian density ${\cal L}_0$ is given by,
\be \label{19.13}
\begin{array}{rcl}
{\cal L}_0 &=&
-  \bar{\psi}
\lbrack \gamma^{\mu} (D_{\mu} -i g_s A_{\mu} ) + m
\rbrack \psi
-\frac{1}{4} \eta^{\mu \rho} \eta^{\nu \sigma}
{\mathbb A}^a_{\mu \nu } {\mathbb A}^a_{ \rho \sigma } \\
&&\\
&& - \frac{1}{16} \eta^{\mu \rho}
\eta^{\nu \sigma} g_{\alpha \beta}
F^{\alpha}_{\mu \nu} F^{\beta}_{\rho \sigma} \\
&&\\
&& - \frac{1}{8} \eta^{\mu \rho}
G^{-1 \nu}_{\beta} G^{-1 \sigma}_{\alpha}
F^{\alpha}_{\mu \nu} F^{\beta}_{\rho \sigma} \\
&&\\
&& + \frac{1}{4} \eta^{\mu \rho}
G^{-1 \nu}_{\alpha} G^{-1 \sigma}_{\beta}
F^{\alpha}_{\mu \nu} F^{\beta}_{\rho \sigma} .
\end{array}
\ee
The full lagrangian density ${\cal L}$ is defined by
\be \label{19.14}
{\cal L} = J(C) {\cal L}_0,
\ee
And the action of the system is
\be \label{19.15}
S  = \int {\rm d}^4 x {\cal L} = \int {\rm d}^4 x J(C) {\cal L}_0.
\ee
\\

Now, we discus symmetry of the system. Under $SU(N)$ gauge
transformations, gravitational gauge field $C_{\mu}(x)$
is kept unchanged. Therefore, $F_{\mu\nu}^{\alpha}$,
$G^{\alpha}_{\mu}$, $G^{-1 \mu}_{\alpha}$ , $D_{\mu}$
and $J(C)$
are not changed under local $SU(N)$ gauge transformations.
Other fields and operators transform as
\be \label{19.16}
\psi  \to \psi' =  (U(x) \psi),
\ee
\be \label{19.17}
A_{\mu} \to A'_{\mu} = U(x) A_{\mu} U^{-1}(x)
- \frac{1}{ig_s} U(x) (D_{\mu} U^{-1}(x)) ,
\ee
\be \label{19.18}
A_{\mu\nu} \to A'_{\mu\nu} = U(x) A_{\mu\nu} U^{-1}(x)
+ \frac{g}{ig_s} F_{\mu\nu}^{\sigma}
U(x) (\partial_{\sigma} U^{-1}(x)) ,
\ee
\be \label{19.19}
{\mathbb A}_{\mu\nu} \to {\mathbb A}'_{\mu\nu}
= U(x) {\mathbb A}_{\mu\nu} U^{-1}(x).
\ee
Using all these relations, we can prove that the lagrangian
density ${\cal L}_0$ does not change under local $SU(N)$
gauge transformations
\be \label{19.20}
{\cal L}_0 \to {\cal L}'_0 = {\cal L}_0.
\ee
Because both integration measure ${\rm d}^4 x$ and $J(C)$ are not
changed under non-Abelian $SU(N)$ gauge transformation, the action
is invariant under $SU(N)$ gauge transformation. Therefore,
the system has local $SU(N)$ gauge symmetry.\\

Under local gravitational gauge transformation, the
transformations of various fields and operators are
\be \label{19.21}
\psi \to \psi ' = (\ehat \psi),
\ee
\be \label{19.22}
\bar\psi \to \bar\psi ' = (\ehat \bar\psi),
\ee
\be \label{19.23}
A_{\mu} \to A'_{\mu} = \ehat A_{\mu} \ehat^{-1},
\ee
\be \label{19.24}
C_{\mu} \to C'_{\mu} = \ehat C_{\mu} \ehat^{-1}
- \frac{1}{ig} \ehat (\partial_{\mu}  \ehat^{-1}),
\ee
\be \label{19.241}
g_{\alpha\beta} \to g'_{\alpha\beta}
= \Lambda_{\alpha}^{~\alpha_1} \Lambda_{\beta}^{~\beta_1}
(\ehat g_{\alpha_1 \beta_1} ),
\ee
\be \label{19.25}
{\mathbb D}_{\mu} \to {\mathbb D}'_{\mu}
= \ehat {\mathbb D}_{\mu} \ehat^{-1},
\ee
\be \label{19.26}
A_{\mu\nu} \to A'_{\mu\nu} = \ehat A_{\mu\nu} \ehat^{-1},
\ee
\be \label{19.27}
F_{\mu\nu}^{\sigma} \to F_{\mu\nu}^{\prime\sigma}
= \Lambda^{\sigma}_{~\rho} \ehat F_{\mu\nu}^{\rho} \ehat^{-1},
\ee
\be \label{19.28}
G^{\alpha}_{\mu} \to G^{\prime\alpha}_{\mu}
= \Lambda^{\alpha}_{~\beta} \ehat G^{\beta}_{\mu} \ehat^{-1},
\ee
\be \label{19.29}
G_{\alpha}^{-1\mu} \to G_{\alpha}^{\prime -1 \mu}
= \Lambda_{\alpha}^{~\beta} \ehat G_{\beta}^{-1\mu} \ehat^{-1},
\ee
\be \label{19.30}
{\mathbb A}_{\mu\nu} \to {\mathbb A}'_{\mu\nu}
= \ehat {\mathbb A}_{\mu\nu} \ehat^{-1},
\ee
\be \label{19.31}
J(C) \to J'(C') = J \cdot \ehat J(C) \ehat^{-1},
\ee
where $J$ is the Jacobian of the corresponding transformation.
Using all these relations and the following relation
\be \label{19.3102}
\int {\rm d}^4 x J(\ehat f(x)) = \int {\rm d}^4 x f(x),
\ee
where $f(x)$ is an arbitrary function, we can prove that
\be \label{19.32}
{\cal L}_0 \to {\cal L}'_0 =( \ehat {\cal L}_0 ),
\ee
\be \label{19.33}
{\cal L} \to {\cal L}' = J( \ehat {\cal L} ),
\ee
\be \label{19.34}
S \to S' = S.
\ee
Therefore, the system has local gravitational gauge symmetry.\\

Combining above results on local $SU(N)$ gauge transformations and local
gravitational gauge transformations, we know that under general
$GSU(N)$ gauge transformation $g(x)$, transformations of various
fields and operators are
\be \label{19.35}
\psi \to \psi ' = (g(x) \psi),
\ee
\be \label{19.36}
\bar\psi \to \bar\psi ' = (\ehat (\bar\psi U^{\dag}(x) )),
\ee
\be \label{19.37}
A_{\mu} \to A'_{\mu} = g(x) \left\lbrack A_{\mu}
- \frac{1}{ig_s} (D_{\mu} U^{-1}(x)) U(x)
\right\rbrack g^{-1}(x),
\ee
\be \label{19.38}
C_{\mu} \to C'_{\mu} = \ehat C_{\mu} \ehat^{-1}
- \frac{1}{ig} \ehat (\partial_{\mu}  \ehat^{-1}),
\ee
\be \label{19.39}
{\mathbb D}_{\mu} \to {\mathbb D}'_{\mu}
= g(x) {\mathbb D}_{\mu} g^{-1}(x),
\ee
\be \label{19.40}
A_{\mu\nu} \to A'_{\mu\nu} = g(x) \left\lbrack  A_{\mu\nu}
+\frac{g}{i g_s} F_{\mu\nu}^{\sigma}
(\partial_{\sigma} U^{-1}(x) ) U(x) \right\rbrack g^{-1}(x),
\ee
\be \label{19.41}
F_{\mu\nu}^{\sigma} \to F_{\mu\nu}^{\prime\sigma}
= \Lambda^{\sigma}_{~\rho} g(x) F_{\mu\nu}^{\rho} g^{-1}(x),
\ee
\be \label{19.42}
G^{\alpha}_{\mu} \to G^{\prime\alpha}_{\mu}
= \Lambda^{\alpha}_{~\beta} g(x) G^{\beta}_{\mu} g^{-1}(x),
\ee
\be \label{19.43}
G_{\alpha}^{-1\mu} \to G_{\alpha}^{\prime -1 \mu}
= \Lambda_{\alpha}^{~\beta} g(x) G_{\beta}^{-1\mu} g^{-1}(x),
\ee
\be \label{19.431}
g_{\alpha\beta} \to g'_{\alpha\beta}
= \Lambda_{\alpha}^{~\alpha_1} \Lambda_{\beta}^{~\beta_1}
\cdot g(x)  g_{\alpha_1 \beta_1} g^{-1}(x),
\ee
\be \label{19.44}
{\mathbb A}_{\mu\nu} \to {\mathbb A}'_{\mu\nu}
= g(x) {\mathbb A}_{\mu\nu} g^{-1}(x),
\ee
\be \label{19.45}
J(C) \to J'(C') = J \cdot g(x) J(C) g^{-1}(x).
\ee
Using all above relations, we can prove that the
action $S$ is invariant local $GSU(N)$ gauge transformation.  \\

The lagrangian density ${\cal L}$ can also be separated into two
parts: the free lagrangian density ${\cal L}_F$ and interaction
lagrangian density ${\cal L}_I$
\be \label{19.46}
{\cal L} = {\cal L}_F + {\cal L}_I,
\ee
where
\be \label{19.47}
\ba{rcl}
{\cal L}_F &=& - \bar{\psi} ( \gamma^{\mu}
\partial_{\mu}  + m ) \psi
- \frac{1}{4} \eta^{\mu \rho} \eta^{\nu \sigma}
A^a_{0 \mu \nu} A^a_{0 \rho \sigma} \\
&&\\
&& - \frac{1}{16} \eta^{\mu \rho} \eta^{\nu \sigma} \eta_{\alpha \beta }
F_{0 \mu \nu}^{\alpha} F_{0 \rho \sigma}^{\beta}
- \frac{1}{8} \eta^{\mu \rho}
F^{\alpha}_{0 \mu \beta} F^{\beta}_{0 \rho \alpha}
+ \frac{1}{4} \eta^{\mu \rho}
F^{\alpha}_{0 \mu \alpha} F^{\beta}_{0 \rho \beta}  ,
\ea
\ee
\be \label{19.48}
\ba{rcl}
{\cal L}_I &=& - ( J(C) - 1 )
 \bar{\psi} ( \gamma^{\mu} \partial_{\mu}  + m ) \psi  \\
 &&\\
&&  - \frac{1}{4} \eta^{\mu \rho} \eta^{\nu \sigma}
(J(C) - 1 )
A^a_{0 \mu \nu} A^a_{0 \rho \sigma} \\
&&\\
&& + g J(C) \bar\psi \gamma^{\mu} (\partial_{\alpha} \psi)
C_{\mu}^{\alpha}
+ i g_s J(C) \bar\psi \gamma^{\mu} T_a \psi
A^a_{\mu}  \\
&&\\
&& + g  \eta^{\mu\rho} \eta^{\nu\sigma} J(C)
(\partial_{\mu} A^a_{\nu} -  \partial_{\nu} A^a_{\mu})
C_{\rho}^{\alpha} (\partial_{\alpha} A^a_{\sigma} )\\
&&\\
&& - \frac{g_s}{2}  \eta^{\mu\rho} \eta^{\nu\sigma} J(C)
f_{abc} A^b_{\rho} A^c_{\sigma}
(\partial_{\mu} A^a_{\nu} -  \partial_{\nu} A^a_{\mu}) \\
&&\\
&& - \frac{g}{2} \eta^{\mu\rho} \eta^{\nu\sigma} J(C)
G^{-1 \lambda}_{\alpha} A_{\lambda}^a
A^a_{\mu\nu} F_{\rho\sigma}^{\alpha} \\
&&\\
&& -\frac{g_s^2}{4} \eta^{\mu\rho} \eta^{\nu\sigma} J(C)
f_{abc} f_{ab_1c_1} A^b_{\mu} A^c_{\nu}
A^{b_1}_{\rho} A_{\sigma}^{c_1} \\
&&\\
&&+ g g_s \eta^{\mu\rho} \eta^{\nu\sigma} J(C)
f_{abc} A^b_{\mu} A^c_{\nu} C_{\rho}^{\alpha}
(\partial_{\alpha} A^a_{\sigma} ) \\
&&\\
&& -\frac{g^2}{2} \eta^{\mu\rho} \eta^{\nu\sigma} J(C)
(C_{\mu}^{\alpha} \partial_{\alpha} A^a_{\nu}
-  C_{\nu}^{\alpha} \partial_{\alpha} A^a_{\mu})
C_{\rho}^{\beta} (\partial_{\beta} A^a_{\sigma} ) \\
&&\\
&&  - \frac{g^2}{4} \eta^{\mu\rho} \eta^{\nu\sigma} J(C)
G^{-1 \lambda}_{\alpha} G^{-1 \kappa}_{\beta}
A^a_{\lambda} A^a_{\kappa}
F_{\mu\nu}^{\alpha} F_{\rho\sigma}^{\beta}\\
&&\\
&& + { self ~interaction~ terms ~of ~Gravitational~gauge~field}.
\ea
\ee
In above relations, $A^a_{0\mu\nu}$ is defined by
\be \label{19.49}
A^a_{0\mu\nu} = (\partial_{\mu} A^a_{\nu})
- (\partial_{\nu} A^a_{\mu}).
\ee
The explicite expression for $J(C)$ is
\be
J(C) = 1 + \sum_{m=1}^{\infty} \frac{1}{m!}
\left( \sum_{n=1}^{\infty}
\frac{g^n}{n}{\rm tr} (C^n)
\right)^m
\label{19.501}
\ee
From eq.(\ref{19.47}), we can write out propagators of Dirac field,
$SU(N)$ non-Abelian gauge field and gravitational gauge field. From
eq.(\ref{19.48}), we can write out Feynman rules for various interaction
vertexes and calculate Feynman diagrams for various interaction
precesses. We can also see that, because of the influence of the
factor $J(C)$,  matter fields can directly
couple to arbitrary number of gravitational gauge field, which
is important for the renormalization of the theory.
 \\

The equation of motion of Dirac field is
\be \label{19.50}
(\gamma^{\mu} {\mathbb D}_{\mu} + m  ) \psi =0.
\ee
The equation of motion of $SU(N)$ gauge field is
\be \label{19.51}
\partial^{\mu} {\mathbb A}^a_{\mu\nu}
= - g_s \eta_{\nu\sigma} J^{\sigma}_{a},
\ee
where
\be \label{19.52}
\ba{rcl}
J_a^{\nu} &=& i \bar\psi \gamma^{\nu} T_a \psi
+ \eta^{\nu\rho} \eta^{\nu_1 \sigma} f_{abc}
A^c_{\nu_1} {\mathbb A}^b_{\rho\sigma}
- \frac{g}{g_s} \eta^{\mu_1 \rho} \eta^{\nu\sigma}
\partial_{\mu} (C_{\mu_1}^{\mu} {\mathbb A}^a_{\rho\sigma} )\\
&&\\
&& - \frac{g}{2 g_s} \eta^{\mu_1 \rho} \eta^{\nu_1 \sigma}
G^{-1 \nu}_{\sigma_1} F^{\sigma_1}_{\mu_1 \nu_1}
{\mathbb A}^a_{\rho\sigma}
 + \frac{g}{g_s} \eta^{\mu\rho} \eta^{\nu\sigma}
G^{-1 \lambda}_{\tau} (D_{\mu} C_{\lambda}^{\tau} )
{\mathbb A}_{\rho\sigma}^a.
\ea
\ee
$J_a^{\nu}$ is a conserved current,
\be \label{19.53}
\partial_{\nu} J_a^{\nu} = 0.
\ee
When gravitational gauge field vanishes, the above current
$J_a^{\nu}$ returns to the conventional current in traditional
non-Abelian $SU(N)$ gauge field theory, which is
\be \label{19.54}
J_a^{\nu} = i \bar\psi \gamma^{\nu} T_a \psi
+ \eta^{\nu\rho} \eta^{\mu \sigma} f_{abc}
A^c_{\mu}  A^b_{\rho\sigma}.
\ee
But if gravitational gauge field does not vanish,
because of the influence from gravitational gauge field,
the conventional current eq.(\ref{19.54}) is no longer a
conserved current.
\\

The equation of motion of gravitational gauge field is
\be
\ba{rl}
\partial_{\mu} ( & \frac{1}{4} \eta^{\mu \rho}
\eta^{\nu \sigma} g_{\alpha \beta}
F_{\rho \sigma}^{\beta}
- \frac{1}{4} \eta^{\nu \rho} F^{\mu}_{\rho\alpha}
+ \frac{1}{4} \eta^{\mu \rho} F^{\nu}_{\rho\alpha} \\
&\\
& - \frac{1}{2} \eta^{\mu\rho} \delta^{\nu}_{\alpha}
F^{\beta}_{\rho\beta}
+ \frac{1}{2} \eta^{\nu\rho} \delta^{\mu}_{\alpha}
F^{\beta}_{\rho\beta})  = - g T_{g \alpha}^{\nu},
\label{19.55}
\ea
\ee
where $T_{g\alpha}^{\nu}$ is the gravitational energy-momentum
tensor
\be \label{19.56}
\ba{rcl}
T_{g\alpha}^{\nu}&=& \bar\psi \gamma^{\nu} \partial_{\alpha} \psi
- \eta^{\mu\rho} \eta^{\nu\sigma} {\mathbb A}^a_{\rho\sigma}
(\partial_{\alpha} A^a_{\mu})  \\
&& - G^{-1 \nu}_{\alpha}  \bar{\psi}
\lbrack \gamma^{\mu} (D_{\mu} -i g_s A_{\mu} ) + m
\rbrack \psi \\
&& - \frac{1}{4} G^{-1 \nu}_{\alpha}
\eta^{\mu \rho} \eta^{\lambda \sigma}
{\mathbb A}^a_{\mu \lambda } {\mathbb A}^a_{ \rho \sigma }\\
&&+ \eta^{\mu_1 \rho} \eta^{\nu \sigma}
\partial_{\mu} ( G^{-1 \lambda}_{\alpha} G^{\mu}_{\mu_1}
A^a_{\lambda} {\mathbb A}^a_{\rho\sigma} ) \\
&& + g \eta^{\mu\rho} \eta^{\nu \sigma}
G^{-1 \lambda_1}_{\tau} (D_{\mu} C^{\tau}_{\lambda_1} )
G^{-1 \lambda}_{\alpha} A^a_{\lambda} {\mathbb A}^a_{\rho\sigma}\\
&& - \frac{g}{2} \eta^{\mu_1 \rho} \eta^{\nu_1 \sigma}
G^{-1 \nu}_{\sigma_1} G^{-1 \lambda}_{\alpha}
A^a_{\lambda} {\mathbb A}^a_{\rho\sigma}
F^{\sigma_1}_{\mu_1 \nu_1} \\
&&- g \eta^{\mu_1 \rho} \eta^{\nu \sigma}
G^{-1 \lambda }_{\sigma_1} A^a_{\lambda}
(\partial_{\alpha} C_{\mu_1}^{\sigma_1} )
{\mathbb A}^a_{\rho\sigma}\\
&& + \frac{1}{4} \eta^{\nu\rho} \eta^{\lambda\sigma}
g_{\beta\gamma} F^{\beta}_{\rho\sigma}
(\partial_{\alpha} C_{\lambda}^{\gamma}) \\
&& + \frac{1}{4} \eta^{\nu\rho} G^{-1 \lambda}_{\beta}
G^{-1 \sigma}_{\gamma} F^{\beta}_{\rho\sigma}
(\partial_{\alpha} C_{\lambda}^{\gamma}) \\
&& - \frac{1}{2} \eta^{\nu\rho} G^{-1 \lambda}_{\gamma}
G^{-1 \sigma}_{\beta} F^{\beta}_{\rho\sigma}
(\partial_{\alpha} C_{\lambda}^{\gamma}) \\
&& - \frac{1}{4} \eta^{\lambda\rho} G^{-1 \nu}_{\beta}
G^{-1 \sigma}_{\gamma} F^{\beta}_{\rho\sigma}
(\partial_{\alpha} C_{\lambda}^{\gamma}) \\
&& + \frac{1}{2} \eta^{\lambda\rho} G^{-1 \nu}_{\gamma}
G^{-1 \sigma}_{\beta} F^{\beta}_{\rho\sigma}
(\partial_{\alpha} C_{\lambda}^{\gamma}) \\
&& + \frac{1}{4} \eta^{\lambda\rho} \eta^{\nu\sigma}
g_{\alpha\beta}
G^{-1 \kappa}_{\gamma} F^{\beta}_{\rho\sigma}
(D_{\lambda} C_{\kappa}^{\gamma}) \\
&& - \frac{1}{4} \eta^{\nu\rho} G^{-1 \sigma}_{\alpha}
G^{-1 \kappa}_{\gamma} F^{\mu}_{\rho\sigma}
(\partial_{\mu} C_{\kappa}^{\gamma}) \\
&& + \frac{1}{2} \eta^{\nu\rho} G^{-1 \sigma}_{\beta}
G^{-1 \kappa}_{\gamma} F^{\beta}_{\rho\sigma}
(\partial_{\alpha} C_{\kappa}^{\gamma}) \\
&& + \frac{1}{4} \eta^{\lambda\rho} G^{-1 \nu}_{\beta}
G^{-1 \sigma}_{\alpha}
G^{-1 \kappa}_{\gamma} F^{\beta}_{\rho\sigma}
(D_{\lambda} C_{\kappa}^{\gamma}) \\
&& - \frac{1}{2} \eta^{\lambda\rho} G^{-1 \nu}_{\alpha}
G^{-1 \sigma}_{\beta}
G^{-1 \kappa}_{\gamma} F^{\beta}_{\rho\sigma}
(D_{\lambda} C_{\kappa}^{\gamma}) \\
&& - \frac{1}{4} \eta^{\lambda\rho} \eta^{\nu\sigma}
\partial_{\mu} (g_{\alpha \beta} C^{\mu}_{\lambda}
 F^{\beta}_{\rho\sigma}) \\
&& - \frac{1}{4} \eta^{\nu\rho}
\partial_{\beta} (C^{\sigma}_{\lambda} G^{-1 \lambda}_{\alpha}
 F^{\beta}_{\rho\sigma}) \\
&&  + \frac{1}{2} \eta^{\nu\rho}
\partial_{\alpha} (C^{\sigma}_{\lambda} G^{-1 \lambda}_{\beta}
 F^{\beta}_{\rho\sigma}) \\
&& + \frac{1}{4g} \eta^{\lambda\rho}
\partial_{\mu} \lbrack ( G^{-1 \nu}_{\beta} G^{-1 \sigma}_{\alpha}
G^{\mu}_{\lambda} - \delta^{\nu}_{\beta} \delta^{\sigma}_{\alpha}
\delta^{\mu}_{\lambda}) F^{\beta}_{\rho\sigma}
\rbrack  \\
&& - \frac{1}{2 g} \eta^{\lambda\rho}
\partial_{\mu} \lbrack ( G^{-1 \nu}_{\alpha} G^{-1 \sigma}_{\beta}
G^{\mu}_{\lambda} - \delta^{\nu}_{\alpha} \delta^{\sigma}_{\beta}
\delta^{\mu}_{\lambda}) F^{\beta}_{\rho\sigma}
\rbrack  \\
&& - \frac{1}{4} \eta^{\kappa \rho} G^{-1 \nu}_{\beta}
G^{-1 \lambda}_{\alpha} G^{-1 \sigma}_{\gamma}
F^{\beta}_{\rho\sigma} F^{\gamma}_{\kappa\lambda}  \\
&& + \frac{1}{2} \eta^{\kappa \rho} G^{-1 \nu}_{\gamma}
G^{-1 \lambda}_{\alpha} G^{-1 \sigma}_{\beta}
F^{\beta}_{\rho\sigma} F^{\gamma}_{\kappa\lambda}  \\
&& - \frac{1}{8} \eta^{\mu \rho} \eta^{\lambda \sigma}
g_{\alpha \gamma} G^{-1 \nu}_{\beta}
F^{\beta}_{\rho\sigma} F^{\gamma}_{\mu\lambda}  \\
&& - \frac{1}{16} \eta^{\mu \rho} \eta^{\lambda \sigma}
g_{\beta \gamma} G^{-1 \nu}_{\alpha}
F^{\beta}_{\rho\sigma} F^{\gamma}_{ \mu\lambda}  \\
&& - \frac{1}{8} \eta^{\mu \rho} G^{-1 \nu}_{\alpha}
G^{-1 \lambda}_{\beta} G^{-1 \sigma}_{\gamma}
F^{\beta}_{\rho\sigma} F^{\gamma}_{\mu\lambda}  \\
&& + \frac{1}{4} \eta^{\mu \rho} G^{-1 \nu}_{\alpha}
G^{-1 \lambda}_{\gamma} G^{-1 \sigma}_{\beta}
F^{\beta}_{\rho\sigma} F^{\gamma}_{\mu\lambda}.
\ea
\ee
The global gravitational gauge symmetry of the system gives out
another energy-momentum tensor which is called inertial
energy-momentum tensor,
\be \label{19.57}
\ba{rcl}
T^{\mu}_{i \alpha} &= J(C) \left(  \right.&
\bar\psi \gamma^{\nu} G^{\mu}_{\nu} (\partial_{\alpha} \psi)
+ \eta^{\mu_1 \rho} \eta^{\nu \sigma}
G^{\mu}_{\mu_1} {\mathbb A}^a_{\rho\sigma}
(\partial_{\alpha} A^a_{\nu})    \\
&&\\
&& - \delta^{\mu}_{\alpha}  \bar{\psi}
\lbrack \gamma^{\nu} (D_{\nu} -i g_s A_{\nu} ) + m
\rbrack \psi \\
&&\\
&& - \frac{1}{4} \delta^{ \mu}_{\alpha}
\eta^{\nu \rho} \eta^{\lambda \sigma}
{\mathbb A}^a_{\nu \lambda } {\mathbb A}^a_{ \rho \sigma }\\
&&\\
&& + g\eta^{\mu_1 \rho} \eta^{\nu \sigma} G^{-1 \lambda}_{\beta}
G^{\mu}_{\mu_1} A^a_{\lambda} {\mathbb A}^a_{\rho\sigma}
(\partial_{\alpha} C^{\beta}_{\nu}) \\
&&\\
&&+ \frac{1}{4} \eta^{\lambda \rho}
\eta^{\nu \sigma} g_{\beta \gamma}
G^{\mu}_{\lambda} F_{\rho \sigma}^{\gamma}
(\partial_{\alpha} C_{\nu}^{\beta}) \\
&&\\
&&- \frac{1}{4} \eta^{\nu\rho} G^{-1 \sigma}_{\beta}
F_{\rho\sigma}^{\mu}
(\partial_{\alpha} C_{\nu}^{\beta}) \\
&&\\
&& + \frac{1}{4} \eta^{\lambda \rho} G^{-1 \nu}_{\gamma}
G^{-1 \sigma}_{\beta} G^{\mu}_{\lambda}
F_{\rho \sigma}^{\gamma}
(\partial_{\alpha} C_{\nu}^{\beta}) \\
&&\\
&& - \frac{1}{2} \eta^{\lambda \rho} G^{-1 \nu}_{\beta}
G^{-1 \sigma}_{\gamma} G^{\mu}_{\lambda}
F_{\rho \sigma}^{\gamma}
(\partial_{\alpha} C_{\nu}^{\beta}) \\
&&\\
&& + \frac{1}{2} \eta^{\nu\rho} \delta^{\mu}_{\beta}
G^{-1 \sigma}_{\gamma} F_{\rho \sigma}^{\gamma}
(\partial_{\alpha} C_{\nu}^{\beta}) \\
&&\\
&& - \frac{1}{16} \delta^{\mu}_{\alpha}
\eta^{\lambda \rho} \eta^{\nu\sigma}
g_{\beta\gamma}
F^{\gamma}_{\lambda\nu} F^{\beta}_{\rho\sigma} \\
&&\\
&& - \frac{1}{8} \delta^{\mu}_{\alpha}
\eta^{\lambda \rho} G^{-1 \nu}_{\beta}
G^{-1 \sigma}_{\gamma}
F^{\gamma}_{\lambda\nu} F^{\beta}_{\rho\sigma} \\
&&\\
&& \left .
+ \frac{1}{4} \delta^{\mu}_{\alpha}
\eta^{\lambda \rho} G^{-1 \nu}_{\gamma}
G^{-1 \sigma}_{\beta} F^{\gamma}_{\lambda\nu} F^{\beta}_{\rho\sigma}
\right ).
\ea
\ee
Compare eq.(\ref{19.56}) with eq.(\ref{19.57}), we can see that the
inertial energy-momentum tensor is not equivalent to the
gravitational energy-momentum tensor. In this case, they
are not equivalent even when gravitational field vanishes.
When gravitational field vanishes, the gravitational
energy-momentum tensor becomes $T^{\nu}_{0 g \alpha}$,
\be \label{19.58}
T_{0g\alpha}^{\nu} = \bar\psi \gamma^{\nu} \partial_{\alpha} \psi
- \eta^{\mu\rho}\eta^{\nu\sigma} A^a_{\rho\sigma}
(\partial_{\alpha} A^a_{\mu})
+ \eta^{\nu}_{1\alpha} {\cal L}_0
+ \eta^{\nu\sigma} \partial^{\mu}
(A^a_{\alpha} A^a_{\mu\sigma} ),
\ee
while inertial energy-momentum tensor becomes $T^{\nu}_{0 i \alpha}$
\be \label{19.59}
T_{0i\alpha}^{\nu} = \bar\psi \gamma^{\nu} \partial_{\alpha} \psi
- \eta^{\mu\rho}\eta^{\nu\sigma} A^a_{\rho\sigma}
(\partial_{\alpha} A^a_{\mu})
+ \delta^{\nu}_{\alpha} {\cal L}_0.
\ee
Therefore, we have
\be \label{19.60}
T_{0g\alpha}^{\nu} = T^{\nu}_{0i\alpha}
+ \eta^{\nu\sigma} \partial^{\mu}
(A^a_{\alpha} A^a_{\mu\sigma} ).
\ee
But this difference has no contribution on energy-momentum. The spatial
integration of time component of energy-momentum tensor gives out
energy-momentum of the system. The inertial energy-momentum
$P_{0 i \alpha}$ is
\be \label{19.61}
P_{0 i \alpha} = \int {\rm d}^3 x T^{0}_{0i\alpha},
\ee
and the gravitational energy-momentum $P_{0 g \alpha}$ is
\be \label{19.62}
P_{0 g \alpha} = \int {\rm d}^3 x T^{0}_{0g\alpha}.
\ee
So, their difference is
\be \label{19.63}
P_{0 g \alpha} - P_{0 i \alpha} = - \int {\rm d}^3 x
\partial_i (A^a_{\alpha} A^a_{i 0} )=0.
\ee
It means that, when gravitational gauge field vanishes, equivalence
principle holds. \\

\section{Unification of Fundamental Interactions}
\setcounter{equation}{0}

It is know that QED, QCD and unified electroweak theory are
all gauge theories. In this chapter, we will
discuss how to unify these gauge theories with gravitational
gauge theory, and how to unify four different kinds of
fundamental interactions formally. \\

First, let's discuss QED theory\cite{12}. As an example, let's discuss
electromagnetic interactions of Dirac field. The traditional
electromagnetic interactions between Dirac field $\psi$ and
electromagnetic field $A_{\mu}$ is
\be
- \frac{1}{4} \eta^{\mu \rho} \eta^{\nu \sigma}
A_{\mu \nu} A_{\rho \sigma}
- \bar{\psi}
( \gamma^{\mu} ( \partial_{\mu} - i e A_{\mu}  ) + m ) \psi.
\label{8.1}
\ee
The Lagrangian that describes gravitational gauge interactions
between gravitational gauge field and Dirac field or electromagnetic
field and describes electromagnetic interactions between Dirac field
and electromagnetic field is
\be
\ba{rcl}
{\cal L}_0 &=&
- \bar{\psi} (\gamma^{\mu} (D_{\mu} - i e A_{\mu}  ) + m ) \psi
- \frac{1}{4} \eta^{\mu \rho} \eta^{\nu \sigma}
{\mathbb A}_{\mu \nu} {\mathbb A}_{\rho \sigma}\\
&&\\
&& - \frac{1}{16} \eta^{\mu \rho}
\eta^{\nu \sigma} g_{\alpha \beta}
F^{\alpha}_{\mu \nu} F^{\beta}_{\rho \sigma} \\
&&\\
&& - \frac{1}{8} \eta^{\mu \rho}
G^{-1 \nu}_{\beta} G^{-1 \sigma}_{\alpha}
F^{\alpha}_{\mu \nu} F^{\beta}_{\rho \sigma} \\
&&\\
&& + \frac{1}{4} \eta^{\mu \rho}
G^{-1 \nu}_{\alpha} G^{-1 \sigma}_{\beta}
F^{\alpha}_{\mu \nu} F^{\beta}_{\rho \sigma},
\ea
\label{8.2}
\ee
where $D_{\mu}$ is the gravitational gauge covariant derivative
which is given by eq.(\ref{4.4}) and the strength of electromagnetic
field $A_{\mu}$ is
\be
{\mathbb A}_{\mu \nu} = A_{\mu \nu} + g G^{-1 \lambda}_{\alpha}
A_{\lambda} F_{\mu \nu}^{\alpha},
\label{8.3}
\ee
where $A_{\mu \nu}$ is given by eq.(\ref{7.4}) and $G^{-1}$ is given by
eq.(\ref{4.703}). The full Lagrangian density and the action of the
system are respectively given by,
\be
{\cal L} = J(C) {\cal L}_0,
\label{8.4}
\ee
\be
S = \int {\rm d}^4 x ~ {\cal L}.
\label{8.5}
\ee
\\

The system given by above Lagrangian has both $U(1)$ gauge
symmetry and gravitational gauge symmetry. Under $U(1)$ gauge
transformations,
\be
\psi(x) \to \psi'(x) = e^{-i \alpha(x)} \psi(x),
\label{8.6}
\ee
\be
A_{\mu}(x) \to A'_{\mu}(x) = A_{\mu}(x) - \frac{1}{e} D_{\mu}
\alpha(x), \label{8.7}
\ee
\be
C_{\mu}^{\alpha}(x) \to C^{\prime \alpha}_{\mu}(x) =
C_{\mu}^{\alpha}(x). \label{8.8}
\ee
It can be proved that the Lagrangian ${\cal L}$ is invariant
under $U(1)$ gauge transformation.
Under gravitational gauge transformations,
\be
\psi(x) \to \psi'(x) = (\ehat \psi(x)),
\label{8.9}
\ee
\be
A_{\mu}(x) \to A'_{\mu}(x) = (\ehat A_{\mu}(x)),
\label{8.10}
\ee
\be
C_{\mu}(x) \to  C'_{\mu}(x) = \ehat (x) C_{\mu} (x) \ehat^{-1} (x)
+ \frac{i}{g} \ehat (x) (\partial_{\mu} \ehat^{-1} (x)).
\label{8.11}
\ee
Using these relations, we can prove that the
action $S$ given by eq.(\ref{8.4}) is invariant under gravitational
gauge transformation. \\

Lagrangian ${\cal L}$ can be separated into free Lagrangian ${\cal L}_F$
and interaction Lagrangian ${\cal L}_I$,
\be
{\cal L} = {\cal L}_F + {\cal L}_I,
\label{8.12}
\ee
where
\be
\ba{rcl}
{\cal L}_F &=& - \frac{1}{4} \eta^{\mu \rho} \eta^{\nu
\sigma} A_{0 \mu \nu} A_{0 \rho \sigma}
- \bar{\psi} ( \gamma^{\mu}
\partial_{\mu}  + m ) \psi \\
&&\\
&&- \frac{1}{16} \eta^{\mu \rho} \eta^{\nu \sigma} \eta_{\alpha \beta }
F_{0 \mu \nu}^{\alpha} F_{0 \rho \sigma}^{\beta}
- \frac{1}{8} \eta^{\mu \rho}
F^{\alpha}_{0 \mu \beta} F^{\beta}_{0 \rho \alpha}
+ \frac{1}{4} \eta^{\mu \rho}
F^{\alpha}_{0 \mu \alpha} F^{\beta}_{0 \rho \beta},
\ea
\label{8.13}
\ee
\be
\begin{array}{rcl}
{\cal L}_I &=&
 ( J(C) - 1 ) \cdot \left\lbrack
- \frac{1}{4} \eta^{\mu \rho} \eta^{\nu
\sigma} A_{0 \mu \nu} A_{0 \rho \sigma}
- \bar{\psi} ( \gamma^{\mu}
\partial_{\mu}  + m ) \psi
 \right\rbrack \\
 &&\\
&& + i e \cdot J(C) \bar{\psi} \gamma^{\mu} \psi A_{\mu} \\
&&\\
&&+ g J(C) \bar\psi \gamma^{\mu} \partial_{\alpha} \psi
C_{\mu}^{\alpha}
 + g J(C) \eta^{\mu \rho} \eta^{\nu \sigma}
A_{0 \mu \nu} C_{\rho}^{\alpha} \partial_{\alpha} A_{\sigma} \\
&&\\
&&- \frac{g}{2} J(C) \eta^{\mu \rho} \eta^{\nu \sigma} A_{\mu
\nu} G^{-1 \lambda }_{\alpha} A_{\lambda} F_{\rho \sigma}^{\alpha}\\
&&\\
&& -\frac{g^2}{4} J(C) \eta^{\mu \rho} \eta^{\nu \sigma} G^{-1
\kappa}_{\alpha} G^{-1 \lambda }_{\beta} A_{\kappa} A_{\lambda}
F_{\mu \nu}^{\alpha} F_{\rho \sigma}^{\beta} \\
&&\\
&& - \frac{g^2}{2} J(C) \eta^{\mu \rho} \eta^{\nu \sigma}
\left( C_{\mu}^{\alpha} C_{\rho}^{\beta} (\partial_{\alpha} A_{\nu} )
(\partial_{\beta} A_{\sigma} )
- C_{\nu}^{\alpha} C_{\rho}^{\beta} (\partial_{\alpha} A_{\mu} )
(\partial_{\beta} A_{\sigma} ) \right )  \\
&&\\
&&  + ~ self ~interaction~ terms ~of ~Gravitational~gauge~field.
\end{array}
\label{8.14}
\ee
\\

The traditional Lagrangian for QCD is
\be
- \sum_n \bar{\psi}_n \lbrack \gamma^{\mu} (\partial_{\mu} -i g_c
A^i_{\mu } \frac{\lambda_i}{2} ) + m_n \rbrack \psi_n -\frac{1}{4}
\eta^{\mu \rho} \eta^{\nu \sigma} A^i_{\mu \nu } A^i_{\rho \sigma
},
\label{8.15}
\ee
where $\psi_n$ is the quark color triplet of the $n$th
flavor, $A_{\mu}^i $ is the color gauge vector
potential, $A_{\mu \nu }^i$ is the color gauge
covariant field strength tensor, $g_c$ is the strong
coupling constant, $\lambda_{\alpha}$ is the Gell-Mann
matrix and $m_n$ is the quark mass of the $n$th flavor.
In gravitational gauge theory, this Lagrangian should be
changed into
\be
\begin{array}{rcl}
{\cal L}_0 &=&
- \sum_n \bar{\psi}_n
\lbrack \gamma^{\mu} (D_{\mu} -i g_c A^i_{\mu}
\frac{\lambda^i}{2} ) + m_n
\rbrack \psi_n  \\
&&\\
&& -\frac{1}{4} \eta^{\mu \rho} \eta^{\nu \sigma}
{\mathbb A}^i_{\mu \nu } {\mathbb A}^i_{ \rho \sigma } \\
&&\\
&& - \frac{1}{16} \eta^{\mu \rho}
\eta^{\nu \sigma} g_{\alpha \beta}
F^{\alpha}_{\mu \nu} F^{\beta}_{\rho \sigma} \\
&&\\
&& - \frac{1}{8} \eta^{\mu \rho}
G^{-1 \nu}_{\beta} G^{-1 \sigma}_{\alpha}
F^{\alpha}_{\mu \nu} F^{\beta}_{\rho \sigma} \\
&&\\
&& + \frac{1}{4} \eta^{\mu \rho}
G^{-1 \nu}_{\alpha} G^{-1 \sigma}_{\beta}
F^{\alpha}_{\mu \nu} F^{\beta}_{\rho \sigma},
\end{array}
\label{8.16}
\ee
where
\be
{\mathbb A}^i_{\mu \nu } = A^i_{\mu \nu} + g G^{-1
\lambda}_{\sigma} A^i_{\lambda} F^{\sigma}_{\mu\nu},
\label{8.17}
\ee
\be
A^i_{\mu \nu } =
D_{\mu} A^i_{\nu } - D_{\nu} A^i_{ \mu }
+ g_c f_{i j k}
A^j_{ \mu } A^k_{\nu }.
\label{8.18}
\ee
It can be proved that this system has both $SU(3)_c$
gauge symmetry and gravitational gauge symmetry.
The unified electroweak model can be discussed in
similar way. \\

Now, let's try to construct a theory which can describe
all kinds of fundamental interactions in Nature\cite{14}. First we
know that the fundamental particles that we know are
fundamental fermions(such as leptons and quarks),
gauge bosons(such as photon, gluons, gravitons and
intermediate gauge bosons $W^{\pm}$ and $Z^0$), and
possible Higgs bosons. According to the Standard Model,
leptons form left-hand doublets and right-hand singlets.
Let's denote
\be
\psi^{(1)}_L =\left (
\begin{array}{c}
\nu_e  \\
e
\end{array}
\right )_L
~~~,~~~
\psi^{(2)}_L =\left (
\begin{array}{c}
\nu_{\mu}  \\
\mu
\end{array}
\right )_L
~~~,~~~
\psi^{(3)}_L =\left (
\begin{array}{c}
\nu_{\tau}  \\
\tau
\end{array}
\right )_L ,
\label{8.19}
\ee
\be
\psi^{(1)}_R=e_R
~~~,~~~
\psi^{(2)}_R=\mu_R
~~~,~~~
\psi^{(3)}_R=\tau_R.
\label{8.20}
\ee
Neutrinos have no right-hand singlets. The weak hypercharge
for left-hand doublets $\psi^{(i)}_L$ is $-1$ and for right-hand
singlet $\psi^{(i)}_R$ is $-2$. All leptons carry no color
charge. In order to define the wave
function for quarks, we have to introduce
Kabayashi-Maskawa mixing matrix first, whose general form is,
\be
K =
\left (
\begin{array}{ccc}
c_1 & s_1 c_3 & s_1 s_3 \\
-s_1 c_2 & c_1 c_2 c_3 - s_2 s_3 e^{i \delta}
& c_1 c_2 s_3 + s_2 c_3 e^{i \delta}   \\
s_1 s_2 & -c_1 s_2 c_3 -c_2 s_3 e^{i \delta}
& -c_1 s_2 s_3 +c_2 c_3 e^{i \delta}
\end{array}
\right )
\label{8.21}
\ee
where
\be
c_i = {\rm cos} \theta_i ~~,~~~~ s_i = {\rm sin} \theta_i ~~~(i=1,2,3)
\label{8.22}
\ee
and $\theta_i$ are generalized Cabibbo angles. The mixing between
three different quarks $d,s$ and $b$ is given by
\be
\left (
\begin{array}{c}
d_{\theta} \\
s_{\theta}  \\
b_{\theta}
\end{array}
\right )
= K
\left (
\begin{array}{c}
d\\
s\\
b
\end{array}
\right ).
\label{8.23}
\ee
Quarks also form left-hand doublets and right-hand singlets,
\be
q_L^{(1)a} =
\left (
\begin{array}{c}
u_L^a \\
d_{\theta L}^a
\end{array}
\right ) ,~~
q_L^{(2)a} =
\left (
\begin{array}{c}
c_L^a \\
s_{\theta L}^a
\end{array}
\right ),~~
q_L^{(3)a} =
\left (
\begin{array}{c}
t_L^a \\
b_{\theta L}^a
\end{array}
\right ) ,
\label{8.24}
\ee
\be
\begin{array}{ccc}
q_u^{(1)a}= u_R^a
& q_u^{(2)a}= c_R^a
& q_u^{(3)a}= t_R^a \\
q_{ \theta d}^{(1)a}= d_{\theta R}^a
& q_{ \theta d}^{(2)a}= s_{\theta R}^a
& q_{ \theta d}^{(3)a}= b_{\theta R}^a,
\end{array}
\label{8.25}
\ee
where index $a$ is color index.
It is known that left-hand doublets have weak isospin
$\frac{1}{2}$ and weak hypercharge  $\frac{1}{3}$,
right-hand singlets  have no weak isospin, $q_u^{(j)a}$s
have weak hypercharge $\frac{4}{3}$ and $q_{\theta d}^{(j)a}$s
have weak hypercharge $ - \frac{2}{3}$. \\

For gauge bosons, gravitational gauge field is also denoted
by $C_{\mu}^{\alpha}$. The gluon field is denoted $A_{\mu}$,
\be
A_{\mu} =
A_{\mu}^i \frac{\lambda^i}{2}.
\label{8.26}
\ee
The color gauge covariant field strength tensor is also
given by eq.(\ref{8.18}). The $U(1)_Y$ gauge field is denoted
by $B_{\mu}$ and $SU(2)$ gauge field is denoted by
$F_{\mu}$
\be
F_{\mu} =
F^n_{\mu} \frac{\sigma_n}{2},
\label{8.27}
\ee
where $\sigma_n$ is the Pauli matrix. The $U(1)_Y$ gauge field strength
tensor is given by
\be
{\mathbb B}_{\mu\nu} = B_{\mu\nu} + g G^{-1 \lambda}_{\alpha}
B_{\lambda} F^{\alpha}_{\mu\nu},
\label{8.28a}
\ee
where
\be B_{ \mu \nu} = D_{\mu} B_{ \nu} - D_{\nu} B_{\mu},
\label{8.28b}
\ee
and the $SU(2)$ gauge field strength tensor is given by
\be
{\mathbb F}^n_{\mu\nu} = F^n_{\mu\nu} + g G^{-1 \lambda}_{\alpha}
F^n_{\lambda} F^{\alpha}_{\mu\nu},
\label{8.29a}
\ee
\be
F_{\mu \nu}^n = D_{\mu} F_{\nu}^n - D_{\nu} F_{\mu}^n + g_w
\epsilon _{lmn} F_{\mu}^l    F_{\nu}^m,
\label{8.29b}
\ee
where $g_w$ is the coupling constant for $SU(2)$ gauge interactions
and the coupling constant for $U(1)_Y$ gauge interactions is $g'_w$.
\\

If there exist Higgs particles in Nature, the Higgs fields is
represented by a complex scalar $SU(2)$ doublet,
\be
\phi =
\left (
\begin{array}{c}
\phi^{\dagger} \\
\phi^0
\end{array}
\right ).
\label{8.30}
\ee
The hypercharge of Higgs field $\phi$ is $1$. \\

The Lagrangian ${\cal L}_0$ that describes four kinds of fundamental
interactions is given by
\be
\begin{array}{rcl}
{\cal L}_0 &=&
-\sum_{j=1}^{3} \overline{\psi}_L^{(j)} \gamma ^{\mu}
(D_{\mu}+ \frac{i}{2} g'_w  B_{\mu} -ig_w F_{\mu} ) \psi_L^{(j)} \\
&&\\
&&- \sum_{j=1}^{3}\overline{e}_R^{(j)} \gamma ^{\mu}
(D_{\mu}+ ig'_w  B_{\mu} ) e_R^{(j)}  \\
&&\\
&&-\sum_{j=1}^{3} \overline{q}_L^{(j)a} \gamma^{\mu}
\left( (D_{\mu}-ig_w F_{\mu}- \frac{i}{6}g'_w B_{\mu} )\delta_{ab}
-i g_c A^k_{\mu} (\frac{\lambda^k}{2})_{ab} \right) q_L^{(j)b} \\
&&\\
&&-\sum_{j=1}^{3} \overline{q}_u^{(j)a} \gamma^{\mu}
\left( (D_{\mu}-i \frac{2}{3} g'_w B_{\mu} )\delta_{ab}
-i g_c A^k_{\mu} (\frac{\lambda^k}{2})_{ab} \right) q_u^{(j)b}  \\
&&\\
&& -\sum_{j=1}^{3} \overline{q}_{\theta d}^{(j)a} \gamma^{\mu}
\left( (D_{\mu} + i \frac{1}{3} g'_w B_{\mu} )\delta_{ab}
-i g_c A^k_{\mu} (\frac{\lambda^k}{2})_{ab} \right) q_{\theta d}^{(j)b} \\
&& \\
&&-\frac{1}{4}  \eta^{\mu \rho} \eta^{\nu \sigma}
{\mathbb F}^{n}_{ \mu \nu} {\mathbb F}^n_{\rho \sigma}
-\frac{1}{4} \eta^{\mu \rho} \eta^{\nu \sigma}
{\mathbb B}_{\mu \nu} {\mathbb B}_{\rho \sigma}
 -\frac{1}{4} \eta^{\mu \rho} \eta^{\nu \sigma}
{\mathbb A}^i_{\mu \nu } {\mathbb A}^i_{ \rho \sigma } \\
&&\\
&& - \frac{1}{16} \eta^{\mu \rho}
\eta^{\nu \sigma} g_{\alpha \beta}
F^{\alpha}_{\mu \nu} F^{\beta}_{\rho \sigma}
 - \frac{1}{8} \eta^{\mu \rho}
G^{-1 \nu}_{\beta} G^{-1 \sigma}_{\alpha}
F^{\alpha}_{\mu \nu} F^{\beta}_{\rho \sigma} \\
&&\\
&& + \frac{1}{4} \eta^{\mu \rho}
G^{-1 \nu}_{\alpha} G^{-1 \sigma}_{\beta}
F^{\alpha}_{\mu \nu} F^{\beta}_{\rho \sigma} \\
&&\\
&& -\left\lbrack (D_{\mu}- \frac{i}{2}
g'_w  B_{\mu} -ig_w F_{\mu}) \phi \right\rbrack ^{\dagger}
\cdot \left\lbrack (D_{\mu}- \frac{i}{2}
g'_w  B_{\mu} -ig_w F_{\mu}) \phi \right\rbrack \\
&&\\
&&   - \mu^2 \phi^{\dagger} \phi
+ \lambda (\phi^{\dagger} \phi)^2  \\
&&\\
&& - \sum_{j=1}^{3} f^{(j)}
\left(\overline{e}_R^{(j)} \phi^{\dag} \psi_L^{(j)}
+\overline{\psi}_L^{(j)} \phi  e_R^{(j)}\right)  \\
&&\\
&& -\sum_{j=1}^{3} \left( f_u^{(j)} \overline{q}_L^{(j)a}
\overline{\phi}
q_u^{(j)a} + f_u^{(j) \ast} \overline{q}_u^{(j)a}
\overline{\phi}^{\dag} q_L^{(j)a} \right)   \\
&&\\
&&-\sum_{j,k=1}^{3} \left( f_d^{(jk)} \overline{q}_L^{(j)a} \phi
q_{\theta d}^{(k)a}
+ f_d^{(jk) \ast} \overline{q}_{\theta d}^{(k)a}
\phi^{\dag} q_L^{(j)a} \right),
\end{array}
\label{8.31}
\ee
where
\be
\overline{\phi} = i \sigma_{2} \phi^{\ast} =
\left (
\begin{array}{c}
\phi^{0 \dag} \\
- \phi
\end{array}
\right ).
\label{8.32}
\ee
The full Lagrangian is given by
\be
{\cal L} = J(C) {\cal L}_0
\label{8.33}
\ee
and the action of the system is
\be  \label{8.34}
S = \int {\rm d}^4 x {\cal L} .
\ee
This Lagrangian describes four kinds of  fundamental interactions
in Nature. \\

Now, let's discuss the symmetry of the system. First, let's study
$SU(3)_c$ symmetry. Denote the $SU(3)_c$ transformation matrix
as $U_3$. Under $SU(3)_c$ transformation, transformations of
various fields and operators are:
\be  \label{8.3401}
\psi_L^{(j)} \to \psi_L^{\prime (j)} = \psi_L^{(j)},
\ee
\be  \label{8.3402}
e_R^{(j)} \to e_R^{\prime (j)} = e_R^{(j)},
\ee
\be  \label{8.3403}
q_L^{(j)a} \to q_L^{\prime (j)a} = U_{3 ab} ~q_L^{(j)b},
\ee
\be  \label{8.3404}
q_u^{(j)a} \to q_u^{\prime (j)a} = U_{3 ab} ~q_u^{(j)b},
\ee
\be  \label{8.3405}
q_{\theta d}^{(j)a} \to q_{\theta d}^{\prime (j)a}
= U_{3 ab} ~q_{\theta d}^{(j)b},
\ee
\be  \label{8.3406}
C_{\mu}^{\alpha}  \to  C_{\mu}^{\prime \alpha}
 = C_{\mu}^{\alpha}
\ee
\be  \label{8.3407}
D_{\mu}  \to D'_{\mu} = D_{\mu},
\ee
\be  \label{8.35}
A_{\mu}  \to A'_{\mu} = U_3 A_{\mu} U_3^{-1}
- \frac{1}{i g_c} U_3 ( D_{\mu} U_3^{-1}),
\ee
\be  \label{8.36}
B_{\mu}  \to B'_{\mu} = B_{\mu},
\ee
\be  \label{8.37}
W_{\mu}  \to W'_{\mu} = W_{\mu},
\ee
\be  \label{8.38}
\phi  \to \phi' = \phi,
\ee
\be  \label{8.39}
J(C)  \to J'(C') = J(C).
\ee
According to above transformation rules, gauge field strength
tensors transform as
\be  \label{8.40}
{\mathbb W}_{\mu\nu} \to {\mathbb W}'_{\mu\nu}
= {\mathbb W}_{\mu\nu},
\ee
\be  \label{8.41}
{\mathbb B}_{\mu\nu} \to {\mathbb B}'_{\mu\nu}
= {\mathbb B}_{\mu\nu},
\ee
\be  \label{8.42}
F_{\mu\nu}^{\sigma} \to F_{\mu\nu}^{\prime\sigma}
= F_{\mu\nu}^{\sigma},
\ee
\be  \label{8.43}
A_{\mu\nu} \to A'_{\mu\nu} = U_3 A_{\mu\nu} U_3^{-1}
- \frac{ig}{g_c} F_{\mu\nu}^{\sigma} U_3
(\partial_{\sigma} U_3^{-1} ),
\ee
\be  \label{8.44}
{\mathbb A}_{\mu\nu} \to {\mathbb A}'_{\mu\nu}
= U_3 {\mathbb A}_{\mu\nu} U_3^{-1}.
\ee
Using all these transformation rules, we can prove that the lagrangian
density ${\cal L}$ is invariant. Therefore, the system has strict
local $SU(3)_c$ gauge symmetry. \\

Denote the transformations matrix of $SU(2)_L$ gauge transformation
as $U_2$.Under $SU(2)_L$ transformation, transformations of
various fields are:
\be  \label{8.45}
\psi_L^{(j)} \to \psi_L^{\prime (j)} =  U_2 \psi_L^{(j)},
\ee
\be  \label{8.46}
e_R^{(j)} \to e_R^{\prime (j)} = e_R^{(j)},
\ee
\be  \label{8.47}
q_L^{(j)a} \to q_L^{\prime (j)a} = U_{2} ~q_L^{(j)a},
\ee
\be  \label{8.48}
q_u^{(j)a} \to q_u^{\prime (j)a} =   q_u^{(j)a},
\ee
\be  \label{8.49}
q_{\theta d}^{(j)a} \to q_{\theta d}^{\prime (j)a}
=  q_{\theta d}^{(j)a},
\ee
\be  \label{8.50}
C_{\mu}^{\alpha}  \to  C_{\mu}^{\prime \alpha}
 = C_{\mu}^{\alpha}
\ee
\be  \label{8.51}
D_{\mu}  \to D'_{\mu} = D_{\mu},
\ee
\be  \label{8.52}
A_{\mu}  \to A'_{\mu} =   A_{\mu}  ,
\ee
\be  \label{8.53}
B_{\mu}  \to B'_{\mu} = B_{\mu},
\ee
\be  \label{8.54}
W_{\mu}  \to W'_{\mu} = U_2 W_{\mu} U_2^{-1}
- \frac{1}{i g_W} U_2 (D_{\mu} U_2^{-1} ),
\ee
\be  \label{8.55}
\phi  \to \phi' = U_2 \phi,
\ee
\be  \label{8.56}
J(C)  \to J'(C') = J(C).
\ee
According to above transformation rules, gauge field strength
tensors transform as
\be  \label{8.57}
{\mathbb A}_{\mu\nu} \to {\mathbb A}'_{\mu\nu}
= {\mathbb A}_{\mu\nu},
\ee
\be  \label{8.58}
{\mathbb B}_{\mu\nu} \to {\mathbb B}'_{\mu\nu}
= {\mathbb B}_{\mu\nu},
\ee
\be  \label{8.59}
F_{\mu\nu}^{\sigma} \to F_{\mu\nu}^{\prime\sigma}
= F_{\mu\nu}^{\sigma},
\ee
\be  \label{8.60}
W_{\mu\nu} \to W'_{\mu\nu} = U_2 W_{\mu\nu} U_2^{-1}
- \frac{ig}{g_W} F_{\mu\nu}^{\sigma} U_2
(\partial_{\sigma} U_2^{-1} ),
\ee
\be  \label{8.61}
{\mathbb W}_{\mu\nu} \to {\mathbb W}'_{\mu\nu}
= U_2 {\mathbb W}_{\mu\nu} U_2^{-1}.
\ee
Using all above realtions, we can prove that the
action of the system is invariant under
the above $SU(2)_L$ gauge transformations. So, the
model has $SU(2)_L$ gauge symmetry.\\

Under $U(1)_Y$ gauge transformation, transformations of
various fields are:
\be  \label{8.62}
\psi_L^{(j)} \to \psi_L^{\prime (j)}
=  e^{i \alpha(x) /2} ~\psi_L^{(j)},
\ee
\be  \label{8.63}
e_R^{(j)} \to e_R^{\prime (j)} =  e^{i \alpha(x)}~ e_R^{(j)},
\ee
\be  \label{8.64}
q_L^{(j)a} \to q_L^{\prime (j)a}
= e^{-i \alpha(x) /6} ~q_L^{(j)a},
\ee
\be  \label{8.65}
q_u^{(j)a} \to q_u^{\prime (j)a}
= e^{-2i \alpha(x) /3} ~ q_u^{(j)a},
\ee
\be  \label{8.66}
q_{\theta d}^{(j)a} \to q_{\theta d}^{\prime (j)a}
=  e^{i \alpha(x) /3} q_{\theta d}^{(j)a},
\ee
\be  \label{8.67}
C_{\mu}^{\alpha}  \to  C_{\mu}^{\prime \alpha}
 = C_{\mu}^{\alpha}
\ee
\be  \label{8.68}
D_{\mu}  \to D'_{\mu} = D_{\mu},
\ee
\be  \label{8.69}
A_{\mu}  \to A'_{\mu} =   A_{\mu}  ,
\ee
\be  \label{8.70}
B_{\mu}  \to B'_{\mu} = B_{\mu}
- \frac{1}{g'_W} (D_{\mu} \alpha(x)),
\ee
\be  \label{8.71}
W_{\mu}  \to W'_{\mu} =  W_{\mu} ,
\ee
\be  \label{8.72}
\phi  \to \phi' = e^{-i \alpha(x) /2} \phi,
\ee
\be  \label{8.73}
\bar\phi  \to \bar\phi' = e^{ i \alpha(x) /2} \bar\phi,
\ee
\be  \label{8.74}
J(C)  \to J'(C') = J(C).
\ee
According to above transformation rules, gauge field strength
tensors transform as
\be  \label{8.75}
{\mathbb A}_{\mu\nu} \to {\mathbb A}'_{\mu\nu}
= {\mathbb A}_{\mu\nu},
\ee
\be  \label{8.76}
{\mathbb W}_{\mu\nu} \to {\mathbb W}'_{\mu\nu}
= {\mathbb W}_{\mu\nu},
\ee
\be  \label{8.77}
F_{\mu\nu}^{\sigma} \to F_{\mu\nu}^{\prime\sigma}
= F_{\mu\nu}^{\sigma},
\ee
\be  \label{8.78}
B_{\mu\nu} \to B'_{\mu\nu} =  B_{\mu\nu}
+ \frac{ g}{g'_W} F_{\mu\nu}^{\sigma}
(\partial_{\sigma} \alpha(x) ),
\ee
\be  \label{8.79}
{\mathbb B}_{\mu\nu} \to {\mathbb B}'_{\mu\nu}
=   {\mathbb B}_{\mu\nu} .
\ee
Using all above relations, we can also prove that
the action of the system is invariant under
the above $U(1)_Y$ gauge transformations. So, the model has
$U(1)_Y$ gauge symmetry\\

Gravitational gauge transformations of various fields are
\be  \label{8.80}
\psi_L^{(j)} \to \psi_L^{\prime (j)}
=  (\ehat \psi_L^{(j)}),
\ee
\be  \label{8.81}
e_R^{(j)} \to e_R^{\prime (j)} =  (\ehat e_R^{(j)}),
\ee
\be  \label{8.82}
q_L^{(j)a} \to q_L^{\prime (j)a}
= (\ehat q_L^{(j)a}),
\ee
\be  \label{8.83}
q_u^{(j)a} \to q_u^{\prime (j)a}
= (\ehat  q_u^{(j)a}),
\ee
\be  \label{8.84}
q_{\theta d}^{(j)a} \to q_{\theta d}^{\prime (j)a}
=  (\ehat q_{\theta d}^{(j)a}),
\ee
\be  \label{8.85}
C_{\mu}   \to  C_{\mu}^{\prime }
 = \ehat C_{\mu} \ehat^{-1}
 - \frac{1}{ig} \ehat (\partial_{\mu} \ehat^{-1}),
\ee
\be  \label{8.86}
D_{\mu}  \to D'_{\mu} = \ehat D_{\mu} \ehat^{-1},
\ee
\be  \label{8.87}
A_{\mu}  \to A'_{\mu} =  (\ehat A_{\mu})  ,
\ee
\be  \label{8.88}
B_{\mu}  \to B'_{\mu} = (\ehat B_{\mu} ),
\ee
\be  \label{8.89}
W_{\mu}  \to W'_{\mu} = (\ehat  W_{\mu} ),
\ee
\be  \label{8.90}
\phi  \to \phi' = (\ehat \phi) ,
\ee
\be  \label{8.91}
\bar\phi  \to \bar\phi' =(\ehat  \bar\phi),
\ee
\be  \label{8.92}
J(C)  \to J'(C') = J \cdot ( \ehat J(C)),
\ee
\be  \label{8.93}
g^{\alpha\beta}  \to g^{\prime\alpha\beta}
= \Lambda^{\alpha}_{~\alpha_1}
\Lambda^{\beta}_{~\beta_1}
(\ehat g^{\alpha_1 \beta_1}),
\ee
\be  \label{8.94}
g_{\alpha\beta}  \to g^{\prime }_{\alpha\beta}
= \Lambda_{\alpha}^{~\alpha_1} \Lambda_{\beta}^{~\beta_1}
(\ehat g_{\alpha_1 \beta_1}).
\ee
According to above transformation rules, gauge field strength
tensors transform as
\be  \label{8.95}
{\mathbb A}_{\mu\nu} \to {\mathbb A}'_{\mu\nu}
= (\ehat {\mathbb A}_{\mu\nu}),
\ee
\be  \label{8.96}
{\mathbb W}_{\mu\nu} \to {\mathbb W}'_{\mu\nu}
= (\ehat {\mathbb W}_{\mu\nu}),
\ee
\be  \label{8.97}
{\mathbb B}_{\mu\nu} \to {\mathbb B}'_{\mu\nu}
= (\ehat  {\mathbb B}_{\mu\nu}) .
\ee
\be  \label{8.98}
F_{\mu\nu}^{\alpha} \to F_{\mu\nu}^{\prime\alpha}
= \Lambda^{\alpha}_{~\beta} (\ehat F_{\mu\nu}^{\beta}),
\ee
According to these transformations, the lagrangian density
${\cal L}_0$ transforms covariantly,
\be  \label{8.99}
{\cal L}_0 \to {\cal L}'_0 = (\ehat {\cal L_0}).
\ee
So,
\be  \label{8.100}
{\cal L}  \to  {\cal L}' = J \cdot (\ehat {\cal L}).
\ee
Using eq.(\ref{4.27}), we can prove that action $S$ is invariant
under gravitational gauge transformations,
\be  \label{8.101}
S \to S' = S.
\ee
Therefore, the system has gravitational gauge symmetry. \\

Now, as a whole, we discuss
$$
(SU(3)_c \times SU(2)_L \times U(1)_Y) \otimes_s
Gravitational ~Gauge ~Group
$$
gauge symmetry. In order to do this, we need define generator
operators. The generator operators of $SU(3)_c$ group are denoted
by $\hat T_{3j}$. The $SU(3)_c$ transformation operator $\hat U_3$
is defined by
\be  \label{8.102}
\hat U_3 = e^{ -i \alpha^j \hat T_{3j} }.
\ee
Matrix $U_3$ is defined by
\be  \label{8.103}
U_3 = e^{-i \alpha^j \lambda^j /2 }.
\ee
When $\hat T_{3j}$ acts on fields, it will becomes
the corresponding representation matrix of generators. So,
\be  \label{8.104}
\hat T_{3i} \psi_L^{(j)} = 0,
\ee
\be  \label{8.105}
\hat T_{3i} e_R^{(j)} = 0,
\ee
\be  \label{8.106}
\hat T_{3i} q_L^{(j)a} =
\left( \frac{\lambda^i}{2} \right)_{ab}
q_L^{(j)b},
\ee
\be  \label{8.107}
\hat T_{3i} q_u^{(j)a} =
\left( \frac{\lambda^i}{2} \right)_{ab}
q_u^{(j)b},
\ee
\be  \label{8.108}
\hat T_{3i} q_{\theta d}^{(j)a} =
\left( \frac{\lambda^i}{2} \right)_{ab}
q_{\theta d}^{(j)b},
\ee
\be  \label{8.109}
\hat T_{3i} \phi = 0,
\ee
\be  \label{8.110}
\lbrack \hat T_{3i} ~~,~~ C_{\mu}^{\alpha} \rbrack = 0,
\ee
\be  \label{8.111}
\lbrack \hat T_{3i} ~~,~~ A_{\mu} \rbrack
= \left\lbrack \frac{\lambda^i}{2}
~~,~~A_{\mu}  \right\rbrack,
\ee
\be  \label{8.112}
\lbrack \hat T_{3i} ~~,~~ B_{\mu} \rbrack = 0,
\ee
\be  \label{8.113}
\lbrack \hat T_{3i} ~~,~~ W_{\mu} \rbrack = 0,
\ee
\be  \label{8.114}
\lbrack \hat T_{3i} ~~,~~ F_{\mu\nu}^{\sigma} \rbrack = 0,
\ee
\be  \label{8.115}
\lbrack \hat T_{3i} ~~,~~ {\mathbb A}_{\mu\nu} \rbrack
= \left\lbrack \frac{\lambda^i}{2}
~~,~~{\mathbb A}_{\mu\nu}  \right\rbrack,
\ee
\be  \label{8.116}
\lbrack \hat T_{3i} ~~,~~ {\mathbb B}_{\mu\nu} \rbrack =0,
\ee
\be  \label{8.117}
\lbrack \hat T_{3i} ~~,~~ {\mathbb W}_{\mu\nu} \rbrack =0.
\ee
The generator operators of $SU(2)_L$ group are denoted
by $\hat T_{2l}$. The $SU(2)_L$ transformation operator $\hat U_2$
is defined by
\be  \label{8.118}
\hat U_2 = e^{ -i \alpha^l \hat T_{2l} }.
\ee
Matrix $U_2$ is defined by
\be  \label{8.119}
U_2 = e^{-i \alpha^l \sigma_l /2 }.
\ee
When $\hat T_{2l}$ acts on fields, it will becomes
the corresponding representation matrix of generators. So,
\be  \label{8.120}
\hat T_{2l} \psi_L^{(j)} = \frac{\sigma_l}{2} \psi_L^{(j)}
\ee
\be  \label{8.121}
\hat T_{2l} e_R^{(j)} = 0,
\ee
\be  \label{8.122}
\hat T_{2l} q_L^{(j)a} =
 \frac{\sigma_l}{2} q_L^{(j)a},
\ee
\be  \label{8.123}
\hat T_{2l} q_u^{(j)a} = 0,
\ee
\be  \label{8.124}
\hat T_{2l} q_{\theta d}^{(j)a} = 0,
\ee
\be  \label{8.125}
\hat T_{2l} \phi = \frac{\sigma_l}{2} \phi
\ee
\be  \label{8.126}
\lbrack \hat T_{2l} ~~,~~ C_{\mu}^{\alpha} \rbrack = 0,
\ee
\be  \label{8.127}
\lbrack \hat T_{2l} ~~,~~ A_{\mu} \rbrack =0,
\ee
\be  \label{8.128}
\lbrack \hat T_{2l} ~~,~~ B_{\mu} \rbrack = 0,
\ee
\be  \label{8.129}
\lbrack \hat T_{2l} ~~,~~ W_{\mu} \rbrack =
\left\lbrack \frac{\sigma_l}{2} ~~,~~W_{\mu} \right\rbrack,
\ee
\be  \label{8.130}
\lbrack \hat T_{2l} ~~,~~ F_{\mu\nu}^{\sigma} \rbrack = 0,
\ee
\be  \label{8.131}
\lbrack \hat T_{2l} ~~,~~ {\mathbb A}_{\mu\nu} \rbrack =0,
\ee
\be  \label{8.132}
\lbrack \hat T_{2l} ~~,~~ {\mathbb B}_{\mu\nu} \rbrack =0,
\ee
\be  \label{8.133}
\lbrack \hat T_{2l} ~~,~~ {\mathbb W}_{\mu\nu} \rbrack =
\left\lbrack \frac{\sigma_l}{2} ~~,~~
{\mathbb W}_{\mu\nu} \right\rbrack .
\ee
The generator operators of $U(1)_Y$ group are denoted
by $\hat T_1$. $2\hat T_1$ is the hypercharge operator.
The $U(1)_Y$ transformation operator $\hat U_1$
is defined by
\be  \label{8.134}
\hat U_1 = e^{ -i \alpha  \hat T_1 }.
\ee
Matrix $U_1$ is defined by
\be  \label{8.135}
U_1 = e^{-i \alpha Y },
\ee
where $Y$ is the hypercharge of the corresponding field.
When $2 \hat T_1$ acts on fields, it will becomes the hypercharge
of the corresponding fields. So,
\be  \label{8.136}
\hat T_{1} \psi_L^{(j)} = - \frac{1}{2} \psi_L^{(j)}
\ee
\be  \label{8.137}
\hat T_{1} e_R^{(j)} = - e_R^{(j)}
\ee
\be  \label{8.138}
\hat T_{1} q_L^{(j)a} =
 \frac{1}{6} q_L^{(j)a},
\ee
\be  \label{8.139}
\hat T_{1} q_u^{(j)a} = \frac{2}{3}  q_u^{(j)a},
\ee
\be  \label{8.140}
\hat T_{1} q_{\theta d}^{(j)a} = -\frac{1}{3} q_{\theta d}^{(j)a}
\ee
\be  \label{8.141}
\hat T_{1} \phi = \frac{1}{2} \phi
\ee
\be  \label{8.142}
\lbrack \hat T_{1} ~~,~~ C_{\mu}^{\alpha} \rbrack = 0,
\ee
\be  \label{8.143}
\lbrack \hat T_{1} ~~,~~ A_{\mu} \rbrack =0,
\ee
\be  \label{8.144}
\lbrack \hat T_{1} ~~,~~ B_{\mu} \rbrack = B_{\mu},
\ee
\be  \label{8.145}
\lbrack \hat T_{1} ~~,~~ W_{\mu} \rbrack = 0,
\ee
\be  \label{8.146}
\lbrack \hat T_{1} ~~,~~ F_{\mu\nu}^{\sigma} \rbrack = 0,
\ee
\be  \label{8.147}
\lbrack \hat T_{1} ~~,~~ {\mathbb A}_{\mu\nu} \rbrack =0,
\ee
\be  \label{8.148}
\lbrack \hat T_{1} ~~,~~ {\mathbb B}_{\mu\nu} \rbrack
= {\mathbb B}_{\mu\nu} ,
\ee
\be  \label{8.149}
\lbrack \hat T_{1} ~~,~~ {\mathbb W}_{\mu\nu} \rbrack = 0.
\ee
Different generator operators act on different spaces, so they
commute each other,
\be  \label{8.150}
\lbrack \hat T_{1} ~~,~~\hat T_{2l} \rbrack = 0,
\ee
\be  \label{8.151}
\lbrack \hat T_{1} ~~,~~ \hat T_{3i}\rbrack= 0,
\ee
\be  \label{8.152}
\lbrack \hat T_{1} ~~,~~ \hat P_{\alpha} \rbrack= 0,
\ee
\be  \label{8.153}
\lbrack \hat T_{2l} ~~,~~ \hat T_{3i} \rbrack= 0,
\ee
\be  \label{8.154}
\lbrack \hat T_{2l} ~~,~~ \hat P_{\alpha} \rbrack= 0,
\ee
\be  \label{8.155}
\lbrack \hat T_{3i} ~~,~~ \hat P_{\alpha} \rbrack= 0.
\ee
As we have mention before, what generators commute each other
does not means that group elements commute each other. \\

A general element of semi-direct product group
$$
(SU(3)_c \times SU(2)_L \times U(1)_Y) \otimes_s
Gravitational ~Gauge ~Group
$$
is denoted by $g(x)$. It can be proved that the $g(x)$ can be
written into the following form
\be  \label{8.156}
g(x) = \ehat \hat U_1 \hat U_2 \hat U_3.
\ee
Define quark color triplet states,
\be \label{8.157}
q_{L}^{(j)} =
\left (
\begin{array}{c}
 q_{L}^{(j)1} \\
 q_{L}^{(j)2}  \\
 q_{L}^{(j)3}
\end{array}
\right ),
\ee
\be \label{8.158}
q_{u}^{(j)} =
\left (
\begin{array}{c}
 q_{u}^{(j)1} \\
 q_{u}^{(j)2}  \\
 q_{u}^{(j)3}
\end{array}
\right ),
\ee
\be \label{8.159}
q_{\theta d}^{(j)} =
\left (
\begin{array}{c}
 q_{\theta d}^{(j)1} \\
 q_{\theta d}^{(j)2}  \\
 q_{\theta d}^{(j)3}
\end{array}
\right ).
\ee
Then, we have the following relations
\be  \label{8.160}
\hat T_{3j} q_{L}^{(j)}
= \frac{\lambda^j}{2} q_L^{(j)} ,
\ee
\be  \label{8.161}
\hat T_{3j} q_{u}^{(j)}
= \frac{\lambda^j}{2} q_u^{(j)} ,
\ee
\be  \label{8.162}
\hat T_{3j} q_{\theta d}^{(j)}
= \frac{\lambda^j}{2} q_{\theta d}^{(j)} .
\ee
Under gauge transformations of semi-direct product group,
various fields and operators transform as
\be  \label{8.163}
\psi_L^{(j)} \to \psi_L^{\prime (j)}
=  (g(x) \psi_L^{(j)}),
\ee
\be  \label{8.164}
e_R^{(j)} \to e_R^{\prime (j)} =  (g(x) e_R^{(j)}),
\ee
\be  \label{8.165}
q_L^{(j)a} \to q_L^{\prime (j) }
= (g(x) q_L^{(j) }),
\ee
\be  \label{8.166}
q_u^{(j)a} \to q_u^{\prime (j) }
= (g(x)  q_u^{(j) }),
\ee
\be  \label{8.167}
q_{\theta d}^{(j)a} \to q_{\theta d}^{\prime (j) }
=  (g(x) q_{\theta d}^{(j) }),
\ee
\be  \label{8.168}
C_{\mu}   \to  C_{\mu}^{\prime }
 = \ehat C_{\mu} \ehat^{-1}
 - \frac{1}{ig} \ehat (\partial_{\mu} \ehat^{-1}),
\ee
\be  \label{8.169}
D_{\mu}  \to D'_{\mu} = \ehat D_{\mu} \ehat^{-1},
\ee
\be  \label{8.170}
F_{\mu\nu}  \to F'_{\mu\nu}
=  \ehat F_{\mu\nu} \ehat^{-1}  ,
\ee
\be  \label{8.171}
A_{\mu}  \to A'_{\mu}
= g(x) \left\lbrack   A_{\mu} - \frac{1}{i g_c}
(D_{\mu} U_3^{-1}) U_3 \right\rbrack g^{-1} (x) ,
\ee
\be  \label{8.172}
W_{\mu}  \to W'_{\mu}
= g(x) \left\lbrack   W_{\mu} - \frac{1}{i g_W}
(D_{\mu} U_2^{-1}) U_2 \right\rbrack g^{-1} (x) ,
\ee
\be  \label{8.173}
B_{\mu}  \to B'_{\mu}
= g(x) \left\lbrack   B_{\mu} - \frac{1}{i g'_W}
(D_{\mu} U_1^{-1}) U_1 \right\rbrack g^{-1} (x) ,
\ee
\be  \label{8.174}
\phi  \to \phi' = (g(x) \phi) ,
\ee
\be  \label{8.175}
J(C)  \to J'(C') = J \cdot g(x)  J(C) g^{-1}(x),
\ee
\be  \label{8.176}
g^{\alpha\beta}  \to g^{\prime \alpha \beta}
= \Lambda^{\alpha}_{~\alpha_1}
\Lambda^{\beta}_{~\beta_1}
g(x) g^{\alpha_1 \beta_1} g^{-1}(x),
\ee
\be  \label{8.177}
g_{\alpha\beta}  \to g^{\prime }_{\alpha\beta}
= \Lambda_{\alpha}^{~\alpha_1} \Lambda_{\beta}^{~\beta_1}
g(x) \eta_{\alpha_1 \beta_1} g^{-1}(x),
\ee
\be  \label{8.178}
{\mathbb A}_{\mu\nu} \to {\mathbb A}'_{\mu\nu}
= g(x) {\mathbb A}_{\mu\nu} g^{-1}(x),
\ee
\be  \label{8.179}
{\mathbb W}_{\mu\nu} \to {\mathbb W}'_{\mu\nu}
= g(x) {\mathbb W}_{\mu\nu} g^{-1}(x),
\ee
\be  \label{8.180}
{\mathbb B}_{\mu\nu} \to {\mathbb B}'_{\mu\nu}
= g(x)  {\mathbb B}_{\mu\nu} g^{-1}(x) .
\ee
Using all these relations, we can prove
that the action of the system has strict local
$$
(SU(3)_c \times SU(2)_L \times U(1)_Y) \otimes_s
Gravitational ~Gauge ~Group
$$
gauge symmetry.\\

It is known that, $SU(3)_c$ color symmetry and gravitational gauge
symmetry are strict symmetry. $SU(2)_L \times U(1)_Y$ symmetry
are not strict symmetry, which is broken to $U(1)_Q$ symmetry.
Now, let's discuss spontaneously symmetry breaking of the system.
The potential of Higgs field is
\be  \label{8.181}
- \mu^2 \phi^{\dag} \phi
+ \lambda (\phi^{\dag} \phi  )^2.
\ee
If,
\be  \label{8.182}
\mu^2 > 0,~~~~~~
\lambda>0,
\ee
the symmetry of vacuum will be spontaneously broken. Suppose that
the vacuum expectation value of neutral Higgs field is non-zero,
that is
\be  \label{8.183}
\langle \phi \rangle_0 =
\left(
\ba{c}
0\\
\frac{v}{\sqrt{2}}
\ea
\right),
\ee
where,
\be  \label{8.184}
v = \sqrt{\frac{\mu^2}{\lambda}  }.
\ee
After a local $SU(2)_L$ gauge transformation, we can select
the Higgs field $\phi(x)$ as,
\be  \label{8.185}
 \phi (x) = \frac{1}{\sqrt{2}}
\left(
\ba{c}
0\\
v + \varphi(x)
\ea
\right).
\ee
After symmetry breaking, the Higgs potential becomes,
\be  \label{8.186}
V( \varphi ) =
\mu^2 \varphi^2 + \lambda v \varphi^3
+\frac{\lambda}{4} \varphi^4
- \frac{\mu^4}{4 \lambda},
\ee
from which we know that the mass of Higgs field is $2 \mu^2$.\\

Define
\be  \label{8.187}
W_{\mu}^{\pm} \define
\frac{1}{\sqrt{2}} \left(
W_{\mu}^1 \mp i W_{\mu}^2,
\right)
\ee
\be  \label{8.188}
A^e_{\mu} \define \cos \theta_W B_{\mu}
+ \sin \theta_W W_{\mu}^3,
\ee
\be  \label{8.189}
Z_{\mu} \define \sin \theta_W B_{\mu}
- \cos \theta_W W_{\mu}^3,
\ee
where
\be  \label{8.190}
{\rm tg} \theta_W = \frac{g'_W}{g_W}.
\ee
$A^e_{\mu}$ in eq.(\ref{8.188}) is the electromagnetic field.
Define two mass parameters as
\be  \label{8.191}
m_W = \frac{v}{2} g_W ,
\ee
\be  \label{8.192}
m_Z = \frac{v}{2} \sqrt{g_W^2 + g_W^{\prime 2}} .
\ee
It is known that, in the standard model, $m_W$ is the mass of
$W^{\pm}$ bosons and $m_Z$ is the mass of $Z$ bosons.
The coupling constant of electromagnetic interactions is
denoted by e,
\be  \label{8.193}
{\rm e} \define g'_W \cos \theta_W
= \frac{g_W g'_W}{\sqrt{g_W^2 + g_W^{\prime 2} }}.
\ee
The current of strong interactions are denoted by
$J^{\mu}_{c i}$,
\be  \label{8.194}
J^{\mu}_{c i} = i \left(
\bar u \gamma^{\mu} \frac{\lambda^i}{2} u
+ \bar c \gamma^{\mu} \frac{\lambda^i}{2} c
+ \bar t \gamma^{\mu} \frac{\lambda^i}{2} t
+ \bar d \gamma^{\mu} \frac{\lambda^i}{2} d
+ \bar s \gamma^{\mu} \frac{\lambda^i}{2} s
+ \bar b \gamma^{\mu} \frac{\lambda^i}{2} b
\right).
\ee
The current of electromagnetic interactions is
\be  \label{8.195}
\ba{rcl}
J^{\mu}_{em} & = & i \left(
- \bar e \gamma^{\mu} e
- \bar \mu \gamma^{\mu} \mu
- \bar \tau \gamma^{\mu} \tau
+ \frac{2}{3} \bar u \gamma^{\mu} u
+ \frac{2}{3} \bar c \gamma^{\mu} c \right.  \\
&&\\
&&\left.
   + \frac{2}{3} \bar t \gamma^{\mu} t
   - \frac{1}{3} \bar d \gamma^{\mu} d
   - \frac{1}{3} \bar s \gamma^{\mu} s
   - \frac{1}{3} \bar b \gamma^{\mu} b
\right).
\ea
\ee
The currents for weak interactions are
\be  \label{8.196}
\ba{rcl}
J^{\mu +}_{W} & = & \frac{i}{2 \sqrt{2}} \left(
 \bar \nu_e \gamma^{\mu} (1 + \gamma_5) e
+ \bar \nu_{\mu} \gamma^{\mu} (1 + \gamma_5) \mu
+ \bar \nu_{\tau} \gamma^{\mu} (1 + \gamma_5) \tau  \right.\\
&&  \\
&&\left.
   + \bar u \gamma^{\mu} (1 + \gamma_5) d_{\theta}
   + \bar c \gamma^{\mu} (1 + \gamma_5) s_{\theta}
   + \bar t \gamma^{\mu} (1 + \gamma_5) b_{\theta}
\right),
\ea
\ee
\be  \label{8.197}
\ba{rcl}
J^{\mu -}_{W} & = & \frac{i}{2 \sqrt{2}} \left(
  \bar e \gamma^{\mu} (1 + \gamma_5) \nu_e
+ \bar \mu \gamma^{\mu} (1 + \gamma_5) \nu_{\mu}
+ \bar \tau \gamma^{\mu} (1 + \gamma_5) \nu_{\tau}  \right.\\
&&  \\
&&\left.
   + \bar d_{\theta} \gamma^{\mu} (1 + \gamma_5) u
   + \bar s_{\theta} \gamma^{\mu} (1 + \gamma_5) c
   + \bar b_{\theta} \gamma^{\mu} (1 + \gamma_5) t
\right),
\ea
\ee
\be  \label{8.198}
J_z^{\mu} = J_3^{\mu} - \sin ^2 \theta_W J^{\mu}_{em},
\ee
where
\be  \label{8.199}
\ba{rcl}
J^{\mu  }_{3} & = & \frac{i}{2 } \left(
+ \bar \nu_{eL} \gamma^{\mu}  \nu_{eL}
+ \bar \nu_{\mu L} \gamma^{\mu}  \nu_{\mu L}
+ \bar \nu_{\tau L} \gamma^{\mu}  \nu_{\tau L}
- \bar e_L \gamma^{\mu}  e_L
- \bar \mu_L \gamma^{\mu}  \mu_L
- \bar \tau_L \gamma^{\mu}  \tau_L  \right.  \\
&&\\
&& \left.
+ \bar u_L \gamma^{\mu}  u_L
+ \bar c_L \gamma^{\mu}  c_L
+ \bar t_L \gamma^{\mu}  t_L
- \bar d_L \gamma^{\mu}  d_L
- \bar s_L \gamma^{\mu}  s_L
- \bar b_L \gamma^{\mu}  b_L
\right).
\ea
\ee
The current for gravitational interactions is
\be  \label{8.200}
\ba{rcl}
J^{\mu  }_{g \alpha} & = &
 \bar \nu_{eL} \gamma^{\mu} \partial_{\alpha} \nu_{eL}
+ \bar \nu_{\mu L} \gamma^{\mu} \partial_{\alpha}  \nu_{\mu L}
+ \bar \nu_{\tau L} \gamma^{\mu} \partial_{\alpha}  \nu_{\tau L}
+ \bar e_L \gamma^{\mu} \partial_{\alpha}  e_L  \\
&&\\
&&
+ \bar \mu_L \gamma^{\mu} \partial_{\alpha}  \mu_L
+ \bar \tau_L \gamma^{\mu} \partial_{\alpha}  \tau_L
+ \bar u_L \gamma^{\mu} \partial_{\alpha}  u_L
+ \bar c_L \gamma^{\mu} \partial_{\alpha}  c_L  \\
&&\\
&&
+ \bar t_L \gamma^{\mu} \partial_{\alpha}  t_L
+ \bar d_L \gamma^{\mu} \partial_{\alpha}  d_L
+ \bar s_L \gamma^{\mu} \partial_{\alpha}  s_L
+ \bar b_L \gamma^{\mu} \partial_{\alpha}  b_L \\
&&\\
&& + \bar e_R \gamma^{\mu} \partial_{\alpha}  e_R
+ \bar \mu_R \gamma^{\mu} \partial_{\alpha}  \mu_R
+ \bar \tau_R \gamma^{\mu} \partial_{\alpha}  \tau_R
+ \bar u_R \gamma^{\mu} \partial_{\alpha}  u_R  \\
&&\\
&& + \bar c_R \gamma^{\mu} \partial_{\alpha}  c_R
+ \bar t_R \gamma^{\mu} \partial_{\alpha}  t_R
+ \bar d_{\theta R} \gamma^{\mu} \partial_{\alpha}  d_{\theta R}
+ \bar s_{\theta R} \gamma^{\mu} \partial_{\alpha}  s_{\theta R}  \\
&&\\
&& + \bar b_{\theta R} \gamma^{\mu} \partial_{\alpha}  b_{\theta R}
.
\ea
\ee
Denote
\be  \label{8.201}
J_{\varphi} = \frac{1}{v} (
m_e \bar e e
+ m_{\mu} \bar \mu \mu
+ m_{\tau} \bar \tau \tau
+ m_u \bar u u
+ m_c \bar c c
+ m_t \bar t t
+ m_d \bar d d
+ m_s \bar s s
+ m_b \bar b b
),
\ee
where
\be     \label{8.20101}
m_i=\frac{1}{\sqrt{2}} f_i v, ~~~~~
(i=e,\mu,\tau,u,c,t,d,s,b).
\ee
Then lagrangian density ${\cal L}_0$ can be written into the
following form,
\be  \label{8.202}
\ba{rcl}
{\cal L}_0 & = &
- \bar \nu_{eL} \gamma^{\mu} \partial_{\mu} \nu_{eL}
- \bar \nu_{\mu L} \gamma^{\mu} \partial_{\mu}  \nu_{\mu L}
- \bar \nu_{\tau L} \gamma^{\mu} \partial_{\mu}  \nu_{\tau L}
- \bar e  ( \gamma^{\mu} \partial_{\mu} + m_e )  e   \\
&&\\
&&
- \bar \mu  ( \gamma^{\mu} \partial_{\mu} + m_{\mu} ) \mu
- \bar \tau ( \gamma^{\mu} \partial_{\mu} + m_{\tau} )  \tau
- \bar u  (\gamma^{\mu} \partial_{\mu} + m_u ) u
- \bar c  (\gamma^{\mu} \partial_{\mu} + m_c ) c   \\
&&\\
&&
- \bar t  (\gamma^{\mu} \partial_{\mu} + m_t ) t
- \bar d  (\gamma^{\mu} \partial_{\mu} + m_d ) d
- \bar s  (\gamma^{\mu} \partial_{\mu} + m_s ) s
- \bar b  (\gamma^{\mu} \partial_{\mu} + m_b ) b   \\
&&\\
&&
+ g_c J_{c i}^{\mu} A_{\mu}^i
+ g_W J_W^{\mu -} W_{\mu}^+
+ g_W J_W^{\mu +} W_{\mu}^-
- \frac{g_W}{\cos \theta_W} J_z^{\mu} Z_{\mu}
+ {\rm e} J^{\mu}_{em} A^e_{\mu}
+ g J^{\mu}_{g \alpha} C_{\mu}^{\alpha}  \\
&&\\
&&
- \frac{1}{2} \eta^{\mu\nu} (D_{\mu} \varphi)(D_{\nu}\varphi)
- \frac{1}{4} g_W^2 \eta^{\mu\nu}
W_{\mu}^+  W_{\nu}^- (2 v \varphi + \varphi^2)\\
&&\\
&&
-\frac{1}{8} (g_W^2 + g_W^{\prime 2}) \eta^{\mu\nu}
Z_{\mu} Z_{\nu} (2 v \varphi + \varphi^2)
- v(\varphi) - J_{\varphi} \phi  \\
&&\\
&& - \frac{1}{4} \eta^{\mu\rho} \eta^{\nu\sigma}
{\mathbb W}^n_{\mu\nu} {\mathbb W}^n_{\rho\sigma}
- m_W^2 \eta^{\mu\nu} W^+_{\mu} W^-_{\nu}
- \frac{1}{2} m^2_Z \eta^{\mu\nu} Z_{\mu} Z_{\nu} \\
&&\\
&&
- \frac{1}{4} \eta^{\mu\rho} \eta^{\nu\sigma}
{\mathbb B}_{\mu\nu} {\mathbb B}_{\rho\sigma}
- \frac{1}{4} \eta^{\mu\rho} \eta^{\nu\sigma}
{\mathbb A}^i_{\mu\nu} {\mathbb A}^i_{\rho\sigma}
- \frac{1}{16} \eta^{\mu \rho}
\eta^{\nu \sigma} g_{\alpha \beta}
F^{\alpha}_{\mu \nu} F^{\beta}_{\rho \sigma} \\
&&\\
&& - \frac{1}{8} \eta^{\mu \rho}
G^{-1 \nu}_{\beta} G^{-1 \sigma}_{\alpha}
F^{\alpha}_{\mu \nu} F^{\beta}_{\rho \sigma}
+ \frac{1}{4} \eta^{\mu \rho}
G^{-1 \nu}_{\alpha} G^{-1 \sigma}_{\beta}
F^{\alpha}_{\mu \nu} F^{\beta}_{\rho \sigma}.
\ea
\ee
\\

According to above discussions, before spontaneously symmetry
breaking,  the system has
$$
(SU(3)_c \times SU(2)_L \times U(1)_Y) \otimes_s
Gravitational ~Gauge ~Group
$$
gauge symmetry. After spontaneously symmetry breaking, the system
has
$$
(SU(3)_c  \times U(1)_Q) \otimes_s
Gravitational ~Gauge ~Group
$$
gauge symmetry.
Four different kinds of fundamental interactions
are unified in the same lagrangian. So, the lagrangian density
${\cal L}$ which is given by eq.(\ref{8.31}) and eq.(\ref{8.33})
can be used to calculate any fundamental interaction process
in Nature. \\

\section{Classical Limit of Quantum Gauge general Relativity}
\setcounter{equation}{0}

In this chapter, we mainly discuss leading order approximation
of quantum gauge general relativity, which will give out
classical Newton's theory of gravity. \\

When we discuss classical limit of quantum gauge general relativity,
we will qualitatively discuss anther important
problem at the same time. It is
know that, in usual gauge theory, such as QED, the coulomb
force between two objects which carry like electric charges
is always mutual repulsive. Gravitational gauge theory is
also a kind of gauge theory, is the force between two static
massive objects attractive or repulsive?  \\

Suppose that the gravitational field is very weak, so that
$g C_{\mu}^{\alpha}$ is a first order infinitesimal
quantity. In leading order approximation, both inertial
energy-momentum tensor and gravitational energy-momentum tensor
give the same results, which we denoted as $T^{\mu}_{\alpha}$.
The space integration of the time component of the current
$T^{\mu}_{\alpha}$ gives out the energy-momentum of
the system. First, we discuss classical limit of the field
equation of gravitational gauge field. Then, we discuss
classical limit of the Einstein's  field equation.
We will find that two field equations give out the same
classical limit, which is what we expected, for these
two field equations are essentially the same. \\

The field equation (\ref{4.41}) of gravitational
gauge field in the leading order is:
\be
\ba{rcl}
-\frac{1}{4} \eta^{\nu\sigma} \eta_{\alpha\beta}
\partial^{\mu} \partial_{\mu} C_{\sigma}^{\beta}
+\frac{1}{4}  \eta_{\alpha\beta}
\partial^{\nu} \partial^{\mu} C_{\mu}^{\beta}
+\frac{1}{4} \eta^{\nu\rho}
\partial_{\mu} \partial_{\rho} C_{\alpha}^{\mu} &&\\
&&\\
-\frac{1}{4}
\partial^{\mu} \partial_{\mu} C_{\alpha}^{\nu}
+\frac{1}{4}
\partial^{\mu} \partial_{\alpha} C_{\mu}^{\nu}
+\frac{1}{2} \delta^{\nu}_{\alpha}
\partial^{\mu} \partial_{\mu} C_{\beta}^{\beta} &&\\
&&\\
-\frac{1}{2} \delta^{\nu}_{\alpha}
\partial^{\mu} \partial_{\beta} C_{\mu}^{\beta}
-\frac{1}{2}
\partial^{\nu} \partial_{\alpha} C_{\beta}^{\beta}
+\frac{1}{4} \eta^{\nu\rho}
\partial_{\alpha} \partial_{\mu} C_{\rho}^{\mu}
&=&  g T^{\nu}_{\alpha},
\ea
\label{9.5}
\ee
where $\partial^{\mu} = \eta^{\mu\nu} \partial_{\nu}$.
In order to simplify the above field equation, harmonic
gauge will be used. In general relativity,
the harmonic coordinate conditions are represented by
\be
\Gamma^{\gamma} \define
g^{\alpha\beta} \Gamma^{\gamma}_{\alpha\beta} = 0,
\label{9.501}
\ee
where $\Gamma^{\gamma}_{\alpha\beta}$ is given by
(\ref{4.1903}). In leading order approximation, it gives
out the following restriction
\be
\partial_{\mu} C_{\nu}^{\mu} - \partial_{\nu} C_{\mu}^{\mu}
= - \eta_{\lambda\nu} \partial^{\mu} C_{\mu}^{\lambda}.
\label{9.502}
\ee
After the harmonic coordinate conditions are considered,
eq. (\ref{9.5}) is changed into
\be
- \frac{1}{4} \eta^{\nu\sigma} \eta_{\alpha \beta}
\partial^{\mu} \partial_{\mu} C_{\sigma}^{\beta}
- \frac{1}{4}
\partial^{\mu} \partial_{\mu} C_{\alpha}^{\nu}
+ \frac{1}{4} \delta^{\nu}_{\alpha}
\partial^{\mu} \partial_{\mu} C_{\beta}^{\beta}
 =g T^{\nu}_{\alpha}.
\label{9.503}
\ee
Multiply both side of the above equation with $\delta_{\nu}^{\alpha}$,
we can get
\be
\partial^{\mu} \partial_{\mu} C_{\beta}^{\beta}
 =2 g T^{\beta}_{\beta}.
\label{9.504}
\ee
Then, eq. (\ref{9.503}) is changed into
\be
- \frac{1}{4} \eta^{\nu\sigma} \eta_{\alpha \beta}
\partial^{\mu} \partial_{\mu} C_{\sigma}^{\beta}
- \frac{1}{4}
\partial^{\mu} \partial_{\mu} C_{\alpha}^{\nu}
 =g \left( T^{\nu}_{\alpha} - \frac{1}{2} \delta^{\nu}_{\alpha}
 T^{\beta}_{\beta} \right).
\label{9.505}
\ee
As a classical limit approximation,
let's consider static gravitational interactions
between two static objects. In this case,
the leading order component of energy-momentum tensor is $T^0_0$, other
components of energy-momentum tensor is a first order infinitesimal
quantity. So, we only need to consider the field
equation (\ref{9.503}) of $\nu = \alpha = 0$, which now becomes
\be
 \partial_{\mu} \partial^{\mu} C_0^0
= -  g T^0_0.
\label{9.6}
\ee
For static problems, all time derivatives vanish. Therefor,
the above equation is changed into
\be
 \nabla ^2 C_0^0
= - g T^0_0.
\label{9.7}
\ee
This is just the Newton's equation of gravitational field. Suppose
that there is only one point object at the origin of the coordinate
system. Because $T^0_0$ is the negative value of energy density,
we can let
\be
T^0_0 = - M \delta(\svec{x}) .
\label{9.8}
\ee
Applying
\be
\nabla ^2  \frac{1}{r} = - 4 \pi \delta(\svec{x}),
\label{9.9}
\ee
with $r =  |\svec{x} |$, we get
\be
C^0_0  = - \frac{gM}{4 \pi r}.
\label{9.10}
\ee
This is just the gravitational potential which is expected in Newton's
theory of gravity. \\

Suppose that there is another point object at the position of
point $\svec{x}$ with mass $m$. The gravitational potential energy
between these two objects is that
\be
V(r) = \int {\rm d}^3 \svec{y} {\cal H}_I
=  -g \int {\rm d}^3 \svec{y} T^0_{2~0} (\svec{x}) C_0^0,
\label{9.11}
\ee
with $C^0_0$ is the gravitational potential generated by the first
point object, and $T^0_{2~0}$ is the $(0,0)$ component of the
energy-momentum tensor of the second object,
\be
T^0_{2~0} (\svec{y})= - m  \delta(\svec{y} - \svec{x}) .
\label{9.12}
\ee
The final result for gravitational potential energy between two
point objects is
\be
V(r) = - \frac{g^2 M m}{4 \pi r}.
\label{9.13}
\ee
The gravitational potential energy between two point objects is always
negative, which is what expected by Newton's theory of gravity and is
the inevitable result of the attractive nature of gravitational
interactions.
\\

The gravitational force that the first point object acts on the second
point object is
\be
\svec{f} = - \nabla V(r) = - \frac{g^2 M m}{4 \pi r^2} \hat{r},
\label{9.14}
\ee
where $\hat r = \svec{r} / r $. This is the famous formula
of Newton's gravitational force.  Therefore, in the classical limit,
the gravitational gauge theory can return to Newton's theory of
gravity. Besides, from eq.(\ref{9.14}), we can clearly see that the
gravitational interaction force between two point objects is
attractive.
\\

Now, we want to ask a problem: why in QED, the force between
two like electric charges is always repulsive, while in
gravitational gauge theory, the force between two like
gravitational charges can be attractive?
A simple answer to this fundamental problem is that the attractive
nature of the gravitational force is an inevitable result of
the global Lorentz symmetry of the system. Because of the requirement
of global Lorentz symmetry, the Lagrangian
function given by eq.(\ref{4.20}) must use
$g_{\alpha \beta}$  and $\eta^{\mu\rho}$,
 can not use the ordinary $\delta$-function $\delta_{\alpha \beta}$
 or $\delta^{\mu\rho}$.
It can be easily prove that, if we use $\delta_{\alpha \beta}$
or $\delta^{\mu\rho}$
instead of $g_{\alpha \beta}$ or $\eta^{\mu\rho}$
in eq.(\ref{4.20}), the Lagrangian
of pure gravitational gauge field is not invariant under global
Lorentz transformation.  On the other hand,
if we use $\delta_{\alpha \beta}$ or $\delta^{\mu\rho}$
instead of  $g_{\alpha \beta}$ or $\eta^{\mu\rho}$
in eq.(\ref{4.20}), the gravitational force will be repulsive which obviously
contradicts with experiment results. In QED, $\delta_{a b}$ is
used to construct the Lagrangian for electromagnetic fields, therefore,
the interaction force between two like electric charges is
always repulsive.
\\

Another form of the field equation of gravitational gauge
field is the Einstein's field equation (\ref{4.5310})
\be
R_{\alpha\beta} -\frac{1}{2} g_{\alpha\beta} R
= - 8 \pi G T_{\alpha\beta}.
\label{9.15}
\ee
Now, we discuss its classical limit. Multiply both side
of the above equation with $g^{\alpha\beta}$, we get
\be
R =  8 \pi G T_{\gamma}^{\gamma}.
\label{9.16}
\ee
Therefore, eq.(\ref{9.15}) can be changed into
\be
R_{\alpha\beta} = - 8 \pi G S_{\alpha\beta},
\label{9.17}
\ee
where
\be
S_{\alpha\beta} = T_{\alpha\beta}
- \frac{1}{2} g_{\alpha\beta} T^{\gamma}_{\gamma}.
\label{9.18}
\ee
$R_{\alpha\beta}$ is given by eq. (\ref{4.1907}). In first order
approximation, it is
\be
\ba{rcl}
R_{\alpha\beta} & = &
g \partial_{\alpha} \partial_{\beta} C_{\mu}^{\mu}
- \frac{g}{2} \partial_{\alpha} \partial_{\lambda}
C_{\beta}^{\lambda}
- \frac{g}{2} \partial_{\beta} \partial_{\lambda}
C_{\alpha}^{\lambda}  \\
&&\\
&& + \frac{g}{2} \eta^{\lambda\nu} \eta_{\beta\gamma}
\partial_{\nu} \partial_{\lambda}
C_{\alpha}^{\gamma}
+ \frac{g}{2} \eta^{\lambda\nu} \eta_{\alpha\gamma}
\partial_{\nu} \partial_{\lambda}
C_{\beta}^{\gamma}  \\
&&\\
&& - \frac{g}{2} \eta^{\mu\nu} \eta_{\beta\gamma}
\partial_{\nu} \partial_{\alpha}
C_{\mu}^{\gamma}
- \frac{g}{2} \eta^{\mu\nu} \eta_{\alpha\gamma}
\partial_{\nu} \partial_{\beta}
C_{\mu}^{\gamma}
+ o((gC)^2).
\ea
\label{9.19}
\ee
Applying harmonic coordinate conditions (\ref{9.502}), we can
simplify the above equation into
\be
R_{\alpha\beta} =
 \frac{g}{2} \eta^{\lambda\nu} \eta_{\beta\gamma}
\partial_{\nu} \partial_{\lambda}
C_{\alpha}^{\gamma}
+ \frac{g}{2} \eta^{\lambda\nu} \eta_{\alpha\gamma}
\partial_{\nu} \partial_{\lambda}
C_{\beta}^{\gamma} .
\label{9.20}
\ee
For classical limit, we only need to consider the
field equation (\ref{9.17}) with $\alpha=\beta=0$, which is
\be
-g \partial^{\nu} \partial_{\nu} C_0^0
= - 8 \pi G S_{00}
\label{9.21}
\ee
Using the following relation
\be
g^2 = 4 \pi G,
\label{9.22}
\ee
\be
S_{00} = \frac{1}{2} T_{00} = - \frac{1}{2} T^0_0,
\label{9.23}
\ee
we can change eq. (\ref{9.21}) into
\be
\partial^{\nu} \partial_{\nu} C_0^0
=  - g T_{0}^{0},
\label{9.21}
\ee
which is the same as eq.(\ref{9.6}). Therefore, the Einstein's
field equation (\ref{4.5310}) gives out the same classical
limit as that of field equation (\ref{4.41}), which is what
we expected.
\\

\section{Path Integral Quantization of Gravitational\\ Gauge Fields}
\setcounter{equation}{0}

For the sake of simplicity, in this chapter and the next chapter, we only
discuss pure gravitational gauge field. For pure gravitational gauge
field, its Lagrangian function is
\be
\ba{rcl}
{\cal L} & = & - \frac{1}{16} \eta^{\mu \rho}
\eta^{\nu \sigma} g_{\alpha \beta} J(C)
F^{\alpha}_{\mu \nu} F^{\beta}_{\rho \sigma} \\
&&\\
&& - \frac{1}{8} \eta^{\mu \rho}
G^{-1 \nu}_{\beta} G^{-1 \sigma}_{\alpha} J(C)
F^{\alpha}_{\mu \nu} F^{\beta}_{\rho \sigma} \\
&&\\
&& + \frac{1}{4} \eta^{\mu \rho}
G^{-1 \nu}_{\alpha} G^{-1 \sigma}_{\beta} J(C)
F^{\alpha}_{\mu \nu} F^{\beta}_{\rho \sigma}.
\ea
\label{10.1}
\ee
Its space-time integration gives out the action of the system
\be
S = \int {\rm d}^4 x  {\cal L}.
\label{10.2}
\ee
This action has local gravitational gauge symmetry. Gravitational
gauge field $C_{\mu}^{\alpha}$ has $4 \times 4 = 16$ degrees of
freedom. But, if gravitons are massless, the system has only
$2 \times 4 = 8$ degrees of freedom. There are gauge degrees
of freedom in the theory. Because only physical degrees of
freedom can be quantized, in order to quantize the system, we
have to introduce gauge conditions to eliminate un-physical degrees
of freedom. For the sake of convenience, we take temporal gauge
conditions
\be
C_0^{\alpha} = 0, ~~~(\alpha = 0,1,2,3).
\label{10.3}
\ee
\\

In temporal gauge, the generating functional $W\lbrack J \rbrack$ is
given by
\be
W\lbrack J \rbrack = N \int \lbrack {\cal D} C\rbrack
\left(\prod_{\alpha, x} \delta( C_0^{\alpha}(x))\right)
exp \left\lbrace i \int {\rm d}^4 x ( {\cal L}
+ J^{\mu}_{\alpha} C^{\alpha}_{\mu}),
\right\rbrace
\label{10.4}
\ee
where $N$ is the normalization constant, $J^{\mu}_{\alpha}$
is a fixed external source and $ \lbrack {\cal D} C\rbrack $ is the
integration measure,
\be
\lbrack {\cal D} C\rbrack
= \prod_{\mu=0}^{3} \prod_{\alpha= 0}^{3} \prod_j
\left(\varepsilon {\rm d}C_{\mu}^{\alpha} (\tau_j)
/ \sqrt{2 \pi i \hbar} \right).
\label{10.5}
\ee
We use this generation functional as our starting
point of the path integral quantization of gravitational gauge field. \\

Generally speaking, the action of the system has local gravitational gauge
symmetry, but the gauge condition has no local gravitational gauge
symmetry. If we make a local gravitational gauge transformations,
the action of the system is kept unchanged while gauge condition will
be changed. Therefore, through local gravitational gauge transformation,
we can change one gauge condition into another gauge condition. The most
general gauge condition is
\be
f^{\alpha} (C(x)) - \varphi^{\alpha} (x) = 0,
\label{10.6}
\ee
where $\varphi^{\alpha}(x)$ is an arbitrary space-time function.
The Fadeev-Popov determinant $\Delta_f(C)$ \cite{f01} is defined by
\be
\Delta_f^{-1} (C) \equiv
\int \lbrack {\cal D} g \rbrack
\prod_{x, \alpha} \delta \left(f^{\alpha} ( ^gC(x))
- \varphi^{\alpha}(x) \right),
\label{10.7}
\ee
where $g$ is an element of gravitational gauge group, $^gC$ is the gravitational
gauge field after gauge transformation $g$ and
$\lbrack {\cal D} g \rbrack $ is the integration measure on
gravitational gauge group
\be
\lbrack {\cal D} g \rbrack
= \prod_x {\rm d}^4 \epsilon(x),
\label{10.8}
\ee
where $\epsilon (x)$ is the transformation parameter of $\ehat$.
Both $\lbrack {\cal D} g \rbrack $ and $\lbrack {\cal D} C \rbrack $
are not invariant under gravitational gauge transformation. Suppose
that,
\be
\lbrack {\cal D} (gg') \rbrack
= J_1(g') \lbrack {\cal D} g \rbrack,
\label{10.9}
\ee
\be
\lbrack {\cal D} ~^gC \rbrack
= J_2(g) \lbrack {\cal D} C \rbrack.
\label{10.10}
\ee
$J_1(g)$ and $J_2(g)$ satisfy the following relations
\be
J_1(g) \cdot J_1(g^{-1}) = 1,
\label{10.11}
\ee
\be
J_2(g) \cdot J_2(g^{-1}) = 1.
\label{10.12}
\ee
It can be proved that, under gravitational gauge transformations,
the Fadeev-Popov determinant transforms as
\be
\Delta_f^{-1} ( ^{g'}C ) = J_1^{-1}(g') \Delta_f^{-1} (C).
\label{10.13}
\ee
\\

Insert eq.(\ref{10.7}) into eq.(\ref{10.4}), we get
\be
\begin{array}{rcl}
W\lbrack J \rbrack &=& N \int \lbrack {\cal D} g \rbrack
\int \lbrack {\cal D} C\rbrack~~
\left\lbrack \prod_{\alpha, y}
\delta( C_0^{\alpha}(y)) \right\rbrack \cdot
\Delta_f  (C) \\
&&\\
&&\cdot \left\lbrack \prod_{\beta, z}
\delta ( f^{\beta} ( ^gC(z))- \varphi^{\beta}(z) )\right\rbrack
 \cdot exp \left\lbrace i \int {\rm d}^4 x ( {\cal L}
+ J^{\mu}_{\alpha} C^{\alpha}_{\mu})
\right\rbrace .
\end{array}
\label{10.14}
\ee
Make a gravitational gauge transformation,
\be
C(x)~~ \to ~~^{g^{-1}} C(x),
\label{10.15}
\ee
then,
\be
^g C(x)~~ \to
~~^{gg^{-1}} C(x).
\label{10.16}
\ee
After this transformation, the generating functional is changed
into
\be
\begin{array}{rcl}
W\lbrack J \rbrack &=& N \int \lbrack {\cal D} g \rbrack
\int \lbrack {\cal D} C\rbrack~~
J_1(g) J_2(g^{-1}) \cdot \left\lbrack \prod_{\alpha, y}
\delta( ^{g^{-1}}C_0^{\alpha}(y)) \right\rbrack \cdot
\Delta_f  (C) \\
&&\\
&&\cdot \left\lbrack \prod_{\beta, z}
\delta ( f^{\beta} ( C(z))- \varphi^{\beta}(z) )\right\rbrack
 \cdot exp \left\lbrace i \int {\rm d}^4 x ( {\cal L}
+ J^{\mu}_{\alpha} \cdot  ^{g^{-1}} \!\!\! C^{\alpha}_{\mu})
\right\rbrace.
\end{array}
\label{10.17}
\ee
\\

Suppose that the gauge transformation $g_0(C)$ transforms general
gauge condition $f^{\beta}(C) - \varphi^{\beta} = 0$ to temporal
gauge condition $C_0^{\alpha} = 0$, and suppose that this transformation
$g_0(C)$ is unique. Then two $\delta$-functions in eq.(\ref{10.17}) require
that the integration on gravitational gauge group must be in the
neighborhood of $g^{-1}_0(C)$. Therefore eq.(\ref{10.17}) is changed into
\be
\begin{array}{rcl}
W\lbrack J \rbrack &=& N \int \lbrack {\cal D} C\rbrack~~
\Delta_f  (C)  \cdot \left\lbrack \prod_{\beta, z}
\delta ( f^{\beta} ( C(z))- \varphi^{\beta}(z) )\right\rbrack \\
&&\\
&& \cdot exp \left\lbrace i \int {\rm d}^4 x ( {\cal L}
+ J^{\mu}_{\alpha} \cdot  ^{g_0}\!C^{\alpha}_{\mu})
\right\rbrace \\
&&\\
&& \cdot J_1(g_0^{-1}) J_2(g_0) \cdot \int \lbrack {\cal D} g \rbrack
\left\lbrack \prod_{\alpha, y}
\delta( ^{g^{-1}}C_0^{\alpha}(y))\right\rbrack.
\end{array}
\label{10.18}
\ee
The last line in eq.(\ref{10.18}) will cause no trouble in renormalization,
and if we consider the contribution from ghost fields which will
be introduced below, it will becomes a quantity which is independent
of gravitational gauge field. So, we put it into normalization constant
$N$ and still denote the new normalization constant as $N$. We also
change $J^{\mu}_{\alpha} ~  ^{g_0}\!C^{\alpha}_{\mu}$
into $J^{\mu}_{\alpha} C^{\alpha}_{\mu}$, this will cause no
trouble in renormalization. Then we get
\be
\begin{array}{rcl}
W\lbrack J \rbrack &=& N \int \lbrack {\cal D} C\rbrack~~
\Delta_f  (C)  \cdot \lbrack \prod_{\beta, z}
\delta ( f^{\beta} ( C(z))- \varphi^{\beta}(z) )\rbrack \\
&&\\
&& \cdot exp \lbrace i \int {\rm d}^4 x ( {\cal L}
+ J^{\mu}_{\alpha} C^{\alpha}_{\mu}) \rbrace.
\end{array}
\label{10.19}
\ee
In fact, we can use this formula as our start-point of path integral
quantization of gravitational gauge field, so we need not worried
about the influences of the third line in eq.(\ref{10.18}).
\\

Use another functional
\be
exp \left\lbrace - \frac{i}{2 \alpha}
\int {\rm d}^4 x \eta_{\alpha \beta}
\varphi^{\alpha}(x) \varphi^{\beta}(x) \right\rbrace,
\label{10.20}
\ee
times both sides of eq.(\ref{10.19}) and then make functional
integration
$\int \lbrack {\cal D} \varphi  \rbrack$,
we get
\be
W\lbrack J \rbrack  = N \int \lbrack {\cal D} C\rbrack~~
\Delta_f (C) \cdot exp \left\lbrace i \int {\rm d}^4 x ( {\cal L}
- \frac{1}{2 \alpha} \eta_{\alpha \beta} f^{\alpha} f^{\beta}
+ J^{\mu}_{\alpha} C^{\alpha}_{\mu}) \right\rbrace.
\label{10.21}
\ee
Now, let's discuss the contribution from $\Delta_f (C)$ which
is related to the ghost fields. Suppose that $g = \ehat$
is an infinitesimal gravitational gauge transformation. Then
eq.(\ref{4.12}) gives out
\be
^gC_{\mu}^{\alpha} (x)
= C_{\mu}^{\alpha} (x)
- \frac{1}{g} {\mathbf D}_{\mu~\sigma}^{\alpha} \epsilon^{\sigma},
\label{10.22}
\ee
where
\be
{\mathbf D}_{\mu~\sigma}^{\alpha}
=\delta^{\alpha}_{\sigma} \partial_{\mu}
- g \delta^{\alpha}_{\sigma} C_{\mu}^{\beta} \partial_{\beta}
+ g \partial_{\sigma} C_{\mu}^{\alpha}.
\label{10.23}
\ee
In order to deduce eq.(\ref{10.22}), the following relation is used
\be
\Lambda^{\alpha}_{~\beta}
= \delta^{\alpha}_{\beta}
+ \partial_{\beta} \epsilon^{\alpha}
+ o( \epsilon^2).
\label{10.24}
\ee
${\mathbf D}_{\mu}$ can be regarded as the covariant derivative
in adjoint representation, for
\be
{\mathbf D}_{\mu} \epsilon
= \lbrack D_{\mu} ~~~,~~~ \epsilon \rbrack,
\label{10.25}
\ee
\be
({\mathbf D}_{\mu} \epsilon)^{\alpha}
= {\mathbf D}_{\mu~\sigma}^{\alpha} \epsilon^{\sigma}.
\label{10.26}
\ee
Using all these relations,  we have,
\be
f^{\alpha} (^gC(x)) = f^{\alpha} (C)
- \frac{1}{g} \int {\rm d}^4 y
\frac{\delta f^{\alpha}(C(x))}{\delta C_{\mu}^{\beta}(y)}
{\mathbf D}_{\mu~\sigma}^{\beta}(y) \epsilon^{\sigma}(y)
+ o(\epsilon^2).
\label{10.27}
\ee
Therefore, according to eq.(\ref{10.7}) and eq.(\ref{10.6}), we get
\be
\Delta_f^{-1} (C) =
\int \lbrack {\cal D} \epsilon \rbrack
\prod_{x, \alpha}
\delta \left( - \frac{1}{g}     \int {\rm d}^4 y
\frac{\delta f^{\alpha}(C(x))}{\delta C_{\mu}^{\beta}(y)}
{\mathbf D}_{\mu~\sigma}^{\beta}(y) \epsilon^{\sigma}(y) \right).
\label{10.28}
\ee
Define
\be
\begin{array}{rcl}
{\mathbf M}^{\alpha}_{~\sigma}(x,y) &=& -g
\frac{\delta}{\delta \epsilon^{\sigma}(y)}
f^{\alpha}(^gC(x)) \\
&&\\
&=&\int {\rm d}^4 z
\frac{\delta f^{\alpha}(C(x))}{\delta C_{\mu}^{\beta}(z)}
{\mathbf D}_{\mu~\sigma}^{\beta}(z) \delta(z-y) .
\end{array}
\label{10.29}
\ee
Then eq.(\ref{10.28}) is changed into
\be
\begin{array}{rcl}
\Delta_f^{-1} (C) &=&
\int \lbrack {\cal D} \epsilon \rbrack
\prod_{x, \alpha}
\delta \left( - \frac{1}{g} \int {\rm d}^4 y
{\mathbf M}^{\alpha}_{~\sigma}(x,y) \epsilon^{\sigma}(y)
\right)  \\
&&\\
&=& const. \times (det {\mathbf M} )^{-1}.
\end{array}
\label{10.30}
\ee
Therefore,
\be
\Delta_f (C) = const. \times det {\mathbf M}.
\label{10.31}
\ee
Put the above constant into normalization constant, then generating
functional eq.(\ref{10.21}) is changed into
\be
W\lbrack J \rbrack  = N \int \lbrack {\cal D} C\rbrack~~
det {\mathbf M} \cdot exp \left\lbrace i \int {\rm d}^4 x ( {\cal L}
- \frac{1}{2 \alpha} \eta_{\alpha \beta} f^{\alpha} f^{\beta}
+ J^{\mu}_{\alpha} C^{\alpha}_{\mu}) \right\rbrace.
\label{10.32}
\ee
\\

In order to evaluate the contribution from $det {\mathbf M}$,
we introduce ghost fields $\eta^{\alpha}(x)$
and $\bar{\eta}_{\alpha}(x)$. Using the following relation
\be
\int \lbrack {\cal D} \eta \rbrack
\lbrack {\cal D}\bar{\eta} \rbrack
exp \left\lbrace i \int {\rm d}^4 x {\rm d}^4 y ~
\bar{\eta}_{\alpha} (x) {\mathbf M}^{\alpha}_{~\beta}(x,y)
\eta^{\beta} (y) \right\rbrace
 = const. \times det {\mathbf M}
\label{10.33}
\ee
and put the constant into the normalization constant, we can get
\be
W\lbrack J \rbrack  = N \int \lbrack {\cal D} C\rbrack
\lbrack {\cal D} \eta \rbrack
\lbrack {\cal D}\bar{\eta} \rbrack
exp \left\lbrace i \int {\rm d}^4 x ( {\cal L}
- \frac{1}{2 \alpha} \eta_{\alpha \beta} f^{\alpha} f^{\beta}
+ \bar{\eta} {\mathbf M } \eta
+ J^{\mu}_{\alpha} C^{\alpha}_{\mu}) \right\rbrace,
\label{10.34}
\ee
where $\int {\rm d}^4x \bar{\eta} {\mathbf M } \eta$ is a
simplified notation, whose explicit expression is
\be
\int {\rm d}^4x \bar{\eta} {\mathbf M } \eta
= \int {\rm d}^4 x {\rm d}^4 y ~
\bar{\eta}_{\alpha} (x) {\mathbf M}^{\alpha}_{~\beta}(x,y)
\eta^{\beta} (y).
\label{10.35}
\ee
The appearance of the non-trivial ghost fields is a inevitable
result of the non-Able nature of the gravitational gauge group.
\\

Set external source $J^{\mu}_{\alpha}$ to zero, we get,
\be
W\lbrack 0 \rbrack  = N \int \lbrack {\cal D} C\rbrack
\lbrack {\cal D} \eta \rbrack
\lbrack {\cal D}\bar{\eta} \rbrack
exp \left\lbrace i \int {\rm d}^4 x ( {\cal L}
- \frac{1}{2 \alpha} \eta_{\alpha \beta} f^{\alpha} f^{\beta}
+ \bar{\eta} {\mathbf M } \eta ) \right\rbrace,
\label{10.3501}
\ee
Now, let's take Lorentz covariant gauge condition,
\be
f^{\alpha} (C) = \partial^{\mu} C_{\mu}^{\alpha} .
\label{10.36}
\ee
Then
\be
\int {\rm d}^4x \bar{\eta} {\mathbf M } \eta =
- \int {\rm d}^4x \left( \partial^{\mu}
\bar{\eta}_{\alpha} (x) \right)
{\mathbf D}_{\mu~\beta}^{\alpha}(x) \eta^{\beta} (x).
\label{10.37}
\ee
And eq.(\ref{10.3501}) is changed into
\be
W\lbrack 0 \rbrack
 =  N \int \lbrack {\cal D} C\rbrack
\lbrack {\cal D} \eta \rbrack
\lbrack {\cal D}\bar{\eta} \rbrack
exp \left\lbrace i \int {\rm d}^4 x ( {\cal L}
- \frac{1}{2 \alpha}
\eta_{\alpha \beta} f^{\alpha} f^{\beta}
 - (\partial^{\mu}\bar{\eta}_{\alpha} )
{\mathbf D}_{\mu~\sigma}^{\alpha} \eta^{\sigma}
) \right\rbrace.
\label{10.3701}
\ee
For quantum gauge general relativity, the external source
of gravitational gauge field should be introduced in
a special way. Define the generating functional
with external sources as
\be
\begin{array}{rcl}
W\lbrack J, \beta, \bar{\beta} \rbrack
& = & N \int \lbrack {\cal D} C\rbrack
\lbrack {\cal D} \eta \rbrack
\lbrack {\cal D}\bar{\eta} \rbrack
exp \left\lbrace i \int {\rm d}^4 x ( {\cal L}
- \frac{1}{2 \alpha}
\eta_{\alpha \beta} f^{\alpha} f^{\beta} \right. \\
&&\\
&& \left. - (\partial^{\mu}\bar{\eta}_{\alpha} )
{\mathbf D}_{\mu~\sigma}^{\alpha} \eta^{\sigma}
+  C^{\alpha}_{\mu}
\swav{\delta}^{\mu\beta}_{\alpha\nu}(x)
J^{\nu}_{0\beta}
+ \bar{\eta}_{\alpha} \beta^{\alpha}
+ \bar{\beta}_{\alpha} \eta^{\alpha}
) \right\rbrace \\
&&\\
& = & N \int \lbrack {\cal D} C\rbrack
\lbrack {\cal D} \eta \rbrack
\lbrack {\cal D}\bar{\eta} \rbrack
exp \left\lbrace i \int {\rm d}^4 x ( {\cal L}
- \frac{1}{2 \alpha}
\eta_{\alpha \beta} f^{\alpha} f^{\beta} \right. \\
&&\\
&& \left. - (\partial^{\mu}\bar{\eta}_{\alpha} )
{\mathbf D}_{\mu~\sigma}^{\alpha} \eta^{\sigma}
+  C^{\alpha}_{\mu}
J^{\mu}_{\alpha}
+ \bar{\eta}_{\alpha} \beta^{\alpha}
+ \bar{\beta}_{\alpha} \eta^{\alpha}
) \right\rbrace ,
\end{array}
\label{10.38}
\ee
where $\swav{\delta}^{\mu \gamma}_{\alpha\rho} (x)$
is defined by
\be
\swav{\delta}^{\mu \gamma}_{\alpha\rho} (x)
\define \frac{1}{2}
\left (
\swav{\delta}^{\mu}_{\rho}(x)  \swav{\delta}^{\gamma}_{\alpha} (x)
+ \swav{\eta}^{\mu\gamma}(x)  \swav{\eta}_{\alpha\rho}(x)
\right),
\label{10.3801}
\ee
and
\be
J^{\mu}_{\alpha} \define
\swav{\delta}^{\mu\beta}_{\alpha\nu} (x)
J^{\nu}_{0\beta}.
\label{10.3802}
\ee
In the above definition, $\swav{\delta}^{\mu}_{\rho}(x)$,
$\swav{\eta}^{\mu\gamma}(x)$ and $\swav{\eta}_{\mu\gamma}(x)$
are defined by
\be
\swav{\delta}^{\mu}_{\rho}(x)
= \delta^{\mu}_{\rho}
- \frac{\partial^{\mu} \partial_{\rho}}
{\square  + i \epsilon},
\label{10.3803}
\ee
\be
\swav{\eta}^{\mu\gamma}(x)
= \eta^{\mu\gamma}
- \frac{\partial^{\mu} \partial^{\gamma}}
{\square  + i \epsilon},
\label{10.3804}
\ee
\be
\swav{\eta}_{\mu\gamma}(x)
= \eta_{\mu\gamma}
- \frac{\partial_{\mu} \partial_{\gamma}}
{\square  + i \epsilon},
\label{10.3805}
\ee
where
\be
\square \define \partial^2
= \partial^{\mu} \partial_{\mu}
= \eta^{\mu\nu} \partial_{\mu} \partial_{\nu}.
\label{10.3806}
\ee
Using these relations, we can prove that
\be
J^{\mu}_{\alpha} =
\swav{\delta}^{\mu\beta}_{\alpha\nu}
J^{\nu}_{\beta}.
\label{10.3807}
\ee
\\

The effective Lagrangian ${\cal L}_{eff}$
is defined by
\be
{\cal L}_{eff} \equiv
{\cal L} - \frac{1}{2 \alpha}
\eta_{\alpha \beta} f^{\alpha} f^{\beta}
- (\partial^{\mu}\bar{\eta}_{\alpha} )
{\mathbf D}_{\mu~\sigma}^{\alpha} \eta^{\sigma}.
\label{10.39}
\ee
${\cal L}_{eff}$ can separate into free Lagrangian
${\cal L}_F$  and interaction Lagrangian ${\cal L}_I$,
\be
{\cal L}_{eff} = {\cal L}_F + {\cal L}_I,
\label{10.40}
\ee
where
\be
\begin{array}{rcl}
{\cal L}_F &=& - \frac{1}{16} \eta^{\mu \rho} \eta^{\nu \sigma} \eta_{\alpha \beta }
F_{0 \mu \nu}^{\alpha} F_{0 \rho \sigma}^{\beta}
- \frac{1}{8} \eta^{\mu \rho}
F^{\alpha}_{0 \mu \beta} F^{\beta}_{0 \rho \alpha}
+ \frac{1}{4} \eta^{\mu \rho}
F^{\alpha}_{0 \mu \alpha} F^{\beta}_{0 \rho \beta} \\
&&\\
&&  -\frac{1}{2 \alpha} \eta_{\alpha \beta}
(\partial^{\mu} C_{\mu}^{\alpha})
(\partial^{\nu} C_{\nu}^{\beta})
- (\partial^{\mu} \bar{\eta}_{\alpha})
(\partial_{\mu} \eta^{\alpha}),
\end{array}
\label{10.41}
\ee
\be
\begin{array}{rcl}
{\cal L}_I &=&  + g(\partial^{\mu} \bar{\eta}_{\alpha})
C_{\mu}^{\beta} (\partial_{\beta} \eta^{\alpha})
- g(\partial^{\mu} \bar{\eta}_{\alpha})
(\partial_{\sigma} C_{\mu}^{\alpha}) \eta^{\sigma}\\
&&\\
&& + {~ self ~interaction~ terms ~of ~Gravitational~gauge~field}.
\end{array}
\label{10.42}
\ee
From the interaction Lagrangian, we can see that ghost fields
do not couple to $J(C)$. This is the reflection of the fact
that ghost fields are not physical fields, they are virtual fields.
Besides, the gauge fixing term does not couple to $J(C)$ either.
Using effective Lagrangian ${\cal L}_{eff}$, the generating
functional $W\lbrack J, \beta, \bar{\beta} \rbrack$ can be
simplified to
\be
W\lbrack J, \beta, \bar{\beta} \rbrack
=  N \int \lbrack {\cal D} C\rbrack
\lbrack {\cal D} \eta \rbrack
\lbrack {\cal D}\bar{\eta} \rbrack
exp \left\lbrace i \int {\rm d}^4 x ( {\cal L}_{eff}
+ J^{\mu}_{\alpha} C^{\alpha}_{\mu}
+ \bar{\eta}_{\alpha} \beta^{\alpha}
+ \bar{\beta}_{\alpha} \eta^{\alpha}
) \right\rbrace,
\label{10.43}
\ee
\\

Using eq.(\ref{10.41}), we can deduce propagator of gravitational gauge
fields and ghost fields. First, after a partial integration,
 we change the form of
eq. (\ref{10.41}) into
\be
\int {\rm d}^4x {\cal L}_F =
\int {\rm d}^4x \left\lbrace \frac{1}{2}
C_{\mu}^{\alpha} {\mathbb M}^{\mu\nu}_{\alpha\beta} (x) C_{\nu}^{\beta}
+ \bar{\eta}_{\alpha}
\partial^2 \eta^{\alpha} \right\rbrace,
\label{10.44}
\ee
where the operator ${\mathbb M}^{\mu\nu}_{\alpha\beta} (x)$
is defined by
\be
\ba{rcl}
{\mathbb M}^{\mu\nu}_{\alpha\beta} (x) & = &
\frac{1}{4} \eta^{\mu\nu} \eta_{\alpha \beta}
\partial^{\rho} \partial_{\rho}
- \frac{1}{4} \eta_{\alpha \beta} (1 - \frac{4}{\alpha})
\partial^{\mu} \partial^{\nu}
- \frac{1}{4} \delta^{\mu}_{\beta}
\partial^{\nu} \partial_{\alpha} \\
&&\\
&& + \frac{1}{4} \delta^{\mu}_{\beta} \delta^{\nu}_{\alpha }
\partial^{\rho} \partial_{\rho}
- \frac{1}{4} \delta^{\nu}_{\alpha}
\partial^{\mu} \partial_{\beta}
+ \frac{1}{2} \delta^{\nu}_{\beta}
\partial^{\mu} \partial_{\alpha}  \\
&&\\
&& - \frac{1}{4} \eta^{\mu\nu}
\partial_{\alpha} \partial_{\beta}
- \frac{1}{2} \delta^{\mu}_{\alpha} \delta^{\nu}_{\beta}
\partial^{\rho} \partial_{\rho}
+ \frac{1}{2} \delta^{\mu}_{\alpha}
\partial^{\nu} \partial_{\beta}.
\ea
\label{10.4401}
\ee
\\

Denote the propagator of gravitational gauge field as
\be
-i \Delta_{F \mu \nu}^{\alpha \beta} (x),
\label{10.45}
\ee
and denote the propagator of ghost field as
\be
-i \Delta_{F \beta}^{\alpha } (x).
\label{10.46}
\ee
They satisfy the following equation,
\be
- {\mathbb M}^{\mu\nu}_{\alpha\beta} (x)
\Delta_{F \nu \rho}^{\beta \gamma} (x)
= \swav{\delta}^{\mu \gamma}_{\alpha\rho} (x)
\delta(x) ,
\label{10.47}
\ee
\be
- \partial^2 \Delta_{F \beta}^{\alpha } (x)
= \delta_{\beta}^{\alpha } \delta(x),
\label{10.48}
\ee
where $\swav{\delta}^{\mu \gamma}_{\alpha\rho} (x)$
is defined by (\ref{10.3801}).
\\

Make Fourier transformations to momentum space
\be
\Delta_{F \mu \nu}^{\alpha \beta} (x)
= \int \frac{{\rm d}^4k}{(2 \pi)^4}
\swav{\Delta}_{F \mu \nu}^{\alpha \beta}(k) \cdot  e^{ikx},
\label{10.49}
\ee
\be
\Delta_{F \beta}^{\alpha } (x)
= \int \frac{{\rm d}^4k}{(2 \pi)^4}
\swav{\Delta}_{F \beta}^{\alpha }(k) \cdot e^{ikx},
\label{10.50}
\ee
where $\swav{\Delta}_{F \mu \nu}^{\alpha \beta}(k)$
and $\swav{\Delta}_{F \beta}^{\alpha }(k)$ are corresponding
propagators in momentum space. They satisfy the following
equations,
\be
- {\mathbb M}^{\mu\nu}_{\alpha\beta}
\swav{\Delta}_{F \nu \rho}^{\beta \gamma}(k)
= \swav{\delta}^{\mu\gamma}_{\alpha\rho} ,
\label{10.51}
\ee
\be
k^2 \swav{\Delta}_{F \beta}^{\alpha }(k)
= \delta^{\alpha}_{\beta},
\label{10.52}
\ee
where the operator ${\mathbb M}^{\mu\nu}_{\alpha\beta} $
is defined by
\be
\ba{rcl}
{\mathbb M}^{\mu\nu}_{\alpha\beta}
& \define & {\mathbb M}^{\mu\nu}_{\alpha\beta} (k) \\
&&\\
& \define &
\frac{1}{4} \eta^{\mu\nu} \eta_{\alpha \beta} k^2
- \frac{1}{4} \eta_{\alpha \beta} (1 - \frac{4}{\alpha})
k^{\mu} k^{\nu}
- \frac{1}{4} \delta^{\mu}_{\beta}
k^{\nu} k_{\alpha} \\
&&\\
&& + \frac{1}{4} \delta^{\mu}_{\beta} \delta^{\nu}_{\alpha }
k^2
- \frac{1}{4} \delta^{\nu}_{\alpha}
k^{\mu} k_{\beta}
+ \frac{1}{2} \delta^{\nu}_{\beta}
k^{\mu} k_{\alpha}  \\
&&\\
&& - \frac{1}{4} \eta^{\mu\nu}
k_{\alpha} k_{\beta}
- \frac{1}{2} \delta^{\mu}_{\alpha} \delta^{\nu}_{\beta}
k^2
+ \frac{1}{2} \delta^{\mu}_{\alpha}
k^{\nu} k_{\beta},
\ea
\label{10.5201}
\ee
and $\swav{\delta}^{\mu \gamma}_{\alpha\rho}$ is defined by
\be
\swav{\delta}^{\mu \gamma}_{\alpha\rho}
\define  \swav{\delta}^{\mu \gamma}_{\alpha\rho} (k) =
\frac{1}{2}
\left (
\swav{\delta}^{\mu}_{\rho}  \swav{\delta}^{\gamma}_{\alpha}
+ \swav{\eta}^{\mu\gamma}  \swav{\eta}_{\alpha\rho}
\right).
\label{10.5202}
\ee
The operator ${\mathbb M}^{\mu\nu}_{\alpha\beta}$
has the following symmetric property
\be
{\mathbb M}^{\mu\nu}_{\alpha\beta}
={\mathbb M}^{\nu\mu}_{\beta\alpha}.
\ee
In the above relation, $\swav{\delta}^{\mu}_{\rho}$,
$\swav{\eta}^{\mu\gamma}$ and $\swav{\eta}_{\mu\gamma}$
are defined by
\be
\swav{\delta}^{\mu}_{\rho}
= \delta^{\mu}_{\rho}
- \frac{k^{\mu} k_{\rho}}
{k^2  - i \epsilon},
\label{10.5203}
\ee
\be
\swav{\eta}^{\mu\gamma}
= \eta^{\mu\gamma}
- \frac{k^{\mu} k^{\gamma}}
{k^2  - i \epsilon},
\label{10.5204}
\ee
\be
\swav{\eta}_{\mu\gamma}
= \eta_{\mu\gamma}
- \frac{k_{\mu} k_{\gamma}}
{k^2  - i \epsilon}.
\label{10.5205}
\ee
It can be easily proved that $\swav{\delta}^{\mu}_{\rho}$,
$\swav{\eta}^{\mu\gamma}$, $\swav{\eta}_{\mu\gamma}$,
${\delta}^{\mu}_{\rho}$,
${\eta}^{\mu\gamma}$ and ${\eta}_{\mu\gamma}$
satisfy the following relations:
\be
\swav{\eta}^{\mu\gamma}
\swav{\eta}_{\gamma \nu}
= \swav{\delta}^{\mu}_{\nu},
\label{10.5206}
\ee
\be
{\eta}^{\mu\gamma}
\swav{\eta}_{\gamma \nu}
= \swav{\delta}^{\mu}_{\nu},
\label{10.5207}
\ee
\be
\swav{\eta}^{\mu\gamma}
{\eta}_{\gamma \nu}
= \swav{\delta}^{\mu}_{\nu},
\label{10.5208}
\ee
\be
\swav{\delta}^{\mu}_{\gamma}
\swav{\delta}^{\gamma}_{ \nu}
= \swav{\delta}^{\mu}_{\nu},
\label{10.5209}
\ee
\be
{\delta}^{\mu}_{\gamma}
\swav{\delta}^{\gamma}_{ \nu}
= \swav{\delta}^{\mu}_{\nu},
\label{10.5210}
\ee
\be
\swav{\delta}^{\mu}_{\gamma}
{\delta}^{\gamma}_{ \nu}
= \swav{\delta}^{\mu}_{\nu},
\label{10.5211}
\ee
\be
\swav{\delta}^{\mu}_{\gamma}
\swav{\eta}^{\gamma \nu}
= \swav{\eta}^{\mu\nu},
\label{10.5212}
\ee
\be
{\delta}^{\mu}_{\gamma}
\swav{\eta}^{\gamma \nu}
= \swav{\eta}^{\mu\nu},
\label{10.5213}
\ee
\be
\swav{\delta}^{\mu}_{\gamma}
{\eta}^{\gamma \nu}
= \swav{\eta}^{\mu\nu},
\label{10.5214}
\ee
\be
\swav{\delta}^{\gamma}_{\mu}
\swav{\eta}_{\gamma \nu}
= \swav{\eta}_{\mu\nu},
\label{10.5215}
\ee
\be
{\delta}^{\gamma}_{\mu}
\swav{\eta}_{\gamma \nu}
= \swav{\eta}_{\mu\nu},
\label{10.5216}
\ee
\be
\swav{\delta}^{\gamma}_{\mu}
{\eta}_{\gamma \nu}
= \swav{\eta}_{\mu\nu},
\label{10.5217}
\ee
\be
k^{\mu} \swav{\eta}_{\mu \nu}
= 0,
\label{10.5218}
\ee
\be
k_{\mu} \swav{\eta}^{\mu \nu}
= 0,
\label{10.5219}
\ee
\be
k^{\mu} \swav{\delta}_{\mu}^{ \nu}
= 0,
\label{10.5220}
\ee
\be
k_{\mu} \swav{\delta}^{\mu}_{ \nu}
= 0.
\label{10.5221}
\ee
Using all these relations, we can prove that
$\swav{\delta}^{\mu \gamma}_{\alpha\rho}$
satisfies the following relation
\be
\swav{\delta}^{\mu \gamma}_{\alpha\rho}
\cdot \swav{\delta}^{\rho \beta}_{\gamma\nu}
= \swav{\delta}^{\mu \beta}_{\alpha\nu}.
\label{10.5222}
\ee
\\

The solutions to the two propagator equations (\ref{10.51})
and (\ref{10.52}) give out the propagators
in momentum space,
\be
-i \swav{\Delta}_{F \mu \nu}^{\alpha \beta}(k)
= \frac{-i}{k^2 - i \epsilon}
\left\lbrack
\swav{\eta}_{\mu\nu} \swav{\eta}^{\alpha\beta}
+ \swav{\delta}_{\mu}^{\beta} \swav{\delta}^{\alpha}_{\nu}
- \swav{\delta}_{\mu}^{\alpha} \swav{\eta}_{\nu}^{\beta}
\right\rbrack,
\label{10.53}
\ee
\be
-i \swav{\Delta}_{F \beta}^{\alpha }(k) =
\frac{-i}{k^2 - i \epsilon} \delta^{\alpha}_{\beta}.
\label{10.54}
\ee
The forms of these propagators are quite beautiful and symmetric.
$\swav{\Delta}_{F \mu \nu}^{\alpha \beta}(k)$ satisfies
the following relation
\be
\swav{\Delta}_{F \mu \nu}^{\alpha \beta}(k)
\cdot \swav{\delta}^{\nu\gamma}_{\beta\rho}
=\swav{\Delta}_{F \mu \rho}^{\alpha \gamma} (k).
\label{10.54a}
\ee
Its Fourier transformation gives out the following relation
\be
 \swav{\delta}^{\nu\gamma}_{\beta\rho} (x) \cdot
{\Delta}_{F \mu \nu}^{\alpha \beta}(x)
= {\Delta}_{F \mu \rho}^{\alpha \gamma} (x).
\label{10.54b}
\ee
The above solution (\ref{10.53}) is not the most general
solution to equation (\ref{10.51}). The most general solution
to the equation (\ref{10.51}) is
\be
\ba{rcl}
-i \swav{\Delta}_{F \mu \nu}^{\alpha \beta}(k)
& = &\frac{-i}{k^2 - i \epsilon}
\left\lbrack
 c_1 \eta_{\mu\nu} \eta^{\alpha\beta}
 - c_1 \eta^{\alpha\beta}
 \frac{k_{\mu} k_{\nu}}{k^2 - i \epsilon}
 + c_3 \eta_{\mu\nu}
\frac{k^{\alpha} k^{\beta}}{k^2 - i \epsilon}
\right .  \\
&&\\
&&+ c_2 \delta_{\mu}^{\beta} \delta_{\nu}^{\alpha}
- c_2 \delta^{\alpha}_{\nu}
\frac{k_{\mu} k^{\beta}}{k^2 - i \epsilon}
+ c_4 \delta_{\mu}^{\beta}
\frac{k_{\nu} k^{\alpha}}{k^2 - i \epsilon}\\
&&\\
&&-  \delta_{\mu}^{\alpha} \delta_{\nu}^{\beta}
 + \delta^{\beta}_{\nu}
\frac{k_{\mu} k^{\alpha}}{k^2 - i \epsilon}
+  \delta_{\mu}^{\alpha}
\frac{k_{\nu} k^{\beta}}{k^2 - i \epsilon} \\
&&\\
&& \left .
+ c_5 \frac{k_{\mu} k_{\nu} k^{\alpha} k^{\beta}}
{(k^2 - i \epsilon)^2}
\right\rbrack,
\ea
\label{10.5401}
\ee
where parameters $c_1$, $c_2$, $c_3$, $c_4$ and $c_5$
satisfy the following restriction
\be
c_1 + c_2 =2,
\label{10.5402}
\ee
\be
c_3 + c_4 + c_5 =-1.
\label{10.5403}
\ee
Set $c_1=c_2=c_5=1$ and $c_3=c_4=-1$, we got the solution
(\ref{10.53}).
\\

The interaction Lagrangian ${\cal L}_I$ is a function of
gravitational gauge field $C_{\mu}^{\alpha}$ and ghost
fields $\eta^{\alpha}$ and $\bar \eta_{\alpha}$,
\be
{\cal L}_I =
{\cal L}_I ( C, \eta, \bar\eta ).
\label{10.55}
\ee
Then eq.(\ref{10.43}) is changed into,
\be
\begin{array}{rcl}
W\lbrack J, \beta, \bar{\beta} \rbrack
& = & N \int \lbrack {\cal D} C\rbrack
\lbrack {\cal D} \eta \rbrack
\lbrack {\cal D}\bar{\eta} \rbrack
~exp \left\lbrace i \int {\rm d}^4 x
 {\cal L}_I ( C, \eta, \bar\eta ) \right\rbrace \\
&&\\
&&\cdot exp \left\lbrace i \int {\rm d}^4 x ( {\cal L}_F
+ J^{\mu}_{\alpha} C^{\alpha}_{\mu}
+ \bar{\eta}_{\alpha} \beta^{\alpha}
+ \bar{\beta}_{\alpha} \eta^{\alpha}
) \right\rbrace  \\
&&\\
&=& exp \left\lbrace i \int {\rm d}^4 x
 {\cal L}_I ( \frac{1}{i}\frac{\delta}{\delta J},
\frac{1}{i}\frac{\delta}{\delta \bar \beta},
\frac{1}{-i}\frac{\delta}{\delta \beta} ) \right\rbrace
\cdot W_0\lbrack J, \beta, \bar{\beta} \rbrack ,
\end{array}
\label{10.56}
\ee
where
\be
\begin{array}{rcl}
W_0\lbrack J, \beta, \bar{\beta} \rbrack
&=&  N \int \lbrack {\cal D} C\rbrack
\lbrack {\cal D} \eta \rbrack
\lbrack {\cal D}\bar{\eta} \rbrack
exp \left\lbrace i \int {\rm d}^4 x
\left( {\cal L}_F
+ J^{\mu}_{\alpha} C^{\alpha}_{\mu}
+ \bar{\eta}_{\alpha} \beta^{\alpha}
+ \bar{\beta}_{\alpha} \eta^{\alpha}
\right ) \right\rbrace  \\
&&\\
&=&  N \int \lbrack {\cal D} C\rbrack
\lbrack {\cal D} \eta \rbrack
\lbrack {\cal D}\bar{\eta} \rbrack
exp \left\lbrace i \int {\rm d}^4 x
\left( \frac{1}{2} C_{\mu}^{\alpha}
{\mathbb M}^{\mu\nu}_{\alpha\beta} (x) C_{\nu}^{\beta}
+ \bar{\eta}_{\alpha}
\partial^2 \eta^{\alpha}  \right. \right.  \\
&&\\
&& \left.\left.
+ J^{\mu}_{\alpha} C^{\alpha}_{\mu}
+ \bar{\eta}_{\alpha} \beta^{\alpha}
+ \bar{\beta}_{\alpha} \eta^{\alpha}
\right ) \right\rbrace  \\
&&\\
&=&  exp \left\lbrace
i \int\int {\rm d}^4 x {\rm d}^4 y
\left\lbrack  \frac{1}{2} J^{\mu}_{\alpha} (x)
\Delta_{F \mu \nu}^{\alpha \beta} (x-y)
J^{\nu}_{\beta} (y) \right. \right.  \\
&&\\
&& \left.\left.~~+ \bar\beta_{\alpha}(x)
\Delta_{F \beta}^{\alpha } (x-y) \beta^{\beta}(y)
\right\rbrack  \right\rbrace .
\end{array}
\label{10.57}
\ee
In order to obtain the above relation, eq. (\ref{10.3807})
is used.
\\

The interaction Feynman rules for interaction vertices
can be obtained from the interaction Lagrangian ${\cal L}_I$.
For example, the interaction
Lagrangian between gravitational gauge field and ghost field
is
\be
+ g(\partial^{\mu} \bar{\eta}_{\alpha})
C_{\mu}^{\beta} (\partial^{\beta} \eta^{\alpha})
- g(\partial^{\mu} \bar{\eta}_{\alpha})
(\partial_{\sigma} C_{\mu}^{\alpha}) \eta^{\sigma}.
\label{10.58}
\ee
This vertex belongs to
$C_{\mu}^{\alpha}(k) \bar{\eta}_{\beta}(q) \eta^{\delta}(p) $
three body interactions, its Feynman rule is
\be
-i g \delta^{\beta}_{\delta} q^{\mu} p_{\alpha}
+ i g \delta^{\beta}_{\alpha} q^{\mu} k_{\delta}.
\label{10.59}
\ee
Feynman rules  for other interaction vertices can be
obtained similarly. We will not calculate Feynman
rules for all kinds of interaction vertices here.
For interaction vertices which involve in gravitational
gauge field, we can find
that all Feynman rules for  interaction vertices are
proportional to energy-momenta of one or more particles,
which is one of the most important properties of gravitational
interactions. In fact, this interaction property is expected
for gravitational interactions, for energy-momentum
is the source of gravity. \\

\section{Renormalization }
\setcounter{equation}{0}

In the quantum gauge general relativity, the gravitational coupling
constant has the dimensionality of negative powers of mass.
According to traditional theory of power counting law, it seems
that the quantum gauge general relativity is a kind of non-renormalizable
theory. But this result is not correct. The power counting
law does not work here. General speaking, power counting law
does not work when a theory has gauge symmetry. If a theory has
gauge symmetry, the constraints from gauge symmetry will make
some divergence cancel each other\cite{c01,c02,c03,c04,c05,c06}.
In the quantum gauge general relativity,
this mechanism works very well. In this chapter, we will give a
strict formal proof on the renormalization of the quantum gauge general
relativity. We will find that the effect of renormalization is
just a scale transformation of the original theory. Though
there are infinite number of divergent vertexes in the quantum gauge general
relativity, we need not introduce infinite number of interaction
terms that do not exist in the original Lagrangian and infinite
number of parameters. All the divergent vertex can find its
correspondence in the original Lagrangian. Therefor, in
renormalization, what we need to do is not to introduce extra
interaction terms to cancel divergent terms, but to redefine
the fields, coupling constants and some other parameters of
the original theory. Because most of counterterms come
from the factor $J(C)$, this factor is key important for
renormalization. Without this factor, the theory is
non-renormalizable. In a word, the quantum gauge general relativity
is a renormalizable quantum gauge theory. Now, let's start our
discussion on renormalization from the generalized BRST
transformations.
\\

The generalized BRST transformations are
\be
\delta C_{\mu}^{\alpha}
= -  {\mathbf D}_{\mu~\beta}^{\alpha} \eta^{\beta}
\delta \lambda,
\label{11.1}
\ee
\be
\delta \eta^{\alpha} = g \eta^{\sigma}
(\partial_{\sigma}\eta^{\alpha}) \delta \lambda,
\label{11.2}
\ee
\be
\delta \bar\eta_{\alpha} = \frac{1}{\alpha}
\eta_{\alpha \beta} f^{\beta} \delta \lambda,
\label{11.3}
\ee
\be
\delta \eta^{\mu \nu} = 0,
\label{11.4}
\ee
\be
\delta g_{\alpha \beta} = g \left (
g_{\alpha \sigma} (\partial_{\beta} \eta^{\sigma})
+ g_{\sigma \beta} (\partial_{\alpha} \eta^{\sigma})
+\eta^{\sigma} (\partial_{\sigma} g_{\alpha\beta})
\right ) \delta \lambda,
\label{11.5}
\ee
where $\delta \lambda$ is an infinitesimal Grassman constant.
It can be strict proved that the generalized BRST transformations
for fields $C_{\mu}^{\alpha}$ and $\eta^{\alpha}$ are nilpotent:
\be
\delta( {\mathbf D}_{\mu~\beta}^{\alpha} \eta^{\beta} )
=0,
\label{11.6}
\ee
\be
\delta(\eta^{\sigma}(\partial_{\sigma}\eta^{\alpha}))
=0.
\label{11.7}
\ee
It means that all second order variations of fields vanish.\\

Using the above transformation rules, it can be strictly proved that
the generalized BRST transformation for gauge field strength tensor
$F^{\alpha}_{\mu \nu}$ is
\be
\delta F^{\alpha}_{\mu \nu} =
g \left\lbrack
- (\partial_{\sigma} \eta^{\alpha}) F^{\sigma}_{\mu \nu}
+ \eta^{\sigma} (\partial_{\sigma} F^{\alpha}_{\mu \nu})
\right \rbrack \delta \lambda,
\label{11.8}
\ee
and the transformation for the factor $J(C)$ is
\be
\delta J(C) =
g \left \lbrack
(\partial_{\alpha} \eta^{\alpha}) J(C)
+ \eta^{\alpha} (\partial_{\alpha} J(C) )
\right \rbrack \delta \lambda.
\label{11.9}
\ee
Therefore, under generalized BRST transformations,
the Lagrangian ${\cal L}$ given by eq.(\ref{10.1}) transforms as
\be
\delta {\cal L} = g (\partial_{\alpha}
(\eta^{\alpha}  {\cal L})) \delta \lambda.
\label{11.10}
\ee
It is a total derivative term, its space-time integration
vanish, i.e., the action of eq.(\ref{10.2}) is invariant under
generalized BRST transformations,
\be
\delta S = \delta \left( \int {\rm d}^4x {\cal L} \right ) = 0.
\label{11.11}
\ee
On the other hand, it can be strict proved that
\be
\delta \left(
- \frac{1}{2 \alpha}
\eta_{\alpha \beta} f^{\alpha} f^{\beta}
+ \bar{\eta}_{\alpha} \partial^{\mu}
{\mathbf D}_{\mu~\sigma}^{\alpha} \eta^{\sigma}
\right) =0.
\label{11.12}
\ee
\\

The non-renormalized effective Lagrangian  is denoted
as $ {\cal L}_{eff}^{\lbrack 0 \rbrack} $. It is given by
\be
{\cal L}_{eff}^{\lbrack 0 \rbrack} =
{\cal L} - \frac{1}{2 \alpha}
\eta_{\alpha \beta} f^{\alpha} f^{\beta}
+ \bar{\eta}_{\alpha} \partial^{\mu}
{\mathbf D}_{\mu~\sigma}^{\alpha} \eta^{\sigma}.
\label{11.13}
\ee
The effective action is defined by
\be
S_{eff}^{\lbrack 0 \rbrack}
= \int {\rm d}^4x {\cal L}_{eff}^{\lbrack 0 \rbrack}.
\label{11.14}
\ee
Using eqs.(\ref{10.11} - \ref{11.12}), we can prove that this effective
action is invariant under generalized BRST transformations,
\be
\delta S_{eff}^{\lbrack 0 \rbrack} =0.
\label{11.15}
\ee
This is a strict relation without any approximation.
It is known that BRST symmetry plays key  role
in the renormalization of gauge theory, for it ensures
the validity of the Ward-Takahashi identity. \\

Before we go any further, we have to do another important
work, i.e., to prove that the functional integration measure
$\lbrack {\cal D}C \rbrack \lbrack {\cal D}\eta \rbrack
\lbrack {\cal D}\bar\eta \rbrack $ is also generalized BRST
invariant.  We have proved before that the functional integration
measure $\lbrack {\cal D}C \rbrack$ is not a gauge invariant
measure, therefore, it is highly important to prove that
$\lbrack {\cal D}C \rbrack \lbrack {\cal D}\eta \rbrack
\lbrack {\cal D}\bar\eta \rbrack $ is a generalized BRST
invariant measure. BRST transformation is a kind of
transformation which involves both bosonic fields and
fermionic fields. For the sake of simplicity, let's formally
denote all bosonic fields as $B = \lbrace B_i \rbrace$
and denote all fermionic fields as $F = \lbrace F_i \rbrace$.
All fields that are involved in generalized BRST transformation
are simply denoted by $(B,F)$. Then, generalized BRST
transformation is formally expressed as
\be
(B,F) ~~~ \to ~~~(B',F').
\label{11.16}
\ee
The transformation matrix of this transformation is
\be
J =
\left (
\begin{array}{cc}
\frac{\partial B_i}{\partial B'_j} &
\frac{\partial B_i}{\partial F'_l} \\
\frac{\partial F_k}{\partial B'_j} &
\frac{\partial F_k}{\partial F'_l}
\end{array}
\right )
=
\left (
\begin{array}{cc}
a &
\alpha \\
\beta &
b
\end{array}
\right )  ,
\label{11.17}
\ee
where
\be
a = \left( \frac{\partial B_i}{\partial B'_j} \right),
\label{11.18}
\ee
\be
b = \left( \frac{\partial F_k}{\partial F'_l} \right),
\label{11.19}
\ee
\be
\alpha = \left( \frac{\partial B_i}{\partial F'_l} \right),
\label{11.20}
\ee
\be
\beta = \left( \frac{\partial F_k}{\partial B'_j} \right).
\label{11.21}
\ee
Matrixes $a$ and $b$ are bosonic square matrix while
$\alpha$ and $\beta$ generally are not square matrix. In order
to calculate the Jacobian $det(J)$,
the  above transformation (\ref{11.16}) is realized
in two steps. The first step is a bosonic transformation
\be
(B,F) ~~~ \to ~~~(B',F).
\label{11.22}
\ee
The transformation matrix of this transformation is denoted as $J_1$,
\be
J_1 =
\left (
\begin{array}{cc}
a - \alpha b^{-1} \beta &
\alpha b^{-1} \\
0 &
1
\end{array}
\right )  .
\label{11.23}
\ee
Its Jacobian is
\be
det~ J_1 = det( a - \alpha b^{-1} \beta ).
\label{11.24}
\ee
Therefore,
\be
\int \prod_i {\rm d}B_i \prod_k {\rm d}F_k
= \int \prod_i {\rm d}B'_i \prod_k {\rm d}F_k
\cdot det( a - \alpha b^{-1} \beta ).
\label{11.25}
\ee
The second step is a fermionic transformation,
\be
(B',F) ~~~ \to ~~~(B',F').
\label{11.26}
\ee
Its transformation matrix is denoted as $J_2$,
\be
J_2 =
\left (
\begin{array}{cc}
1  &
0  \\
\beta &
b
\end{array}
\right )  .
\label{11.27}
\ee
Its Jacobian is the inverse of the determinant of the transformation
matrix,
\be
(det~ J_2)^{-1} = (det~b)^{-1}.
\label{11.28}
\ee
Using this relation, eq.(\ref{11.25}) is changed into
\be
\int \prod_i {\rm d}B_i \prod_k {\rm d}F_k
= \int \prod_i {\rm d}B'_i \prod_k {\rm d}F'_k
\cdot det( a - \alpha b^{-1} \beta ) (det~b)^{-1} .
\label{11.29}
\ee
For generalized BRST transformation, all non-diagonal matrix elements
are proportional to Grassman constant $\delta \lambda$. Non-diagonal
matrix $\alpha$ and $\beta$ contains only non-diagonal matrix elements,
so,
\be
\alpha b^{-1} \beta  \propto (\delta \lambda)^2 = 0.
\label{11.30}
\ee
It means that
\be
\int \prod_i {\rm d}B_i \prod_k {\rm d}F_k
= \int \prod_i {\rm d}B'_i \prod_k {\rm d}F'_k
\cdot det( a  ) \cdot (det~b)^{-1} .
\label{11.31}
\ee
Generally speaking, $C_{\mu}^{\alpha}$ and
$\partial_{\nu} C_{\mu}^{\alpha}$ are independent degrees of freedom,
so are $\eta^{\alpha}$ and $\partial_{\nu} \eta^{\alpha}$. Using
eqs.(\ref{11.1} - \ref{11.3}), we obtain
\be
\begin{array}{rcl}
(det ~ a^{-1}) & = &
det \left\lbrack
( \delta^{\alpha}_{\beta} +
g (\partial_{\beta} \eta^{\alpha} )
\delta \lambda ) \delta^{\mu}_{\nu}
\right\rbrack   \\
&&\\
&=&
\prod_{x}
( 1 + g (\partial_{\alpha} \eta^{\alpha} )
\delta \lambda)^4.
\end{array}
\label{11.32}
\ee
\be
\begin{array}{rcl}
(det ~ b^{-1}) & = &
det \left\lbrack \left(
 \delta^{\alpha}_{\beta} +
g (\partial_{\beta} \eta^{\alpha} )
\delta \lambda \right)  \delta^{\gamma}_{\delta}  \right\rbrack  \\
&&  \\
& = &
\prod_{x}
( 1 + g (\partial_{\alpha} \eta^{\alpha} )
\delta \lambda)^4 .
\end{array}
\label{11.33}
\ee
In eq.(\ref{11.33}), $\delta^{\gamma}_{\delta}$ comes
from contribution of ghost fields $\bar\eta_{\delta}$.
Using these two relations, we have
\be
det( a  ) \cdot (det~b)^{-1} ~=~ \prod_x~{\rm\bf 1} ~=~1.
\label{11.34}
\ee
Therefore, under generalized BRST transformation, functional
integrational measure
$\lbrack {\cal D}C \rbrack \lbrack {\cal D}\eta \rbrack
\lbrack {\cal D}\bar\eta \rbrack $
is invariant,
\be
\lbrack {\cal D}C \rbrack \lbrack {\cal D}\eta \rbrack
\lbrack {\cal D}\bar\eta \rbrack  ~=~
\lbrack {\cal D}C' \rbrack \lbrack {\cal D}\eta' \rbrack
\lbrack {\cal D}\bar\eta' \rbrack .
\label{11.35}
\ee
Though both $\lbrack {\cal D}C \rbrack$ and
$\lbrack {\cal D}\eta \rbrack$ are not invariant under generalized
BRST transformation, their product is invariant under generalized
BRST transformation. This result is interesting and important. \\

The generating functional
$W^{\lbrack 0 \rbrack} \lbrack J \rbrack $ is
\be
W^{\lbrack 0 \rbrack} \lbrack J \rbrack
=  N \int \lbrack {\cal D} C\rbrack
\lbrack {\cal D} \eta \rbrack
\lbrack {\cal D}\bar{\eta} \rbrack
exp \left\lbrace i \int {\rm d}^4 x
( {\cal L}^{\lbrack 0 \rbrack} _{eff}
+ J^{\mu}_{\alpha} C^{\alpha}_{\mu}
) \right\rbrace,
\label{10.36}
\ee
where the external source $J^{\mu}_{\alpha}$ satisfies
the following restriction
\be
J^{\mu}_{\alpha} = \swav{\delta}^{\mu\beta}_{\alpha\nu}
J^{\nu}_{\beta}.
\label{10.3601}
\ee
Because
\be
\int {\rm d}\eta^{\beta} {\rm d}\bar\eta^{\sigma}
\cdot \bar\eta^{\alpha} \cdot f(\eta,\bar\eta ) = 0,
\label{11.37}
\ee
where $f(\eta,\bar\eta )$ is a bilinear function of
$\eta$ and $\bar\eta$, we have
\be
\int \lbrack {\cal D} C\rbrack
\lbrack {\cal D} \eta \rbrack
\lbrack {\cal D}\bar{\eta} \rbrack
\cdot \bar\eta^{\alpha}(x) \cdot
exp \left\lbrace i \int {\rm d}^4 y
( {\cal L}^{\lbrack 0 \rbrack}_{eff}(y)
+ J^{\mu}_{\alpha}(y) C^{\alpha}_{\mu}(y)
) \right\rbrace = 0.
\label{11.38}
\ee
If all fields are the fields after generalized BRST transformation,
eq.(\ref{11.38}) still holds, i.e.
\be
\int \lbrack {\cal D} C' \rbrack
\lbrack {\cal D} \eta' \rbrack
\lbrack {\cal D}\bar{\eta}' \rbrack
\cdot \bar\eta^{\prime \alpha}(x) \cdot
exp \left\lbrace i \int {\rm d}^4 y
( {\cal L}^{\prime \lbrack 0 \rbrack}_{eff}(y)
+ J^{\mu}_{\alpha}(y) C^{\prime \alpha}_{\mu}(y)
) \right\rbrace = 0,
\label{11.39}
\ee
where ${\cal L}^{\prime \lbrack 0 \rbrack}_{eff} $ is the
effective Lagrangian after generalized BRST transformation.
Both functional integration measure and effective action
$\int {\rm d}^4 y {\cal L}^{\prime \lbrack 0 \rbrack}_{eff}(y)$
are generalized BRST invariant, so, using
eqs.(\ref{11.1} - \ref{11.3}), we get
\be
\begin{array}{ll}
\int \lbrack {\cal D} C \rbrack
\lbrack {\cal D} \eta \rbrack
\lbrack {\cal D}\bar{\eta} \rbrack
& \left\lbrack
\frac{1}{\alpha} f^{\alpha} (C(x)) \delta\lambda
- i \bar\eta^{\alpha} (C(x))
\int {\rm d}^4z ( J^{\mu}_{\beta}(z)
{\mathbf D}_{\mu \sigma}^{\beta}(z)
\eta^{\sigma}(z) \delta\lambda )
\right\rbrack  \\
&\\
&\cdot exp \left\lbrace i \int {\rm d}^4 y
( {\cal L}^{ \lbrack 0 \rbrack}_{eff}(y)
+ J^{\mu}_{\alpha}(y) C^{\alpha}_{\mu}(y)
) \right\rbrace = 0.
\end{array}
\label{11.40}
\ee
This equation will lead to
\be
\frac{1}{\alpha} f^{\alpha}
\left(\frac{1}{i} \frac{\delta}{\delta J(x)} \right)
W^{\lbrack 0 \rbrack} \lbrack J \rbrack
- \int {\rm d}^4y ~ J^{\mu}_{\beta}(y)
{\mathbf D}_{\mu \sigma}^{\beta}
\left( \frac{1}{i} \frac{\delta}{\delta J(x)} \right)
W^{\lbrack 0 \rbrack \sigma \alpha} \lbrack y,x, J \rbrack =0,
\label{11.41}
\ee
where
\be
W^{\lbrack 0 \rbrack\sigma \alpha} \lbrack y,x,J \rbrack
=  N i  \int \lbrack {\cal D} C\rbrack
\lbrack {\cal D} \eta \rbrack
\lbrack {\cal D}\bar{\eta} \rbrack
\bar\eta^{\alpha}(x) \eta^{\sigma}(y)
exp \left\lbrace i \int {\rm d}^4 z
( {\cal L}^{\lbrack 0 \rbrack} _{eff}
+ J^{\mu}_{\alpha} C^{\alpha}_{\mu}
) \right\rbrace.
\label{10.42}
\ee
This is the generalized Ward-Takahashi identity for
generating functional $W^{\lbrack 0 \rbrack} \lbrack J \rbrack$.\\

Now, let's introduce the external sources of ghost fields, then
the generation functional becomes
\be
W^{\lbrack 0 \rbrack}\lbrack J, \beta, \bar{\beta} \rbrack
=  N \int \lbrack {\cal D} C\rbrack
\lbrack {\cal D} \eta \rbrack
\lbrack {\cal D}\bar{\eta} \rbrack
exp \left\lbrace i \int {\rm d}^4 x
( {\cal L}_{eff}^{\lbrack 0 \rbrack}
+ J^{\mu}_{\alpha} C^{\alpha}_{\mu}
+ \bar{\eta}_{\alpha} \beta^{\alpha}
+ \bar{\beta}_{\alpha} \eta^{\alpha}
) \right\rbrace,
\label{11.43}
\ee
In renormalization of the theory, we have to introduce external
sources $K^{\mu}_{\alpha}$ and $L_{\alpha}$
of the following composite operators,
\be
{\mathbf D}_{\mu \beta}^{\alpha} \eta^{\beta}
~~,~~g \eta^{\sigma} (\partial_{\sigma} \eta^{\alpha} ).
\label{11.44}
\ee
Then the effective Lagrangian becomes
\be
\begin{array}{rcl}
{\swav{\cal L}}^{\lbrack 0 \rbrack} (C,\eta,\bar\eta,K,L)
&=& {\cal L} - \frac{1}{2 \alpha}
\eta_{\alpha \beta} f^{\alpha} f^{\beta}
+ \bar{\eta}_{\alpha} \partial^{\mu}
{\mathbf D}_{\mu~\sigma}^{\alpha} \eta^{\sigma}
+K^{\mu}_{\alpha} {\mathbf D}_{\mu \beta}^{\alpha} \eta^{\beta}
+ g L_{\alpha} \eta^{\sigma} (\partial_{\sigma} \eta^{\alpha})  \\
&&\\
&=& {\cal L}_{eff}^{\lbrack 0 \rbrack}
+K^{\mu}_{\alpha} {\mathbf D}_{\mu \beta}^{\alpha} \eta^{\beta}
+ g L_{\alpha} \eta^{\sigma} (\partial_{\sigma} \eta^{\alpha}) .
\end{array}
\label{11.45}
\ee
Then,
\be
\swav{S}^{\lbrack 0 \rbrack} \lbrack C,\eta,\bar\eta,K,L \rbrack
= \int {\rm d}^4x {\cal \swav{L}}^{\lbrack 0 \rbrack}
(C,\eta,\bar\eta,K,L).
\label{11.46}
\ee
It is easy to deduce that
\be
\frac{\delta \swav{S}^{\lbrack 0 \rbrack}}
{\delta K^{\mu}_{\alpha} } =
{\mathbf D}_{\mu \beta}^{\alpha} \eta^{\beta},
\label{11.47}
\ee
\be
\frac{\delta \swav{S}^{\lbrack 0 \rbrack}}
{\delta L_{\alpha} } =
g \eta^{\sigma} (\partial_{\sigma} \eta^{\alpha}).
\label{11.48}
\ee
The generating functional now becomes,
\be
W^{\lbrack 0 \rbrack}\lbrack J, \beta, \bar{\beta},K,L \rbrack
=  N \int \lbrack {\cal D} C\rbrack
\lbrack {\cal D} \eta \rbrack
\lbrack {\cal D}\bar{\eta} \rbrack
exp \left\lbrace i \int {\rm d}^4 x
( \swav{{\cal L}}^{\lbrack 0 \rbrack}
+ J^{\mu}_{\alpha} C^{\alpha}_{\mu}
+ \bar{\eta}_{\alpha} \beta^{\alpha}
+ \bar{\beta}_{\alpha} \eta^{\alpha}
) \right\rbrace.
\label{11.49}
\ee
In previous discussion, we have already proved that
$S_{eff}^{\lbrack 0 \rbrack}$ is generalized BRST
invariant. External sources $K^{\mu}_{\alpha}$
and $L_{\alpha}$ keep unchanged under generalized BRST
transformation. Using nilpotent property of generalized
BRST transformation, it is easy to prove that the two new
terms
$K^{\mu}_{\alpha} {\mathbf D}_{\mu \beta}^{\alpha} \eta^{\beta}$
and
$g L_{\alpha} \eta^{\sigma} (\partial_{\sigma} \eta^{\alpha})$
in ${\swav{\cal L}}^{\lbrack 0 \rbrack}$ are also
generalized BRST invariant,
\be
\delta (K^{\mu}_{\alpha}
{\mathbf D}_{\mu \beta}^{\alpha} \eta^{\beta})
=0,
\label{11.50}
\ee
\be
\delta( g L_{\alpha} \eta^{\sigma}
(\partial_{\sigma} \eta^{\alpha}) ) =0.
\label{11.51}
\ee
Therefore, the action given by (\ref{11.46}) are generalized
BRST invariant,
\be
\delta \swav{S}^{\lbrack 0 \rbrack} = 0.
\label{11.52}
\ee
It gives out
\be
\begin{array}{l}
\int {\rm d}^4x \left\lbrace
 - ({\mathbf D}_{\mu \beta}^{\alpha} \eta^{\beta}(x))
\delta\lambda \frac{\delta}{\delta C_{\mu}^{\alpha}(x)}
+ g \eta^{\sigma}(x) (\partial_{\sigma} \eta^{\alpha}(x))
\delta\lambda  \frac{\delta}{\delta \eta^{\alpha}(x)} \right.  \\
\\
 \left. + \frac{1}{\alpha} f^{\alpha}(C(x)) \delta\lambda
\frac{\delta}{\delta \bar\eta_{\alpha}(x)}
\right\rbrace \swav{S}^{\lbrack 0 \rbrack} = 0.
\end{array}
\label{11.53}
\ee
Using relations (\ref{11.47} - \ref{11.48}), we can get
\be
\int {\rm d}^4x \left\lbrace
\frac{\delta \swav{S}^{\lbrack 0 \rbrack}}
{\delta K^{\mu}_{\alpha}(x) }
\frac{\delta \swav{S}^{\lbrack 0 \rbrack}}
{\delta C_{\mu}^{\alpha}(x) }
+ \frac{\delta \swav{S}^{\lbrack 0 \rbrack}}
{\delta L_{\alpha}(x) }
\frac{\delta \swav{S}^{\lbrack 0 \rbrack}}
{\delta \eta^{\alpha}(x) }
+ \frac{1}{\alpha} f^{\alpha}(C(x))
\frac{\delta \swav{S}^{\lbrack 0 \rbrack}}
{\delta \bar\eta_{\alpha}(x) }
\right\rbrace  = 0.
\label{11.54}
\ee
On the other hand, from (\ref{11.45} - \ref{11.46}), we can obtain that
\be
\frac{\delta \swav{S}^{\lbrack 0 \rbrack}}
{\delta \bar\eta_{\alpha}(x) }
=\partial^{\mu}
\left( {\mathbf D}_{\mu \beta}^{\alpha} \eta^{\beta}(x)\right).
\label{11.55}
\ee
Combine (\ref{11.47}) with (\ref{11.55}), we get
\be
\frac{\delta \swav{S}^{\lbrack 0 \rbrack}}
{\delta \bar\eta_{\alpha}(x) }
= \partial^{\mu} \left( \frac{\delta \swav{S}^{\lbrack 0 \rbrack}}
{\delta K^{\mu}_{\alpha}(x) } \right).
\label{11.56}
\ee
\\

In generation functional
$W^{\lbrack 0 \rbrack}\lbrack J, \beta, \bar{\beta},K,L \rbrack$,
all fields are integrated, so, if we set all fields to the fields
after generalized BRST transformations, the final result should
not be changed, i.e.
\be
\begin{array}{l}
\swav{W}^{\lbrack 0 \rbrack}\lbrack J, \beta, \bar{\beta},K,L \rbrack
=  N \int \lbrack {\cal D} C' \rbrack
\lbrack {\cal D} \eta' \rbrack
\lbrack {\cal D} \bar{\eta}' \rbrack  \\
\\
~~~~~\cdot exp \left\lbrace i \int {\rm d}^4 x
( {\cal L}^{\lbrack 0 \rbrack} (C',\eta',\bar\eta',K,L)
+ J^{\mu}_{\alpha} C^{\prime \alpha}_{\mu}
+ \bar{\eta}'_{\alpha} \beta^{\alpha}
+ \bar{\beta}_{\alpha} \eta^{\prime \alpha}
) \right\rbrace.
\end{array}
\label{11.57}
\ee
Both action (\ref{11.46}) and functional integration
measure $\lbrack {\cal D} C \rbrack
\lbrack {\cal D} \eta \rbrack
\lbrack {\cal D} \bar{\eta} \rbrack$
are generalized BRST invariant, so, the above relation
gives out
\be
\begin{array}{l}
\int \lbrack {\cal D} C \rbrack
\lbrack {\cal D} \eta \rbrack
\lbrack {\cal D} \bar{\eta} \rbrack
\left\lbrace
 i \int {\rm d}^4x \left(
J^{\mu}_{\alpha} \frac{\delta \swav{S}^{\lbrack 0 \rbrack}}
{\delta K^{\mu}_{\alpha}(x) }
- \bar\beta_{\alpha} \frac{\delta \swav{S}^{\lbrack 0 \rbrack}}
{\delta L_{\alpha}(x) }
+ \frac{1}{\alpha} \eta_{\alpha \sigma} f^{\alpha} \beta^{\sigma}
\right)\right\rbrace \\
\\
\cdot exp \left\lbrace i \int {\rm d}^4 y
( {\cal L}^{\lbrack 0 \rbrack} (C,\eta,\bar\eta,K,L)
+ J^{\mu}_{\alpha} C^{\alpha}_{\mu}
+ \bar{\eta}_{\alpha} \beta^{\alpha}
+ \bar{\beta}_{\alpha} \eta^{\alpha}
) \right\rbrace = 0.
\end{array}
\label{11.58}
\ee
On the other hand, because the ghost field $\bar\eta_{\alpha}$
was integrated in
$W^{\lbrack 0 \rbrack}\lbrack J, \beta, \bar{\beta},K,L \rbrack$,
if we use $\bar\eta'_{\alpha}$ in the in functional integration,
it will not change the generating functional. That is
\be
\begin{array}{l}
\swav{W}^{\lbrack 0 \rbrack}\lbrack J, \beta, \bar{\beta},K,L \rbrack
=  N \int \lbrack {\cal D} C \rbrack
\lbrack {\cal D} \eta \rbrack
\lbrack {\cal D} \bar{\eta}' \rbrack  \\
\\
~~~~~\cdot exp \left\lbrace i \int {\rm d}^4 x
( {\cal L}^{\lbrack 0 \rbrack} (C,\eta,\bar\eta',K,L)
+ J^{\mu}_{\alpha} C^{\alpha}_{\mu}
+ \bar{\eta}'_{\alpha} \beta^{\alpha}
+ \bar{\beta}_{\alpha} \eta^{\alpha}
) \right\rbrace.
\end{array}
\label{11.59}
\ee
Suppose that
\be
\bar{\eta}'_{\alpha} = \bar{\eta}_{\alpha}
+ \delta \bar{\eta}_{\alpha}.
\label{11.60}
\ee
Then (\ref{11.59}) and (\ref{11.49}) will gives out
\be
\begin{array}{l}
\int \lbrack {\cal D} C \rbrack
\lbrack {\cal D} \eta \rbrack
\lbrack {\cal D} \bar{\eta} \rbrack
\left\lbrace
\int {\rm d}^4x \delta \bar{\eta}_{\alpha}
( \frac{\delta \swav{S}^{\lbrack 0 \rbrack}}
{\delta \bar\eta_{\alpha}(x) } + \beta^{\alpha} (x)
)\right\rbrace \\
\\
\cdot exp \left\lbrace i \int {\rm d}^4 y
( {\cal L}^{\lbrack 0 \rbrack} (C,\eta,\bar\eta,K,L)
+ J^{\mu}_{\alpha} C^{\alpha}_{\mu}
+ \bar{\eta}_{\alpha} \beta^{\alpha}
+ \bar{\beta}_{\alpha} \eta^{\alpha}
) \right\rbrace = 0.
\end{array}
\label{11.61}
\ee
Because $\delta \bar{\eta}_{\alpha}$ is an arbitrary
variation, from (\ref{11.61}), we will get
\be
\begin{array}{l}
\int \lbrack {\cal D} C \rbrack
\lbrack {\cal D} \eta \rbrack
\lbrack {\cal D} \bar{\eta} \rbrack
\left( \frac{\delta \swav{S}^{\lbrack 0 \rbrack}}
{\delta \bar\eta_{\alpha}(x) } + \beta^{\alpha} (x)
\right)  \\
\\
\cdot exp \left\lbrace i \int {\rm d}^4 y
( {\cal L}^{\lbrack 0 \rbrack} (C,\eta,\bar\eta,K,L)
+ J^{\mu}_{\alpha} C^{\alpha}_{\mu}
+ \bar{\eta}_{\alpha} \beta^{\alpha}
+ \bar{\beta}_{\alpha} \eta^{\alpha}
) \right\rbrace = 0.
\end{array}
\label{11.62}
\ee
\\

The generating functional of connected Green function is given by
\be
\swav{Z}^{\lbrack 0 \rbrack}\lbrack J,
\beta, \bar{\beta},K,L \rbrack
= - i~ {\rm ln}~
\swav{W}^{\lbrack 0 \rbrack}\lbrack J,
\beta, \bar{\beta},K,L \rbrack.
\label{11.63}
\ee
After Legendre transformation, we will get the generating
functional of irreducible vertex
$\swav{\Gamma}^{\lbrack 0 \rbrack}\lbrack C ,
\bar\eta,\eta,K,L \rbrack$,
\be
\begin{array}{rcl}
\swav{\Gamma}^{\lbrack 0 \rbrack}\lbrack C,
\bar\eta,\eta,K,L \rbrack
&=& \swav{Z}^{\lbrack 0 \rbrack}\lbrack J,
\beta, \bar{\beta},K,L \rbrack  \\
&&\\
&& - \int {\rm d}^4 x
\left( J^{\mu}_{\alpha} C^{\alpha}_{\mu}
+ \bar{\eta}_{\alpha} \beta^{\alpha}
+ \bar{\beta}_{\alpha} \eta^{\alpha}
\right) .
\end{array}
\label{11.64}
\ee
Functional derivative of the generating functional
$\swav{Z}^{\lbrack 0 \rbrack}$ gives out the classical fields
$C_{\mu}^{\alpha}$, $\eta^{\alpha}$ and $\bar\eta_{\alpha}$,
\be
C_{\mu}^{\alpha} =
\frac{\delta \swav{Z}^{\lbrack 0 \rbrack}}
{\delta J^{\mu}_{\alpha}},
\label{11.65}
\ee
\be
\eta^{\alpha} =
\frac{\delta \swav{Z}^{\lbrack 0 \rbrack}}
{\delta \bar\beta_{\alpha}},
\label{11.66}
\ee
\be
\bar\eta_{\alpha} =
- \frac{\delta \swav{Z}^{\lbrack 0 \rbrack}}
{\delta \beta^{\alpha}}.
\label{11.67}
\ee
Then, functional derivative of the generating functional
$\swav{\Gamma}^{\lbrack 0 \rbrack}$ gives out external
sources $J^{\mu}_{\alpha}$, $\bar\beta_{\alpha}$ and
$\beta^{\alpha}$,
\be
\frac{\delta \swav{\Gamma}^{\lbrack 0 \rbrack}}
{\delta C_{\mu}^{\alpha}}
= - J^{\mu}_{\alpha},
\label{11.68}
\ee
\be
\frac{\delta \swav{\Gamma}^{\lbrack 0 \rbrack}}
{\delta \eta^{\alpha}}
=  \bar\beta_{\alpha},
\label{11.69}
\ee
\be
\frac{\delta \swav{\Gamma}^{\lbrack 0 \rbrack}}
{\delta \bar\eta_{\alpha}}
=  - \beta^{\alpha}.
\label{11.70}
\ee
Besides, there are two other relations which can be strictly
deduced from (\ref{11.64}),
\be
\frac{\delta\swav{\Gamma}^{\lbrack 0 \rbrack}}
{\delta K^{\mu}_{\alpha}}
= \frac{\delta\swav{Z}^{\lbrack 0 \rbrack}}
{\delta K^{\mu}_{\alpha}},
\label{11.71}
\ee
\be
\frac{\delta\swav{\Gamma}^{\lbrack 0 \rbrack}}
{\delta L_{\alpha}}
= \frac{\delta \swav{Z}^{\lbrack 0 \rbrack}}
{\delta L_{\alpha}}.
\label{11.72}
\ee
\\

It is easy to prove that
\be
\begin{array}{rcl}
&& i \frac{\delta \swav{S}^{\lbrack 0 \rbrack}}
{\delta K^{\mu}_{\alpha}(x) }
exp \left\lbrace i \int {\rm d}^4 y
( {\cal L}^{\lbrack 0 \rbrack} (C,\eta,\bar\eta,K,L)
+ J^{\mu}_{\alpha} C^{\alpha}_{\mu}
+ \bar{\eta}_{\alpha} \beta^{\alpha}
+ \bar{\beta}_{\alpha} \eta^{\alpha}
) \right\rbrace  \\
&&\\
&=& \frac{\delta} {\delta K^{\mu}_{\alpha}(x) }
exp \left\lbrace i \int {\rm d}^4 y
( {\cal L}^{\lbrack 0 \rbrack} (C,\eta,\bar\eta,K,L)
+ J^{\mu}_{\alpha} C^{\alpha}_{\mu}
+ \bar{\eta}_{\alpha} \beta^{\alpha}
+ \bar{\beta}_{\alpha} \eta^{\alpha}
) \right\rbrace,
\end{array}
\label{11.73}
\ee
\be
\begin{array}{rcl}
&& i \frac{\delta \swav{S}^{\lbrack 0 \rbrack}}
{\delta L_{\alpha}(x) }
exp \left\lbrace i \int {\rm d}^4 y
( {\cal L}^{\lbrack 0 \rbrack} (C,\eta,\bar\eta,K,L)
+ J^{\mu}_{\alpha} C^{\alpha}_{\mu}
+ \bar{\eta}_{\alpha} \beta^{\alpha}
+ \bar{\beta}_{\alpha} \eta^{\alpha}
) \right\rbrace  \\
&&\\
&=& \frac{\delta} {\delta L_{\alpha}(x) }
exp \left\lbrace i \int {\rm d}^4 y
( {\cal L}^{\lbrack 0 \rbrack} (C,\eta,\bar\eta,K,L)
+ J^{\mu}_{\alpha} C^{\alpha}_{\mu}
+ \bar{\eta}_{\alpha} \beta^{\alpha}
+ \bar{\beta}_{\alpha} \eta^{\alpha}
) \right\rbrace.
\end{array}
\label{11.74}
\ee
Use these two relations, we can change (\ref{11.58}) into
\be
\begin{array}{l}
\int \lbrack {\cal D} C \rbrack
\lbrack {\cal D} \eta \rbrack
\lbrack {\cal D} \bar{\eta} \rbrack
\left\lbrace
 \int {\rm d}^4x \left(
J^{\mu}_{\alpha}(x) \frac{\delta } {\delta K^{\mu}_{\alpha}(x) }
- \bar\beta_{\alpha}(x) \frac{\delta }{\delta L_{\alpha}(x) }
+ \frac{i}{\alpha} \eta_{\alpha \sigma}
f^{\alpha}(\frac{1}{i}\frac{\delta }{\delta J^{\mu}_{\gamma}(x) } )
\beta^{\sigma}(x)   \right)\right\rbrace \\
\\
\cdot exp \left\lbrace i \int {\rm d}^4 y
( {\cal L}^{\lbrack 0 \rbrack} (C,\eta,\bar\eta,K,L)
+ J^{\mu}_{\alpha} C^{\alpha}_{\mu}
+ \bar{\eta}_{\alpha} \beta^{\alpha}
+ \bar{\beta}_{\alpha} \eta^{\alpha}
) \right\rbrace = 0.
\end{array}
\label{11.75}
\ee
Using relations (\ref{11.68} -  \ref{11.70})
and definition of generating functional
(\ref{11.57}), we can rewrite this equation into
\be
\int {\rm d}^4x \left\lbrace
\frac{\delta \swav{W}^{\lbrack 0 \rbrack} }
{\delta K^{\mu}_{\alpha}(x) }
\frac{\delta \swav{\Gamma}^{\lbrack 0 \rbrack} }
{\delta C_{\mu}^{\alpha}(x) }
+ \frac{\delta \swav{W}^{\lbrack 0 \rbrack} }
{\delta L_{\alpha}(x) }
\frac{\delta \swav{\Gamma}^{\lbrack 0 \rbrack} }
{\delta \eta^{\alpha}(x) }
+ \frac{i}{\alpha} \eta_{\alpha \sigma}
f^{\alpha} \left(\frac{1}{i}
\frac{\delta }{\delta J^{\mu}_{\rho}(x) } \right)
\swav{W}^{\lbrack 0 \rbrack}
\frac{\delta \swav{\Gamma}^{\lbrack 0 \rbrack} }
{\delta \bar\eta_{\sigma}(x) }
\right\rbrace = 0.
\label{11.76}
\ee
Using (\ref{11.63}), we can obtain that
\be
\frac{\delta \swav{W}^{\lbrack 0 \rbrack} }
{\delta K^{\mu}_{\alpha}(x) }
= i \frac{\delta \swav{\Gamma}^{\lbrack 0 \rbrack} }
{\delta K^{\mu}_{\alpha}(x) } \cdot
\swav{W}^{\lbrack 0 \rbrack},
\label{11.77}
\ee
\be
\frac{\delta \swav{W}^{\lbrack 0 \rbrack} }
{\delta L_{\alpha}(x) }
= i \frac{\delta \swav{\Gamma}^{\lbrack 0 \rbrack} }
{\delta L_{\alpha}(x) } \cdot
\swav{W}^{\lbrack 0 \rbrack}.
\label{11.78}
\ee
Then (\ref{11.76}) is changed into
\be
\int {\rm d}^4x \left\lbrace
\frac{\delta \swav{\Gamma}^{\lbrack 0 \rbrack} }
{\delta K^{\mu}_{\alpha}(x) }
\frac{\delta \swav{\Gamma}^{\lbrack 0 \rbrack} }
{\delta C_{\mu}^{\alpha}(x) }
+ \frac{\delta \swav{\Gamma}^{\lbrack 0 \rbrack} }
{\delta L_{\alpha}(x) }
\frac{\delta \swav{\Gamma}^{\lbrack 0 \rbrack} }
{\delta \eta^{\alpha}(x) }
+ \frac{i}{\alpha} \eta_{\alpha \sigma}
f^{\alpha}
\frac{\delta \swav{\Gamma}^{\lbrack 0 \rbrack} }
{\delta \bar\eta_{\sigma}(x) }
\right\rbrace = 0.
\label{11.79}
\ee
Using (\ref{11.56}) and (\ref{11.73}), (\ref{11.62}) becomes
\be
\begin{array}{l}
\int \lbrack {\cal D} C \rbrack
\lbrack {\cal D} \eta \rbrack
\lbrack {\cal D} \bar{\eta} \rbrack
\left\lbrack  - i \partial^{\mu}
\frac{\delta } {\delta K^{\mu}_{\alpha}(x) }
+ \beta^{\alpha} (x)   \right\rbrack  \\
\\
\cdot exp \left\lbrace i \int {\rm d}^4 y
( {\cal L}^{\lbrack 0 \rbrack} (C,\eta,\bar\eta,K,L)
+ J^{\mu}_{\alpha} C^{\alpha}_{\mu}
+ \bar{\eta}_{\alpha} \beta^{\alpha}
+ \bar{\beta}_{\alpha} \eta^{\alpha}
) \right\rbrace = 0.
\end{array}
\label{11.80}
\ee
In above equation, the factor $- i \partial^{\mu}
\frac{\delta } {\delta K^{\mu}_{\alpha}(x) }
+ \beta^{\alpha} (x)$
can  be moved out of functional integration,
then (\ref{11.80}) gives out
\be
\partial^{\mu}
\frac{\delta \swav{\Gamma}^{\lbrack 0 \rbrack}  }
{\delta K^{\mu}_{\alpha}(x) }
= \frac{\delta \swav{\Gamma}^{\lbrack 0 \rbrack}  }
{\delta \bar\eta_{\alpha}(x) }.
\label{11.81}
\ee
In order to obtain this relation, (\ref{11.57}), (\ref{11.77}) and
(\ref{11.70}) are used. \\

Define
\be
\bar{\Gamma}^{\lbrack 0 \rbrack}
\lbrack C, \bar\eta, \eta, K, L \rbrack
= \swav{\Gamma}^{\lbrack 0 \rbrack}
\lbrack C, \bar\eta, \eta, K, L \rbrack
+ \frac{1}{2 \alpha}
\int {\rm d}^4x \eta_{\alpha \beta}
f^{\alpha} f^{\beta}
\label{11.82}
\ee
It is easy to prove that
\be
\frac{\delta \bar{\Gamma}^{\lbrack 0 \rbrack}  }
{\delta K^{\mu}_{\alpha}(x) }
= \frac{\delta \swav{\Gamma}^{\lbrack 0 \rbrack}  }
{\delta K^{\mu}_{\alpha}(x) }
\label{11.83},
\ee
\be
\frac{\delta \bar{\Gamma}^{\lbrack 0 \rbrack}  }
{\delta L_{\alpha}(x) }
= \frac{\delta \swav{\Gamma}^{\lbrack 0 \rbrack}  }
{\delta L_{\alpha}(x) },
\label{11.84}
\ee
\be
\frac{\delta \bar{\Gamma}^{\lbrack 0 \rbrack}  }
{\delta \eta^{\alpha}(x) }
= \frac{\delta \swav{\Gamma}^{\lbrack 0 \rbrack}  }
{\delta \eta^{\alpha}(x) }    ,
\label{11.85}
\ee
\be
\frac{\delta \bar{\Gamma}^{\lbrack 0 \rbrack}  }
{\delta \bar\eta_{\alpha}(x) }
= \frac{\delta \swav{\Gamma}^{\lbrack 0 \rbrack}  }
{\delta \bar\eta_{\alpha}(x) }    ,
\label{11.86}
\ee
\be
\frac{\delta \bar{\Gamma}^{\lbrack 0 \rbrack}  }
{\delta C_{\mu}^{\alpha}(x) }
= \frac{\delta \swav{\Gamma}^{\lbrack 0 \rbrack}  }
{\delta C_{\mu}^{\alpha}(x) }
- \frac{1}{\alpha} \eta_{\alpha \beta}
\partial^{\mu}  f^{\beta}.
\label{11.87}
\ee
Using these relations, (\ref{11.81}) and (\ref{11.79}) are changed into
\be
\partial^{\mu}
\frac{\delta \bar{\Gamma}^{\lbrack 0 \rbrack}  }
{\delta K^{\mu}_{\alpha}(x) }
= \frac{\delta \bar{\Gamma}^{\lbrack 0 \rbrack}  }
{\delta \bar\eta_{\alpha}(x) },
\label{11.88}
\ee
\be
\int {\rm d}^4x \left\lbrace
\frac{\delta \bar{\Gamma}^{\lbrack 0 \rbrack} }
{\delta K^{\mu}_{\alpha}(x) }
\frac{\delta \bar{\Gamma}^{\lbrack 0 \rbrack} }
{\delta C_{\mu}^{\alpha}(x) }
+ \frac{\delta \bar{\Gamma}^{\lbrack 0 \rbrack} }
{\delta L_{\alpha}(x) }
\frac{\delta \bar{\Gamma}^{\lbrack 0 \rbrack} }
{\delta \eta^{\alpha}(x) }
\right\rbrace = 0.
\label{11.89}
\ee
Eqs.(\ref{11.88} - \ref{11.89}) are generalized
Ward-Takahashi identities
of generating functional of regular vertex. It is the foundations
of the renormalization of the quantum general relativity.\\

Generating functional $\swav{\Gamma}^{\lbrack 0 \rbrack}$
is the generating functional of regular vertex with external
sources, which is constructed from the Lagrangian
$\swav{\cal L}_{eff}^{\lbrack 0 \rbrack}$. It is a functional
of physical field, therefore, we can make a functional
expansion
\be
\begin{array}{rcl}
\swav{\Gamma}^{\lbrack 0 \rbrack} &=&
\sum_n \int \frac{\delta^n \swav{\Gamma}^{\lbrack 0 \rbrack}}
{\delta C^{\alpha_1}_{\mu_1}(x_1) \cdots
\delta C^{\alpha_n}_{\mu_n}(x_n) } |_{C=\eta=\bar\eta=0}
C^{\alpha_1}_{\mu_1}(x_1) \cdots C^{\alpha_n}_{\mu_n}(x_n)
{\rm d}^4x_1 \cdots {\rm d}^4x_n  \\
&&\\
&& + \sum_n \int \frac{\delta^2}
{\delta \bar\eta_{\beta_1}(y_1) \delta\eta^{\beta_2}(y_2) }
\frac{\delta^n \swav{\Gamma}^{\lbrack 0 \rbrack}}
{\delta C^{\alpha_1}_{\mu_1}(x_1) \cdots
\delta C^{\alpha_n}_{\mu_n}(x_n) } |_{C=\eta=\bar\eta=0}  \\
&&\\
&&~~~ \cdot \bar\eta_{\beta_1}(y_1) \eta^{\beta_2}(y_2)
C^{\alpha_1}_{\mu_1}(x_1) \cdots C^{\alpha_n}_{\mu_n}(x_n)
{\rm d}^4y_1 {\rm d}^4y_2 {\rm d}^4x_1 \cdots {\rm d}^4x_n\\
&&\\
&& + \cdots.
\end{array}
\label{11.90}
\ee
In this functional expansion, the expansion coefficients are
regular vertexes with external sources. Before renormalization,
these coefficients contain divergences. If we calculate these
divergences in the methods of dimensional regularization,
the form of these divergence will not violate gauge symmetry
of the theory\cite{t01,t02}.
In other words, in the method of dimensional
regularization, gravitational gauge symmetry is not
violated and the generating functional of regular vertex
satisfies Ward-Takahashi identities
(\ref{11.88} - \ref{11.89}). In order to eliminate
the ultraviolet divergences of the theory, we need to introduce
counterterms into Lagrangian. All these counterterms are
formally denoted by $\delta {\cal L}$. Then the renormalized
Lagrangian is
\be
\swav{\cal L}_{eff} =
\swav{\cal L}_{eff}^{\lbrack 0 \rbrack}
+ \delta {\cal L}.
\label{11.91}
\ee
Because $\delta {\cal L}$ contains all counterterms,
$\swav{\cal L}_{eff}$ is the Lagrangian density after
renormalization. The generating functional of regular vertex
which is calculate from $\swav{\cal L}_{eff}$ is denoted
by $\swav{\Gamma}$. The regular vertexes calculated from
this generating functional $\swav{\Gamma}$ contain no
ultraviolet divergence anymore. Then let external sources
$K^{\mu}_{\alpha}$ and $L_{\alpha}$ vanish, we will get
generating functional $\Gamma$ of regular vertex without external
sources,
\be
\Gamma = \swav{\Gamma} |_{K=L=0} .
\label{11.92}
\ee
The regular vertexes which are generated from $\Gamma$ will
contain no ultraviolet divergence either. Therefore, the
S-matrix for all physical process are finite. For a renormalizable
theory, the counterterm $\delta {\cal L}$ only contain finite
unknown parameters which are needed to be determined by
experiments. If conterterm $\delta {\cal L}$ contains infinite
unknown parameters, the theory will lost its predictive power
and it is conventionally regarded as a non-renormalizable
theory. Now, the main task for us is to prove that
the conterterm $\delta {\cal L}$ for the quantum gauge general
relativity only contains a few unknown parameters. If this
goal was reached, we will have proved that the quantum gauge general
relativity is renormalizable quantum gravity. \\

Now, we use inductive method to prove the renormalizability
of the quantum gauge general relativity.
In the previous discussion, we have proved that the generating
functional of regular vertex before renormalization satisfies
Ward-Takahashi identities (\ref{11.88} - \ref{11.89}).
The effective Lagrangian density that contains all counterterms
which cancel all divergences of $l$-loops ($0 \leq l \leq L $)
is denoted by $\swav{\cal  L}^{\lbrack L \rbrack}$.
$\swav{\Gamma}^{\lbrack L \rbrack}$ is the generating functional
of regular vertex which is calculated from
$\swav{\cal  L}^{\lbrack L \rbrack}$. The regular vertex which
is generated by $\swav{\Gamma}^{\lbrack L \rbrack}$ will contain
no divergence if the number of the loops of the diagram is not
greater than $L$. We have proved that the generating functional
$\swav{\Gamma}^{\lbrack L \rbrack}$ satisfies Ward-Takahashi
identities if $L=0$. Hypothesis that Ward-Takahashi identities
are also satisfied when $L=N$, that is
\be
\partial^{\mu}
\frac{\delta \bar{\Gamma}^{\lbrack N \rbrack}  }
{\delta K^{\mu}_{\alpha}(x) }
= \frac{\delta \bar{\Gamma}^{\lbrack N \rbrack}  }
{\delta \bar\eta_{\alpha}(x) },
\label{11.93}
\ee
\be
\int {\rm d}^4x \left\lbrace
\frac{\delta \bar{\Gamma}^{\lbrack N \rbrack} }
{\delta K^{\mu}_{\alpha}(x) }
\frac{\delta \bar{\Gamma}^{\lbrack N \rbrack} }
{\delta C_{\mu}^{\alpha}(x) }
+ \frac{\delta \bar{\Gamma}^{\lbrack N \rbrack} }
{\delta L_{\alpha}(x) }
\frac{\delta \bar{\Gamma}^{\lbrack N \rbrack} }
{\delta \eta^{\alpha}(x) }
\right\rbrace = 0.
\label{11.94}
\ee
Our goal is to prove that  Ward-Takahashi identities
are also satisfied when $L=N+1$.\\

Now, let's introduce a special product which is defined by
\be
A * B \equiv \int {\rm d}^4x \left\lbrace
\frac{\delta A }{\delta K^{\mu}_{\alpha}(x) }
\frac{\delta B }{\delta C_{\mu}^{\alpha}(x) }
+ \frac{\delta A }{\delta L_{\alpha}(x) }
\frac{\delta B }{\delta \eta^{\alpha}(x) }
\right\rbrace .
\label{11.95}
\ee
Then (\ref{11.94}) will be simplified to
\be
\bar{\Gamma}^{\lbrack N \rbrack} *
\bar{\Gamma}^{\lbrack N \rbrack} = 0.
\label{11.96}
\ee
$\bar{\Gamma}^{\lbrack N \rbrack}$ contains all contributions from
all possible diagram with arbitrary loops. The contribution from
$l$-loop diagram is proportional to $\hbar^l$. We can expand
$\bar{\Gamma}^{\lbrack N \rbrack}$ as a power serials of $\hbar^l$,
\be
\bar{\Gamma}^{\lbrack N \rbrack}
= \sum_M \hbar^M \bar{\Gamma}^{\lbrack N \rbrack}_M,
\label{11.97}
\ee
where $\bar{\Gamma}^{\lbrack N \rbrack}_M$ is the contribution
from all $M$-loop diagrams. According to our inductive hypothesis,
all $\bar{\Gamma}^{\lbrack N \rbrack}_M$ are finite is $M \leq N$.
Therefore, divergence first appear in
$\bar{\Gamma}^{\lbrack N \rbrack}_{N+1}$.
Substitute (\ref{11.97}) into (\ref{11.96}), we will get
\be
\sum_{M,L}  \hbar^{M+L}
\bar{\Gamma}_M^{\lbrack N \rbrack} *
\bar{\Gamma}_L^{\lbrack N \rbrack} = 0.
\label{11.98}
\ee
The $(L+1)$-loop contribution of (\ref{11.98}) is
\be
\sum_{M=0}^{N+1}
\bar{\Gamma}_M^{\lbrack N \rbrack} *
\bar{\Gamma}_{N-M+1}^{\lbrack N \rbrack} = 0.
\label{11.99}
\ee
$\bar{\Gamma}_{N+1}^{\lbrack N \rbrack}$ can separate into
two parts: finite part
$\bar{\Gamma}_{N+1,F}^{\lbrack N \rbrack}$
and divergent part
$\bar{\Gamma}_{N+1,div}^{\lbrack N \rbrack}$, that is
\be
\bar{\Gamma}_{N+1}^{\lbrack N \rbrack}
= \bar{\Gamma}_{N+1,F}^{\lbrack N \rbrack}
+ \bar{\Gamma}_{N+1,div}^{\lbrack N \rbrack}.
\label{11.100}
\ee
$\bar{\Gamma}_{N+1,div}^{\lbrack N \rbrack}$ is a divergent
function of $(4-D)$ if we calculate loop diagrams in dimensional
regularization. In other words, all terms in
$\bar{\Gamma}_{N+1,div}^{\lbrack N \rbrack}$
are divergent terms when $(4-D)$ approaches zero.
Substitute (\ref{11.100}) into (\ref{11.99}), if we only
concern divergent  terms, we will get
\be
\bar{\Gamma}_{N+1,div}^{\lbrack N \rbrack} *
\bar{\Gamma}_0^{\lbrack N \rbrack}  +
\bar{\Gamma}_0^{\lbrack N \rbrack} *
\bar{\Gamma}_{N+1,div}^{\lbrack N \rbrack} = 0.
\label{11.101}
\ee
$\bar{\Gamma}_{N+1,F}^{\lbrack N \rbrack}$ has no contribution
to the divergent part. Because $\bar{\Gamma}_0^{\lbrack N \rbrack}$
represents contribution from tree diagram and counterterms has
no contribution to tree diagram, we have
\be
\bar{\Gamma}_0^{\lbrack N \rbrack}
= \bar{\Gamma}_0^{\lbrack 0 \rbrack}.
\label{11.102}
\ee
Denote
\be
\bar{\Gamma}_0
= \bar{\Gamma}_0^{\lbrack N \rbrack}
= \swav{S}^{\lbrack 0 \rbrack}
+ \frac{1}{2 \alpha}
\int {\rm d}^4x \eta_{\alpha \beta}
f^{\alpha} f^{\beta}.
\label{11.103}
\ee
Then (\ref{11.101}) is changed into
\be
\bar{\Gamma}_{N+1,div}^{\lbrack N \rbrack} *
\bar{\Gamma}_0  +
\bar{\Gamma}_0 *
\bar{\Gamma}_{N+1,div}^{\lbrack N \rbrack} = 0.
\label{11.104}
\ee
Substitute (\ref{11.97}) into (\ref{11.93}), we get
\be
\partial^{\mu}
\frac{\delta \bar{\Gamma}_{N+1}^{\lbrack N \rbrack}  }
{\delta K^{\mu}_{\alpha}(x) }
= \frac{\delta \bar{\Gamma}_{N+1}^{\lbrack N \rbrack}  }
{\delta \bar\eta_{\alpha}(x) }.
\label{11.105}
\ee
The finite part $\bar{\Gamma}_{N+1,F}^{\lbrack N \rbrack}$
has no contribution to the divergent part, so we have
\be
\partial^{\mu}
\frac{\delta \bar{\Gamma}_{N+1,div}^{\lbrack N \rbrack}  }
{\delta K^{\mu}_{\alpha}(x) }
= \frac{\delta \bar{\Gamma}_{N+1,div}^{\lbrack N \rbrack}  }
{\delta \bar\eta_{\alpha}(x) }.
\label{11.106}
\ee
\\

The operator $\hat{g}$ is defined by
\be
\begin{array}{rcl}
\hat{g} &\equiv& \int {\rm d}^4x \left\lbrace
\frac{\delta \bar{\Gamma}_0 }{\delta C_{\mu}^{\alpha}(x) }
\frac{\delta  }{\delta K^{\mu}_{\alpha}(x) }
+ \frac{\delta \bar{\Gamma}_0 }{\delta L_{\alpha}(x) }
\frac{\delta  }{\delta \eta^{\alpha}(x) }  \right. \\
&&\\
&& \left.
+ \frac{\delta \bar{\Gamma}_0 }{\delta K^{\mu}_{\alpha}(x) }
\frac{\delta  }{\delta C_{\mu}^{\alpha}(x) }
+ \frac{\delta \bar{\Gamma}_0 }{\delta \eta^{\alpha}(x) }
\frac{\delta  }{\delta  L_{\alpha}(x) }
\right\rbrace .
\end{array}
\label{11.107}
\ee
Using this definition, (\ref{11.104}) simplifies to
\be
\hat{g} \bar{\Gamma}_{N+1,div}^{\lbrack N \rbrack} = 0.
\label{11.108}
\ee
\\

It can be strictly proved that operator $\hat{g}$ is a
nilpotent operator, i.e.
\be
\hat{g}^2=0.
\label{11.109}
\ee
Suppose that $f\lbrack C \rbrack$ is an arbitrary functional
of gravitational gauge field $C_{\mu}^{\alpha}$ which is invariant
under local gravitational gauge transformation.
$f\lbrack C \rbrack$ is invariant under generalized BRST
transformation. The generalized BRST transformation
of $f\lbrack C \rbrack$ is
\be
\delta f\lbrack C \rbrack =
- \int {\rm d}^4x
\frac{\delta f\lbrack C \rbrack}
{\delta C_{\mu}^{\alpha}(x)}
{\mathbf D}_{\mu \beta}^{\alpha} \eta^{\beta} \delta \lambda.
\label{11.110}
\ee
Because $\delta \lambda$ is an arbitrary Grassman variable,
(\ref{11.110}) gives out
\be
\delta f\lbrack C \rbrack =
- \int {\rm d}^4x
\frac{\delta f\lbrack C \rbrack}
{\delta C_{\mu}^{\alpha}(x)}
{\mathbf D}_{\mu \beta}^{\alpha} \eta^{\beta}.
\label{11.111}
\ee
Because $f \lbrack C \rbrack$ is a functional of only gravitational
gauge fields, its functional derivatives to other fields vanish
\be
\frac{\delta f \lbrack C \rbrack }
{\delta K^{\mu}_{\alpha}(x) } =0,
\label{11.112}
\ee
\be
\frac{\delta f \lbrack C \rbrack }
{\delta L_{\alpha}(x) } =0,
\label{11.113}
\ee
\be
\frac{\delta f \lbrack C \rbrack }
{\delta \eta^{\alpha}(x) } =0.
\label{11.114}
\ee
Using these relations, (\ref{11.111}) is changed into
\be
\delta f\lbrack C \rbrack =
\hat{g} f\lbrack C \rbrack.
\label{11.115}
\ee
The generalized BRST symmetry of $f\lbrack C \rbrack$ gives
out the following important property of operator $\hat{g}$,
\be
\hat{g} f\lbrack C \rbrack = 0.
\label{11.116}
\ee
\\

Using two important properties of operator $\hat{g}$ which are
shown in eq.(\ref{11.109})
and eq.(\ref{11.116}), we could see that the solution
of eq.(\ref{11.108}) can be written in the following form
\be
\bar{\Gamma}_{N+1,div}^{\lbrack N \rbrack}
= f\lbrack C \rbrack
+ \hat{g} f'\lbrack C,\eta,\bar\eta,K,L \rbrack,
\label{11.117}
\ee
$f\lbrack C \rbrack$ is a gauge invariant functional and
$f'\lbrack C,\eta,\bar\eta,K,L \rbrack$ is an arbitrary functional
of fields $C_{\mu}^{\alpha}(x)$, $\eta^{\alpha}(x)$,
$\bar\eta_{\alpha}(x)$ and external sources
$K^{\mu}_{\alpha}(x)$ and $L_{\alpha}(x)$.
\\

Now, let's consider constrain from eq.(\ref{11.105}).
Using eq.(\ref{11.112})
and eq.(\ref{11.114}), we can see that $f \lbrack C \rbrack$ satisfies
eq.(\ref{11.105}), so eq.(\ref{11.105}) has no
constrain on $f \lbrack C \rbrack$.
Define a new variable
\be
B^{\mu}_{\alpha}
= K^{\mu}_{\alpha}
- \partial^{\mu} \bar\eta_{\alpha}.
\label{11.118}
\ee
$f_1 \lbrack B \rbrack$ is an arbitrary functional of $B$.
It can be proved that
\be
\frac{\delta f_1 \lbrack B \rbrack }
{\delta B^{\mu}_{\alpha}(x) }
= \frac{\delta f_1 \lbrack B \rbrack }
{\delta K^{\mu}_{\alpha}(x) } ,
\label{11.119}
\ee
\be
\frac{\delta f_1 \lbrack B \rbrack }
{\delta \bar\eta_{\alpha}(x) }
= \partial^{\mu} \frac{\delta f_1 \lbrack B \rbrack }
{\delta B^{\mu}_{\alpha}(x) }.
\label{11.120}
\ee
Combine these two relations, we will get
\be
\frac{\delta f_1 \lbrack B \rbrack }
{\delta \bar\eta_{\alpha}(x) }
= \partial^{\mu} \frac{\delta f_1 \lbrack B \rbrack }
{\delta K^{\mu}_{\alpha}(x) }.
\label{11.121}
\ee
There $f_1 \lbrack B \rbrack$ is a solution to eq.(\ref{11.106}).
Suppose that there is another functional $f_2$ that is
given by,
\be
f_2 \lbrack K,C,\eta,L \rbrack
= \int {\rm d}^4x ~ K^{\mu}_{\alpha}
T_{\mu}^{\alpha}(C,\eta,L),
\label{11.122}
\ee
where $T_{\mu}^{\alpha}$ is a conserved current
\be
\partial^{\mu} T_{\mu}^{\alpha} =0.
\label{11.123}
\ee
It can be easily proved that $f_2 \lbrack K,C,\eta,L \rbrack$
is also a solution of eq.(\ref{11.105}). Because $\bar{\Gamma}_0$ satisfies
eq.(\ref{11.106}) (please see eq.(\ref{11.88})), operator $\hat{g}$ commutes
with $\frac{\delta}{\delta \bar\eta_{\alpha}(x) }
- \partial^{\mu} \frac{\delta }{\delta K^{\mu}_{\alpha}(x) } $.
It means that functional $f'\lbrack C,\eta,\bar\eta,K,L \rbrack$
in eq.(\ref{11.117}) must satisfy eq.(\ref{11.106}). According to these
discussion, the solution of $f'\lbrack C,\eta,\bar\eta,K,L \rbrack$
has the following form,
\be
f'\lbrack C,\eta,\bar\eta,K,L \rbrack
= f_1 \lbrack C,\eta,
K^{\mu}_{\alpha} - \partial^{\mu} \bar\eta_{\alpha}, L \rbrack
+ \int {\rm d}^4x ~ K^{\mu}_{\alpha}
T_{\mu}^{\alpha}(C,\eta,L).
\label{11.124}
\ee
\\

In order to determine $f'\lbrack C,\eta,\bar\eta,K,L \rbrack$,
we need to study dimensions of various fields and external sources.
Set the dimensionality of mass to $1$, i.e.
\be
D \lbrack \hat{P}_{\mu} \rbrack =1.
\label{11.125}
\ee
 Then we have
\be
D \lbrack C_{\mu}^{\alpha} \rbrack =1,
\label{11.126}
\ee
\be
D \lbrack {\rm d}^4x \rbrack =-4,
\label{11.127}
\ee
\be
D \lbrack D_{\mu} \rbrack =1,
\label{11.128}
\ee
\be
D \lbrack \eta \rbrack  = D \lbrack \bar\eta \rbrack =1,
\label{11.129}
\ee
\be
D \lbrack K \rbrack  = D \lbrack L \rbrack =2,
\label{11.130}
\ee
\be
D \lbrack g \rbrack   = - 1,
\label{11.131}
\ee
\be
D \lbrack \bar{\Gamma}_{N+1,div}^{\lbrack N \rbrack} \rbrack
= D \lbrack S \rbrack = 0.
\label{11.132}
\ee
Using these relations, we can prove that
\be
D \lbrack \hat{g} \rbrack   =1,
\label{11.133}
\ee
\be
D \lbrack f' \rbrack   =-1.
\label{11.134}
\ee
Define virtual particle number $N_g$ of ghost field
$\eta$ is 1, and that of ghost field $\bar\eta$ is -1, i.e.
\be
N_g \lbrack \eta \rbrack = 1,
\label{11.135}
\ee
\be
N_g \lbrack \bar\eta \rbrack = - 1.
\label{11.136}
\ee
The virtual particle number is a additive conserved quantity, so
Lagrangian and action carry no virtual particle number,
\be
N_g \lbrack S \rbrack =  N_g \lbrack {\cal L} \rbrack    =  0.
\label{11.137}
\ee
The virtual particle number $N_g$ of other fields and external
sources are
\be
N_g \lbrack C \rbrack = N_g \lbrack D_{\mu} \rbrack =  0,
\label{11.138}
\ee
\be
N_g \lbrack g \rbrack =  0,
\label{11.139}
\ee
\be
N_g \lbrack \bar{\Gamma} \rbrack =  0,
\label{11.140}
\ee
\be
N_g \lbrack K \rbrack =  -1,
\label{11.141}
\ee
\be
N_g \lbrack L \rbrack =  -2.
\label{11.142}
\ee
Using all these relations, we can determine the virtual particle
number $N_g$ of $\hat g$ and $f'$,
\be
N_g \lbrack \hat g \rbrack =  1,
\label{11.143}
\ee
\be
N_g \lbrack f' \rbrack =  -1.
\label{11.144}
\ee
According to eq.(\ref{11.134}) and eq.(\ref{11.144}), we know that the
dimensionality of $f'$ is $-1$ and its virtual particle
number is also $-1$. Besides, $f'$ must be a Lorentz scalar.
Combine all these results, the only two possible solutions of
$f_1 \lbrack C,\eta, K^{\mu}_{\alpha} - \partial^{\mu}
\bar\eta_{\alpha}, L \rbrack$ in eq.(\ref{11.124}) are
\be
(K^{\mu}_{\alpha} - \partial^{\mu} \bar\eta_{\alpha}  )
C_{\mu}^{\alpha},
\label{11.145}
\ee
\be
\bar\eta^{\alpha} L_{\alpha}.
\label{11.146}
\ee
The only possible solution of $T_{\mu}^{\alpha}$ is
$C_{\mu}^{\alpha}$. But in general gauge conditions,
$C_{\mu}^{\alpha}$ does not satisfy the conserved
equation  eq.(\ref{11.123}). Therefore, the solution to
$f'\lbrack C,\eta,\bar\eta,K,L \rbrack$
is linear combination of (\ref{11.145}) and (\ref{11.146}), i.e.
\be
f'\lbrack C,\eta,\bar\eta,K,L \rbrack =
\int {\rm d}^4x \left\lbrack
\beta_{N+1}(\varepsilon) (K^{\mu}_{\alpha} - \partial^{\mu}
\bar\eta_{\alpha}  ) C_{\mu}^{\alpha}
+ \gamma_{N+1}(\varepsilon)
\bar\eta^{\alpha} L_{\alpha}
\right\rbrack,
\label{11.147}
\ee
where $\varepsilon = (4-D)$, $\beta_{N+1}(\varepsilon)$
and $\gamma_{N+1}(\varepsilon)$ are divergent parameters when
$\varepsilon$ approaches zero.
Then using the definition of $\hat g$,
we can obtain the following result,
\be
\begin{array}{rcl}
\hat g f'\lbrack C,\eta,\bar\eta,K,L \rbrack
&=& - \beta_{N+1}(\varepsilon)
\int {\rm d}^4x \left\lbrack
\frac{\delta \bar{\Gamma}_0 }{\delta C_{\mu}^{\alpha}(x) }
C_{\mu}^{\alpha}(x)
+ \frac{\delta \bar{\Gamma}_0 }{\delta K^{\mu}_{\alpha}(x) }
K^{\mu}_{\alpha}(x)
- \bar\eta_{\alpha} \partial^{\mu}
{\mathbf D}_{\mu \beta}^{\alpha} \eta^{\beta}
\right\rbrack  \\
&&\\
&&- \gamma_{N+1}(\varepsilon)
\int {\rm d}^4x \left\lbrack
\frac{\delta \bar{\Gamma}_0 }{\delta L_{\alpha}(x) }
L_{\alpha}(x)
+ \frac{\delta \bar{\Gamma}_0 }{\delta \eta^{\alpha}(x) }
\eta^{\alpha}(x)
\right\rbrack.
\end{array}
\label{11.148}
\ee
The  solution to $f \lbrack C \rbrack$ of eq.(\ref{11.117})
which is constructed from  pure gravitational gauge fields
is the action $S\lbrack C \rbrack$ of gauge fields.
Therefore, the most general solution of
$\bar{\Gamma}_{N+1,div}^{\lbrack N \rbrack}$ is
\be
\begin{array}{rcl}
\bar{\Gamma}_{N+1,div}^{\lbrack N \rbrack}
&=& \alpha_{N+1}(\varepsilon)  S\lbrack C \rbrack
- \int {\rm d}^4x \left\lbrack
\beta_{N+1}(\varepsilon) \frac{\delta \bar{\Gamma}_0 }
{\delta C_{\mu}^{\alpha}(x) } C_{\mu}^{\alpha}(x)
+ \beta_{N+1}(\varepsilon) \frac{\delta \bar{\Gamma}_0 }
{\delta K^{\mu}_{\alpha}(x) } K^{\mu}_{\alpha}(x)
\right. \\
&&\\
&& \left.
+ \gamma_{N+1}(\varepsilon) \frac{\delta \bar{\Gamma}_0 }
{\delta  L_{\alpha}(x) } L_{\alpha}(x)
+ \gamma_{N+1}(\varepsilon) \frac{\delta \bar{\Gamma}_0 }
{\delta  \eta^{\alpha}(x) } \eta^{\alpha}(x)
- \beta_{N+1}(\varepsilon) \bar\eta_{\alpha} \partial^{\mu}
{\mathbf D}_{\mu \beta}^{\alpha} \eta^{\beta}
\right\rbrack.
\end{array}
\label{11.149}
\ee
\\

In fact, the action $S\lbrack C \rbrack$ is a functional
of pure gravitational gauge field. It also contains gravitational
coupling constant $g$. So, we can denote it as
$S\lbrack C,g \rbrack$. From eq.(\ref{4.20}),
eq.(\ref{4.24}) and eq.(\ref{4.25}),
we can prove that the action $S\lbrack C,g \rbrack$ has
the following important properties,
\be
S\lbrack gC,1 \rbrack = g^2 S\lbrack C,g \rbrack.
\label{11.150}
\ee
Differentiate both sides of eq.(\ref{11.150}) with respect to coupling
constant $g$, we can get
\be
S\lbrack C,g \rbrack
= \frac{1}{2} \int {\rm d}^4x
\frac{\delta S\lbrack C,g \rbrack }{\delta C_{\mu}^{\alpha}(x) }
C_{\mu}^{\alpha}(x)
- \frac{1}{2} g \frac{\partial S\lbrack C,g \rbrack}{\partial g}.
\label{11.151}
\ee
\\

It can be easily proved that
\be
\begin{array}{l}
\int {\rm d}^4x C_{\mu}^{\alpha}(x)
\frac{\delta  }{\delta C_{\mu}^{\alpha}(x) }
\left\lbrack
\int {\rm d}^4y \bar\eta_{\alpha}(y) \partial^{\mu}
{\mathbf D}_{\mu \beta}^{\alpha} \eta^{\beta}(y)
\right\rbrack  \\
\\
= \int {\rm d}^4x \left\lbrack
(\partial^{\mu} \bar\eta_{\beta}(x))
(\partial_{\mu} \eta^{\beta}(x))
+ \bar\eta_{\alpha}(x) \partial^{\mu}
{\mathbf D}_{\mu \beta}^{\alpha} \eta^{\beta}(x)
\right\rbrack,
\end{array}
\label{11.152}
\ee
\be
\begin{array}{l}
\int {\rm d}^4x C_{\mu}^{\alpha}(x)
\frac{\delta  }{\delta C_{\mu}^{\alpha}(x) }
\left\lbrack
\int {\rm d}^4y K^{\mu}_{\alpha}(y)
{\mathbf D}_{\mu \beta}^{\alpha} \eta^{\beta}(y)
\right\rbrack  \\
\\
= - \int {\rm d}^4x \left\lbrack
K^{\mu}_{\alpha}(x) \partial_{\mu} \eta^{\alpha}(x)
- K^{\mu}_{\alpha}(x) {\mathbf D}_{\mu \beta}^{\alpha}
\eta^{\beta}(x)
\right\rbrack,
\end{array}
\label{11.153}
\ee
\be
\begin{array}{l}
g \frac{\partial  }{\partial g }
\left\lbrack
\int {\rm d}^4x \bar\eta_{\alpha}(x) \partial^{\mu}
{\mathbf D}_{\mu \beta}^{\alpha} \eta^{\beta}(x)
\right\rbrack  \\
\\
= \int {\rm d}^4x \left\lbrack
(\partial^{\mu} \bar\eta_{\beta}(x))
(\partial_{\mu} \eta^{\beta}(x))
+ \bar\eta_{\alpha}(x) \partial^{\mu}
{\mathbf D}_{\mu \beta}^{\alpha} \eta^{\beta}(x)
\right\rbrack,
\end{array}
\label{11.154}
\ee
\be
\begin{array}{l}
g \frac{\partial  }{\partial g }
\left\lbrack
\int {\rm d}^4x K^{\mu}_{\alpha}(x)
{\mathbf D}_{\mu \beta}^{\alpha} \eta^{\beta}(x)
\right\rbrack  \\
\\
= - \int {\rm d}^4x \left\lbrack
K^{\mu}_{\alpha}(x) \partial_{\mu} \eta^{\alpha}
- K^{\mu}_{\alpha}(x) {\mathbf D}_{\mu \beta}^{\alpha}
\eta^{\beta}
\right\rbrack,
\end{array}
\label{11.155}
\ee
\be
\begin{array}{l}
g \frac{\partial  }{\partial g }
\left\lbrack
\int {\rm d}^4x ~g L_{\alpha}(x) \eta^{\beta}(x)
(\partial_{\beta} \eta^{\alpha}(x) )
\right\rbrack  \\
\\
=
\int {\rm d}^4x~ g L_{\alpha}(x) \eta^{\beta}(x)
(\partial_{\beta} \eta^{\alpha}(x) ),
\end{array}
\label{11.156}
\ee

Using eqs.(\ref{11.152} - \ref{11.153}),
eq.(\ref{11.103}) and eq.(\ref{11.45}),
we can prove that
\be
\begin{array}{rcl}
\int {\rm d}^4x
\frac{\delta S\lbrack C,g \rbrack  }{\delta C_{\mu}^{\alpha}(x) }
C_{\mu}^{\alpha}(x)  &=&
\int {\rm d}^4x
\frac{\delta \bar{\Gamma}_0  }{\delta C_{\mu}^{\alpha}(x) }
C_{\mu}^{\alpha}(x)
+ \int {\rm d}^4x \left\lbrack
-(\partial^{\mu} \bar\eta_{\alpha}(x))
(\partial_{\mu} \eta^{\alpha}(x)) \right. \\
&&\\
&&\left. - \bar\eta_{\alpha}(x) \partial^{\mu}
{\mathbf D}_{\mu \beta}^{\alpha} \eta^{\beta}(x)
+K^{\mu}_{\alpha}(x) \partial_{\mu} \eta^{\alpha}(x)
- K^{\mu}_{\alpha}(x) {\mathbf D}_{\mu \beta}^{\alpha}
\eta^{\beta}(x)
\right\rbrack.
\end{array}
\label{11.157}
\ee
Similarly, we can get,
\be
\begin{array}{rcl}
g \frac{\partial S\lbrack C,g \rbrack }{\partial g }
&=&  g \frac{\partial \bar{\Gamma}_0 }{\partial g }
+  \int {\rm d}^4x \left\lbrack
- (\partial^{\mu} \bar\eta_{\alpha}(x))
(\partial_{\mu} \eta^{\alpha}(x))
 - \bar\eta_{\alpha}(x) \partial^{\mu}
{\mathbf D}_{\mu \beta}^{\alpha} \eta^{\beta}(x)
+K^{\mu}_{\alpha}(x) \partial_{\mu} \eta^{\alpha}(x) \right.  \\
&&\\
&& \left.
- K^{\mu}_{\alpha}(x) {\mathbf D}_{\mu \beta}^{\alpha}
\eta^{\beta}(x)
- g L_{\alpha}(x) \eta^{\beta}(x)
(\partial_{\beta} \eta^{\alpha}(x) )
\right\rbrack.
\end{array}
\label{11.158}
\ee
Substitute eqs.(\ref{11.157} -\ref{11.158})
into eq.(\ref{11.151}), we will get
\be
S\lbrack C,g \rbrack
= \frac{1}{2} \int {\rm d}^4x
\frac{\delta \bar{\Gamma}_0 }{\delta C_{\mu}^{\alpha}(x) }
C_{\mu}^{\alpha}(x)
- \frac{1}{2} g \frac{\partial \bar{\Gamma}_0}{\partial g}
+ \int {\rm d}^4x \left\lbrace
\frac{1}{2} g L_{\alpha}(x) \eta^{\beta}(x)
(\partial_{\beta} \eta^{\alpha}(x) \right\rbrace.
\label{11.159}
\ee
Substitute eq.(\ref{11.159}) into eq.(\ref{11.149}). we will get
\be
\begin{array}{rcl}
\bar{\Gamma}_{N+1,div}^{\lbrack N \rbrack}
&=& \int {\rm d}^4x \left\lbrack
(\frac{\alpha_{N+1}(\varepsilon)}{2}
- \beta_{N+1}(\varepsilon) ) C_{\mu}^{\alpha}(x)
\frac{\delta \bar{\Gamma}_0 }{\delta C_{\mu}^{\alpha}(x) }
 \right. \\
&&\\
&&+ (\frac{\alpha_{N+1}(\varepsilon)}{2}
- \gamma_{N+1}(\varepsilon) ) L_{\alpha}(x)
\frac{\delta \bar{\Gamma}_0 }
{\delta  L_{\alpha}(x) }   \\
&&\\
&& + \gamma_{N+1}(\varepsilon) \eta^{\alpha}(x)
\frac{\delta \bar{\Gamma}_0 } {\delta  \eta^{\alpha}(x) }
+ \beta_{N+1}(\varepsilon) \bar\eta_{\alpha}(x)
\frac{\delta \bar{\Gamma}_0 } {\delta  \bar\eta_{\alpha}(x) } \\
&&\\
&& \left. + \beta_{N+1}(\varepsilon) K^{\mu}_{\alpha}(x)
\frac{\delta \bar{\Gamma}_0 }{\delta K^{\mu}_{\alpha}(x) }
\right\rbrack - \frac{\alpha_{N+1}(\varepsilon)}{2}
g \frac{\partial \bar{\Gamma}_0 }{\partial g }
\end{array}
\label{11.160}
\ee
\\

On the other hand, we can prove the following relations
\be
\begin{array}{l}
\int {\rm d}^4x  \eta^{\alpha}(x)
\frac{\delta  }{\delta \eta^{\alpha}(x) }
\left\lbrack
\int {\rm d}^4y ~ \bar\eta_{\beta}(y) \partial^{\mu}
{\mathbf D}_{\mu \sigma}^{\beta} \eta^{\sigma}(y)
\right\rbrack
= \int {\rm d}^4x~
 \bar\eta_{\beta}(x) \partial^{\mu}
{\mathbf D}_{\mu \sigma}^{\beta} \eta^{\sigma}(x),
\end{array}
\label{11.161}
\ee
\be
\begin{array}{l}
\int {\rm d}^4x  \eta^{\alpha}(x)
\frac{\delta  }{\delta \eta^{\alpha}(x) }
\left\lbrack
\int {\rm d}^4y ~ K^{\mu}_{\beta}(y)
{\mathbf D}_{\mu \sigma}^{\beta} \eta^{\sigma}(y)
\right\rbrack
=  \int {\rm d}^4x ~
 K^{\mu}_{\beta}(x)
{\mathbf D}_{\mu \sigma}^{\beta} \eta^{\sigma}(x),
\end{array}
\label{11.162}
\ee
\be
\begin{array}{l}
\int {\rm d}^4x  \eta^{\alpha}(x)
\frac{\delta  }{\delta \eta^{\alpha}(x) }
\left\lbrack
\int {\rm d}^4y ~  g L_{\beta}(y)
\eta^{\sigma}(y) (\partial_{\sigma} \eta^{\beta}(y) )
\right\rbrack
= 2  \int {\rm d}^4x~
 g L_{\beta}(x)
\eta^{\sigma}(x) (\partial_{\sigma} \eta^{\beta}(x) ),
\end{array}
\label{11.163}
\ee

\be
\begin{array}{l}
\int {\rm d}^4x  \eta^{\alpha}(x)
\frac{\delta \bar\Gamma_0  }{\delta \eta^{\alpha}(x) }
= \int {\rm d}^4x \left\lbrace
 \bar\eta_{\beta}(x) \partial^{\mu}
{\mathbf D}_{\mu \sigma}^{\beta} \eta^{\sigma}(x) \right. \\
\\
  \left.
+  K^{\mu}_{\beta}(x)
{\mathbf D}_{\mu \sigma}^{\beta} \eta^{\sigma}(x)
+ 2  g L_{\beta}(x)
\eta^{\sigma}(x) (\partial_{\sigma} \eta^{\beta}(x))
\right\rbrace.
\end{array}
\label{11.164}
\ee
Substitute eqs.(\ref{11.161} - \ref{11.163})
into eq.(\ref{11.164}), we will get
\be
\int {\rm d}^4x  \left\lbrace
- \eta^{\alpha}
\frac{\delta \bar\Gamma_0  }{\delta \eta^{\alpha} }
+  \bar\eta_{\alpha}
\frac{\delta \bar\Gamma_0  }{\delta \bar\eta_{\alpha} }
+ K^{\mu}_{\alpha}
\frac{\delta \bar\Gamma_0  }{\delta K^{\mu}_{\alpha} }
+ 2 L_{\alpha}
\frac{\delta \bar\Gamma_0  }{\delta L_{\alpha} }
\right\rbrace = 0.
\label{11.165}
\ee
Eq.(\ref{11.165}) times $\frac{\gamma_{N+1}-\beta_{N+1}}{2}$, then
add up this results and eq.(\ref{11.160}), we will get
\be
\begin{array}{rcl}
\bar{\Gamma}_{N+1,div}^{\lbrack N \rbrack}
&=& \int {\rm d}^4x \left\lbrack
(\frac{\alpha_{N+1}(\varepsilon)}{2}
- \beta_{N+1}(\varepsilon) )\left( C_{\mu}^{\alpha}(x)
\frac{\delta \bar{\Gamma}_0 }{\delta C_{\mu}^{\alpha}(x) }
+  L_{\alpha}(x)
\frac{\delta \bar{\Gamma}_0 }
{\delta  L_{\alpha}(x) } \right) \right.  \\
&&\\
&& \left.
\frac{\beta_{N+1}(\varepsilon)+\gamma_{N+1}(\varepsilon)}{2}
 \left( \eta^{\alpha}(x)
\frac{\delta \bar\Gamma_0  }{\delta \eta^{\alpha}(x) }
+ \bar\eta_{\alpha}(x)
\frac{\delta \bar{\Gamma}_0 } {\delta  \bar\eta_{\alpha}(x) }
 +  K^{\mu}_{\alpha}(x)
\frac{\delta \bar{\Gamma}_0 }{\delta K^{\mu}_{\alpha}(x) }
\right)
\right\rbrack \\
&&\\
&&  - \frac{\alpha_{N+1}(\varepsilon)}{2}
g \frac{\partial \bar{\Gamma}_0 }{\partial g }.
\end{array}
\label{11.166}
\ee
This is the most general form of
$\bar{\Gamma}_{N+1,div}^{\lbrack N \rbrack}$
which satisfies Ward-Takahashi identities. \\

According to minimal subtraction, the counterterm that cancel
the divergent part of $\bar{\Gamma}_{N+1}^{\lbrack N \rbrack}$
is just $-\bar{\Gamma}_{N+1,div}^{\lbrack N \rbrack}$, that is
\be
\swav{S}^{\lbrack N+1 \rbrack}
= \swav{S}^{\lbrack N \rbrack}
- \bar{\Gamma}_{N+1,div}^{\lbrack N \rbrack}
+ o(\hbar^{N+2}),
\label{11.167}
\ee
where the term of $o(\hbar^{N+2})$ has no contribution to
the divergences of $(N+1)$-loop diagrams. Suppose that
$\bar{\Gamma}_{N+1}^{\lbrack N+1 \rbrack}$ is the generating
functional of regular vertex which is calculated from
$\swav{S}^{\lbrack N+1 \rbrack}$. It can be easily proved that
\be
\Gamma_{N+1}^{\lbrack N+1 \rbrack}
= \Gamma_{N+1}^{\lbrack N \rbrack}
- \bar{\Gamma}_{N+1,div}^{\lbrack N \rbrack}.
\label{11.168}
\ee
Using eq.(\ref{11.100}), we can get
\be
\Gamma_{N+1}^{\lbrack N+1 \rbrack}
= \bar{\Gamma}_{N+1,F}^{\lbrack N \rbrack}.
\label{11.169}
\ee
$\Gamma_{N+1}^{\lbrack N+1 \rbrack}$ contains no divergence
which is just what we expected. \\

Now, let's try to determine the form of
$\swav{S}^{\lbrack N+1 \rbrack}$. Denote the non-renormalized
action of the system as
\be
\swav{S}^{\lbrack 0 \rbrack}
= \swav{S}^{\lbrack 0 \rbrack}
\lbrack
C_{\mu}^{\alpha}, \bar\eta_{\alpha}, \eta^{\alpha},
K^{\mu}_{\alpha}, L_{\alpha}, g, \alpha
\rbrack.
\label{11.170}
\ee
As one of the inductive hypothesis, we suppose that the
action of the system after $\hbar^N$ order renormalization is
\be
\begin{array}{l}
\swav{S}^{\lbrack N \rbrack}\lbrack
C_{\mu}^{\alpha}, \bar\eta_{\alpha}, \eta^{\alpha},
K^{\mu}_{\alpha}, L_{\alpha}, g, \alpha \rbrack  \\
\\
= \swav{S}^{\lbrack 0 \rbrack}
\lbrack
\sqrt{Z_1^{\lbrack N \rbrack}}C_{\mu}^{\alpha},
\sqrt{Z_2^{\lbrack N \rbrack}}\bar\eta_{\alpha},
\sqrt{Z_3^{\lbrack N \rbrack}}\eta^{\alpha},
\sqrt{Z_4^{\lbrack N \rbrack}}K^{\mu}_{\alpha},
\sqrt{Z_5^{\lbrack N \rbrack}}L_{\alpha},
      Z_g^{\lbrack N \rbrack} g,
      Z_{\alpha}^{\lbrack N \rbrack} \alpha
\rbrack.
\end{array}
\label{11.171}
\ee
Substitute eq(\ref{11.166}) and
eq.(\ref{11.171}) into eq.(\ref{11.167}),
we obtain
\be
\begin{array}{rcl}
\swav{S}^{\lbrack N+1 \rbrack} &=&
\swav{S}^{\lbrack 0 \rbrack}
\lbrack
\sqrt{Z_1^{\lbrack N \rbrack}}C_{\mu}^{\alpha},
\sqrt{Z_2^{\lbrack N \rbrack}}\bar\eta_{\alpha},
\sqrt{Z_3^{\lbrack N \rbrack}}\eta^{\alpha},
\sqrt{Z_4^{\lbrack N \rbrack}}K^{\mu}_{\alpha},
\sqrt{Z_5^{\lbrack N \rbrack}}L_{\alpha},
      Z_g^{\lbrack N \rbrack} g,
      Z_{\alpha}^{\lbrack N \rbrack} \alpha
\rbrack  \\
&&\\
&& - \int {\rm d}^4x \left\lbrack
(\frac{\alpha_{N+1}(\varepsilon)}{2}
- \beta_{N+1}(\varepsilon) )\left( C_{\mu}^{\alpha}(x)
\frac{\delta \bar{\Gamma}_0 }{\delta C_{\mu}^{\alpha}(x) }
+  L_{\alpha}(x)
\frac{\delta \bar{\Gamma}_0 }
{\delta  L_{\alpha}(x) } \right) \right.  \\
&&\\
&& \left.
+ \frac{\beta_{N+1}(\varepsilon)+\gamma_{N+1}(\varepsilon)}{2}
 \left( \eta^{\alpha}(x)
\frac{\delta \bar\Gamma_0  }{\delta \eta^{\alpha}(x) }
+ \bar\eta_{\alpha}(x)
\frac{\delta \bar{\Gamma}_0 } {\delta  \bar\eta_{\alpha}(x) }
 +  K^{\mu}_{\alpha}(x)
\frac{\delta \bar{\Gamma}_0 }{\delta K^{\mu}_{\alpha}(x) }
\right)
\right\rbrack \\
&&\\
&&  + \frac{\alpha_{N+1}(\varepsilon)}{2}
g \frac{\partial \bar{\Gamma}_0 }{\partial g }
 + o(\hbar^{N+2}).
\end{array}
\label{11.172}
\ee
Using eq.(\ref{11.103}), we can prove that
\be
\int {\rm d}^4x C_{\mu}^{\alpha}(x)
\frac{\delta \bar{\Gamma}_0}{\delta C_{\mu}^{\alpha}(x)}
= \int {\rm d}^4x C_{\mu}^{\alpha}(x)
\frac{\delta \swav{S}^{\lbrack 0 \rbrack}}
{\delta C_{\mu}^{\alpha}(x)}
+ 2 \alpha
\frac{\partial  \swav{S}^{\lbrack 0 \rbrack}}
{\partial \alpha},
\label{11.173}
\ee
\be
\int {\rm d}^4x L_{\alpha}(x)
\frac{\delta \bar{\Gamma}_0}{\delta L_{\alpha}(x)}
= \int {\rm d}^4x L_{\alpha}(x)
\frac{\delta \swav{S}^{\lbrack 0 \rbrack}}
{\delta L_{\alpha}(x)},
\label{11.174}
\ee
\be
\int {\rm d}^4x \bar\eta_{\alpha}(x)
\frac{\delta \bar{\Gamma}_0}{\delta \bar\eta_{\alpha}(x)}
= \int {\rm d}^4x   \bar\eta_{\alpha}(x)
\frac{\delta \swav{S}^{\lbrack 0 \rbrack}}
{\delta \bar\eta_{\alpha}(x)},
\label{11.175}
\ee
\be
\int {\rm d}^4x \eta^{\alpha}(x)
\frac{\delta \bar{\Gamma}_0}{\delta \eta^{\alpha}(x)}
= \int {\rm d}^4x   \eta^{\alpha}(x)
\frac{\delta \swav{S}^{\lbrack 0 \rbrack}}
{\delta \eta^{\alpha}(x)},
\label{11.176}
\ee
\be
\int {\rm d}^4x K^{\mu}_{\alpha}(x)
\frac{\delta \bar{\Gamma}_0}{\delta K^{\mu}_{\alpha}(x)}
= \int {\rm d}^4x K^{\mu}_{\alpha}(x)
\frac{\delta \swav{S}^{\lbrack 0 \rbrack}}
{\delta K^{\mu}_{\alpha}(x)},
\label{11.177}
\ee
\be
g \frac{\partial \bar{\Gamma}_0}{\partial g}
= g \frac{\partial \swav{S}^{\lbrack 0 \rbrack}}{\partial g}.
\label{11.178}
\ee
Using these relations, eq.(\ref{11.172}) is changed into,
\be
\begin{array}{rcl}
\swav{S}^{\lbrack N+1 \rbrack} &=&
\swav{S}^{\lbrack 0 \rbrack}
\lbrack
\sqrt{Z_1^{\lbrack N \rbrack}}C_{\mu}^{\alpha},
\sqrt{Z_2^{\lbrack N \rbrack}}\bar\eta_{\alpha},
\sqrt{Z_3^{\lbrack N \rbrack}}\eta^{\alpha},
\sqrt{Z_4^{\lbrack N \rbrack}}K^{\mu}_{\alpha},
\sqrt{Z_5^{\lbrack N \rbrack}}L_{\alpha},
      Z_g^{\lbrack N \rbrack} g,
      Z_{\alpha}^{\lbrack N \rbrack} \alpha
\rbrack  \\
&&\\
&& - \int {\rm d}^4x \left\lbrack
(\frac{\alpha_{N+1}(\varepsilon)}{2}
- \beta_{N+1}(\varepsilon) )\left( C_{\mu}^{\alpha}(x)
\frac{\delta  \swav{S}^{\lbrack 0 \rbrack} }
{\delta C_{\mu}^{\alpha}(x) }
+  L_{\alpha}(x)
\frac{\delta  \swav{S}^{\lbrack 0 \rbrack} }
{\delta  L_{\alpha}(x) } \right) \right.  \\
&&\\
&& \left.
+ \frac{\beta_{N+1}(\varepsilon)+\gamma_{N+1}(\varepsilon)}{2}
 \left( \eta^{\alpha}(x)
\frac{\delta  \swav{S}^{\lbrack 0 \rbrack} }
{\delta \eta^{\alpha} }
+ \bar\eta_{\alpha}(x)
\frac{\delta  \swav{S}^{\lbrack 0 \rbrack} }
{\delta  \bar\eta_{\alpha}(x) }
 +  K^{\mu}_{\alpha}(x)
\frac{\delta  \swav{S}^{\lbrack 0 \rbrack} }
{\delta K^{\mu}_{\alpha}(x) }
\right)
\right\rbrack \\
&&\\
&&  + \frac{\alpha_{N+1}(\varepsilon)}{2}
g \frac{\partial  \swav{S}^{\lbrack 0 \rbrack} }{\partial g }
- 2 (\frac{\alpha_{N+1}(\varepsilon)}{2}
- \beta_{N+1}(\varepsilon) )
\alpha \frac{\partial \swav{S}^{\lbrack 0 \rbrack} }
{\partial \alpha} + o(\hbar^{N+2})
\end{array}
\label{11.179}
\ee
We can see that this relation has just the form of first order functional
expansion. Using this relation, we can determine the form of
$\swav{S}^{\lbrack N+1 \rbrack}$. It is
\be
\begin{array}{l}
\swav{S}^{\lbrack N+1 \rbrack} \lbrack
C_{\mu}^{\alpha}, \bar\eta_{\alpha}, \eta^{\alpha},
K^{\mu}_{\alpha}, L_{\alpha}, g, \alpha
\rbrack   \\
\\
= \swav{S}^{\lbrack 0 \rbrack}
\lbrack
\sqrt{Z_1^{\lbrack N+1 \rbrack}}C_{\mu}^{\alpha},
\sqrt{Z_2^{\lbrack N+1 \rbrack}}\bar\eta_{\alpha},
\sqrt{Z_3^{\lbrack N+1 \rbrack}}\eta^{\alpha},
\sqrt{Z_4^{\lbrack N+1 \rbrack}}K^{\mu}_{\alpha},
\sqrt{Z_5^{\lbrack N+1 \rbrack}}L_{\alpha},
      Z_g^{\lbrack N+1 \rbrack} g,
      Z_{\alpha}^{\lbrack N+1 \rbrack}  \alpha
\rbrack,
\end{array}
\label{11.180}
\ee
where
\be
\sqrt{Z_1^{\lbrack N+1 \rbrack}}
= \sqrt{Z_5^{\lbrack N+1 \rbrack}}
= \sqrt{Z_{\alpha}^{\lbrack N+1 \rbrack}}
= 1- \sum_{L=1}^{N+1}
\left( \frac{\alpha_L (\varepsilon)}{2}
-\beta_L (\varepsilon) \right),
\label{11.181}
\ee
\be
\sqrt{Z_2^{\lbrack N+1 \rbrack}}
= \sqrt{Z_3^{\lbrack N+1 \rbrack}}
= \sqrt{Z_4^{\lbrack N+1 \rbrack}}
= 1- \sum_{L=1}^{N+1}
\frac{\beta_L (\varepsilon)+\gamma_L  (\varepsilon) }{2} ,
\label{11.182}
\ee
\be
 Z_g^{\lbrack N+1 \rbrack}
= 1 + \sum_{L=1}^{N+1}
\frac{\alpha_L (\varepsilon)}{2}.
\label{11.183}
\ee
Denote
\be
\sqrt{Z^{\lbrack N+1 \rbrack}}
\define \sqrt{Z_1^{\lbrack N+1 \rbrack}}
= \sqrt{Z_5^{\lbrack N+1 \rbrack}}
= \sqrt{Z_{\alpha}^{\lbrack N+1 \rbrack}},
\label{11.184}
\ee
\be
  \sqrt{\swav{Z}^{\lbrack N+1 \rbrack}}
\define \sqrt{Z_2^{\lbrack N+1 \rbrack}}
= \sqrt{Z_3^{\lbrack N+1 \rbrack}}
= \sqrt{Z_4^{\lbrack N+1 \rbrack}}.
\label{11.185}
\ee
The eq.(\ref{11.180}) is changed into
\be
\begin{array}{l}
\swav{S}^{\lbrack N+1 \rbrack} \lbrack
C_{\mu}^{\alpha}, \bar\eta_{\alpha}, \eta^{\alpha},
K^{\mu}_{\alpha}, L_{\alpha}, g, \alpha
\rbrack   \\
\\
= \swav{S}^{\lbrack 0 \rbrack}
\lbrack
\sqrt{Z^{\lbrack N+1 \rbrack}}C_{\mu}^{\alpha},
\sqrt{\swav{Z}^{\lbrack N+1 \rbrack}}\bar\eta_{\alpha},
\sqrt{\swav{Z}^{\lbrack N+1 \rbrack}}\eta^{\alpha},
\sqrt{\swav{Z}^{\lbrack N+1 \rbrack}}K^{\mu}_{\alpha},
\sqrt{Z^{\lbrack N+1 \rbrack}}L_{\alpha},
      Z_g^{\lbrack N+1 \rbrack} g,
      Z^{\lbrack N+1 \rbrack} \alpha
\rbrack.
\end{array}
\label{11.186}
\ee
Using eq.(\ref{11.186}), we can easily prove that
\be
\begin{array}{l}
\bar{\Gamma}^{\lbrack N+1 \rbrack} \lbrack
C_{\mu}^{\alpha}, \bar\eta_{\alpha}, \eta^{\alpha},
K^{\mu}_{\alpha}, L_{\alpha}, g, \alpha
\rbrack   \\
\\
= \bar{\Gamma}^{\lbrack 0 \rbrack}
\left\lbrack
\sqrt{Z^{\lbrack N+1 \rbrack}}C_{\mu}^{\alpha},
\sqrt{\swav{Z}^{\lbrack N+1 \rbrack}}\bar\eta_{\alpha},
\sqrt{\swav{Z}^{\lbrack N+1 \rbrack}}\eta^{\alpha},
\sqrt{\swav{Z}^{\lbrack N+1 \rbrack}}K^{\mu}_{\alpha},
\sqrt{Z^{\lbrack N+1 \rbrack}}L_{\alpha},
        Z_g^{\lbrack N+1 \rbrack} g,
    Z^{\lbrack N+1 \rbrack} \alpha
\right\rbrack.
\end{array}
\label{11.187}
\ee
\\

Now, we need to prove that all inductive hypotheses hold at $L=N+1$. The main
inductive hypotheses which is used in the above proof are the following three:
when $L=N$, the following three hypotheses hold,
\begin{enumerate}

\item the lowest divergence of $\bar\Gamma^{\lbrack N \rbrack}$ appears
in the $(N+1)$-loop diagram;

\item $\bar\Gamma^{\lbrack N \rbrack}$ satisfies Ward-Takahashi identities
eqs.(\ref{11.93}-\ref{11.94});

\item after $\hbar^N th$ order renormalization, the action of the system
has the form of eq.(\ref{11.171}).

\end{enumerate}

First, let's see the first hypothesis. According to eq.(\ref{11.169}), after
introducing $(N+1) th$ order counterterm, the $(N+1)$-loop diagram
contribution of $\bar\Gamma^{\lbrack N+1 \rbrack}$ is finite. It means
that the lowest order divergence of $\bar\Gamma^{\lbrack N+1 \rbrack}$
appears in the $(N+2)$-loop diagram. So, the first inductive
hypothesis hold when $L=N+1$.\\

It is known that the
non-renormalized generating functional of regular vertex
\be
\bar{\Gamma}^{\lbrack 0 \rbrack} =
\bar{\Gamma}^{\lbrack 0 \rbrack}
\lbrack C, \bar\eta, \eta, K, L ,g,\alpha \rbrack
\label{11.216}
\ee
satisfies Ward-Takahsshi identities
eqs.(\ref{11.88}-\ref{11.89}). If we define
\be
\bar{\Gamma} ' =
\bar{\Gamma}^{\lbrack 0 \rbrack}
\lbrack C', \bar\eta', \eta', K', L' ,g',\alpha',
 \rbrack,
\label{11.217}
\ee
then, it must satisfy the following Ward-Takahashi identities
\be
\partial^{\mu}
\frac{\delta \bar{\Gamma}'  }
{\delta K^{\prime\mu}_{\alpha}(x) }
= \frac{\delta \bar{\Gamma} '  }
{\delta \bar\eta'_{\alpha}(x) },
\label{11.218}
\ee
\be
\int {\rm d}^4x \left\lbrace
\frac{\delta \bar{\Gamma} ' }
{\delta K^{\prime\mu}_{\alpha}(x) }
\frac{\delta \bar{\Gamma} ' }
{\delta C_{\mu}^{\prime\alpha}(x) }
+ \frac{\delta \bar{\Gamma} ' }
{\delta L'_{\alpha}(x) }
\frac{\delta \bar{\Gamma} ' }
{\delta \eta^{\prime\alpha}(x) }
\right\rbrace = 0.
\label{11.219}
\ee
Set,
\be
C_{\mu}^{\prime\alpha}
= \sqrt{Z^{\lbrack N+1 \rbrack}} C_{\mu}^{\alpha},
\label{11.222}
\ee
\be
K^{\prime\mu}_{\alpha}
= \sqrt{\swav{Z}^{\lbrack N+1 \rbrack}} K^{\mu}_{\alpha},
\label{11.223}
\ee
\be
L'_{\alpha}
= \sqrt{Z^{\lbrack N+1 \rbrack}} L_{\alpha},
\label{11.224}
\ee
\be
\eta^{\prime\alpha}
= \sqrt{\swav{Z}^{\lbrack N+1 \rbrack}} \eta^{\alpha},
\label{11.225}
\ee
\be
\bar\eta'_{\alpha}
= \sqrt{\swav{Z}^{\lbrack N+1 \rbrack}} \bar\eta_{\alpha},
\label{11.226}
\ee
\be
g' = Z_g^{\lbrack N+1 \rbrack} g,
\label{11.227}
\ee
\be
\alpha' = Z^{\lbrack N+1 \rbrack} \alpha.
\label{11.228}
\ee
In this case, we have
\be
\begin{array}{rcl}
\bar\Gamma' &=& \bar{\Gamma}^{\lbrack 0 \rbrack}
\left\lbrack
\sqrt{Z^{\lbrack N+1 \rbrack}}C_{\mu}^{\alpha},
\sqrt{\swav{Z}^{\lbrack N+1 \rbrack}}\bar\eta_{\alpha},
\sqrt{\swav{Z}^{\lbrack N+1 \rbrack}}\eta^{\alpha},\right.\\
&&\\
&&~~~~~~~\left.
\sqrt{\swav{Z}^{\lbrack N+1 \rbrack}}K^{\mu}_{\alpha},
\sqrt{Z^{\lbrack N+1 \rbrack}}L_{\alpha},
        Z_g^{\lbrack N+1 \rbrack} g,
    Z^{\lbrack N+1 \rbrack} \alpha,
    ,\eta_1,\eta_2
\right\rbrack\\
&&\\
&=& \bar\Gamma^{\lbrack N+1 \rbrack}.
\end{array}
\label{11.231}
\ee
Then eq.(\ref{11.218}) is changed into
\be
\frac{1}{\sqrt{\swav{Z}^{\lbrack N+1 \rbrack}}}
\partial^{\mu}
\frac{\delta \bar{\Gamma}^{\lbrack N+1 \rbrack}  }
{\delta K^{\mu}_{\alpha}(x) }
= \frac{1}{\sqrt{\swav{Z}^{\lbrack N+1 \rbrack}}}
\frac{\delta \bar{\Gamma}^{\lbrack N+1 \rbrack}  }
{\delta \bar\eta_{\alpha}(x) }.
\label{11.232}
\ee
Because $\frac{1}{\sqrt{\swav{Z}^{\lbrack N+1 \rbrack}}}$ does not
vanish, the above equation gives out
\be
\partial^{\mu}
\frac{\delta \bar{\Gamma}^{\lbrack N+1 \rbrack}  }
{\delta K^{\mu}_{\alpha}(x) }
=\frac{\delta \bar{\Gamma}^{\lbrack N+1 \rbrack}  }
{\delta \bar\eta_{\alpha}(x) }.
\label{11.233}
\ee
Eq.(\ref{11.219}) gives out
\be
\int {\rm d}^4x \left\lbrace
\frac{1}{\sqrt{\swav{Z}^{\lbrack N+1 \rbrack}}
\sqrt{Z^{\lbrack N+1 \rbrack}}}
\left\lbrack
\frac{\delta \bar{\Gamma}^{\lbrack N+1 \rbrack} }
{\delta K^{\mu}_{\alpha}(x) }
\frac{\delta \bar{\Gamma}^{\lbrack N+1 \rbrack} }
{\delta C_{\mu}^{\alpha}(x) }
+ \frac{\delta \bar{\Gamma}^{\lbrack N+1 \rbrack} }
{\delta L_{\alpha}(x) }
\frac{\delta \bar{\Gamma}^{\lbrack N+1 \rbrack} }
{\delta \eta^{\alpha}(x) }\right\rbrack
\right\rbrace = 0.
\label{11.234}
\ee
Because $\frac{1}{\sqrt{\swav{Z}^{\lbrack N+1 \rbrack}}
\sqrt{Z^{\lbrack N+1 \rbrack}}}$ does not vanish,
we can obtain
\be
\int {\rm d}^4x \left\lbrace
\frac{\delta \bar{\Gamma}^{\lbrack N+1 \rbrack} }
{\delta K^{\mu}_{\alpha}(x) }
\frac{\delta \bar{\Gamma}^{\lbrack N+1 \rbrack} }
{\delta C_{\mu}^{\alpha}(x) }
+ \frac{\delta \bar{\Gamma}^{\lbrack N+1 \rbrack} }
{\delta L_{\alpha}(x) }
\frac{\delta \bar{\Gamma}^{\lbrack N+1 \rbrack} }
{\delta \eta^{\alpha}(x) }
 \right\rbrace = 0.
\label{11.235}
\ee
Eq.(\ref{11.233}) and eq.(\ref{11.235})
are just the Ward-Takahashi identities
for $L=N+1$. Therefore, the second inductive hypothesis holds when
$L=N+1$. \\

The third inductive hypothesis has already been proved which is shown
in eq.(\ref{11.186}). Therefore, all three inductive hypothesis hold when
$L=N+1$. According to inductive principle, they will hold when $L$ is
an arbitrary non-negative number, especially they hold when $L$
approaches infinity. \\

In above discussions, we have proved that, if we suppose that
when $L=N$ eq.(\ref{11.171}) holds, then it also holds when $L=N+1$.
According to inductive principle, we know that
eq.(\ref{11.186}  -  \ref{11.187})
hold for any positive integer $N$, that is
\be
\begin{array}{l}
\swav{S} \lbrack
C_{\mu}^{\alpha}, \bar\eta_{\alpha}, \eta^{\alpha},
K^{\mu}_{\alpha}, L_{\alpha}, g, \alpha
\rbrack   \\
\\
= \swav{S}^{\lbrack 0 \rbrack}
\left\lbrack
\sqrt{Z}C_{\mu}^{\alpha},
\sqrt{\swav{Z}}\bar\eta_{\alpha},
\sqrt{\swav{Z}}\eta^{\alpha},
\sqrt{\swav{Z}}K^{\mu}_{\alpha},
\sqrt{Z}L_{\alpha},
      Z_g g,
      Z \alpha
\right\rbrack,
\end{array}
\label{11.188}
\ee
\be
\begin{array}{l}
\bar{\Gamma} \lbrack
C_{\mu}^{\alpha}, \bar\eta_{\alpha}, \eta^{\alpha},
K^{\mu}_{\alpha}, L_{\alpha}, g, \alpha
\rbrack   \\
\\
= \bar{\Gamma}^{\lbrack 0 \rbrack}
\left\lbrack
\sqrt{Z}C_{\mu}^{\alpha},
\sqrt{\swav{Z}}\bar\eta_{\alpha},
\sqrt{\swav{Z}}\eta^{\alpha},
\sqrt{\swav{Z}}K^{\mu}_{\alpha},
\sqrt{Z}L_{\alpha},
      Z_g  g,
      Z \alpha
\right\rbrack,
\end{array}
\label{11.189}
\ee
where
\be
\sqrt{Z} \define  \lim_{N \to \infty}
 \sqrt{Z^{\lbrack N \rbrack}}
= 1- \sum_{L=1}^{\infty}
\left( \frac{\alpha_L (\varepsilon)}{2}
-\beta_L (\varepsilon) \right),
\label{11.190}
\ee
\be
\sqrt{\swav{Z}}
\define \lim_{N \to \infty} \sqrt{\swav{Z}^{\lbrack N \rbrack}}
= 1- \sum_{L=1}^{\infty}
\frac{\beta_L (\varepsilon)+\gamma_L  (\varepsilon) }{2} ,
\label{11.191}
\ee
\be
Z_g \define \lim_{N \to \infty}
Z_g^{\lbrack N \rbrack}
= 1 + \sum_{L=1}^{\infty}
\frac{\alpha_L (\varepsilon)}{2}.
\label{11.192}
\ee
$\swav{S} \lbrack C_{\mu}^{\alpha}, \bar\eta_{\alpha},
\eta^{\alpha}, K^{\mu}_{\alpha}, L_{\alpha}, g, \alpha \rbrack$
and $\bar{\Gamma} \lbrack C_{\mu}^{\alpha}, \bar\eta_{\alpha},
\eta^{\alpha}, K^{\mu}_{\alpha}, L_{\alpha}, g, \alpha \rbrack$
are renormalized action and generating functional of regular vertex.
The generating functional of regular vertex $\bar{\Gamma}$ contains
no divergence. All kinds of vertex that generated from $\bar{\Gamma}$
are finite. From eq.(\ref{11.188})
and eq.(\ref{11.189}), we can see that we only
introduce three known parameters which are $\sqrt{Z}$, $\sqrt{\swav{Z}}$
and $Z_g$. Therefore, quantum gauge general relativity is a renormalizable
theory. \\

From eq.(\ref{11.189}) and eqs.(\ref{11.88} - \ref{11.89}),
we can deduce that the renormalized
generating functional of regular vertex satisfies Ward-Takahashi
identities,
\be
\partial^{\mu}
\frac{\delta \bar{\Gamma}  }
{\delta K^{\mu}_{\alpha}(x) }
= \frac{\delta \bar{\Gamma}  }
{\delta \bar\eta_{\alpha}(x) },
\label{11.193}
\ee
\be
\int {\rm d}^4x \left\lbrace
\frac{\delta \bar{\Gamma} }
{\delta K^{\mu}_{\alpha}(x) }
\frac{\delta \bar{\Gamma} }
{\delta C_{\mu}^{\alpha}(x) }
+ \frac{\delta \bar{\Gamma} }
{\delta L_{\alpha}(x) }
\frac{\delta \bar{\Gamma} }
{\delta \eta^{\alpha}(x) }
\right\rbrace = 0.
\label{11.194}
\ee
It means that the renormalized theory has the structure of gauge symmetry.
If we define
\be
C_{0 \mu}^{\alpha} = \sqrt{Z} C_{\mu}^{\alpha},
\label{11.195}
\ee
\be
\eta_0^{\alpha} = \sqrt{\swav{Z}} \eta^{\alpha},
\label{11.196}
\ee
\be
\bar\eta_{0 \alpha} = \sqrt{\swav{Z}} \bar\eta_{\alpha},
\label{11.197}
\ee
\be
K^{\mu}_{0 \alpha} = \sqrt{\swav{Z}} K^{\mu}_{\alpha},
\label{11.198}
\ee
\be
L_{0 \alpha} = \sqrt{Z} L_{\alpha},
\label{11.199}
\ee
\be
g_0 = Z_g g,
\label{11.200}
\ee
\be
\alpha_0 = Z \alpha.
\label{11.201}
\ee
Therefore, eqs.(\ref{11.188} - \ref{11.189}) are changed into
\be
\swav{S} \lbrack
C_{\mu}^{\alpha}, \bar\eta_{\alpha}, \eta^{\alpha},
K^{\mu}_{\alpha}, L_{\alpha}, g, \alpha
\rbrack
= \swav{S}^{\lbrack 0 \rbrack}
\lbrack
C_{0 \mu}^{\alpha}, \bar\eta_{0 \alpha}, \eta_0^{\alpha},
K^{\mu}_{0 \alpha}, L_{0 \alpha}, g_0, \alpha_0
\rbrack,
\label{11.202}
\ee
\be
\bar{\Gamma} \lbrack
C_{\mu}^{\alpha}, \bar\eta_{\alpha}, \eta^{\alpha},
K^{\mu}_{\alpha}, L_{\alpha}, g, \alpha
\rbrack
= \bar{\Gamma}^{\lbrack 0 \rbrack}
\lbrack
C_{0 \mu}^{\alpha}, \bar\eta_{0 \alpha}, \eta_0^{\alpha},
K^{\mu}_{0 \alpha}, L_{0 \alpha}, g_0, \alpha_0
\rbrack.
\label{11.203}
\ee
$C_{0 \mu}^{\alpha}$, $\bar\eta_{0 \alpha}$ and
$\eta_0^{\alpha}$ are renormalized wave function,
$K^{\mu}_{0 \alpha}$ and $L_{0 \alpha}$ are renormalized
external sources, and $g_0$ is the renormalized gravitational
coupling constant. \\

The action $\swav{S}$ which is given by eq.(\ref{11.202}) is invariant
under the following generalized BRST transformations,
\be
\delta C_{0\mu}^{\alpha}
= -  {\mathbf D}_{0\mu~\beta}^{\alpha} \eta_0^{\beta}
\delta \lambda,
\label{11.252}
\ee
\be
\delta \eta_0^{\alpha} = g_0 \eta_0^{\sigma}
(\partial_{\sigma}\eta_0^{\alpha}) \delta \lambda,
\label{11.253}
\ee
\be
\delta \bar\eta_{0\alpha} = \frac{1}{\alpha_0}
\eta_{\alpha \beta} f_0^{\beta} \delta \lambda,
\label{11.254}
\ee
\be
\delta \eta^{\mu \nu} = 0,
\label{11.255}
\ee
where,
\be
{\mathbf D}^{\alpha}_{0 \mu \beta} =
\delta_{\beta}^{\alpha} \partial_{\mu}
- g_0 \delta^{\alpha}_{\beta} C^{\sigma}_{0 \mu} \partial_{\sigma}
+ g_0 (\partial_{\beta} C^{\alpha}_{0\mu}),
\label{11.258}
\ee
\be
f_0^{\alpha} = \partial^{\mu} C_{0 \mu}^{\alpha}.
\label{11.259}
\ee
Therefore, the normalized action has generalized BRST symmetry, which
means that the normalized theory has the structure of gauge theory.\\

\section{Classical Tests of Quantum Gauge general Relativity}
\setcounter{equation}{0}

In the above chapters, the quantum gravity is formulated
in the traditional framework of quantum field theory, i.e.,
the physical space-time is always flat and the space-time
metric is always selected to be the Minkowski metric. In this
picture, gravity is treated as physical interactions
in flat physical space-time. Our gravitational gauge
transformation does not act on physical space-time
coordinates, but act on
physical fields, so gravitational gauge transformation
does not affect the structure of physical space-time.
This is the physics picture of gravity.\\

According to above discussions, there is another picture
of gravity which is widely used in Einstein's general
relativity, i.e., the geometrical picture of gravity.
For gravitational gauge theory, if we treat
$G^{\alpha}_{\mu}$ ( or $G^{-1 \mu}_{\alpha}$ )
as a fundamental physical input of the
theory, we can also set up the geometrical picture
of gravity\cite{c44}. For gravitational gauge theory, the geometrical
nature of gravity essentially originates from the
geometrical nature of the gravitational gauge transformation.
In the geometrical picture of gravity, gravity is not
treated as a kind of physical interactions, but is
treated as the geometry of space-time. So, there
is no physical gravitational interactions in space-time
and space-time is curved if
gravitational field does not vanish. In this case, the
space-time metric is not Minkowski metric, but
$g^{\alpha\beta}$ (or $g_{\alpha\beta}$). In other words,
Minkowski metric is the space-time metric if we discuss
gravity in physics picture of gravity while metric tensor
$g^{\alpha\beta}$ ( or $g_{\alpha\beta}$) is space-time
metric if we discuss gravity in geometrical picture of gravity.
So, if we use Minkowski metric in our discussion,
it means that we are in physics picture of
gravity; if we use metric tensor
$g^{\alpha\beta}$ (or $g_{\alpha\beta}$) in our discussion,
it means that we are in geometrical picture of gravity. \\

In one picture, gravity is treated as physical
interactions, while in another picture, gravity
is treated as geometry of space-time. But we know that
there is only one physics for gravity, so two
pictures of gravity must be equivalent. This
equivalence means that the nature of gravity is
physics-geometry duality, i.e., gravity is a kind
of physical interactions which has the characteristics
of geometry; it is also a geometry of space-time
which is essentially a kind of physical interactions.
\\

When we discuss quantization and renormalization of
quantum gauge general relativity, it is better to perform all
discussions in physics picture of gravity; but when we discuss
classical effects of gravity, it's convenient for us to
perform our discussions in geometrical picture of  gravity.
In this chapter, we will discuss classical tests of gravity
and our discussions are in geometrical picture of gravity.
\\

In quantum gauge general relativity, the field equation of gravitational
gauge field is the same as that of the Einstein's field equation.
It is known that, the Einstein's field equation for empty space
has Schwarzschild solution, it is expected that the field equation
of gravitational gauge field in empty space should also have
Schwarzschild solution. In spherical coordinate system
($t,r,\theta,\phi$),
the simplest form of gravitational gauge field which has
spherical symmetry is
\be
C_{\mu}^{\alpha} = \left (
\ba{cccc}
u(r) &0&0&0\\
0 & v(r) & 0 & 0 \\
0&0&0&0  \\
0&0&0&0
\ea
\right).
\label{14.1}
\ee
We can prove that, when we select
\be
g u(r) = 1-
\frac{1}{\sqrt{1- \frac{2 G M}{r}}},
\label{14.2}
\ee
\be
g v(r) = 1-
\sqrt{1- \frac{2 G M}{r}},
\label{14.3}
\ee
gravitational gauge field $C_{\mu}^{\alpha}$ given by
(\ref{14.1}) is just the solution of field equation (\ref{4.41}) for
pure gravitational gauge field. (\ref{14.1}) is just the
Schwarzschild solution in quantum gauge general relativity.
Using relation ({\ref{4.707}}), we can get
\be
g_{\alpha \beta} = \left(
\ba{cccc}
-1 + \frac{2 G M}{r} & 0 & 0& 0  \\
0 & \frac{1}{1- \frac{2GM}{r}} & 0&0 \\
0&0& r^2 & 0 \\
0&0&0&r^2 {\rm sin}^2 \theta
\ea
\right ),
\label{14.4}
\ee
which is the standard Schwarzschild solution in general
relativity. We can easily check that (\ref{14.4}) is
the solution to the field equation (\ref{4.5310}) when
$T^{\alpha \beta}$ vanishes.\\

As we have stated above,
in the geometrical picture of gravity,
gravity is not treated as physical interactions
in space-time.  In the geometrical picture of
gravity, if there is no other
physical interactions in space time, any mass point
can not feel any physical forces when it moves in
space-time. So, it must move along the curve which
has the least length.
Suppose that a particle is moving from point
A to point B along a curve. Define
\be  \label{14.5}
T_{BA} = \int_A^B
\sqrt{ - g_{\mu\nu} \frac{{\rm d}x^{\mu}}{{\rm d}p}
\frac{{\rm d}x^{\nu}}{{\rm d}p} } {\rm d} p,
\ee
where $p$ is a parameter that describe the orbit.
The real curve that the particle moving along
corresponds to the minimum of $T_{AB}$. The
minimum of $T_{AB}$ gives out the following
geodesic curve equation
\be  \label{14.6}
\frac{{\rm d}^2 x^{\mu}}{{\rm d}p^2}
+ \Gamma^{\mu}_{\nu\lambda}
\frac{{\rm d}x^{\nu}}{{\rm d}p}
\frac{{\rm d}x^{\lambda}}{{\rm d}p} =0,
\ee
where $\Gamma^{\mu}_{\nu\lambda}$ is the affine
connection defined by equation (\ref{4.1901}).
Eq.(\ref{14.6}) gives out the curve that a free particle
moves along in curved space-time if we discuss physics
in the geometrical picture of gravity.
\\

Combine Schwarzschild solution (\ref{14.4}) with
the geodesic curve equation (\ref{14.6}), theoretical
predictions on classical tests of gravity can be obtained.
Details on this calculation can be found in literature
\cite{r01,r02} and many other standard textbook on
general relativity. Here, I only list the final results.
For the deflection of light by  sun, the theoretical
prediction is
\be
\Delta \varphi =
\frac{4 G M}{r_0},
\label{14.7}
\ee
where $M$ is the mass of the sun and $r_0$ is the distance
of closest approach to the sun. For the procession of perihelia,
the theoretical prediction is
\be
\Delta \varphi =
\frac{6 \pi G M}{L},
\label{14.8}
\ee
where $L$ is the semilatus rectum. For the time delay
of radar echo, the theoretical prediction is
\be
(\Delta t)_{max} \simeq
4 GM \left \lbrace
1 + {\rm ln} \left(
\frac{4 r_1 r_2}{R^2_0}
\right) \right\rbrace,
\label{14.9}
\ee
where $r_1$ and $r_2$ are the distances from the center
of the sun to the radar antenna on earth, and
the point on Mercury's surface closest to the
earth, and $R_0$ is the radius of the sun. \\

For the precession of orbiting gyroscope, quantum gauge general
relativity should also give out the same theoretical
predictions as that of Einstein's general relativity.
In fact, quantum gauge general relativity and Einstein's
general relativity should have the same classical
effects of gravitational interactions, and they should
give out the same theoretical predictions on any
phenomenon which belongs to the classical effects
of gravitational interactions, they have the same
post-Newtonian approximation. \\

Robertson-Walker metric is a solution of Einstein's
field equation, so it is also the solution of the
field equation of gravitational gauge field in quantum
gauge general relativity. It is known that, in general relativity,
the cosmological model is dominated by classical
gravitational interactions. Because quantum gauge general
relativity has the fundamental laws of classical
gravitational interactions as those of general
relativity, so it will give out the same cosmological
model as that of general relativity. All traditional
calculations on cosmological model can be copied
to the present model without any changes on the
final results. \\

\section{Quantum Effects of Gravitational Interactions}
\setcounter{equation}{0}

Quantum gauge general relativity can not only explain classical
effects of gravitational interactions, but also explain
quantum effects of gravitational interactions. Up to now,
we know that there are two quantum effects of gravitational
interactions which had already been detected by experiments,
that is, gravitational phase effects and gravitational
shielding effects. In this chapter, we will discuss these
two effects. \\

First, let's discuss gravitational phase effects. According
to above discussions, the equation of motion of a Dirac particle
in gravitational field is
\be \label{15.1}
( \gamma^{\mu} D_{\mu} + m ) \psi = 0.
\ee
If electromagnetic interactions are considered, this
equation of motion will be changed into
\be \label{15.2}
\lbrack \gamma^{\mu} (D_{\mu} - i e A_{\mu})
+ m \rbrack \psi = 0.
\ee
Its non-relativistic limit gives out
\be \label{15.3}
\ba{rcl}
i D_0 \Psi &=& \left\lbrack
\frac{1}{2m} (-i \svec{D} - e \svec{A})^2
+ m g C_0^0 - e A_0  \right.\\
&&\\
&& \left.
- \frac{g}{2} \svec{\Sigma} \cdot \svec{B}^0
+ \frac{e}{2 m} \svec{\Sigma} \cdot \svec{B}_e
+ \frac{i g}{2}  \svec{\alpha} \cdot \svec{E}^0
- \frac{i e }{ 2 m}\svec{\alpha} \cdot \svec{E}_e
 \right\rbrack \Psi,
\ea
\ee
where
\be \label{15.4}
\alpha^i = - \gamma^0 \gamma^i,
\ee
\be \label{15.5}
E_i = F_{0i} = E_i^{\alpha} \hat P_{\alpha},
\ee
\be \label{15.6}
E_{ei} = A_{0i},
\ee
\be \label{15.7}
B_i = - \frac{1}{2} \varepsilon_{ijk} F_{jk} =
 B_i^{\alpha} \hat P_{\alpha},
\ee
\be \label{15.8}
B_{ei} = - \frac{1}{2} \varepsilon_{ijk} A_{jk}.
\ee
In equation ({\ref{15.3}), the term $mg C_0^0$ gives
out the gravitational phase effects detected by
COW experiments\cite{q01,q02,q03}. For COW experiments,
the phase difference given this term is
\be
\delta \phi = - \left \lbrack
\frac{4 \pi \lambda g m^2 }{h^2} L^2 {\rm tan} \theta
\right \rbrack {\rm sin} \alpha,
\label{15.9}
\ee
where $\lambda$ is the wave length, $g$ gravitational
acceleration, $m$ is the mass of test particle (for COW
experiments, $m$ is the mass of neutron), $2L$ is the
distance between the first and the last crystal plate,
$\theta$ is the Bragg angle of the crystal and $\alpha$
is the angle between the neutron interferometer and the
horizontal plane. This result is strictly confirmed
by COW  experiments\cite{q01,q02,q03}. \\

Another quantum effect of gravitational interactions
is gravitational shielding effects which is first detected
by E.E.Podkletnov and R.Nieminen in 1992. In Podkletnov
experiments, a specially made ceramic superconductor with
composite structure has revealed weak shielding properties
against gravitational force. From the point of view
of quantum gauge theory of gravity, the nature of
gravitational shielding effects is exponentially decreasing
of gravitational field in inhomogeneous vacuum of scalar field
after spontaneous symmetry breaking\cite{q07}.  \\

In gauge theory of gravity, the lagrangian that describes
the  gravitational gauge interactions of complex scalar
field is
\be
\ba{rcl}
{\cal L}_0 & = &
- \eta^{\mu\nu} (D_{\mu} \phi) (D_{\nu} \phi)^* - V(\phi)  \\
&&\\
&& - \frac{1}{16} \eta^{\mu \rho}
\eta^{\nu \sigma} g_{\alpha \beta}
F^{\alpha}_{\mu \nu} F^{\beta}_{\rho \sigma} \\
&&\\
&& - \frac{1}{8} \eta^{\mu \rho}
G^{-1 \nu}_{\beta} G^{-1 \sigma}_{\alpha}
F^{\alpha}_{\mu \nu} F^{\beta}_{\rho \sigma} \\
&&\\
&& + \frac{1}{4} \eta^{\mu \rho}
G^{-1 \nu}_{\alpha} G^{-1 \sigma}_{\beta}
F^{\alpha}_{\mu \nu} F^{\beta}_{\rho \sigma},
\ea
\label{15.10}
\ee
where $V(\phi)$ is the potential of scalar field
\be
V(\phi) = \frac{1}{2} \mu^2 | \phi |^2
+ \frac{\lambda}{4} |\phi|^4,
~~~~~~(\lambda > 0).
\label{15.11}
\ee
The full Lagrangian of the system is
\be
{\cal L} = J(C) \cdot {\cal L}_0 ,
\label{15.12}
\ee
\\

When $\mu^2 <0$, the origin $\phi = 0$ will become a local
maximum and the symmetry of the system will be spontaneously
broken. After symmetry breaking, the physical vacuum moves
to $\phi_0$
\be
| \phi_0 | =  \sqrt{-\frac{\mu^2}{\lambda}}.
\label{15.13}
\ee
$\phi_0$ is the vacuum of the complex scalar field, generally
speaking, it is required that $\phi_0$ is even distributed
in microscopic scale, so that there is no quantum excitation,
otherwise $\phi_0$ is not a classical state, but a quantum
state. But on the other hand, $|\phi_0|^2$ represents
Cooper pairs density in superconductor, and generally speaking,
Cooper pairs density is not a constant in macroscopic scale,
so $\phi_0$ can not be a constant in macroscopic scale. So,
in this paper, our basic hypothesis on the physical vacuum
$\phi_0$ is that it is a classical state, it is even
distributed or a constant in microscopic scale and it
is a function of space-time coordinates in macroscopic
scale. In a word,
\be
\phi_0 = \phi_0(x).
\label{15.14}
\ee
$\phi_0$ is a classical state, it has no quantum excitation,
though it is space-time dependent. After symmetry breaking,
the scalar field becomes
\be
\phi(x) = \varphi(x) + \phi_0.
\label{15.15}
\ee
$\varphi(x)$ is the quantum scalar field after symmetry
breaking, which represents the small perturbations around
physical vacuum $\phi_0$. \\

After symmetry breaking, lagrangian becomes
\be
\ba{rcl}
{\cal L}_0 & = &
- \eta^{\mu\nu} (D_{\mu} \varphi) (D_{\nu} \varphi)^*
- \eta^{\mu\nu} (D_{\mu} \varphi) (D_{\nu} \phi_0)^*
- \eta^{\mu\nu} (D_{\mu} \varphi)^*  (D_{\nu} \phi_0) \\
&&\\
&&
- \eta^{\mu\nu} (D_{\mu} \phi_0) (D_{\nu} \phi_0)^*
- V(\varphi)
- \frac{1}{16} \eta^{\mu \rho}
\eta^{\nu \sigma} g_{\alpha \beta}
F^{\alpha}_{\mu \nu} F^{\beta}_{\rho \sigma} \\
&&\\
&& - \frac{1}{8} \eta^{\mu \rho}
G^{-1 \nu}_{\beta} G^{-1 \sigma}_{\alpha}
F^{\alpha}_{\mu \nu} F^{\beta}_{\rho \sigma}
 + \frac{1}{4} \eta^{\mu \rho}
G^{-1 \nu}_{\alpha} G^{-1 \sigma}_{\beta}
F^{\alpha}_{\mu \nu} F^{\beta}_{\rho \sigma},
\ea
\label{15.16}
\ee
where $V(\varphi)$ is the potential of scalar field
\be
V(\varphi)  =  \frac{\lambda}{4} |\varphi|^4
+ \frac{\lambda}{2} |\varphi|^2
(\varphi^* \phi_0 + \varphi \phi_0^*)
- \frac{\mu^2}{2} |\varphi|^2
 + \frac{\lambda}{4}
(\varphi^{* 2} \phi_0^2 + \varphi^2 \phi_0^{*2} )
- \frac{\mu^4}{4 \lambda}.
\label{15.17}
\ee
There are two important properties which could be seen from
the above lagrangian. Firstly, when the vacuum $\phi_0$
is space-time dependent, it can directly couple to
quantum field $\varphi$. In other words, in the place where
$\partial_{\mu} \phi_0$ does not vanish, vacuum state $\phi_0$
can directly excite quantum scalar state, so it can be
considered to be an external source of scalar field.
Secondly, when the vacuum $\phi_0$ is space-time dependent,
it will couple to gravitational gauge field. But this interaction
is completely in the classical level. In other words, it is also
an external source of gravitational gauge field.
\\

The field equation of gravitational gauge field in the
superconductor is
\be
\ba{rl}
\partial_{\mu} ( & \frac{1}{4} \eta^{\mu \rho}
\eta^{\nu \sigma} g_{\alpha \beta}
F_{\rho \sigma}^{\beta}
- \frac{1}{4} \eta^{\nu \rho} F^{\mu}_{\rho\alpha}
+ \frac{1}{4} \eta^{\mu \rho} F^{\nu}_{\rho\alpha} \\
&\\
& - \frac{1}{2} \eta^{\mu\rho} \delta^{\nu}_{\alpha}
F^{\beta}_{\rho\beta}
+ \frac{1}{2} \eta^{\nu\rho} \delta^{\mu}_{\alpha}
F^{\beta}_{\rho\beta})  = - g T_{g \alpha}^{\nu},
\label{15.18}
\ea
\ee
where $T_{g \alpha}^{\nu}$ is the gravitational energy-momentum
tensor
\be
\ba{rcl}
T_{g \alpha}^{\nu} & = &
\eta^{\mu\nu} (D_{\mu} \varphi^*) \partial_{\alpha} \varphi
+ \eta^{\mu\nu} (D_{\mu} \phi^*_0) \partial_{\alpha} \varphi
+ \eta^{\mu\nu} (D_{\mu} \varphi) \partial_{\alpha} \varphi^*
+ \eta^{\mu\nu} (D_{\mu} \phi_0) \partial_{\alpha} \varphi^*  \\
&&\\
&& + \eta^{\mu\nu} (D_{\mu} \varphi^*) \partial_{\alpha} \phi_0
+ \eta^{\mu\nu} (D_{\mu} \phi_0^*) \partial_{\alpha} \phi_0
+ \eta^{\mu\nu} (D_{\mu} \varphi) \partial_{\alpha} \phi_0^*
+ \eta^{\mu\nu} (D_{\mu} \phi_0) \partial_{\alpha} \phi_0^*  \\
&&\\
&& - \frac{\partial {\cal L}_0}{\partial D_{\nu} C_{\mu}^{\beta}}
\partial_{\alpha} C_{\mu}^{\beta}
+ G_{\alpha}^{-1 \nu} {\cal L}_0
- G^{-1 \lambda}_{\beta} (\partial_{\mu} C_{\lambda}^{\beta} )
\frac{\partial {\cal L}_0}
{\partial \partial_{\mu} C_{\nu}^{\alpha}} \\
&&\\
&& - \frac{1}{4} \eta^{\lambda\rho} \eta^{\nu\sigma}
\partial_{\mu} (g_{\alpha \beta} C^{\mu}_{\lambda}
 F^{\beta}_{\rho\sigma})
 - \frac{1}{4} \eta^{\nu\rho}
\partial_{\beta} (C^{\sigma}_{\lambda} G^{-1 \lambda}_{\alpha}
 F^{\beta}_{\rho\sigma})
 + \frac{1}{2} \eta^{\nu\rho}
\partial_{\alpha} (C^{\sigma}_{\lambda} G^{-1 \lambda}_{\beta}
 F^{\beta}_{\rho\sigma}) \\
 &&\\
&& + \frac{1}{4g} \eta^{\lambda\rho}
\partial_{\mu} \lbrack ( G^{-1 \nu}_{\beta} G^{-1 \sigma}_{\alpha}
G^{\mu}_{\lambda} - \delta^{\nu}_{\beta} \delta^{\sigma}_{\alpha}
\delta^{\mu}_{\lambda}) F^{\beta}_{\rho\sigma}
\rbrack  \\
&&\\
&& - \frac{1}{2 g} \eta^{\lambda\rho}
\partial_{\mu} \lbrack ( G^{-1 \nu}_{\alpha} G^{-1 \sigma}_{\beta}
G^{\mu}_{\lambda} - \delta^{\nu}_{\alpha} \delta^{\sigma}_{\beta}
\delta^{\mu}_{\lambda}) F^{\beta}_{\rho\sigma}
\rbrack  \\
&&\\
&& - \frac{1}{4} \eta^{\kappa \rho} G^{-1 \nu}_{\beta}
G^{-1 \lambda}_{\alpha} G^{-1 \sigma}_{\gamma}
F^{\beta}_{\rho\sigma} F^{\gamma}_{\kappa\lambda}
 + \frac{1}{2} \eta^{\kappa \rho} G^{-1 \nu}_{\gamma}
G^{-1 \lambda}_{\alpha} G^{-1 \sigma}_{\beta}
F^{\beta}_{\rho\sigma} F^{\gamma}_{\kappa\lambda}  \\
&&\\
&& - \frac{1}{8} \eta^{\mu \rho} \eta^{\lambda \sigma}
g_{\alpha \gamma} G^{-1 \nu}_{\beta}
F^{\beta}_{\rho\sigma} F^{\gamma}_{\mu\lambda}.
\ea
\label{15.19}
\ee
In fact, when we deduce the field equation of gravitational
gauge field from least action principle, the original field
equation that we obtained is
\be
\ba{rcl}
J(C) \cdot \partial_{\mu} \left (  \frac{1}{4} \eta^{\mu \rho}
\eta^{\nu \sigma} g_{\alpha \beta}
F_{\rho \sigma}^{\beta}
- \frac{1}{4} \eta^{\nu \rho} F^{\mu}_{\rho\alpha} \right . &&\\
&&\\
\left .+ \frac{1}{4} \eta^{\mu \rho} F^{\nu}_{\rho\alpha}
 - \frac{1}{2} \eta^{\mu\rho} \delta^{\nu}_{\alpha}
F^{\beta}_{\rho\beta}
+ \frac{1}{2} \eta^{\nu\rho} \delta^{\mu}_{\alpha}
F^{\beta}_{\rho\beta} \right )  & = &  - g J(C)  T_{g \alpha}^{\nu},
\label{15.20}
\ea
\ee
Because $J(C)$ does not vanish, we can be eliminate it from
the above equation to obtain the field equation eq.(\ref{15.18}).
Now, in order to obtain correct gravitational shielding
effects, we need to start our discussions directly from
eq.(\ref{15.20}), which can be changed into another form
\be
\ba{rcl}
&&  \partial_{\mu} \left (  \frac{1}{4} \eta^{\mu \rho}
\eta^{\nu \sigma} g_{\alpha \beta}
F_{\rho \sigma}^{\beta}
- \frac{1}{4} \eta^{\nu \rho} F^{\mu}_{\rho\alpha}
 + \frac{1}{4} \eta^{\mu \rho} F^{\nu}_{\rho\alpha}
 - \frac{1}{2} \eta^{\mu\rho} \delta^{\nu}_{\alpha}
 F^{\beta}_{\rho\beta}
+ \frac{1}{2} \eta^{\nu\rho} \delta^{\mu}_{\alpha}
F^{\beta}_{\rho\beta} \right )  \\
&& \\
& = &   - g J(C)  T_{g \alpha}^{\nu}
+ (1-J(C)) \partial_{\mu} \left (  \frac{1}{4} \eta^{\mu \rho}
\eta^{\nu \sigma} g_{\alpha \beta}
F_{\rho \sigma}^{\beta}
- \frac{1}{4} \eta^{\nu \rho} F^{\mu}_{\rho\alpha}
 + \frac{1}{4} \eta^{\mu \rho} F^{\nu}_{\rho\alpha} \right . \\
 && \\
&& \left .
 - \frac{1}{2} \eta^{\mu\rho} \delta^{\nu}_{\alpha}
F^{\beta}_{\rho\beta}
+ \frac{1}{2} \eta^{\nu\rho} \delta^{\mu}_{\alpha}
F^{\beta}_{\rho\beta} \right )
\label{15.21}
\ea
\ee
The equation of motion of complex scalar field is
\be
\ba{rcl}
\eta^{\mu\nu} D_{\mu} D_{\nu} \varphi - \frac{\mu^2}{2} \varphi
&=& - \eta^{\lambda \nu} (\partial_{\mu} G^{\mu}_{\lambda})
D_{\nu} \varphi - \eta^{\mu\nu} D_{\mu}D_{\nu} \phi_0  \\
&&\\
&& - \eta^{\lambda \nu} (\partial_{\mu} G_{\lambda}^{\mu})
D_{\nu} \phi_0
- g \eta^{\mu\nu} G^{-1 \kappa}_{\alpha}
(D_{\mu} C_{\kappa}^{\alpha}) D_{\nu} \varphi  \\
&&\\
&& - g \eta^{\mu\nu} G^{-1 \kappa}_{\alpha}
(D_{\mu} C_{\kappa}^{\alpha}) D_{\nu} \phi_0
- \frac{\lambda}{2} \varphi^* (\varphi + \phi_0)^2
- \frac{\lambda}{2} \varphi^2 \phi^*_0.
\ea
\label{15.22}
\ee
This field equation can be changed into another form
\be
\left (\eta^{\mu\nu} \partial_{\mu} \partial_{\nu}
- \frac{\mu^2}{2} \right ) \varphi =
-\eta^{\mu\nu} D_{\mu}D_{\nu} \phi_0
- \eta^{\lambda\nu} (\partial_{\mu} G^{\mu}_{\lambda})
D_{\nu} \phi_0
- g \eta^{\mu\nu} G^{-1 \kappa}_{\alpha}
(D_{\mu} C_{\kappa}^{\alpha}) D_{\nu} \phi_0 + \cdots.
\label{15.23}
\ee
From the right hand side of the above equation, we can see that
when the vacuum of the complex scalar field is not a constant,
it will become source of quantum scalar field $\varphi$. In other
words, inhomogeneous vacuum will excite quantum states. \\

After some complicated calculations, the field equation (\ref{15.21})
can be changed into the following form
\be
\ba{rcl}
&&  \partial_{\mu} \left (  \frac{1}{4} \eta^{\mu \rho}
\eta^{\nu \sigma} g_{\alpha \beta}
F_{\rho \sigma}^{\beta}
- \frac{1}{4} \eta^{\nu \rho} F^{\mu}_{\rho\alpha}
 + \frac{1}{4} \eta^{\mu \rho} F^{\nu}_{\rho\alpha}
 - \frac{1}{2} \eta^{\mu\rho} \delta^{\nu}_{\alpha}
 F^{\beta}_{\rho\beta}
+ \frac{1}{2} \eta^{\nu\rho} \delta^{\mu}_{\alpha}
F^{\beta}_{\rho\beta} \right )  \\
&& \\
& = &    - g N^{\nu}_{\alpha}
+ g^2 M^{\nu\mu}_{\alpha \beta} C_{\mu}^{\beta} + \cdots,
\ea
\label{15.24}
\ee
where
\be
\ba{rcl}
N^{\nu}_{\alpha} & = &
\delta^{\nu}_{\alpha} \lbrack
- \eta^{\mu\lambda} (\partial_{\mu} \phi_0)
(\partial_{\lambda} \phi_0^* )+ \frac{\mu^4}{4 \lambda}
\rbrack  \\
&&\\
&& + \eta^{\mu\nu} (\partial_{\mu} \phi_0)
(\partial_{\alpha} \phi_0^* )
+ \eta^{\mu\nu} (\partial_{\mu} \phi_0^*)
(\partial_{\alpha} \phi_0 ),
\ea
\label{15.25}
\ee

\be
\ba{rcl}
M^{\nu\mu}_{\alpha\beta} & = &
(\delta^{\mu}_{\alpha} \delta^{\nu}_{\beta}
+  \delta^{\mu}_{\beta} \delta^{\nu}_{\alpha})
\lbrack \eta^{\rho\lambda} (\partial_{\rho} \phi_0)
(\partial_{\lambda} \phi_0^*)
- \frac{\mu^4}{4 \lambda}  \rbrack \\
&&\\
&& + (\eta^{\mu\nu} \delta^{\lambda}_{\beta}
- \eta^{\lambda \nu} \delta^{\mu}_{\beta})
\lbrack (\partial_{\lambda} \phi_0) (\partial_{\alpha} \phi_0^*)
+ (\partial_{\lambda} \phi_0^*) (\partial_{\alpha} \phi_0) \rbrack \\
&&\\
&& - \delta^{\nu}_{\alpha} \eta^{\mu\lambda}
\lbrack (\partial_{\lambda} \phi_0) (\partial_{\beta} \phi_0^*)
+ (\partial_{\lambda} \phi_0^*) (\partial_{\beta} \phi_0) \rbrack.
\ea
\label{15.26}
\ee
From above expressions, all terms in $N^{\nu}_{\alpha}$
and $M^{\nu\mu}_{\alpha\beta}$ are classical quantities,
so $N^{\nu}_{\alpha}$ and $M^{\nu\mu}_{\alpha\beta}$
themselves are classical quantities. Therefore, the first two
terms of the right hand side of eq.(\ref{15.24}) do not
represent interaction terms. The first term is the source
of gravity in superconductor, and the second term is
mass  term of gravitational gauge field. In fact, from
eq.(\ref{15.25}), we can see that $N^{\nu}_{\alpha}$
is just the energy-momentum tensor of the vacuum of complex
scalar field, which is the source of gravitational gauge
field. But, because the coupling constant of gravitational
interactions is extremely small and the total energy in
a superconductor is finite, the magnitude of total gravitational
gauge field generated by energy-momentum of the superconductor
is many orders smaller than that of the gravitational gauge
field generated by the earth. In experiments, we can neglect
the gravity generated by the superconductor itself. So,
$N^{\nu}_{\alpha}$ has no contribution to gravitational
shielding effects. We neglect this term for the moment. \\

Let's turn to the second term of the right hand side of
eq.(\ref{15.24}). It is a mass term, but it is not a
constant mass term, so it is a variable mass term or
local mass term. At different positions,
$M^{\nu\mu}_{\alpha\beta}$ has different values. It is
known that, in vacuum space, graviton is massless and
gravitational force are long range force. But in
superconductor, gravitational gauge field obtain a small
mass term, so in superconductor, gravitational force
decreased exponentially, which is the nature of gravitational
shielding effects. \\

Selecting harmonic gauge (\ref{9.502}), equation (\ref{15.24})
can be changed into
\be
 \frac{1}{4} \eta^{\nu\sigma} \eta_{\alpha \beta}
\partial^{\mu} \partial_{\mu} C_{\sigma}^{\beta}
+ \frac{1}{4}
\partial^{\mu} \partial_{\mu} C_{\alpha}^{\nu}
- \frac{1}{4} \delta^{\nu}_{\alpha}
\partial^{\mu} \partial_{\mu} C_{\beta}^{\beta}
 = g^2 M^{\nu\mu}_{\alpha \beta} C_{\mu}^{\beta} + \cdots.
\label{15.27}
\ee
For earth's gravitational gauge field, in post-Nowtonian
approximation, we have the following leading order
results
\be
\ba{l}
g C^0_0 = \phi_e = - \frac{G M}{ r} \\
\\
g C^i_j = - \delta^i_j \phi_e ,
\ea
\label{15.28}
\ee
where $\phi_e$ is the gravitational potential generated
by earth.
For Podkletnov experiments, the dominant component of earth's
gravitational gauge field is $C^0_0$, which corresponds
to classical Newton's gravity. So, in order to explain
gravitational shielding effects quantitatively, we need
to study the propagation of gravitational gauge field
$C_0^0$ in superconductor. For earth's gravitational
field $C_{\mu}^{\alpha}$, it is static, so we can set
all time derivative of $C_{\mu}^{\alpha}$ to zero.
From field equation (\ref{15.27}), the following field
equation can be obtained
\be
\nabla^2  \phi_e
=  g^2 ( M^{00}_{00} - M^{0i}_{0i})  \phi_e ,
\label{15.29}
\ee
where
\be
M^{00}_{00} = 2 | \nabla \phi_0 |^2 + 2 V(\phi_0),
\label{15.30}
\ee
\be
M^{0i}_{0i} =  | \nabla \phi_0 |^2  - |\tdot{\phi}_0|^2  +  V(\phi_0),
\label{15.31}
\ee
\be
M^{00}_{00} - M^{0i}_{0i} = | \nabla \phi_0 |^2
+  |\tdot{\phi}_0|^2 +  V(\phi_0),
\label{15.32}
\ee
and the contribution from $N^0_0$ is neglected.
Now, the mass term becomes very simple form.
In ordinary superconductor, the Cooper pair density is almost
a constant and space gradient of $\phi_0$ is almost zero.
In this case,
\be
M^{00}_{00} =  V(\phi_0) = - \frac{\mu^4}{4 \lambda}.
\label{15.33}
\ee
After omitting contribution from $N^0_0$, the
field equation (\ref{15.29}) becomes
\be
\nabla^2 \phi_e = - \frac{g^2 \mu^4}{4 \lambda} \phi_e.
\label{15.34}
\ee
Selecting spherical coordinate system. For the earth's gravitational
field, $\phi_e$ is approximately a function of $r$ coordinate. In this
case, the general solution of eq. (\ref{15.34}) is
\be
\phi_e (r)=
\frac{c_1}{r} {\rm cos}
\left (\sqrt{\frac{g^2 \mu^4}{4 \lambda}} (r-r_0) \right)
+ \frac{c_2}{r} {\rm sin}
\left (\sqrt{\frac{g^2 \mu^4}{4 \lambda}} (r-r_0) \right),
\label{15.35}
\ee
where $r_0$ is the position of the lower surface of superconductor.
Generally speaking, $\frac{g^2 \mu^4}{4 \lambda}$ is a extremely
small quantity, in a small region of ordinary superconductor, we have
\be
\phi_e (r) \approx \frac{c_1}{r}.
\label{15.36}
\ee
That is, ordinary superconductor almost have no effects on
earth's gravitational field, or it shows no gravitational
shielding effects. But for Podkletnov experiment, $\phi_0$
is not a constant, $|\nabla \phi_0|^2$ and $|\tdot{\phi}_0|^2$
are many orders larger than $2 V(\phi_0)$, so the dominant
contribution of $M^{00}_{00}$ is
$ |\nabla \phi_0|^2 + |\tdot{\phi}_0|^2$ and
it becomes a positive quantity. In this case,
$ |\nabla \phi_0|^2 + |\tdot{\phi}_0|^2$
 is not a constant. Denote $g^2( M^{00}_{00}-M^{0i}_{0i})$
by $m_g^2$
\be
m_g^2 = g^2 ( M^{00}_{00} - M^{0i}_{0i})
= g^2 \left ( |\nabla \phi_0|^2 + |\tdot{\phi}_0|^2 \right ) ,
\label{15.37}
\ee
and omit the influence from $N_0^0$, then the above field
equation (\ref{15.27}) becomes
\be
\nabla^2 \phi_e = m_g^2 \phi_e.
\label{15.38}
\ee
The general solution of the above equation is
\be
\phi_e (r) =
c_1 \frac{e^{-m_g (r-r_0)}}{r}
+ c_2 \frac{e^{m_g (r-r_0)}}{r}.
\label{15.39}
\ee
In Podkletnov experiment, $c_2$ vanishes and earth's gravitational
field decrease exponentially
\be
\phi_e (r) =
c_1 \frac{e^{-m_g (r-r_0)}}{r}.
\label{51.40}
\ee
That is, the earth's gravitational field decrease exponentially
in inhomogeneous superconductor. Suppose that the thickness
of the gravitational shielding region is $\Delta r_0$ and $m_g$
denotes the average value of the mass of graviton in this region,
then the relative gravity loss in the upper surface of
 the superconducting disk is
\be
\varepsilon \approx m_g  \Delta r_0.
\label{15.41}
\ee
\\

The Newton's gravitational constant is
\be
G = \frac{g^2}{ 4 \pi},
\label{15.42}
\ee
so
\be
m_g^2 =  4 \pi G ( |\nabla \phi_0|^2
+|\tdot{\phi}_0|^2).
\label{15.43}
\ee
Obviously, increase space gradient or un-stability of $\phi_0$ will increase
$m_g$, and therefore increase gravitational shielding effects.
According to this spirit, the key structure of the ceramic
superconductor disk in Podkletnov experiment is that is has
three zones with different crystal structure and the gravitational
shielding effects mainly come from the transition part of the disk,
which is consists of randomly oriented grains with typical
sizes between 5 and 15 $\mu m$. The key step in the experiment
is the rotation of the disk and strong disturbance from
outside high frequency magnetic field.
In the experiment, the upper
layer of the disk is superconducting, while the lower layer
is not. There is a transition region between the tow layers.
The upper part of the transition region is partially in
superconducting,  some part of the  transition region is
critical, while the lower part of the transition region is not
superconducting. Because  of the granular structure of the lower
part, the Cooper pair density is strongly inhomogeneous,
the supercurrent is strongly unstable and the
space gradient of $\phi_0$ will be large. The
supercurrent is disturbed by a high frequency magnetic field,
which will increase the time derivative of $\phi_0$. A rough
estimation shows that the present model can semi-quantitatively
explain the gravitational shielding effects in Podkletnov
experiments\cite{q07}.
\\

\section{Discussions}
\setcounter{equation}{0}

In this paper, a renormalizable quantum gauge general relativity
is formulated in the framework of quantum gauge theory
of gravity, where gravity is treated as a kind
of physical interactions and space-time is always
kept flat. This treatment satisfies the fundamental
spirit of traditional quantum field theory, and
go along this way, four different kinds of fundamental
interactions can be unified on the same fundamental
baseline\cite{12,13,14}. The most advantage of this approach is that
the renormalizability of the quantum gravity is easy
to be proved. Its transcendental  foundation is gauge
principle, not equivalence principle and the
principle of general covariance.
Gravitational gauge interactions is
determined by gauge symmetry. In other words, the
Lagrangian of the system is  determined by
gauge symmetry. Using the langurage of Cartan tetrad,
we set up the geometrical formulation of this new
quantum gauge general relativity and to study its
geometrical foundation. So, gravity theory has both
physical picture and geometrical picture, which is
the reflection of the physics-geometry duality of
gravity.  Finally, we give a simple summary to the
whole theory.

\begin{enumerate}

\item In leading order approximation, the quantum gauge general
relativity gives out classical Newton's theory of gravity.

\item In quantum gauge general relativity, the field equation
of the gravitational gauge field is just the  Einstein's field
equation. In classical level, gauge general relativity is
the same as that of the Einstein's general relativity.

\item For classical tests of gravity, such as the gravitational
red-shift, the perihelion procession, deflection of light,
time delay of radar echo and the procession of a gyroscope,
the gauge general relativity gives out the same theoretical
predictions as that of the Einstein's general relativity. It
also gives out the same cosmological model as that of the
Einstein's general relativity.

\item quantum gauge general relativity is a perturbatively
renormalizable quantum gravity.

\item Four different kinds of fundamental interactions can be easily
unified in a simple and beautiful way in quantum gauge general
relativity.

\item Quantum gauge general relativity can also explain quantum
effects of gravitational interactions, such as gravitational
phase effects and gravitational shielding effects.

\end{enumerate}

{\lbrack \bf acknowledgement \rbrack}
The authors would like to express 
his sincere gratitude to Prof. Dahua Zhang for some
constructive discussions on this topic.
\\

\end{document}